\documentclass[hyper,12pt,letterpaper]{JHEP3}

\usepackage{latexsym,amsmath,amsfonts,amssymb}
\usepackage[dvips]{graphicx}
\usepackage{bbm}
\usepackage{bm}
\usepackage{subfigure}
\usepackage{paralist}
\newcommand{\vol}{\mathrm{vol}}

\newcommand{\kq}{/\!/}

\numberwithin{equation}{section}

\newcommand{\nn}{\nonumber}
\newcommand{\mat}[1]{\begin{pmatrix} #1 \end{pmatrix}}
\newcommand{\be}{\begin{equation}} \newcommand{\ee}{\end{equation}}
\newcommand{\bea}{\begin{equation} \begin{aligned}} \newcommand{\eea}{\end{aligned} \end{equation}}

\newcommand{\cA}{\mathcal{A}}

\newcommand{\cC}{\mathcal{C}}

\newcommand{\cE}{\mathcal{E}}
\newcommand{\cF}{\mathcal{F}}
\newcommand{\cG}{\mathcal{G}}

\newcommand{\cI}{\mathcal{I}}

\newcommand{\cM}{\mathcal{M}}
\newcommand{\cN}{\mathcal{N}}
\newcommand{\cO}{\mathcal{O}}

\newcommand{\cQ}{\mathcal{Q}}

\newcommand{\cZ}{\mathcal{Z}}

\newcommand{\bC}{\mathbb{C}}
\newcommand{\bF}{\mathbb{F}}

\newcommand{\bP}{\mathbb{P}}

\newcommand{\bR}{\mathbb{R}}

\newcommand{\bZ}{\mathbb{Z}}

\newcommand{\CP}{\bC\bP}

\newcommand{\eps}{\epsilon}

\newcommand{\Ypq}{Y^{p,\,\bm{q}}}

\title{Toric Fano varieties and Chern-Simons quivers}

\let\CC\spadesuit

\let\BB\clubsuit

\author{Cyril Closset$^\CC$, Stefano Cremonesi$^\BB$\\

$^\CC$ Department of Particle Physics and Astrophysics \\
Weizmann Institute of Science, Rehovot 76100, Israel. \\
\vspace{-5pt}

$^\BB$ Theoretical Physics Group, Imperial College London,\\
Prince Consort Road, London, SW7 2AZ, UK\\
}
\preprint{Imperial/TP/12/SC/01\\ WIS/01/12-JAN-DPPA}

\abstract{In favourable cases the low energy dynamics of a stack of M2-branes at a toric Calabi-Yau fourfold singularity can be described by an $\cN=2$ supersymmetric Chern-Simons quiver theory, but there still does not exists an ``inverse algorithm'' going from the toric data of the $CY_4$ to the CS quiver. We make progress in that direction by deriving CS quiver theories for M2-branes probing cones over a large class of geometries $\Ypq(B_4)$, which are $S^3/\bZ_p$ bundles over toric Fano varieties $B_4$.
We rely on the type IIA understanding of CS quivers, giving a firm string theory footing to our CS theories. In particular we give a derivation of some previously conjectured CS quivers in the case $B_4= \CP^1\times \CP^1$, as field theories dual to M-theory backgrounds with nontrivial torsion $G_4$ fluxes.
}

\begin{document}

\section{Introduction}
\label{sec: intro}

The $AdS_4$/CFT$_3$ correspondence has been the object of intense study in recent years, especially in its maximally supersymmetric version ($\cN=6$ or $8$ in three dimension) embodied in the ABJM proposal \cite{Aharony:2008ug}, where the three dimensional conformal field theory is a Chern-Simons (CS) quiver gauge theory. Many more instances of  $AdS_4$/CFT$_3$ dualities can be proposed if the number of supersymmetries is lowered. A particularly interesting field of study is the 3d $\cN=2$ supersymmetric case, the minimal number that allows holomorphy. The corresponding $AdS$/CFT duality arises from a decoupling limit \cite{Maldacena:1997re} on $N$ M2-branes located at a Calabi-Yau (CY) fourfold conical singularity \cite{Acharya:1998db}.

The first thing to understand in order to study those dualities explicitly is the low energy theory on the worldvolume of $N$ M2-branes at a $CY_4$ singularity. This is an interesting problem even independently from the $AdS$/CFT motivation. Soon after the ABJM proposal, numerous examples of $\cN=2$ quiver gauge theories were proposed to describe theories which might flow to the correct M2-brane theories in the infrared (IR) \cite{Jafferis:2008qz, Martelli:2008si, Hanany:2008cd, Franco:2008um, Ueda:2008hx, Imamura:2008qs, Hanany:2008fj}. Unlike the higher supersymmetric cases where the field theories could be deduced from brane constructions, however, many of those proposals were lacking a convincing motivation, apart from their reproducing a $CY_4$ geometry as their classical moduli space. This state of affairs started to change with the proposal of \cite{Aganagic:2009zk}: the idea is to use the M-theory/type IIA duality mapping M2-branes on a CY fourfold to D2-branes on a Calabi-Yau threefold together with Ramond-Ramond (RR) background fluxes accounting for the non-trivial M-theory fibration. This proposal was sharpened and developed in \cite{Jafferis:2009th, Benini:2009qs, Benini:2011cma}:
from the string theory side it was understood that generically D6-branes are present in the type IIA reduction, while from the field theory side progress was made to include in the analysis of the moduli space quantum corrections  due to D2-D6 open string modes.

We should note that many of the theories derived%
\footnote{More precisely, some theories had been proposed before by an inspired guess, such as the equal rank $\bC^3/\bZ_3$ quiver for $Y^{p,q}(\CP^2)$ first proposed in \cite{Martelli:2008si} and derived from string theory in \cite{Benini:2011cma}. That particular theory (including some subtle corrections introduced in \cite{Benini:2011cma}) was nicely checked recently at the level of the $S^3$ partition function and of the superconformal index  \cite{Gang:2011jj}.}
 that way have passed non-trivial independent checks by matching their large $N$ partition function on $S^3$ \cite{Jafferis:2010un, Hama:2010av} or/and their superconformal index \cite{Imamura:2011su} to the expectation from supergravity \cite{Cheon:2011vi,Martelli:2011qj,Cheon:2011th,  Jafferis:2011zi, Amariti:2011uw, Gang:2011jj}.

The type IIA framework allows to connect $\cN=2$ Chern-Simons quiver theories to the well studied setup of D-branes on $CY_3$ cones. In this work we develop this perspective in the case of \emph{toric} Calabi-Yau threefolds, which can be studied in terms of ``toric quivers'', also known as brane tilings \cite{Hanany:2005ve,Franco:2005rj}.

We study M2-branes on toric $CY_4$ cones over seven dimensional manifolds that we dub $\Ypq(B_4)$. Those $\Ypq$ spaces are $S^3/\bZ_p$ bundles over complex two dimensional Fano varieties $B_4$, which generalise the $Y^{p,\,q}$ geometries studied in \cite{Gauntlett:2004hh, Martelli:2008rt}.%
\footnote{We are not looking for explicit metrics on those $SE_7$ spaces, we only need that they exist, as guaranteed in the toric case by \cite{Futaki:2006cc}.}
The relevant type IIA dual setup corresponds to D2-branes on the $CY_3$
\be\label{Ytilde first appearance}
\tilde{Y} \cong  \cO_{B_4}(K)\, ,
\ee
the canonical bundle over the Fano variety $B_4$. Importantly, there are also D6-branes wrapped on the compact 4-cycle $B_4$ in the IIA reduction. The presence of M5-branes on torsion 3-cycles in $\Ypq$ (corresponding to turning on torsion $G_4$ flux through $\Ypq$ in the $AdS_4\times \Ypq$ background) can also lead to D4-branes on 2-cycles in type IIA. Such a situation has been studied in detail in the simplest case $B_4 = \CP^2$ \cite{Benini:2011cma}. In the present work we generalise the results of \cite{Benini:2011cma} to any toric Fano variety $B_4$. There are only 16 such varieties, including the five  smooth del Pezzo surfaces. Much of the corresponding $CY_3$ cones (\ref{Ytilde first appearance}) were studied from the brane tiling point of view in the literature, and the corresponding brane tilings are fully classified  \cite{Hanany:2012hi}.

This paper is organised as follows.
In section \ref{section: toric quivers and dictionaries}, we introduce a formalism which allows an efficient study of fractional branes on any of the spaces (\ref{Ytilde first appearance}). We review the crucial notion of $\theta$-stability \cite{King1994} and some of its applications to $CY_3$ quivers.

In section \ref{sec: Mtheory to IIA} we explain which KK reduction we perform to obtain a useful IIA background from the conical geometry $C(\Ypq)$ in M-theory. Several interesting details on the $\Ypq$ geometry and the type IIA reduction are relegated to Appendix \ref{sec: Appendix topology}.

In section \ref{section: IIA to quiver} we explain how to translate the type IIA data into a Chern-Simons quiver gauge theory and we show that the field theory reproduces the type IIA geometry as its semiclassical moduli space by cnstruction.%
\footnote{As explained in \cite{Benini:2011cma}, additional Chern-Simons interactions coupling the central $U(1)$ factors of the $U(N_i)$ gauge groups are often needed to cancel a $\bZ_2$ global anomaly. While their precise type IIA origin is not yet understood, it is always possible to add them without changing the result of the semiclassical analysis of the moduli space. We leave this subtlety aside in this paper.}
 We also briefly comment on the inclusion of monopole operators, and how they naturally fit into the language of GIT quotient for quiver moduli spaces.

In section \ref{sec:_examples} we use the type IIA approach to give a first principle derivation of $\cN=2$ CS quiver gauge theories dual to any of the $\Ypq(B_4)$ geometries. For simplicity we focus on the case where the M-theory background contains no $G_4$ torsion flux.

In section \ref{sec: par resol and Higgsing} we study Higgsing and the resulting RG flow between the field theories of section \ref{sec:_examples}, showing in numerous examples that they match with the geometric process of partial resolution.

Finally, in section \ref{sec:torsion flux for F0} we consider adding $G_4$ discrete torsion flux to the $Y^{p,\, q_1,\, q_2}(\bF_0)$ geometry, showing that the dual theory is a CS quiver with generic ranks and Chern-Simons levels. In particular we show that the $U(N)^4$ quiver gauge theories of \cite{Tomasiello:2010zz}, in the absence of Romans mass in IIA, are dual to M-theory backgrounds with specific nontrivial torsion $G_4$ fluxes; we generalise the duality to allow for generic torsion $G_4$ fluxes in $Y^{p, \,q}(\bF_0)$, \emph{i.e.} when $q_1=q_2=q$.
An interesting ingredient in this analysis are the 3d Seiberg dualities for chiral quivers studied in \cite{Benini:2011mf, Closset:2012eq}.

Most of the computations of this paper have been algorithmised using \emph{Mathematica} \cite{Mathematica7}, building on a package developed by Jurgis Pasukonis for \cite{Davey:2009bp}.

%%%%%%%%%%%%%%%%%%%%%%%%%%%%%%%%%%%%%%%%%%%%%%%%%%%%%%
\section{Toric quivers, $CY_3$ cones and fractional brane charges}\label{section: toric quivers and dictionaries}
D-branes at conical Calabi-Yau singularities have been extensively studied in the past \cite{Douglas:1996sw, Diaconescu:1997br, Lawrence:1998ja, Klebanov:1998hh, Morrison:1998cs, Beasley:1999uz, Feng:2000mi, Feng:2002fv}. At the singularity, a transverse D-brane decays into a marginal bound state of so-called fractional branes. The low energy physics on the bound state of fractional branes is described by a supersymmetric quiver gauge theory with four supercharges. When the cone $Y$ is a \emph{toric} CY threefold, the quiver gauge theory has a very convenient description as a \emph{brane tiling}: a bipartite graph on the torus which encodes the quiver and superpotential data \cite{Hanany:2005ve,Franco:2005rj}. We refer to \cite{Kennaway:2007tq} for a review.

Quiver gauge theories are best known to describe D3-branes at CY$_3$ singularities $Y$ in type IIB, but they arise more generally for any lower dimensional D-branes transverse to $Y$. In this work we are interested in D2-branes in type IIA string theory, giving rise to three dimensional $\cN=2$ supersymmetric gauge theories.

If there are only D2-branes transverse to the singularity, the quiver gauge theory has a gauge group $U(N)^G$ with equal ranks and $G$ the number of quiver nodes. In this paper we will consider a more general set of D2-branes together with D4- and D6-branes wrapped on compact 2- and 4-cycles in $Y$, respectively, giving rise to rather generic field theories with $\prod_i U(N_i)$ gauge groups. We thus need to know the translation between the number of compactly supported D-branes and the quiver ranks. We will find matrices $Q^\vee$ such that
\be
\bm{N}=\bm{Q}_{\text{brane}}(Q^{\vee})^{-1} \,  ,
\ee
where $\bm{N}=(N_i)$ are the quiver ranks, and $\bm{Q}_{\text{brane}}$ is a vector encoding the D-brane Page charges \cite{Marolf:2000cb} of the supergravity background. The $G\times G$ matrix $Q^{\vee}$ containing the brane charges sourced by fractional branes will be called the \emph{dictionary}.

Although most of what follows is valid more generally, our main focus will be on $Y$ a complex cone over a two-complex-dimensional Fano variety. In this case we have an explicit algorithm to determine $Q^{\vee}$. Along the way we will review various important results about quivers and their moduli spaces, which will be useful when we turn to M2-brane theories.

\subsection{Toric quiver and crepant resolutions of $Y$}\label{subsec: toric quiver and resol}
\begin{figure}[t]
\begin{center}
\subfigure[\small $PdP_2$ quiver.]{
\includegraphics[height=4cm]{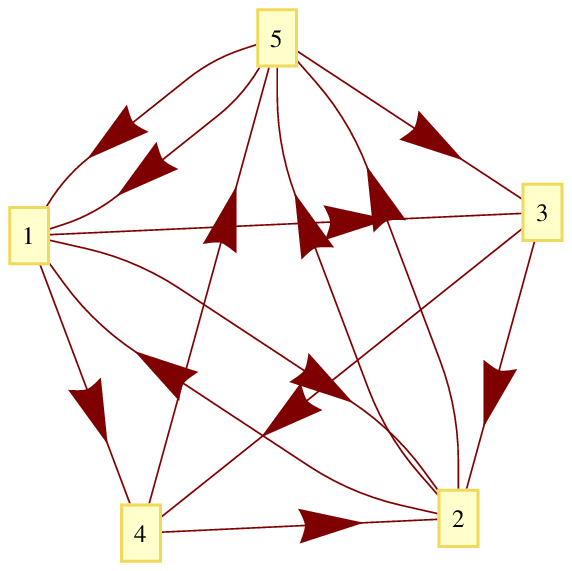}
\label{fig: PdP2 quiver0}}
\,
\subfigure[\small Brane tiling.]{
\includegraphics[height=4cm]{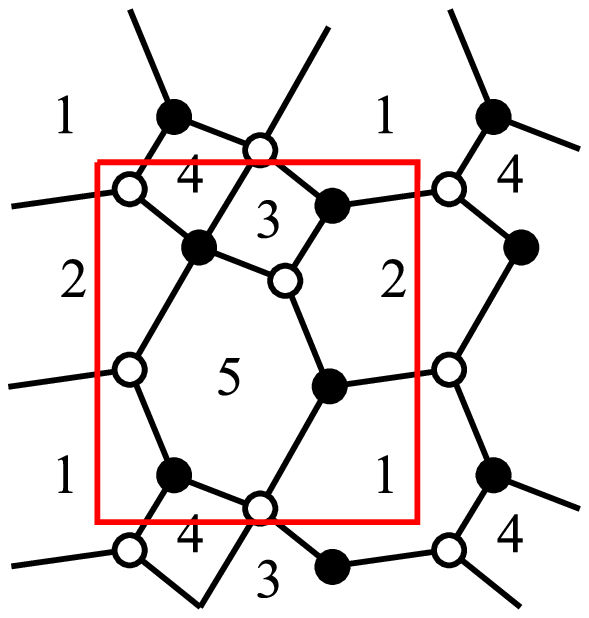}
\label{fig: PdP2 tiling}}
\,
\subfigure[\small Toric diagram.]{
\includegraphics[height=4cm]{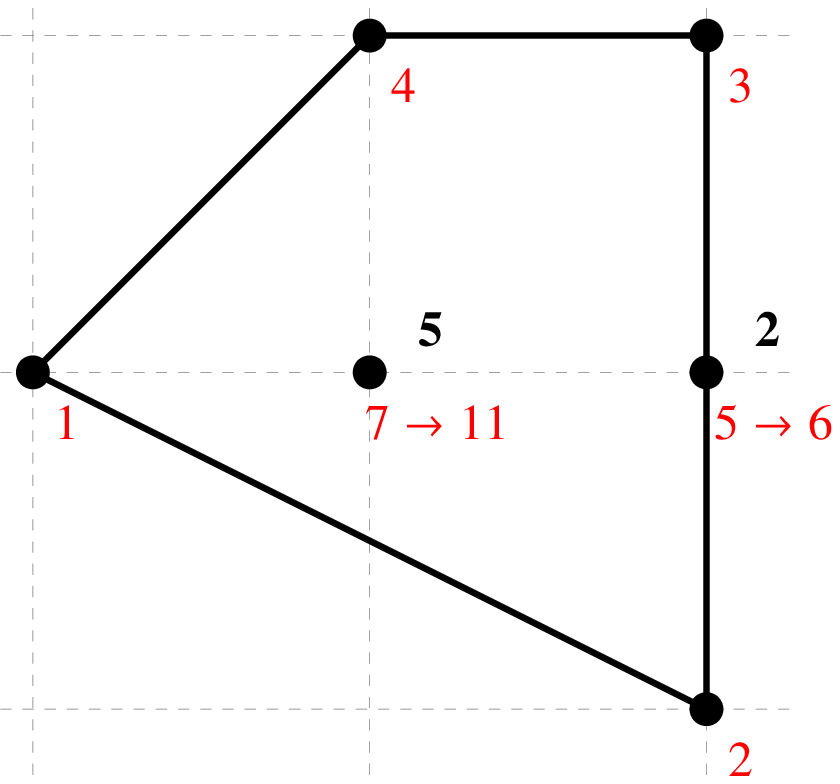}
\label{fig: PdP2 toric diag}}
\caption{\small Quiver, brane tiling and toric diagram for the complex cone over $PdP_2$. Black numbers above toric points are the multiplicities of these points (when they are larger than one), red numbers are the names of the corresponding perfect matchings. The strictly internal point conventionally has coordinates $(0,0)$.}
\end{center}
\end{figure}

To any toric $CY_3$ $Y$ we can associate at least one toric quiver with superpotential $\cQ$. Toric quiver theories admit a description in term of a brane tiling (see Figure \ref{fig: PdP2 tiling} for an example): each quiver node becomes a face, each arrow becomes an edge and each superpotential term becomes a white or black vertex in the brane tiling, depending on its sign.
A \textit{dimer} is a distinguished edge in a brane tiling. A \textit{perfect matching} $p_k$ is a configuration of dimers such that every vertex is touched exactly once. We define the perfect matching matrix $\bm{P}_{a k}$ as
\be\label{definition pm matrix}
 \bm{P}_{a k } = \left\{ \begin{array}{ll}
                  	1 &\quad \mbox{if the perfect matching $p_k$ contains the fields $X_{a}$}\\
			0 & \quad \mbox{otherwise}
                  \end{array}\right.
\ee
A dimer model is a brane tiling together with its perfect matchings.
Efficient ``inverse algorithms'' exist to go from the geometry to $\cQ$ \cite{Hanany:2005ss, Gulotta:2008ef}.

The low energy worldvolume theory of a single D2-brane transverse to the cone $Y$ is a 3d $\cN=2$ quiver gauge theory $\cQ$ with Abelian gauge group $\cG= U(1)^G$.
Indeed the variety $Y$ probed by the D2-brane is reproduced as the vacuum moduli space of the Abelian quiver.
Resolving the cone $Y$ to $\tilde{Y}$ corresponds to turning on Fayet-Ilioupoulos (FI) terms in the quiver gauge theory.
The FI parameters $\bm{\xi}=(\xi_i)$ affect D-term equations, leading to non-zero levels for the moment maps of $\cG$ in the K\"ahler quotient description of the moduli space,
\be\label{Kahl description 101}
 \cM(\cQ ; \bm{\xi})_{K} \equiv \{X_a \, | \, \partial W =0 \} //_{\bm{\xi}}\, \cG\,  \, .
\ee
We have that $\tilde{Y}\cong    \cM(\cQ ; \bm{\xi})_{K} $. The moment maps correspond to the D-terms
\be\label{abelian moment maps}
\mu_i \equiv  \sum_{X_a= X_{ij}} |X_a|^2 -\sum_{X_a= X_{j i}} |X_a|^2 = \xi_i\, .
\ee
The great advantage of toric quiver gauge theories is that we can trade the F-term equations $\partial W=0$ in (\ref{Kahl description 101}) for D-term equations in some auxiliary gauged linear sigma model (GLSM) with no superpotential.
The idea is to trivialise the relations $\partial W=0$ by introducing new variables $p_k$, $k=1, \cdots, c$  \cite{Feng:2000mi}. The solution of the F-term equations is given explicitly in term of so-called perfect matching (p.m.) variables $p_k$. We have
\be\label{X as pm var}
X_a = \prod_k (p_k)^{\bm{P}_{a k}}\, ,
\ee
where $\bm{P}$ is the perfect matching matrix (\ref{definition pm matrix}) \cite{Franco:2005rj, Franco:2006gc}.
To eliminate the redundancy in the description (\ref{X as pm var}) one introduces a spurious $U(1)^{c-G-2}$ gauge symmetry. Let $Q_D$ be the charge matrix of the variables  $p_k$ under the original gauge group $U(1)^{G}$ and $Q_F$ the charge matrix under the spurious gauge symmetry. The GLSM is conveniently summarised by its charge matrix together with its FI parameters:
\be\label{def GLSM toric VMS with D and F}
\begin{tabular}{l|ccc|c}
 & $p_1$ &$\cdots$ & $p_c$ & FI \\
 \hline
    $U(1)_F^l$  &$(Q_F)^l_1$ &$\cdots$ & $(Q_F)^l_c$&  $0$\\
  $U(1)_D^i$  &$(Q_D)_1^i$ &$\cdots$ & $(Q_D)_c^i$ & $\xi^i$
\end{tabular}
\ee
Remark that one should not introduce FI parameters for the spurious gauge symmetries. This connects to the GLSM description of toric varieties. Each $p_k$ corresponds to a point in the toric diagram $\Gamma$ of $Y$, and moving in FI parameter space allows to describe various (complete or partial) resolutions of $Y$. Importantly, the GLSM (\ref{def GLSM toric VMS with D and F}) never probes non-geometric phases of $Y$. This is possible because any internal point in the toric diagram is associated to several variables $p_k$ (the number of $p_k$'s associated to a toric point is called its \emph{multiplicity}).%
\footnote{A note on terminology. The toric diagram $\Gamma$ of $Y$ is a convex lattice polygon. We call a lattice point which is either inside  $\Gamma$ or inside some of its edges \emph{internal}, a point inside  $\Gamma$ \emph{strictly internal}, and an internal point of an external edge \emph{internal-external}. Finally a point which is not internal (i.e. one of the vertices of  $\Gamma$) is called \emph{strictly external}.}
 For any choice of $\xi_i$, we can always choose a unique variable $p_{0;\, r}$ for each internal point $w_r$ such that all other variables $p_{k;\, r}$ associated to $w_r$ are written in term of $p_{0;\, r}$ for any field value, making these $p_{k;\, r}$ redundant. More precisely, the D-term equations of the GLSM relate the modulus according to $|p_{k;\, r}|^2 = |p_{0;\, r}|^2 +\xi_{k,0} $, with $\xi_{k,0}$ some positive combination of the FI parameters; the phase of $p_{k;\, r}$ is fixed by the gauge symmetry.

\begin{figure}[t]
\begin{center}
\includegraphics[height=3.2cm]{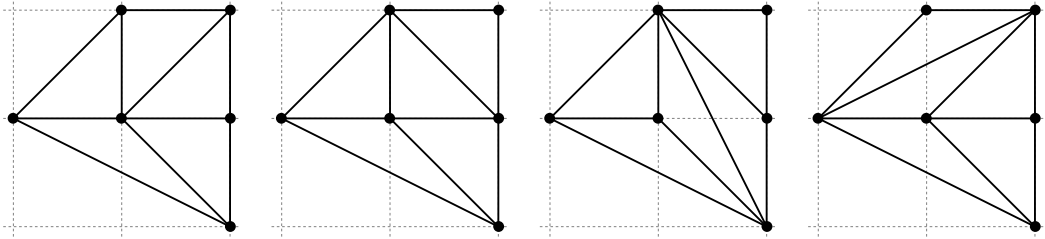}
\caption{\small The four possible triangulations $T_{\Gamma}^{(1)}, \cdots, T_{\Gamma}^{(4)}$ of the toric diagram of $C_{\bC}(PdP_2)$. %We denote them $T_{\Gamma}^{(1)}, \cdots, T_{\Gamma}^{(4)}$, from left to right.
}\label{fig: PdP2tianguluations}
\end{center}
\end{figure}

Going the other way, any choice of a single p.m. variable per point in $\Gamma$ determines an \emph{open string K\"ahler chamber} in FI parameter space, denoted
\be\label{def KC in term of internal pm}
KC = \{p_{r_1}, \cdots, p_{r_{n_I}}  \}\, ,
\ee
where $n_I$ denote the number of internal points in $\Gamma$. Such a choice determines a wedge in FI space by requiring that all other p.m.  variables from internal points can be solved for in term of the variables (\ref{def KC in term of internal pm}) --- see the example below.
This leads us to a minimal GLSM for $\tilde{Y}$,
\be\label{reduced GLSM}
\begin{tabular}{l|cc|c}
 $CY_3$ $\tilde{Y}$& $p_{s}$ & $p_{r}$ & FI \\
 \hline
  $ U(1)_{\alpha}$  &$Q^{\alpha}_s$ & $Q^{\alpha}_r$&  $\chi_{\alpha}(\xi)$
\end{tabular}
\ee
where $p_s$ and $p_r$ are strictly external and internal points in $\Gamma$, respectively, and the resolution parameters $\chi_{\alpha}$ are the K\"ahler volumes of a basis of 2-cycles $\cC_{\alpha}$, which depend linearly on the FI parameters in such a way that (\ref{reduced GLSM}) is always in a geometric phase.
%%%%%%%%%%%%%%%%%%%%%%%%%%%%%%%%%%%%%%%%%%%%%
%%%%%%%%%%%%%%%%%%%%%%%%%%%%%%%%%%%
\paragraph{Example:  The $PdP_2$ quiver.} As an example which contains all the complications of the general case, consider the quiver of Figure \ref{fig: PdP2 quiver0}, with superpotential
\bea \label{W_PdP2}
W_{PdP_2} &= X_{13}X_{34}X_{45}X_{51}^2 +X_{21}X_{14}X_{42}+X_{51}^1X_{12}X_{25}^2+X_{53}X_{32}X_{25}^1 +\\
& -X_{13}X_{32}X_{21}-X_{14}X_{45}X_{51}^1 -X_{51}^2X_{12}X_{25}^1 -X_{53}X_{34}X_{42}X_{25}^2\, .
\eea
The brane tiling is shown in Fig. \ref{fig: PdP2 tiling}. It describes D-branes transverse to the complex cone over the pseudo-del Pezzo surface $PdP_2$, whose toric diagram is shown in Figure \ref{fig: PdP2 toric diag}. The perfect matchings, as collections of dimers in the brane tiling, are
\bea
&p_1 =   \left\{X_{21}^{},X_{53}^{},X_{12}^{},X_{45}^{}\right\} \, ,
&\qquad & p_7= \left\{X_{14}^{},X_{53}^{},X_{12}^{},X_{13}^{}\right\}\, , \\
&p_2 =  \left\{X_{14}^{},X_{32}^{},X_{25}^2,X_{51}^2\right\} \, ,
&\qquad & p_8= \left\{X_{14}^{},X_{32}^{},X_{12}^{},X_{34}^{}\right\} \, ,\\
&p_3 =   \left\{X_{42}^{},X_{25}^1,X_{51}^1,X_{13}^{}\right\}\, ,
 &\qquad & p_9=\left\{X_{42}^{},X_{32}^{},X_{12}^{},X_{45}^{}\right\}\, , \\
&p_4 = \left\{X_{21}^{},X_{25}^1,X_{51}^1,X_{34}^{}\right\}  \, ,
&\qquad & p_{10}= \left\{X_{21}^{},X_{25}^1,X_{25}^2,X_{45}^{}\right\}\, , \\
&p_5 =   \left\{X_{14}^{},X_{25}^1,X_{25}^2,X_{13}^{}\right\} \, ,
 &\qquad  & p_{11}= \left\{X_{21}^{},X_{53}^{},X_{51}^1,X_{51}^2\right\}\, ,  \\
&p_6 =   \left\{X_{42}^{},X_{32}^{},X_{51}^1,X_{51}^2\right\} \,    &
\eea
From this we can read the perfect matching matrix $\bm{P}$ and express the 13 fields $X_{ij}$ in term of p.m. variables $p_k$, according to (\ref{X as pm var}). The GLSM (\ref{def GLSM toric VMS with D and F}) is
\be\label{GLSM PdP2 with mult}
\begin{tabular}{l|ccccccccccc|c}
 & $p_1$ & $p_2$ & $p_3$& $p_4$& $p_5$& $p_6$& $p_7$& $p_8$& $p_9$& $p_{10}$& $p_{11}$& FI \\
 \hline
   $U(1)^F_1$ & $0$ & $ -1$ & $ 0$ & $ -1$ & $ 1$ & $ 0$ & $ -1$ & $ 1$ & $ 0$ & $ 0$ & $ 1$ & $ 0$\\
  $U(1)^F_2$ & $-1$ & $ 0$ & $ 0$ & $ 0$ & $ -1$ & $ 0$ & $ 1$ & $ 0$ & $ 0$ & $ 1$ & $ 0$ & $ 0$\\
   $U(1)^F_3$ & $-1$ & $ 0$ & $ -1$ & $ 1$ & $ 0$ & $ 0$ & $ 1$ & $ -1$ & $ 1$ & $ 0$ & $ 0$ & $ 0$\\
   $U(1)^F_4$ & $ 0$ & $ -1$ & $ -1$ & $ 0$ & $ 1$ & $ 1$ & $ 0$ & $ 0$ & $ 0$ & $ 0$ & $ 0$ & $ 0 $ \\
   $U(1)^D_1$ & $ 0$ & $ -1$ & $ 0$ & $ -1$ & $ 1$ & $ 0$ & $ 0$ & $ 1$ & $ 0$ & $ 0$ & $ 0$ & $ \xi_1 $ \\
   $U(1)^D_2$ & $ 0$ & $ 0$ & $ -1$ & $ 1$ & $ 1$ & $ 0$ & $ 0$ & $ -1$ & $ 0$ & $ 0$ & $ 0$ & $ \xi_2$  \\
   $U(1)^D_3$ & $ 0$ & $ 0$ & $ 0$ & $ 0$ & $ 0$ & $ 0$ & $ -1$ & $ 1$ & $ 0$ & $ 0$ & $ 0$ & $ \xi_3$ \\
   $U(1)^D_4$ & $ 1$ & $ 0$ & $ 1$ & $ -1$ & $ 0$ & $ 0$ & $ -1$ & $ 0$ & $ 0$ & $ 0$ & $ 0$ & $ \xi_4$  \\
\end{tabular}
\ee
The last line of $Q_D$ (for the fifth gauge group) is omitted because it is redundant. There are 10 K\"ahler chambers, corresponding to choosing one of the 5 p.m. variables $(p_7, \cdots, p_{11})$ for the toric point $(0,0)$ and one of the 2 variables $(p_5, p_6)$ for the point $(1,0)$.
Indeed, the D-term equations of the GLSM (\ref{GLSM PdP2 with mult}) can be massaged into
\bea\label{PdP2:Dterms for KCs}
& |p_5|^2-|p_6|^2= \xi_1+\xi_2\,
 , \qquad |p_7|^2-|p_{11}|^2= \xi_1\, , \qquad |p_8|^2-|p_{11}|^2= \xi_1+\xi_3\, ,  \\
& |p_9|^2-|p_{11}|^2= \xi_1+\xi_3+\xi_4\, , \qquad |p_{10}|^2-|p_{11}|^2= \xi_1+\xi_2+\xi_3+\xi_4\, ,
\eea
together with three more equations.
Consider for instance the choice $\{p_5,p_7\}$. We can solve the D-terms (\ref{PdP2:Dterms for KCs}) in term of  $p_5$ as $|p_6|^2=|p_5|^2-\xi_1-\xi_2$ as long as $\xi_1+\xi_2\leq 0$, and similarly for $p_8, \cdots, p_{11}$ in term of $p_7$. Proceeding that way, we find 10 Kahler chambers $\{p_{(1,0)}, p_{(0,0)}\}$ with the conditions
\be\label{KC PdP2: choice for A1 CP1}
p_{(1,0)} = \begin{cases} p_5 \, : \quad \xi_1+\xi_2\leq 0\,  \\ p_6 \, : \quad \xi_1+\xi_2\geq 0\, \end{cases}\, ,
\ee
and
\be\label{KC PdP2: choice for internal pm}
p_{(0,0)} = \begin{cases} p_7   :\, \quad  \xi_3 \geq 0, \,  \xi_3 + \xi_4 \geq 0, \,  \xi_2 + \xi_3 + \xi_4 \geq 0, \,  \xi_1 \leq 0 \\
                          p_8   :\,\quad \xi_3 \leq 0, \,  \xi_4 \geq 0, \,  \xi_2 + \xi_4 \geq 0, \,  \xi_1 + \xi_3 \leq 0\, \,  \\
                          p_9   : \, \quad  \xi_3 + \xi_4 \leq 0, \,  \xi_4 \leq 0, \,  \xi_2 \geq 0, \,  \xi_1 + \xi_3 + \xi_4 \leq 0\\
                          p_{10}  :\quad \xi_2 + \xi_3 + \xi_4 \leq 0, \,  \xi_2 + \xi_4 \leq 0, \,  \xi_2 \leq 0, \,  \xi_1 + \xi_2 + \xi_3 + \xi_4 \leq 0  \\
                          p_{11}   :  \quad  \xi_1 \geq 0, \,  \xi_1 + \xi_3 \geq 0, \,  \xi_1 + \xi_3 + \xi_4 \geq 0, \,  \xi_1 + \xi_2 + \xi_3 + \xi_4 \geq 0\end{cases}
\ee
These conditions subdivide the FI parameter space $\bm{\xi}\cong \bR^4$ into 10 wedges. For a generic choice of $\bm{\xi}$, the singularity $Y= C_{\bC}(PdP_2)$ is fully resolved, consisting of one of the four possible triangulations of the toric diagram shown in Figure \ref{fig: PdP2tianguluations}.
%%%%%%%%%%%%%%%%%%5
%%%%%%%%%%%%%%%%%%%%%%%%%%%%%%
\\

We will discuss an elegant way of recovering these K\"ahler chambers in the following.
Let us note already that this notion of K\"ahler chamber stems from the quiver $\cQ$ and is ultimately related to the complexified K\"ahler moduli space seen by the type II string, including $\alpha'$ corrections. On the other hand, given a toric $CY_3$ $Y$ with toric diagram $\Gamma$, the classical geometric notion of (partial) resolution corresponds to a (partial) triangulation of the toric diagram. The space $\tilde{Y}$ is completely smooth if it is described by a simplicial fan, corresponding to a complete triangulation of $\Gamma$ such as in Fig. \ref{fig: PdP2tianguluations}. For a given K\"ahler chamber (\ref{def KC in term of internal pm}) only some of the triangulations of $\Gamma$ might be allowed.%
\footnote{Any internal edge $(r_1, r_2)$ in the triangulated toric diagram corresponds to a curve $D_{r_1}\cap D_{r_2}$ of positive volume $\chi$, where $D_r$ is the toric divisor associated to the toric point $w_r$. The volume $\chi$ is a linear combination of the $\chi_{\alpha}$'s in (\ref{reduced GLSM}), which in turn depend on the FI parameters in a specific way in each wedge in FI space (K\"ahler chamber). It might thus happen that $\chi$ is never positive in that K\"ahler chamber.}
We will return to this point below after introducing more powerful tools.

\subsection{GIT quotient and moduli space}

In the absence of FI parameters, i.e. when all D-term equations give vanishing moment maps on the space of constant fields $\{X_a\}$,  the moduli space can be recovered algebraically by ignoring the D-terms and quotienting the space of F-flatness solutions by the complexified gauge group. We recover the cone $Y$ as an affine algebraic variety,
\be\label{GIT quotient 101}
Y \cong \{X_a \, | \, \partial W =0 \} /\cG_{\bC}\;.
\ee
This is because whenever $\xi_i=0$ there always exists a unique solution of (\ref{abelian moment maps}) in the closure of each complexified gauge orbit \cite{Luty:1995sd}. Such quotient is called a GIT (Geometric Invariant Theory) quotient, and it is very intuitive from the physics point of view: we just consider the classical chiral ring of holomorphic gauge invariant operators.

There is a natural GIT generalization of (\ref{GIT quotient 101}) to allow for partial resolution of $Y$. Let $\cZ= \{X_a |\partial W=0 \} = \text{Spec}\, \bC[X_a]/(\partial W)$ be the set of solutions of the F-term equations, also known as the \emph{master space}, and $z_a$ some affine coordinates on $\cZ$. In our toric Abelian theory, this can also be described as
\be
\cZ = \text{Spec}\; \bC[p_1, \cdots, p_c]^{\cG_{\bC}^{F}},
\ee
in term of polynomials in the p.m. variables invariant under the spurious gauge symmetry $\cG^{F}= U(1)^{c-G-2}$. Consider a trivial line bundle $\cZ\times \bC$, with $t$ the coordinate on $\bC$, and pick some integers $(\theta_i)\equiv \bm{\theta} \in \bZ^G$ (such that $ \sum_i\theta_i=0$). The choice of $\bm{\theta}$ determines a one-dimensional representation $\chi_{\bm{\theta}}$  of $\cG_{\bC} \cong (\bC^{*})^G$ on the $\bC$ fibre,
\be\label{action of character theta}
(z_a, \, t)\mapsto  (\lambda\cdot z_a,\, \chi_{\bm{\theta}}(\lambda)\, t)\, , \qquad \mathrm{with}\qquad \chi_{\bm{\theta}} = \prod_{i=1}^{G} \lambda_i^{\theta_i}\, ,
\ee
for  $\lambda=(\lambda_1, \cdots, \lambda_G) \, \in (\bC^{*})^G$. Let $\cG_{\bC}(\bm{\theta})$ denote the action of the gauge group $\cG_{\bC}$ on $\cZ\times \bC$. The GIT quotient is given by
\be\label{GIT construction 00}
\cM(\cQ; \bm{\theta})_{GIT} = \text{Proj} \;\; \bC[\cZ\times \bC]^{\cG_{\bC}(\bm{\theta})} \, .
\ee
We refer to \cite{Martelli:2008cm} for some background on this construction in the present context. A crucial result is the Kempf-Ness theorem stating the equivalence of GIT quotient and K\"ahler quotient,
\be
\cM(\cQ; \bm{\theta})_{GIT} \cong \cM(\cQ; \bm{\xi})_{K} \, ,\qquad \mathrm{with}\quad \bm{\xi} = \bm{\theta} \, .
\ee
The parameters $\bm{\theta}$ are discretised FI parameters, which determine discretised K\"ahler classes of the underlying $CY_3$ cone $Y$.

\subsection{Quiver representations, $\theta$-stability and K\"ahler chambers}
From a mathematician's point of view, a quiver is nothing but an oriented graph consisting of nodes $i=1, \cdots, G$ and arrows $a$ connecting the nodes. Let us denote by $a= a_{ij}$ an arrow that goes from  $i$ to node $j$. The arrows generate a non-commutative algebra $\bC\cQ$ consisting of all the paths in the quiver, where the multiplication operation is the obvious concatenation of paths. If we associate to each node the trivial path $e_i$, the algebra $\bC\cQ$ has an identity element $\sum_i e_i$.

The quivers we study are also equipped with a superpotential which is a formal sum of quiver loops $l=\{i_1, i_2, \cdots, i_{n_l}\}$:
\be\label{abstract W}
W= \sum_{l \in L_0} \pm\, a_{i_1 i_2}\cdots a_{i_{n_l-1}i_{n_l}},
\ee
where $L_0$ denotes some subset of all the closed loops in $\cQ$. Formal derivation with respect to the arrows $a_{ij}$ leads to relations between the paths according to $\partial_{a} W =0$. The fundamental algebraic object associated to the quiver $\cQ$ is the path algebra obtained after quotienting by superpotential relations,
\be\label{define path algebra}
\cA\equiv \bC\cQ/(\partial W)\, .
\ee
For a toric quiver every arrow $X_a$ appears twice in (\ref{abstract W}), with opposite signs.

A \emph{quiver representation}  $R$ is a choice of  vector space $V_i$ for each node $i$ and of linear map $X_a$ for each arrow $a$, with the linear maps satisfying the superpotential relations:
\be
R: \begin{cases}i \; \mapsto V_i\cong \bC^{N_i}\\  a_{ij} \mapsto X_{ij}  \qquad \text{such that}\quad \partial_{X} W=0\, .  \end{cases}
\ee
The vector
\be
\bm{N} \equiv (N_1, \cdots, N_G) \equiv \text{dim}R
\ee
is the \emph{dimension vector} of  $R$.
In physical terms, a quiver representation is a choice of gauge group
\be\label{generic gauge group of Q}
\cG = U(N_1)\times \cdots \times U(N_G)
\ee
for a supersymmetric quiver gauge theory, together with a choice of VEVs for the chiral superfields $X_{ij}$. The dimension vector $\bm{N}$ gives the ranks of the gauge group.

Given two quiver representations $R$ and $R'$, a \emph{morphism} $\phi: R\rightarrow R'$ is a set of linear maps $\phi_i : V_i \rightarrow V_i'$ such that
\be\label{def morphism}
\phi_i X_a' = X_a \phi_j \, , \qquad \forall\, a=a_{ij}\, .
\ee
If $\phi$ is injective,  $R$ is a called a \emph{subrepresentation} of $R'$. Two representations $R$ and $R'$ are isomorphic if there exists a bijective morphism between them. As long as we consider holomorphic quantities, the gauge group of the supersymmetric quiver theory is effectively complexified to $\cG_{\bC} = \prod_i GL(N_i , \bC)$. Isomorphic quiver representations are simply gauge equivalent supersymmetric vacua in a given quiver gauge theory, which are physically identified. Isomorphism classes of $\cQ$ representations can also be understood as $\cA$-modules, i.e. representations of the path algebra (\ref{define path algebra}).

The moduli space of quiver representations of dimension $\bm{N}= \bm{\alpha} \equiv (1, 1, \cdots, 1)$ is our space $Y$, seen algebraically:
\be\label{moduli space of quiver w/o stab}
\cM(\cQ, \bm{\alpha}) = \{R_{\bm{\alpha}}\}/\cG_{\bC}\,  \cong Y \, .
\ee
We are after a similar description of partial resolutions $\pi : \tilde{Y}\rightarrow Y$, in parallel to the discussion of K\"ahler and GIT quotients in section \ref{subsec: toric quiver and resol}. We need some notion that adds some additional $\cG_{\bC}$-orbits to (\ref{moduli space of quiver w/o stab}). The crucial notion to do so is $\theta$-stability \cite{King1994}.

\paragraph{Definition: $\theta$-stability.} Consider a quiver $\cQ$ with $G$ nodes. Given a vector $\bm{\theta} \in \bZ^G$, a quiver representation $R$ of dimension $\bm{N}$ is $\theta$-stable (resp. semi-stable) if and only if $\bm{\theta}\bm{N}=0$ and for any proper subrepresentation $R'$ of dimension $\bm{N}'$ we have $\bm{\theta}\bm{N}'<0$ (resp. $\bm{\theta}\bm{N}'\leq 0$).\\

The main result of \cite{King1994} is that the moduli space of $\theta$-semistable quiver representations of a given dimension $\bm{N}$ can be obtained by a GIT quotient. In particular for $\bm{N}=\bm{\alpha}$,
\be\label{VMS as semistab reps}
\cM(\cQ, \bm{\alpha}; \bm{\theta}) \equiv  \{R_{\bm{\alpha}}\}^{ss} /\cG_{\bC}\,  \cong\, \cM(\cQ; \bm{\theta})_{GIT}\, ,
\ee
with $ \cM(\cQ; \bm{\theta})_{GIT}$ defined in (\ref{GIT construction 00}).

One can reformulate the considerations of section \ref{subsec: toric quiver and resol} about K\"ahler chambers in this quiver language \cite{2009arXiv0908.3475M, 2009arXiv0909.2013B}.
%For a given $\bm{\theta}\cong \bm{\xi}$ such that $\sum_i \theta_i =0$, we need to know which are the semi-stable $\bm{\alpha}$-reps (quiver representations of dimension $\bm{\alpha}$).
Consider any $\bm{\alpha}$-rep ($\cQ$ representation of dimension $\bm{\alpha}$) $R_{\alpha}\cong \{\bm{\alpha}, X_a\}$. Let $\cQ_{R_{\alpha}}$ be the quiver obtained from $\cQ$ by deleting any arrow $a$ such that $X_a=0$ in $R_{\alpha}$. Any subrepresentation of $R_{\alpha}$ is also a representation of $\cQ_{R_{\alpha}}$, obviously. A representation $R'_{\beta}\cong \{\bm{\beta}, X'_a\}$, with $\bm{\beta}\leq \bm{\alpha}$, is a subrepresentation of $R_{\alpha}$ if and only if for any $\beta_i \neq 0$ there exists a non-zero complex number $\phi_i$ such that (\ref{def morphism}) holds. Denote by $I_0$ the set of nodes $\{i_0\}$ such that $\beta_{i_0}=0$; we thus have $\phi_i=0$ if $i\in I_0$ and $\phi_i \neq 0$ otherwise. From (\ref{def morphism}), we have
\be
0= \phi_j X_{a} \, , \qquad \forall a=a_{j \, i_0}\,, \quad \forall i_0\in I_0\, ,
\ee
which holds if and only if $j \in I_0$ too. We thus showed that $R_{\beta}$ is a suprepresentation of $R_{\alpha}$ only if the dimension vector $\bm{\beta}$ is such that for any $i_0 \in I_0$, all nodes $j$ connected to $i_0$ \emph{from the left} in the auxilliary quiver $\cQ_{R_{\alpha}}$ (i.e. nodes such that there is a $X_{j i_0}\neq 0$ in $R_{\alpha}$) are also in $I_0$.

Consequently, any representation $R_{\alpha}$ such that $\cQ_{R_{\alpha}}$ is strongly connected%
\footnote{A quiver is called \emph{strongly connected} if for any pair of nodes $i,j$ there exists a quiver path $\mathsf{p}: i\rightarrow j$.}
has no subrepresentations and is therefore stable for \emph{any} $\bm{\theta}$. This is what happens for generic $R_\alpha$ representations, corresponding to a D2-brane probing the resolved cone $\tilde{Y}$ away from the exceptional locus $B_4$ (and away from any singularity that might remain in $\tilde{Y}$).

On the other hand, the $R_{\alpha}$ representations corresponding to the exceptional locus $\pi^{-1}(0)\subset \tilde{Y}$ are all representations of a quiver $\cQ_{B_4}$ with no closed loop \cite{Hanany:2006nm} and therefore \emph{not} strongly connected; we call such quiver a ``pseudo-Beilinson quiver'' for $B_4$. Choosing a K\"ahler chamber in the toric quiver in the sense of (\ref{def KC in term of internal pm}), the exceptional locus $B_{4}$ is obtained as a compact toric divisor $\{p_0=0\}$, with $p_0 \in \{p_r\}$ corresponding to the strictly internal point $w_0$. Therefore the pseudo-Beilinson  quiver $\cQ_{B_4}$  is obtained from $\cQ$ by setting to zero any field $X_a$ appearing in the perfect matching $p_0$.

More generally, for any perfect matching or collection of perfect matchings $\{p_k\}$, we define a $\cQ$-representation of dimension $\bm{\alpha}$ \cite{2009arXiv0909.2013B}%
\footnote{It turns out that for perfect matchings this is a good representation \cite{2009arXiv0909.2013B}, which is not completely obvious due to the superpotential relations. }
\be
R_{\{p_k\}}\, : \; X_{a}= \begin{cases}0\, , \qquad a\in \{p_0\}  \\ 1\, , \qquad a\notin \{p_k\}  \end{cases}\, ,
\ee
and a pseudo-Beilinson quiver $\cQ_{R_{\{p_k\}}}$. The collection $\{p_k\}$ is called $\bm{\theta}$-stable if $R_{\{p_k\}}$ is $\theta$-stable.

For any choice of K\"ahler chamber (\ref{def KC in term of internal pm}), we need that every perfect matching $p_r\in KC$ is $\theta$-stable. This gives inequalities on $\bm{\theta}\cong \bm{\xi}$, reproducing the same result as in section \ref{subsec: toric quiver and resol}.
Given such a K\"ahler chamber, we further specify a triangulation $T_{\Gamma}$ of the toric diagram as a collection of pairs of perfect matchings, $T_{\Gamma}=\{ (p_{r_1},p_{r_2})\}$, according to the edges in $T_{\Gamma}$.  This triangulation is allowed only for $\bm{\theta}$ such that every pair $(p_{r_1},p_{r_2})$ in $T_{\Gamma}$ is $\theta$-stable as well.

When implemented on a computer, this gives an algorithm to find all K\"ahler chambers in FI parameter space which runs in about the same time as the algorithm described in section \ref{subsec: toric quiver and resol}. On the other hand, the present method is much more efficient to discuss triangulations of the toric diagram and how they depend on the FI parameters of the quiver, besides being more elegant conceptually.

%%%%%%%%%%%%%%%%%%%%%%%
\begin{figure}[t]
\begin{center}
\subfigure[\small $\cQ_{p_5}$.]{
\includegraphics[height=2cm]{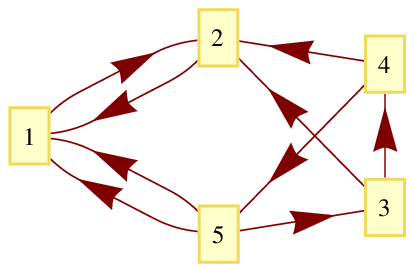}
\label{fig: PdP2 Bel quiv p5}}
\quad
\subfigure[\small $\cQ_{p_6}$.]{
\includegraphics[height=2cm]{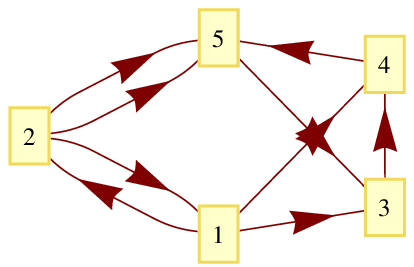}
\label{fig: PdP2 Bel quiv p6}}
\quad
\subfigure[\small $\cQ_{p_7}$.]{
\includegraphics[height=2cm]{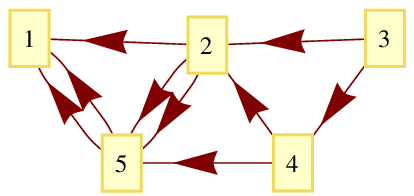}
\label{fig: PdP2 Bel quiv p7}}
\caption{\small Examples of pseudo-Beilinson quivers obtained from the toric quiver for $PdP_2$, for the perfect matchings $p_5$, $p_6$ and $p_7$ respectively.}\label{fig: PBelquivers for PdP2}
\end{center}
\end{figure}
%%%%%%%%%%%%%%%%%%%%%%%%%
\paragraph{The $PdP_2$ example.} Consider the $PdP_2$ quiver introduced before. A K\"ahler chamber $KC \cong \{p_{(1,0)}, p_{(0,0)}\}$ is such that these two perfect matchings are $\theta$-stable for any $\bm{\theta}$ in the chamber (and $\theta$-semistable on the walls of the chamber). The pseudo-Beilinson quivers associated to some perfect matchings are shown in Figure \ref{fig: PdP2 Bel quiv p5}. In the case $p_{(1,0)}=p_5$, we easily see from the quiver in Fig. \ref{fig: PdP2 Bel quiv p5} that the only subrepresentation of the $\bm{\alpha}$-rep $R_{p_5}$ has dimension vector $\bm{\beta}=(1,1,0,0,0)$, leading to the $\theta$-stability condition $\theta_1+\theta_2< 0$. Similarly, from Fig. \ref{fig: PdP2 Bel quiv p6} we find that the only subrepresentation of $R_{p_6}$ has dimension vector $(0,0,1,1,1)$, so that we should have $\theta_3+\theta_4+\theta_5= -\theta_1-\theta_2 < 0$. This reproduces (\ref{KC PdP2: choice for A1 CP1}). Similarly for the choice of strictly internal perfect matching: from Fig. \ref{fig: PdP2 Bel quiv p7} we see that the subrepresentations of $R_{p_7}$ are $(1, 0, 0, 0, 0)$, $(1, 0, 0, 0, 1)$, $(1, 1, 0, 0, 1)$ and $(1, 1, 0, 1, 1)$, giving the conditions in the first line of (\ref{KC PdP2: choice for internal pm}).

Consider the four triangulations $T_{\Gamma}^{(1)}, \cdots, T_{\Gamma}^{(4)}$ in Figure \ref{fig: PdP2tianguluations}. Any edge between the internal point $(0,0)$ and a strictly external point is $\theta$-stable whenever $p_{(0,0)}$ is $\theta$-stable. On the other hand any other edge gives a new non-trivial constraint. We find that the choice $p_{(1,0)}= p_6$ (i.e. a choice of FI parameters with $\xi_1+\xi_2>0$) does not allow any of the four complete triangulations. On the other hand, for $p_{(1,0)}= p_5$, the five K\"ahler chambers $\{p_5, p_{(0,0)}\}$ are compatible with some of the triangulations according to:
\be
\begin{tabular}{l|ccccc}
Compatible ? &  $\{p_5, p_7\}$&  $\{p_5, p_8\}$ &  $\{p_5, p_9\}$ &  $\{p_5, p_{10}\}$ &  $\{p_5, p_{11}\}$\\
\hline
$T_{\Gamma}^{(1)}$ &Yes &Yes&Yes&Yes&Yes \\
$T_{\Gamma}^{(2)}$ &Yes &Yes&No&Yes&Yes \\
$T_{\Gamma}^{(3)}$ &No &Yes&No&Yes&Yes \\
$T_{\Gamma}^{(4)}$ &Yes &No&Yes&Yes&Yes \\
\end{tabular}
\ee

\subsection{Tilting collection of line bundles from toric quiver}\label{subsec: tilting collection of line bdl}
Branes of type II string theory wrapping holomorphic cycles in $Y$ are described mathematically as objects in the B-brane category of $Y$, also known as derived category of coherent sheaves $D(\text{Coh}\, \tilde{Y})$. These objects are the branes of the topological B-model, and for that reason they are independent of the K\"ahler structure, allowing us to probe singular geometries where  large $\alpha'$ corrections are expected.

Given a toric variety $\tilde{Y}$, suppose that we can find a collection of line bundles
\be\label{notation cE}
\cE= \{ \mathsf{L}_1, \cdots , \mathsf{L}_G  \}
\ee
such that
\be
H^n(\tilde{Y}, \mathsf{L}_j\otimes  \mathsf{L}_i^*)=0 \, , \qquad \forall\, n>0 \, , \; \forall\, i,j\, ,
\ee
and which generates $D(\text{Coh}\, \tilde{Y})$. Such an object (\ref{notation cE}) is called a \emph{tilting collection} of line bundles and gives a particularly nice generating set of the B-branes on $Y$.
It also has the right properties to be associated to a quiver: each $\mathsf{L_i}$  corresponds to a node of the quiver $\cQ$ with path algebra (\ref{define path algebra}) given by
\be
\cA^{\text{op}} \cong  \oplus_{i,j}\, \text{Hom}(\mathsf{L_i},\mathsf{L_j}) = \oplus_{i,j}\, H^0(\tilde{Y}, \mathsf{L}_j\otimes  \mathsf{L}_i^*)\, .
\ee
See in particular \cite{Aspinwall:2008jk} for more background on this construction.%
\footnote{See also \cite{Closset:2012eq}. As reviewed there, the line bundles $\mathsf{L_i}$ (or sheaves $\mathsf{P_i}$ in the notation of \cite{Closset:2012eq}) correspond to the projective (right) $\cA$-modules $\cA e_i$ in the quiver language. }
$\cA$ and $\cA^{\text{op}}$ are related by reversing the orientation of every path, and $\cA$ defines the quiver $\cQ$ (with superpotential relations) implicitly.

Given a toric quiver $\cQ$ corresponding to a toric $CY_3$ $Y$, we can construct a tilting collection on any of the partial resolutions $\tilde{Y}$  following  \cite{Hanany:2006nm, 2009arXiv0909.2013B}.
A \emph{weak path} $\mathsf{p}$ is a path in $\cQ$ using both the arrows $a$ and their inverse $a^{-1}$:
\be\label{def weak path}
\mathsf{p}\,:\quad a_1^{\epsilon_1}a_2^{\epsilon_2}\cdots a_l^{\epsilon_l}\, , \qquad\quad \epsilon = \pm 1\, .
\ee
To any arrow $a$, the $\Psi$-map associates the formal sum of perfect matchings in which $X_a$ appears, $\Psi(a)= \sum_k \bm{P}_{ak} p_k$. This extends to  any weak path  linearly.
Consider the resolved space $\tilde{Y}_{\theta}$ associated to some $\bm{\theta}$ parameter, and more generally to some chamber (\ref{def KC in term of internal pm}). We denote by $\Psi_{\theta}$ the $\Psi$-map with range restricted to $\theta$-stable perfect matchings. Since the latter are associated to rays in the toric fan of $\tilde{Y}_{\theta}$, thus to toric divisors, $\Psi_{\theta}$ is really mapping paths to  divisors of $\tilde{Y}_{\theta}$. For a weak path (\ref{def weak path}), we have
\be\label{def psi map with div}
\Psi_{\theta}(\mathsf{p}) = \sum_{n=1}^l \epsilon_n \sum_{k |p_k\, \\ \text{ $\theta$-stab}} \,  \bm{P}_{a_n k} D_k\, ,
\ee
where $D_k \cong \{p_k= 0\}$.
Choose some conventional ``first node'' $i=1$ in $\cQ$, and associate to every node $i$ a weak path $\mathsf{p}_i : 1\rightarrow i$ (with $\mathsf{p}_1$ the trivial path). We can associate a line bundle over $\tilde{Y}_{\theta}$ to every quiver node according to
\be\label{line bundle from Psi theta map}
\mathsf{L}_i = \cO(  \Psi_{\theta}(\mathsf{p}_i)  )\, .
\ee
It was proven in Theorem 4.2 of \cite{2009arXiv0909.2013B} that the collection $\{\mathsf{L}_i\}_{i=1}^G$ obtained in this way is a tilting collection.

\paragraph{Example.} In our $PdP_2$ example (see Fig. \ref{fig: PdP2 quiver0}), we can take for instance the weak paths $\mathsf{p_1}= e_1$, $\mathsf{p_2}= a_{12}$, $\mathsf{p_2}= a_{13}$, $\mathsf{p_3}= a_{13}$, $\mathsf{p_4}= a_{14}$ and $\mathsf{p_5}= (a_{51}^1)^{-1}$. The $\Psi$-map gives us
\bea\label{result Psi map PdP2 example}
&\Psi(\mathsf{p_1})= 0\, ,& \qquad & \Psi(\mathsf{p_2})= p_1+p_7+p_9\, ,\\
&\Psi(\mathsf{p_3})= p_3+p_5+p_7\, ,& \qquad & \Psi(\mathsf{p_4})= p_2+p_5+p_7+p_8 \, ,\\
&\Psi(\mathsf{p_5})= -p_3-p_4-p_6-p_{11}\, .& \qquad & \\
\eea
In any K\"ahler chamber $KC\cong \{p_{(1,0)}, p_{(0,0)}\}$, the $\Psi_{\theta}$ map restricts (\ref{result Psi map PdP2 example}) to the $\theta$-stable perfect matchings, and to the corresponding tilting collections of line bundles (\ref{line bundle from Psi theta map}). For instance
\bea\label{examples of BM collections PdP2}
&KC\cong \{p_5, p_7\}\, &:& \, \qquad \begin{cases}\cO, \, \cO(D_1+D_{(0,0)}), \, \cO(D_3+D_{(1,0)}+D_{(0,0)}), \,\\ \cO(D_2+D_{(1,0)}+D_{(0,0)} ), \, \cO(-D_3-D_4), \, \end{cases} \\
&KC\cong \{p_5, p_8\}\, &:& \, \qquad \begin{cases}\cO, \, \cO(D_1), \, \cO(D_3+D_{(1,0)}), \,\\ \cO(D_2+D_{(1,0)}+D_{(0,0)} ), \, \cO(-D_3-D_4), \, \end{cases}
\eea
and so on for the 10 open string K\"ahler chambers of this $PdP_2$ quiver.

\subsection{2d toric Fano varieties and brane charge dictionaries}\label{subsec:_Fano}

%%%%%%%%%%%%%%%%toric diagrams for 16 Fanos:%%%%%%%%%%%%%%%%%%%%%%%%%%%%%
\begin{figure}[h]
\begin{center}
\subfigure[\small $dP_0$]{\includegraphics[height=2.7cm]{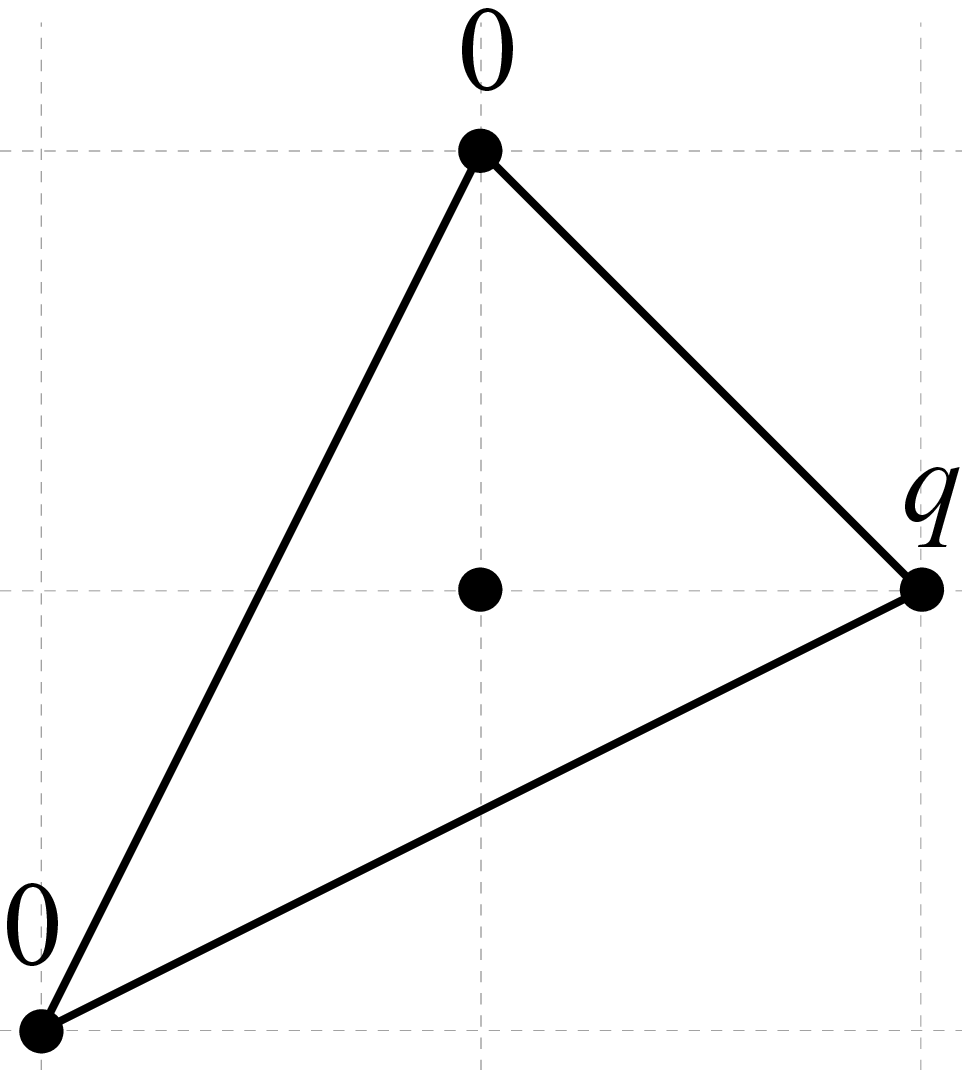}\label{fig: torDiag mod 01}}\qquad
\subfigure[\small $F_0$]{\includegraphics[height=2.7cm]{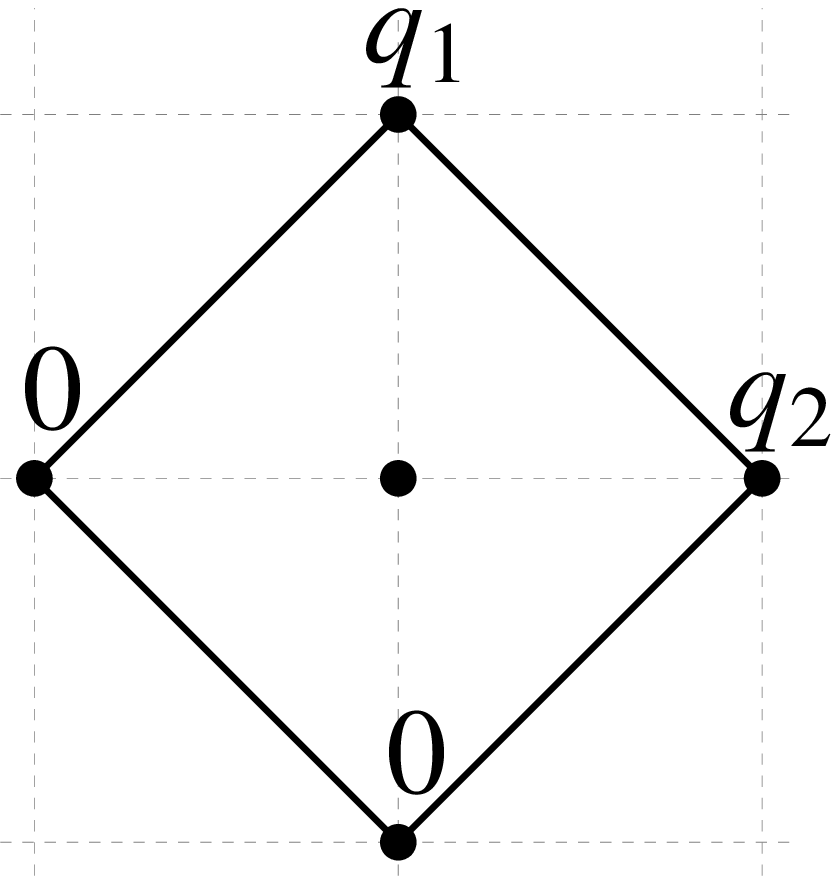}\label{fig: torDiag mod 02}}\qquad
\subfigure[\small $dP_1$]{\includegraphics[height=2.7cm]{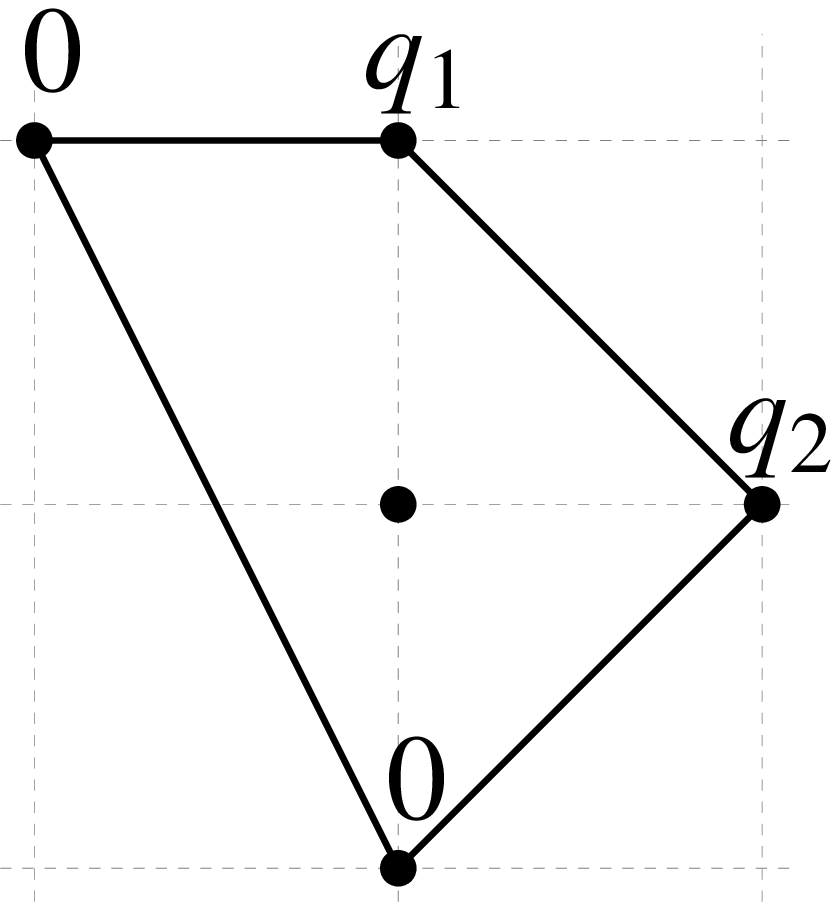}\label{fig: torDiag mod 03}}\qquad
\subfigure[\small $W\CP^2_{[1,1,2]}$ ]{\includegraphics[height=2.7cm]{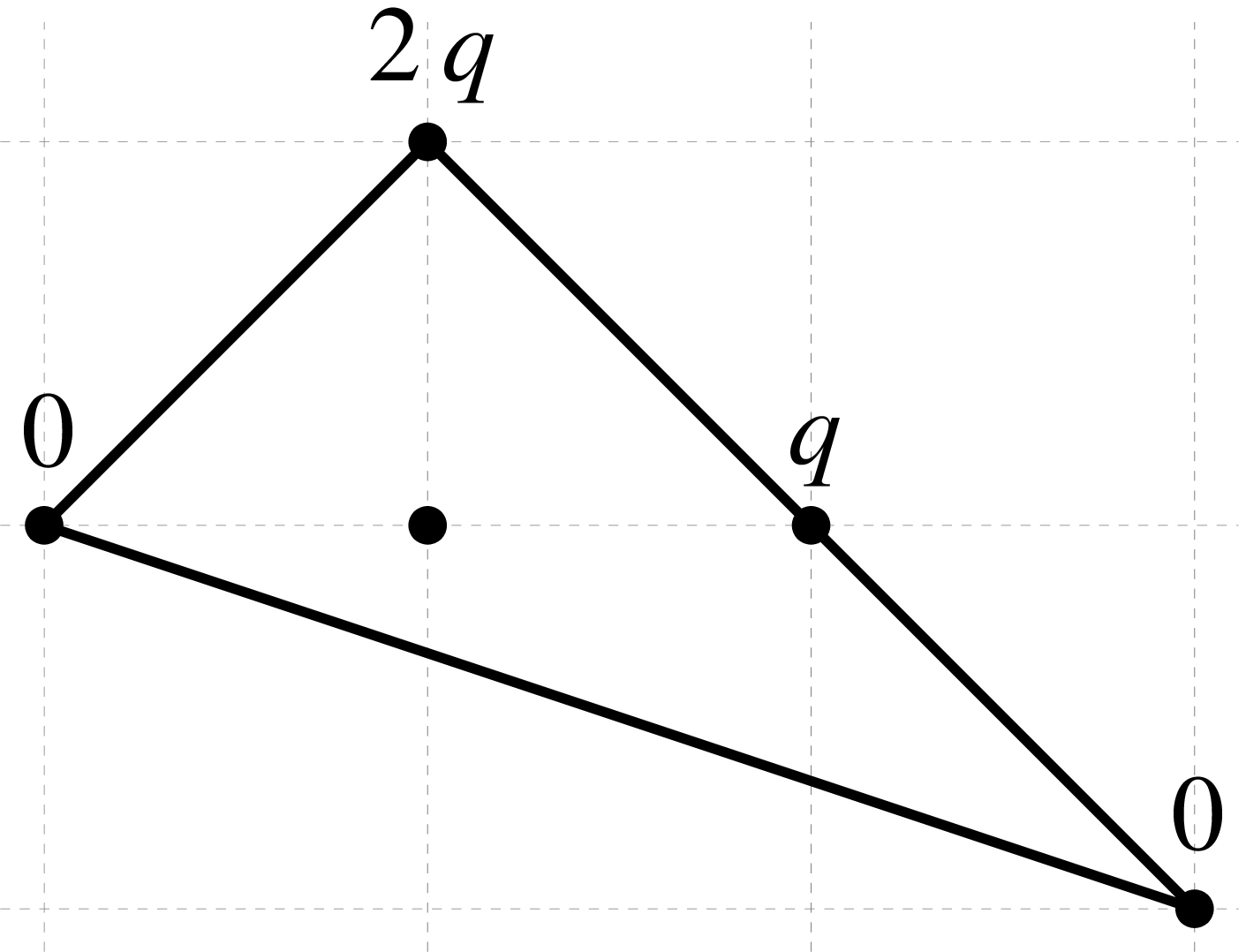}\label{fig: torDiag mod 04}}\\
\subfigure[\small $dP_2$]{\includegraphics[height=2.7cm]{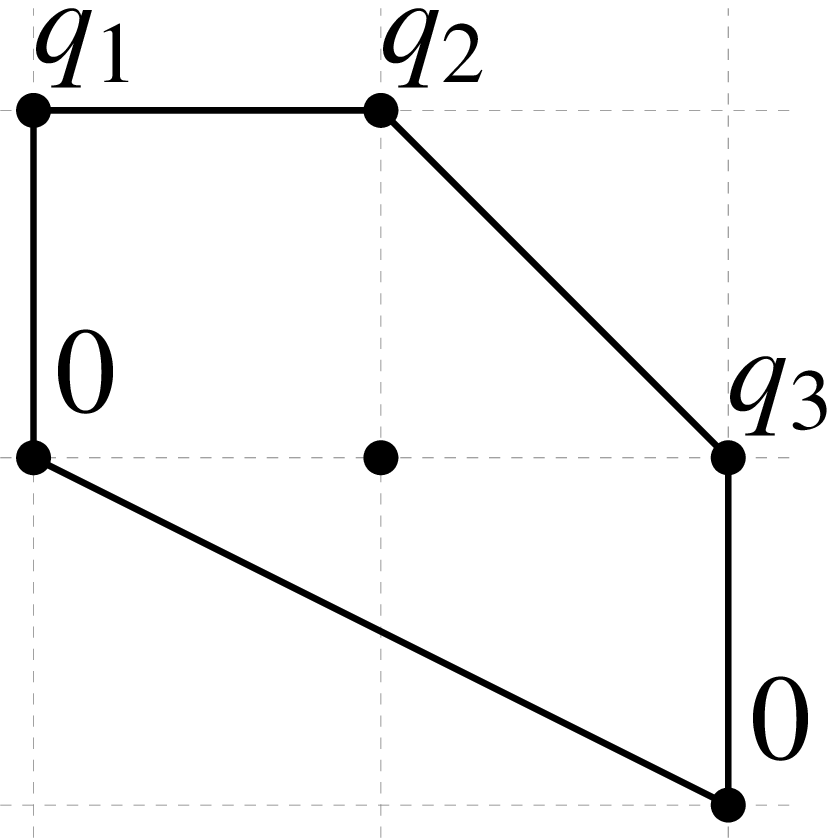}\label{fig: torDiag mod 05}}\qquad
\subfigure[\small $PdP_2$]{\includegraphics[height=2.7cm]{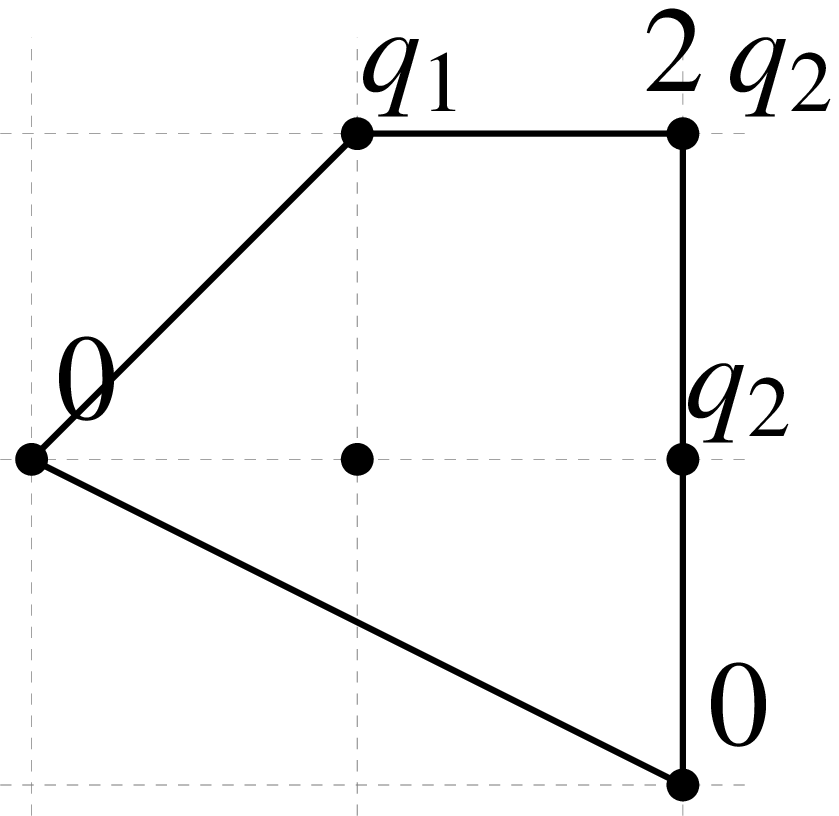}\label{fig: torDiag mod 06}}\qquad
\subfigure[\small $dP_3$]{\includegraphics[height=2.7cm]{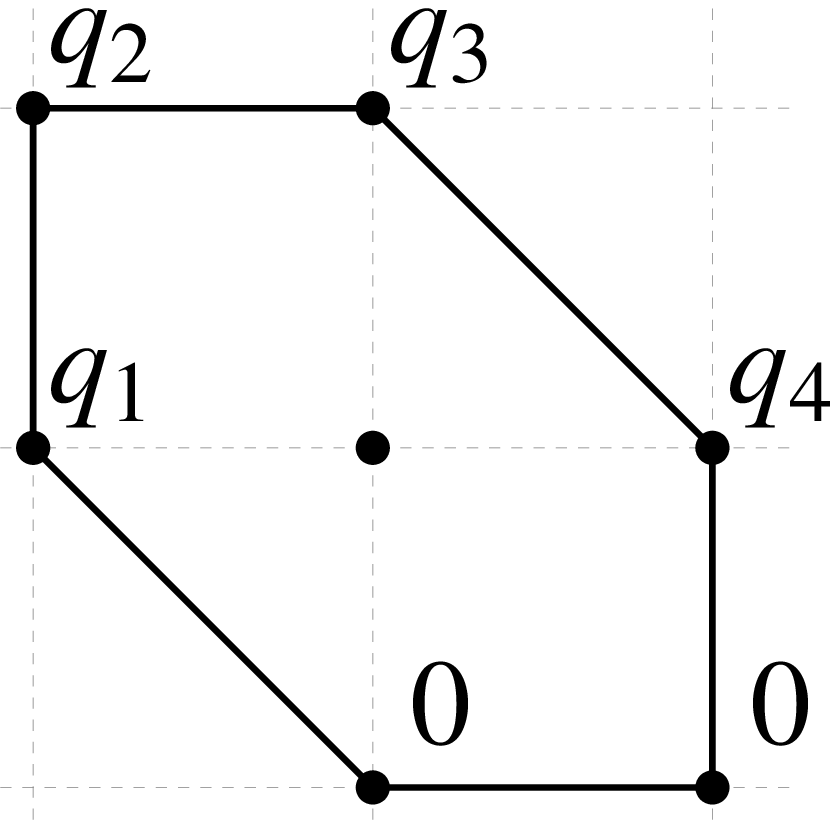}\label{fig: torDiag mod 07}}\qquad
\subfigure[\small $PdP_{3b}$]{\includegraphics[height=2.7cm]{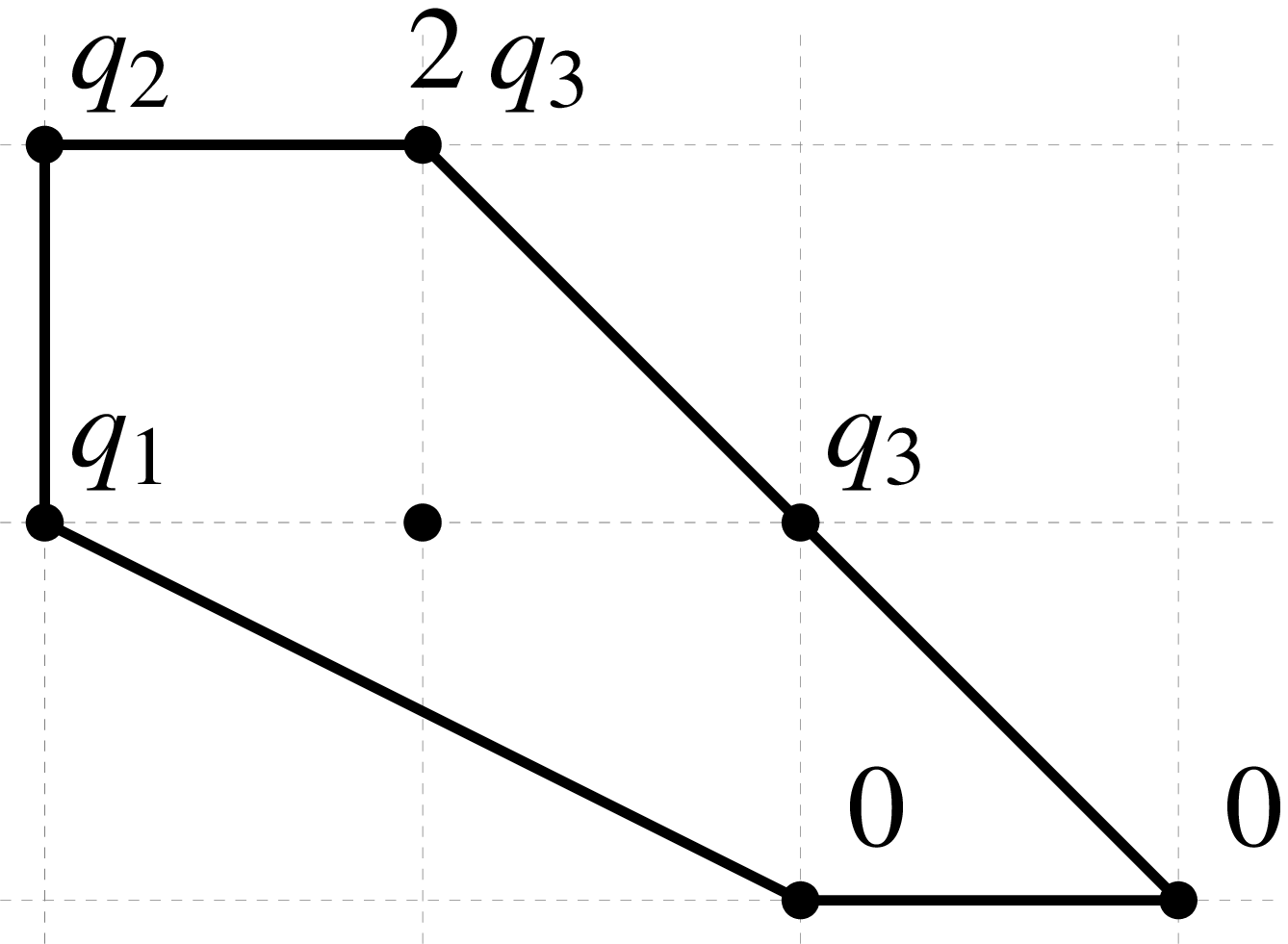}\label{fig: torDiag mod 08}}\\
\subfigure[\small $PdP_{3c}$]{\includegraphics[height=2.7cm]{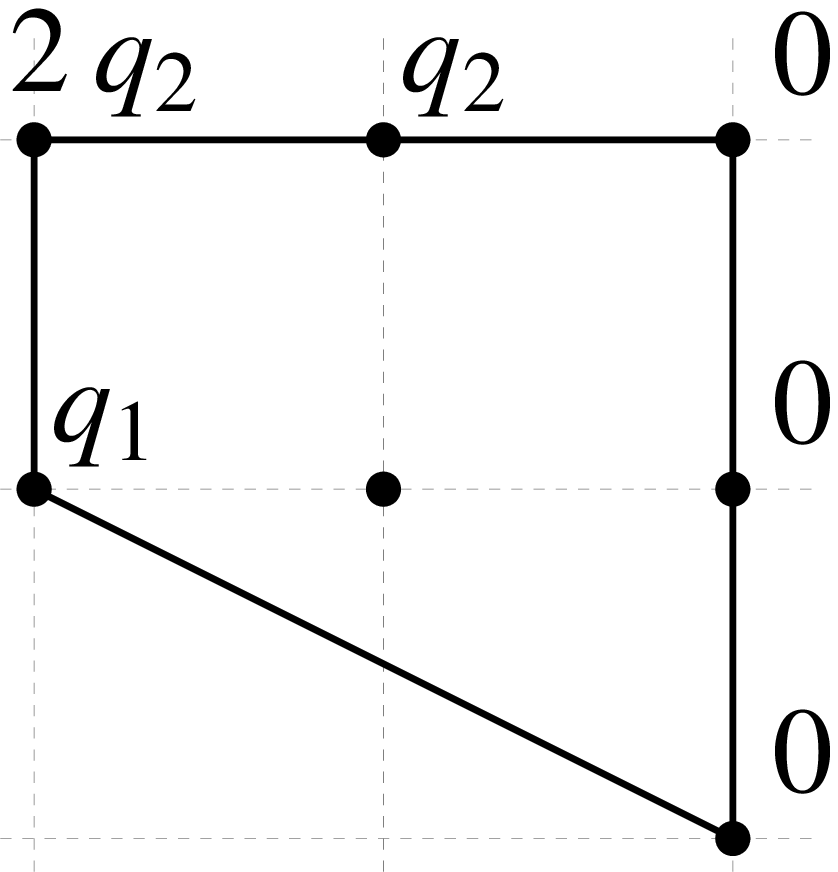}\label{fig: torDiag mod 09}}\qquad
\subfigure[\small $W\CP^2_{[1,2,3]}$ ]{\includegraphics[height=3.2cm]{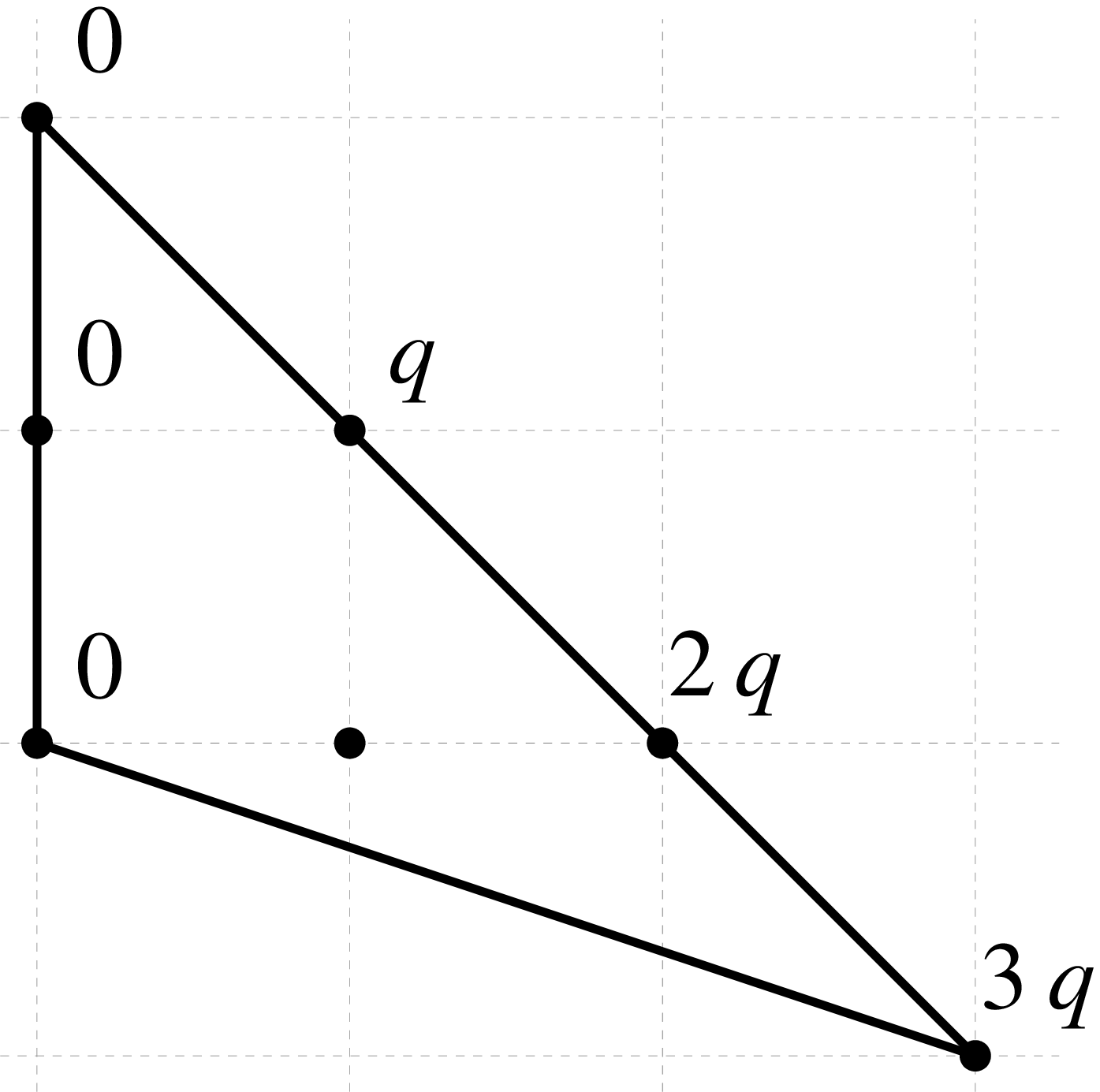}\label{fig: torDiag mod 10}}\qquad
\subfigure[\small $PdP_4$]{\includegraphics[height=2.7cm]{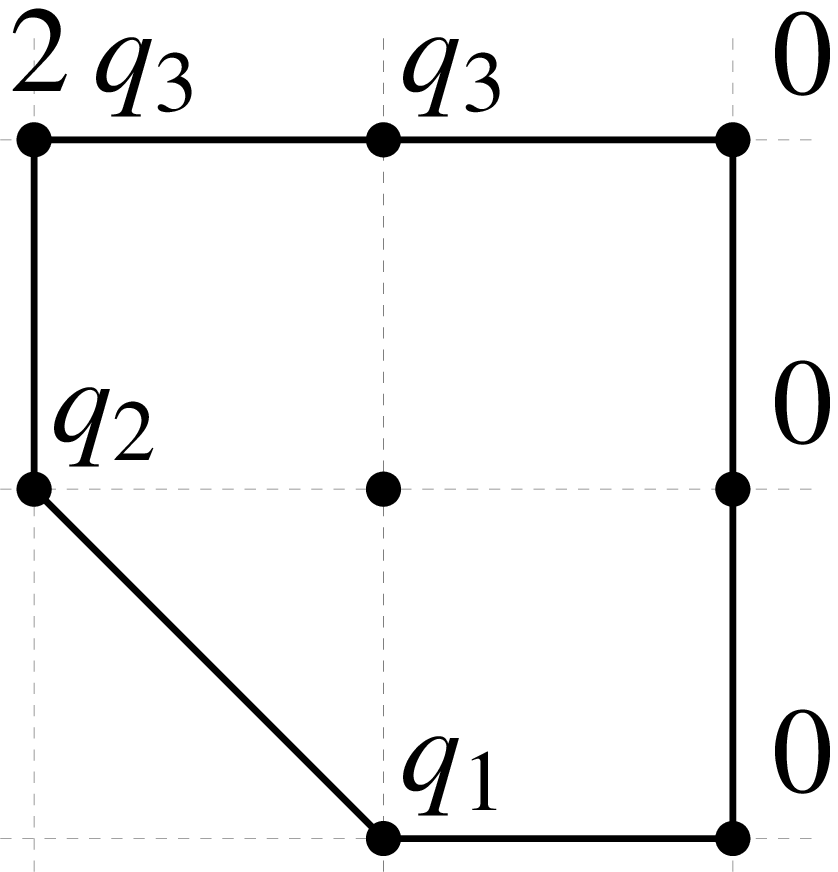}\label{fig: torDiag mod 11}}\qquad
\subfigure[\small $PdP_{4b}$ ]{\includegraphics[height=3.2cm]{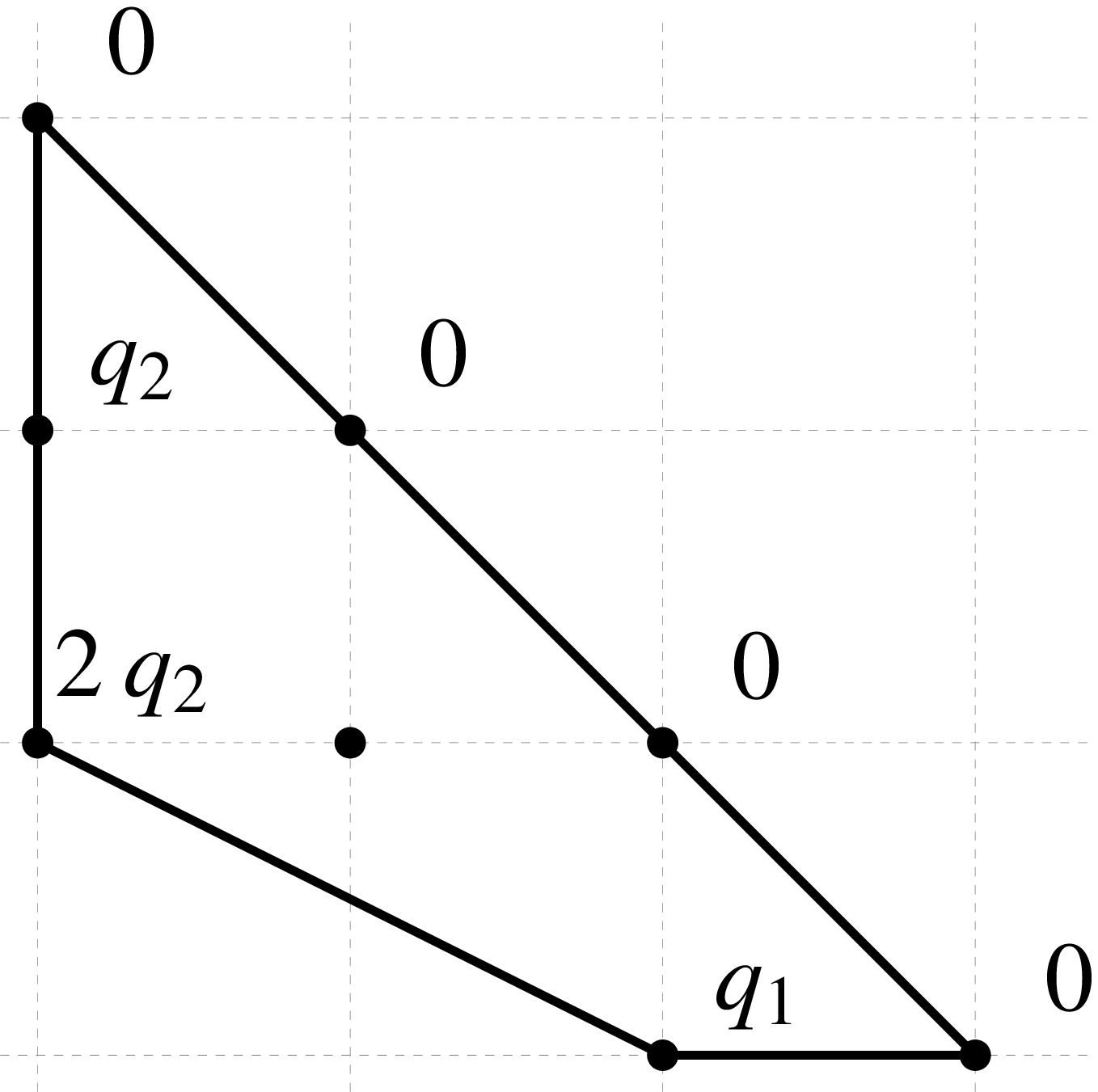}\label{fig: torDiag mod 12}}\\
\subfigure[\small $PdP_5$]{\includegraphics[height=2.7cm]{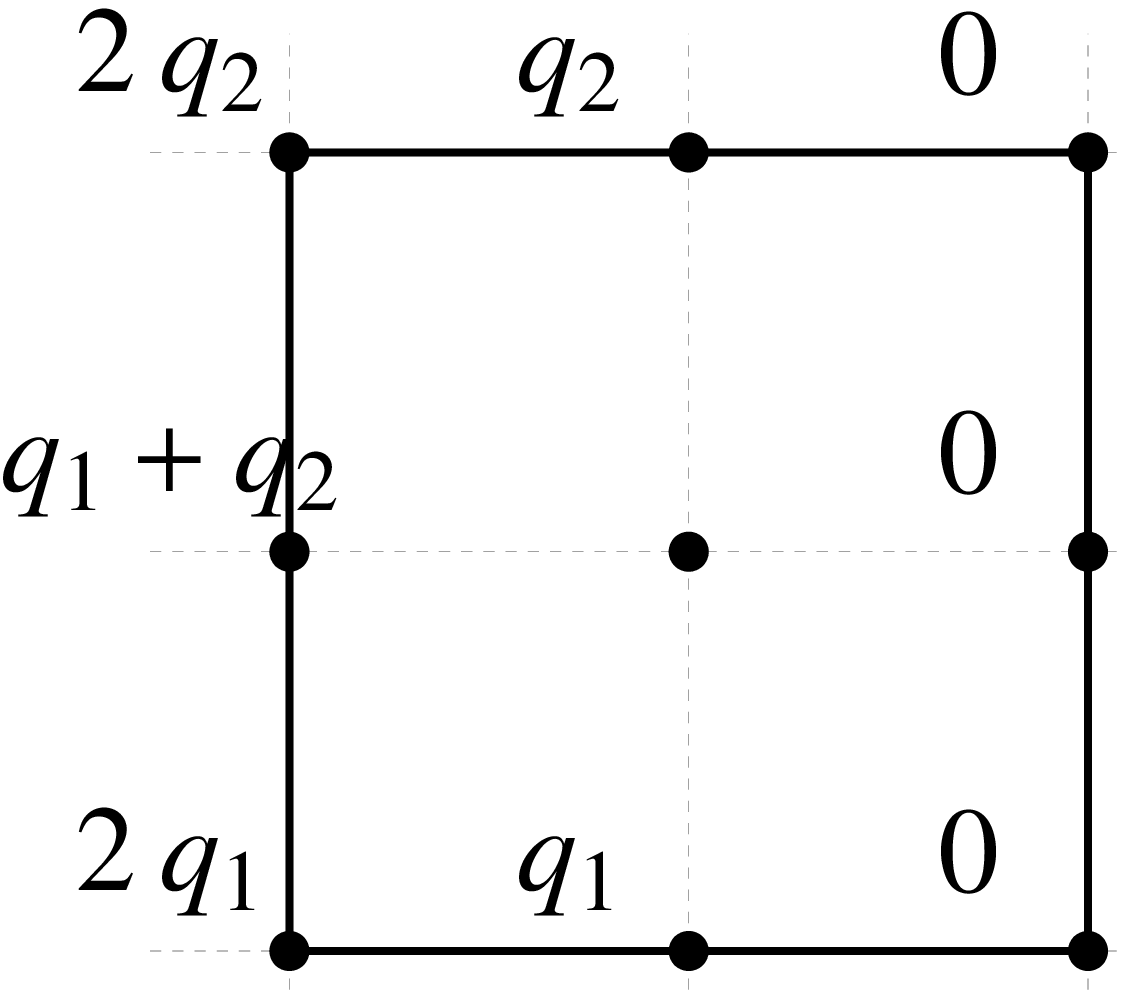}\label{fig: torDiag mod 13}}\,
\subfigure[\small $PdP_{5b}$]{\includegraphics[height=2.7cm]{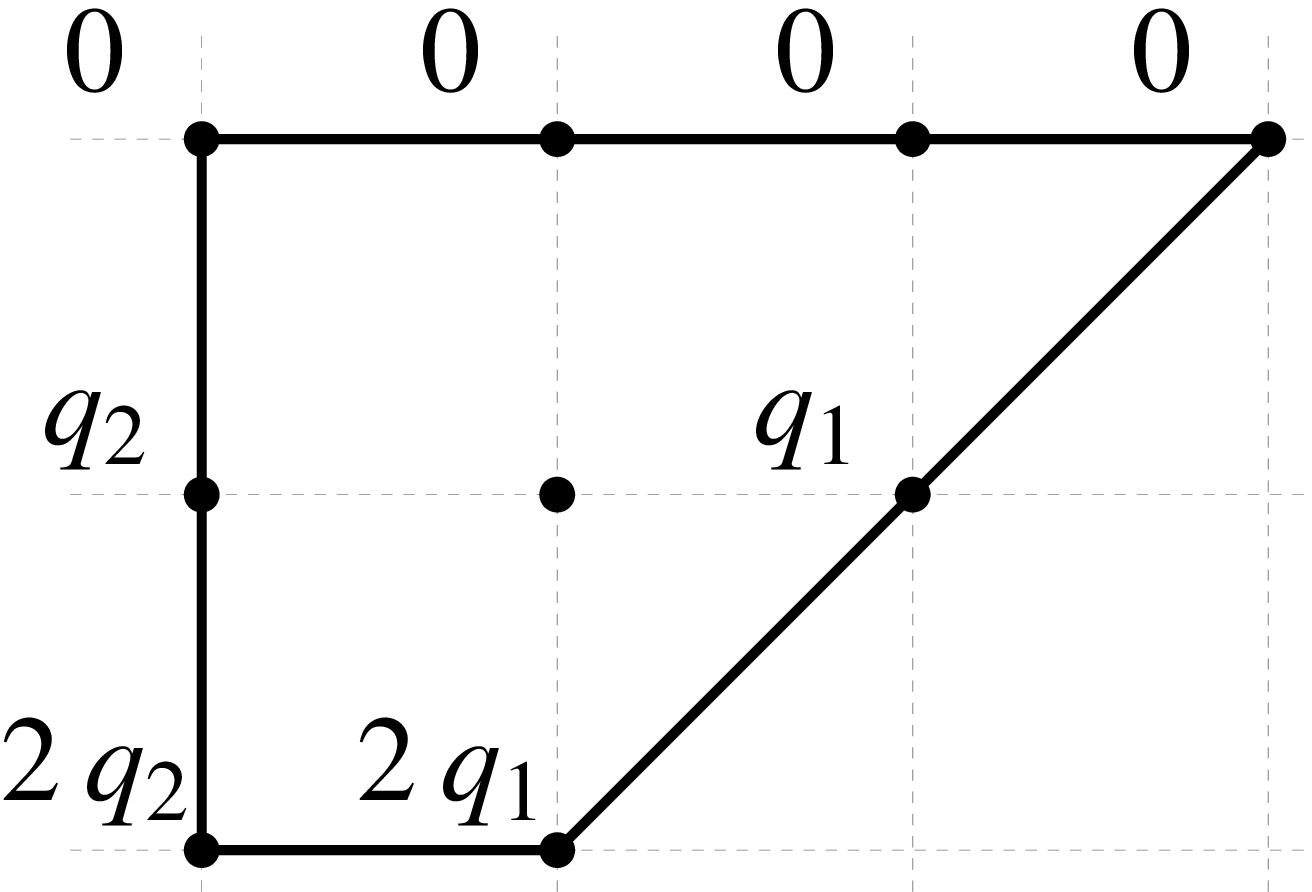}\label{fig: torDiag mod 14}}\,
\subfigure[\small $W\CP^2_{[2,2,4]}$ ]{\includegraphics[height=2.5cm]{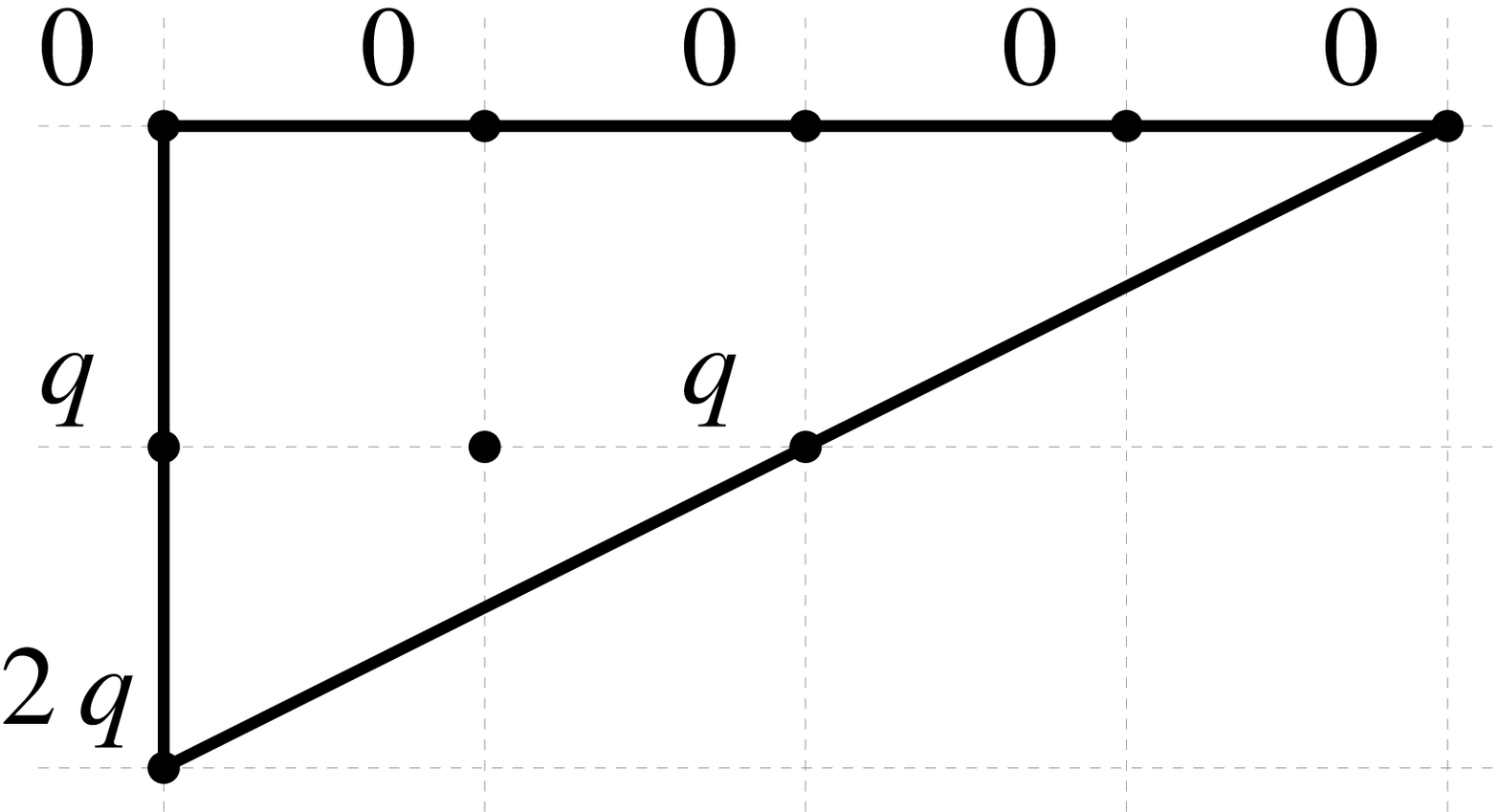}\label{fig: torDiag mod 15}}\,
\subfigure[\small $W\CP^2_{[3,3,3]}$ ]{\includegraphics[height=3.2cm]{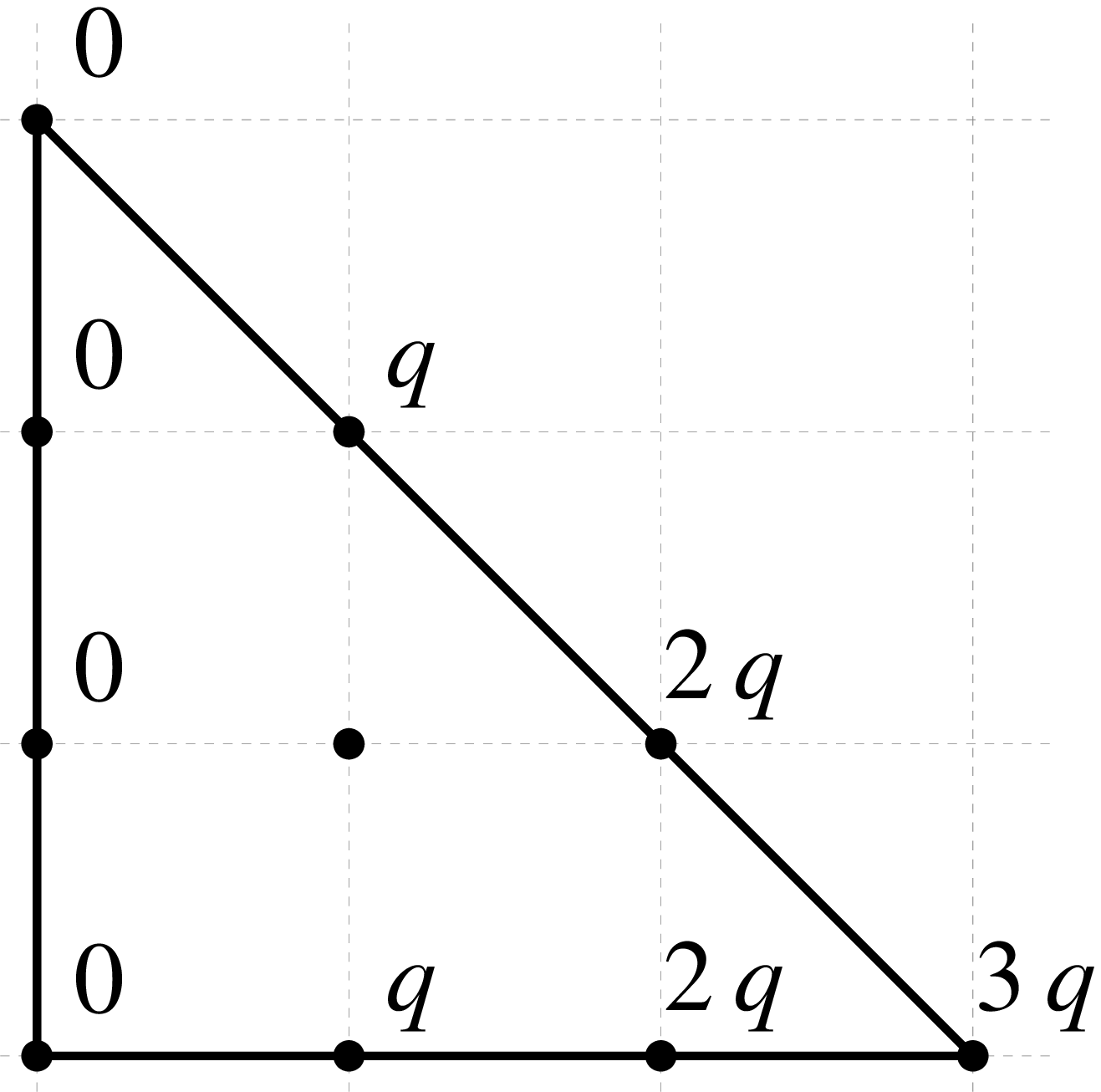}\label{fig: torDiag mod 16}}
\caption{\small Toric diagrams for the 16 two-dimensional toric Fano varieties. The labels over the external points give the height of the point in the corresponding 3d toric diagrams which we shall introduce later on.}\label{Figs: the 16 toric diagrams}
\end{center}
\end{figure}
%%%%%%%%%%%%%%%%%%%%%%%%%%%%%%%%%%%
To any K\"ahler chamber (\ref{def KC in term of internal pm})  we associate a tilting collection of line bundles (\ref{notation cE}). However, the objects $\mathsf{L_i}$ are \emph{not} the physical fractional branes. The fractional branes are --- loosely speaking --- branes wrapping compact cycles. For the algorithm of section \ref{section: IIA to quiver} below, we will need to know their brane charges, a piece of information which is not so easy to extract in general. In this paper we restrict ourselves to the case in which $Y$ is a complex cone over a  \emph{toric Fano variety}. There are 16 of them, with toric diagrams that are reflexive polygons, having a single strictly internal point \cite{Cox:book}.

Adding the interior point to the toric fan corresponds to a partial resolution
\be
\pi: \tilde{Y}\rightarrow Y \, , \qquad \tilde{Y}\cong \cO_{B_4}(K)\, ,
\ee
with $\cO_{B_4}(K)$ the canonical bundle over the Fano variety $B_4$. The toric fan of $B_4$ is obtained from the toric diagram $\Gamma$ by taking the strictly interior point as the origin and drawing a toric vector to every strictly external point. When there are internal-external points, $B_4$ has orbifold singularities, which can be resolved by adding the corresponding toric vectors to the fan. The resulting smooth manifold is denoted $\tilde{B_4}$.

For all practical purposes we can restrict our attentions to the B-branes on $B_4$, which naturally lift to B-branes  of $\tilde{Y}$.  Let us denote by $\{p_r^{EI}, p_0\}$ a K\"ahler chamber, where the perfect matchings $p_r^{EI}$ correspond to external-internal points and $p_0$ correspond to the strictly internal point. To any such chamber we associate a smooth space $\tilde{Y}$ and a pseudo-Beilinson quiver $\cQ_{B_4}$ obtained from $\cQ$ by removing all arrows appearing in $p_0$. We then associate a collection of line bundles over $B_4$ to this K\"ahler chamber by the $\Psi_{\theta}$-map,  for any $\bm{\theta}$ inside the chamber. We write the divisors (\ref{def psi map with div}) in term of the toric divisors $D_k$ for the external points only (using $D_0 = -\sum_k D_k$ in homology, with $D_0\cong B_4$ the divisor from the internal point), and these naturally restrict to divisors $D_k$ on $B_4$. The resulting collection
\be\label{cE coll on B4}
\cE = \{ \mathsf{E}_1, \cdots, \mathsf{E}_G\}
\ee
is also a tilting collection for $D(\text{Coh}\, B_4)$, the B-brane category on $B_4$ \cite{Carqueville:2009xu}.%
\footnote{See Proposition C.1. of that paper; we thank the authors of \cite{Carqueville:2009xu} for bringing our attention to their result. (Remark that it can also be checked explicitly (using the \emph{SAGE} \cite{sage} package) that in all the examples we studied $H^{n}(\tilde{B_4}, \mathsf{E}_j\otimes \mathsf{E}_j^*)=0$ for $n>0$. We thank Noppadol Mekareeya for helping us with that computation.)}

The fractional branes we are looking for are related to the line bundles (\ref{cE coll on B4}), but they are somewhat more complicated objects in $D(\text{Coh}\, B_4)$. A generic object in the B-brane category is a chain complex of sheaves,
\be
\label{generic element of D(B4) 00}
\mathsf{E}\;=\; \cdots \rightarrow E_{(-2)} \rightarrow E_{(-1)} \rightarrow E_{(0)} \rightarrow E_{(1)} \rightarrow \cdots \;.
\ee
In this work we will only be concerned with the \emph{charges} of the branes, so that we can ignore the subtleties of the derived category formalism. The charge of a B-brane (\ref{generic element of D(B4) 00}) is a K-theory class \cite{Minasian:1997mm}, but for our purpose it suffices to define the charge as the Chern character
\be\label{Chern character of B brane}
ch(\mathsf{E})=\sum_n (-1)^n \,ch(E_{(n)}) \;,
\ee
where $ch(E_{(n)})$ are the Chern characters of the individual sheaves. To discuss the most general brane charges we consider the full resolution $\pi : \tilde{B}_4\rightarrow B_4$, which is what the B-model probes. Let us denote
\be\label{def 2 cycles and cI}
\cC_{\alpha} \in H^2(\tilde{B}_4, \bZ)\, , \qquad \alpha= 1, \cdots, m\, , \qquad \quad\quad \cC_{\alpha}\cdot\cC_{\beta}= \cI_{\alpha\beta}\, .
\ee
a primitive basis of 2-cycles, with $\cI$ the intersection matrix. We will generally choose the 2-cycles $\cC_{\alpha}$ to coincide with $m$ of the $m+2$ toric divisors: $\cC_{\alpha}\cong D_{\alpha}$ with $\{D_{\alpha}\} \subset \{D_k\}$.

There are $G\equiv m+2$ charges for the compactly supported branes (branes wrapping $B_4$, $\cC_{\alpha}$ or a point), and we denote the Chern character (\ref{Chern character of B brane}) of a generic B-brane $\mathsf{E}$ by the covector
\be\label{ch as vector}
\bm{Q}_{\text{branes}}(\mathsf{E}) \,=\, ch(\mathsf{E})\,=\, (rk(\mathsf{E}), c_1(\mathsf{E}), ch_2(\mathsf{E}))\, .
\ee
A natural pairing on the space of charges is given by the Euler character%
\footnote{We refer to \cite{Aspinwall:2004jr} for a physical introduction to $\text{Ext}$ groups. For the purpose of this paper one could as well take (\ref{pairing of E, F}) as the primary definition.}
\be
\chi(\mathsf{E}_i, \mathsf{E}_j)\,\equiv\, \sum_q (-1)^q \, \mathrm{dim}\, \mathrm{Ext}^q(\mathsf{E}_i, \mathsf{E}_j) \;,
\ee
which can be computed by the Riemann-Roch theorem:
\bea\label{pairing of E, F}
\chi(\mathsf{E}_i, \mathsf{E}_j) \,=\, \int_{\tilde{B}_4} \,ch(\mathsf{E}_i^*)\, ch(\mathsf{E}_j)\, Td(\tilde{B}_4) \;.
\eea
We can conveniently rewrite this in matrix notation,
\be
\label{definition of X}
\chi(\mathsf{E}_i,\mathsf{E}_j) = ch(\mathsf{E}_i) \, \mathbf{X}_{\tilde{B}_4} \, ch(\mathsf{E}_j)^T \;, \qquad \quad \text{with}\qquad
\mathbf{X}_{\tilde{B}_4} = \mat{ 1 & \frac12 c_1 & 1 \\ - \frac12 c_1 & - \cI & 0 \\ 1 & 0 & 0 }\, ,
\ee
where $\cI$ is defined in (\ref{def 2 cycles and cI}) and $c_1= c_1(\tilde{B}_4)$; thus the matrix $\mathbf{X}_{\tilde{B}_4}$ is intrinsic to the geometry we consider.
Given the tilting collection (\ref{cE coll on B4}) on $B_4$, we define
\be\label{definition S}
S_{ij} \equiv \chi(\mathsf{E}_i, \mathsf{E}_j)= \mathrm{dim}\, \text{Hom}(\mathsf{E}_i, \mathsf{E}_j)  \;.
\ee
The matrix element $S_{ij}$ is the number of independent open paths from $j$ to $i$ in the pseudo-Beilinson quiver $\cQ_{B_4}$. It is equal to the number of global sections of the line bundle $\mathsf{E}_j\otimes\mathsf{E}_i^*$,
\be
S_{ij}=  \mathrm{dim}\, H^0(\tilde{B}_4, \mathsf{E}_j\otimes\mathsf{E}_i^*)\, ,
\ee
which is easily computed by toric methods. The fractional branes form a collection
\be\label{cEvee coll on B4}
\cE^{\vee} = \{ \mathsf{E}_1^{\vee}, \cdots, \mathsf{E}_G^{\vee}\}\,
\ee
which is dual to (\ref{cE coll on B4}) with respect to the Euler character:
\be\label{E Ev Chi-dual}
\chi(\mathsf{E}_i, \mathsf{E}_j^{\vee})= \delta_{ij}\, .
\ee
This also implies $\chi(\mathsf{E}^{\vee}_j, \mathsf{E}_i^{\vee}) = S^{-1}_{ij}$.
We introduce the two $G\times G$ matrices
\be\label{definition Q Qv}
Q = \mat{ ch(\mathsf{E}_1) \\ \vdots \\ ch(\mathsf{E}_G) } \;,\qquad\qquad
    Q^\vee = \mat{ ch(\mathsf{E}_1^\vee) \\ \vdots \\ ch(\mathsf{E}_G^\vee) } \;.
\ee
whose rows are the charges of the B-branes in $\cE$ and $\cE^{\vee}$, respectively.
In term of these charge matrices we can rewrite (\ref{E Ev Chi-dual}) and (\ref{definition S}) in a compact way:
\be
Q^{\vee T }= (\mathbf{X}_{\tilde{B}_4})^{-1}\, Q^{-1} \, , \qquad \qquad S= Q\, \mathbf{X}_{\tilde{B}_4}\,  Q^T\, .
\ee
The antisymmetric adjacency matrix $A$ of the complete quiver $\cQ$ can be found from $S$, according to
\be\label{adj matrix from S}
A\equiv S^{-1 T}- S^{-1}\, .
\ee
Remark that we did not need to give a concrete definition of the fractional branes $\mathsf{E}^{\vee}$ in order to extract their brane charges $Q^{\vee}$. Instead we will just conjecture that there exist objects in $D(\text{Coh}\,B_4)$ with the right properties.%
\footnote{If $\cE^{\vee}$ is a complete strongly exceptional collection (corresponding to $S$ an upper-triangular matrix), the fractional branes can be obtained from the line bundles $\mathsf{E}_i$ by explicit mutations \cite{Herzog:2004qw}.}
We call the matrix $Q^{\vee}$ a \emph{dictionary}. It allows to translate between the brane charge basis (\ref{ch as vector}) and the fractional brane basis, namely the quiver ranks:
\be\label{rel N and Qbrane}
\bm{Q}_{\text{branes}} = \bm{N} \, Q^{\vee}\, .
\ee

\paragraph{Example.} Consider the total resolution $\tilde{PdP_2}$ of $PdP_2$, whose toric fan looks like the triangulation $T_{\Gamma}^{(1)}$ from Figure \ref{fig: PdP2tianguluations}. We take our homology basis (\ref{def 2 cycles and cI}) to be
\be\label{choice H2 basis PdP2}
\{\cC_{\alpha}\} \cong \{D_3, D_4, D_{(1,0)}  \}\, , \qquad \quad \cI = \mat{-1& 1& 1\\ 1& -1& 0\\ 1& 0& -2 }\, .
\ee
where the divisors of $\tilde{PdP_2}$ are inherited from the divisors of $\tilde{Y}$. We have the homology relations $D_1= 2D_3+D_4+D_{(1,0)}$ and $D_2=D_3+D_4$; we also have the relation $D_{(0,0)}=-4D_3-3D_4-2D_{(0,0)}$ in $\tilde{Y}$. The tilting collection (\ref{cE coll on B4}) for $\tilde{PdP_2}$ is directly obtained from (\ref{examples of BM collections PdP2}) by using these homology relations. Let us focus on the first K\"ahler chamber for definiteness. We have:
\be\nn
\cE_{\{ p_5, p_7 \} } = \{\cO(0,0,0), \, \cO(-2,-2,-1) , \, \cO(-3,-3,-1), \, \cO(-3,-2,-1), \, \cO(-1,-1,0)  \}
\ee
in the basis (\ref{choice H2 basis PdP2}).
The charge matrix for these line bundles is
\be
Q=\mat{1 & 0 & 0& 0&  0\\1& -2&-2&-1&1 \\1& -3&-3&-1&2\\1&-3 &-2&-1&\frac32\\ 1& -1&-1&0&0}\, ,
\ee
and we have
\be
\mathbf{X}_{\tilde{B}_4}=\mat{1&\frac12 &\frac12 &0&1\\-\frac12 &1&-1&-1&0\\\frac12 &-1&1&0&0\\0 &-1&0&2&0\\1&0&0&0&0}\, , \qquad S=\mat{1&0&0&0&0\\4&1&0&0&2\\6&2&1&1&4\\5&1&0&1&3\\2&0&0&0&1}\, ,
\ee
from which we can compute the dictionary $Q^{\vee}= Q^{-1T}\mathbf{X}_{\tilde{B}_4}^{-1T}$. The actual dictionary we will use will contain some additional half-integer shift of the charges due to the Freed-Witten anomaly \cite{Freed:1999vc}.

\subsection{K\"ahler moduli space and quiver locus}
Most B-branes on $\tilde{Y}$ are not physical D-branes, because they do not lift to half-BPS objects in physical string theory. The D-brane spectrum at any given value of the K\"ahler moduli is given by the spectrum of $\Pi$-stable B-branes \cite{Douglas:2000ah}. In the regime of interest to us, $\Pi$-stability reduces to $\theta$-stability of quiver representations \cite{Douglas:2000ah, Aspinwall:2004mb}.

To any compactly supported D-brane $\mathsf{E}^{\vee}$ on $Y$ one associates a complex central charge $Z(\mathsf{E}^{\vee})$, which determines which half of the supersymmetry of the closed string background it preserves. Two BPS D-branes $\mathsf{E}^{\vee}_1$ and $\mathsf{E}^{\vee}_2$ are mutually BPS if and only if  their central charges are aligned,
\be
\arg(Z(\mathsf{E}^{\vee}_1)) =\arg(Z(\mathsf{E}^{\vee}_2))\, .
\ee
The central charge depends on the closed string K\"ahler moduli. Our space $\tilde{Y}$ has $m$ complexified Kahler moduli, corresponding to
\be
t_{\alpha}\,\equiv \, \int_{\cC_{\alpha}}  (B+ i\, J)  \, \equiv \, b_{\alpha} + i\, \chi_{\alpha}\, ,
\ee
where the 2-cycles $\cC_{\alpha}$, $\alpha= 1, \cdots, m$, were defined in (\ref{def 2 cycles and cI}). The \emph{quiver locus} $\cM_{\cQ}$ is the locus in Kahler moduli space $\cM_K$ where the $G=m+2$ fractional branes (\ref{cEvee coll on B4}) are mutually BPS \cite{Aspinwall:2004vm}. It has codimension $m+1$ in $\cM_K$. Since the fractional branes are mutually BPS at the tip of the cone, we expect the quiver locus to be located at
\be
\chi_{\alpha} = 0\, .
\ee
There is one more constraint on the $m$ B-field periods $b_{\alpha}$, which we denote by $\chi_0=0$. Let us define a vector
\be\label{def big chi}
\bm{\chi} \equiv (\chi_0, \chi_{\alpha}, 0)\, ,
\ee
corresponding to the directions transverse to $\cM_{\cQ}$ in $\cM_K$. As long as the central charges $Z(\mathsf{E}^{\vee}_i)$ are almost aligned, the quiver $\cQ$ is a good description of D-brane physics. The closed string modes (\ref{def big chi}) couple to the fractional branes as FI parameters $\bm{\xi}= (\xi_i)$  \cite{Douglas:1996sw}, and stability of D-branes corresponds to $\theta$-stability of quiver representations. One can show that the FI parameters are related to the K\"ahler moduli (\ref{def big chi}) by the dictionary $Q^{\vee}$:
\be\label{dico xi to chi 0}
\bm{\xi} \, = \, Q^{\vee}\, \bm{\chi}\, .
\ee
The $m-1$ directions along the quiver locus correspond to marginal gauge couplings for the so-called ``non-anomalous'' fractional branes, which are D4-branes wrapped on 2-cycles in $\tilde{B}_4$ dual to non-compact divisors in $\tilde{Y}$. In the quiver regime, the central charges of the fractional branes are given by
\be
Z(\mathsf{E}^{\vee}_i) \approx \frac{1}{g_i^2} + i \xi_i\, .
\ee
Whenever the inverse squared gauge coupling of a non-anomalous fractional brane becomes negative, one should change basis of fractional branes, leading to a Seiberg dual quiver (which might be a self-similar quiver, or a new quiver in a different ``toric phase'').

\subsection{Dictionaries, monodromies and Freed-Witten anomaly}\label{subsec: dic, monodrom, FW}
A quiver with its FI parameters $\bm{\xi}$ describes the fractional branes near a particular point in the quiver locus, probing $\cM_K$ in directions transverse to $\cM_{\cQ}$. We have seen how the quiver can probe numerous K\"ahler chambers $\{p_r^{EI}, p_0\}$, related to the multiplicities of points in the toric diagram. Going from one K\"ahler chamber to the next one in FI parameter space corresponds to crossing a wall of marginal stability in $\cM_K$ (a codimension 1 wall where two fractional branes become mutually BPS).

To each of these K\"ahler chamber we associated a dictionary $Q^{\vee}$. However, there are some ambiguities to this procedure, which corresponds physically to the fact that the very concept of brane charge is not well defined on K\"ahler moduli space, but only on its universal cover (its Teichm\"uller space).

The central charge of a generic D-brane $\mathsf{E}$ with brane charge (\ref{ch as vector}) can be written
\be\label{exact central charge Z}
Z(\mathsf{E}) \, = \bm{Q}_{\text{brane}}\cdot \bm{\Pi} =  \, r(\mathsf{E})\, \Pi_6 + c_1^{\alpha}(\mathsf{E}^{\vee}) \,\Pi_{4, \alpha} + ch_2(\mathsf{E}^{\vee}) \,\Pi_2\, ,
\ee
where $\Pi_4$, $\Pi_{2, \alpha}$ and $\Pi_0$ are so-called \emph{periods} associated to the states with Chern characters $(1,0,0)$, $(0,\delta^{\alpha}_{\beta},0)$ and $(0,0,1)$, respectively.  These periods are not single valued functions on $\cM_{K}$, but instead suffer from \textit{monodromies} around various singular loci. On the other hand the central charge of any physical state is invariant under such monodromies. Denoting by $M$ the monodromy matrix acting on the periods, there is a corresponding action on the brane charges:
\be
\bm{\Pi}\rightarrow M \bm{\Pi}\, , \qquad \quad \bm{Q}_{\text{brane}}\rightarrow \bm{Q}_{\text{brane}} M^{-1}\, .
\ee
The best understood monodromies are the monodromies around the large volume limit in $\cM_K$. At large volume, up to instanton corrections to $\Pi_6$, the periods are given by
\bea
&\Pi_6 & &\simeq \quad \frac{1}{2}\int_{\tilde{B}_4} (B+iJ)^2 \, + \frac{1}{24}\chi(\tilde{B}_4) \\
&\Pi_{4, \alpha}  & &=   \quad t_{\alpha}    \\
&\Pi_2 & &=  \quad 1\;.
\eea
The large volume monodromies (LVM) corresponds to the shift of the B-field by some cohomomology class in $H^2(\tilde{B}_4)$,
\be
B\rightarrow B + \sum_{\alpha} m_{\alpha} [D_\alpha]\, ,
\ee
for $m_{\alpha}\in \bZ$. Its action on  the periods $\bm{\Pi}= (\Pi_6,\Pi_{4, \alpha} , \Pi_2)$ is given by
\be
M_{\infty}(\bm{m} ) = \mat{1 &\bm{m} &\; \frac12 \bm{m}\cI\bm{m} \\ 0 & \bm{1} & \cI\bm{m}  \\0  &0 &1}\,
\ee
with $\bm{m}= (m_{\alpha})$ and $\cI$ the intersection matrix.
In the algorithm described above to find $Q^{\vee}$, a different choice of ``first node'' to construct the line bundles (\ref{cE coll on B4}) corresponds to such a large volume monodromy. Fixing the order of the nodes once and for all, we are still free to perform any LVM, generating new dictionaries which are valid for different values of the background B-field. A generic dictionary takes the form
\be\label{generic Qv from LVM}
Q^{\vee}[KC, \bm{m}] = Q^{\vee}[KC, \bm{0}] \, M_{\infty}(\bm{m})^{-1},
\ee
with $KC\cong \{p_r^{EI}, p_0\}$ a K\"ahler chamber. We have to fix some convention on what we call $Q^{\vee}[KC, \bm{0}]$. In every example studied in this paper we choose convenient conventions which are kept implicit. Instead we state the actual dictionary $Q^\vee$ whenever needed.

When the manifold $B_4$ is not spin, we cannot wrap a D6 over it without turning as well $F=1/2$ units of worldvolume flux%
\footnote{More precisely we have to introduce some (ill-defined) line bundle $\sqrt{L}$ such that $\sqrt{TB_4} \otimes \sqrt{L}$ is a well defined spin$^c$ bundle. Heuristically $F= \frac12$ is the first Chern class of $\sqrt{L}$.} \cite{Freed:1999vc}, and this results in half-integral shift of the Page charges --- see Appendix \ref{subsec: D6s and FW}. To take this Freed-Witten anomaly into account in our dictionaries, we need to shift them according to
\be
Q^{\vee}[KC, \bm{0}]\; \rightarrow\;  Q^{\vee}[KC, \bm{0}] \, M_{\infty}(\frac{\bm{s}}{2})\, ,
\ee
where the Freed-Witten anomaly parameters $\bm{s}= (s_{\alpha})$ are defined in (\ref{wv flux for FW and def of s}).

There are further monodromies apart from the LVM's, generically called quantum monodromies because they arise in the region of $\cM_K$ which suffers large $\alpha'$ corrections. We will have little to say about them, but we should keep in mind that there can exist more dictionaries than those of (\ref{generic Qv from LVM}). In section \ref{sec:torsion flux for F0} we will encounter an instance of such extra monodromies which are related to Seiberg duality of quivers \cite{Herzog:2004qw, Closset:2012eq}.

%%%%%%%%%%%%%%%%%%%%%%%%%%%%%%%%%%%%%%%%%%%%%%%%%%%%%%
\section{Reduction of M-theory on $\bR^{1,2}\times CY_4$ to type IIA}\label{sec: Mtheory to IIA}

In this section we describe the first step of the stringy derivation of the theories on M2 branes probing the toric $CY_4$ cone over $\Ypq(B_4)$, with $B_4$ a 2-complex-dimensional toric Fano surface, namely the reduction of the M-theory background $\bR^{1,2}\times C(\Ypq(B_4))$ to a type IIA background.
In the next section this type IIA background will be used to deduce the field theory on D2-brane probes: its low energy limit is the M2-brane theory we are interested in. We will streamline the presentation, referring the reader to \cite{Benini:2011cma} for more background on this kind of computations.
We start by discussing aspects of the reduction for general toric $CY_4$ cones, before applying it to the $C(\Ypq(B_4))$ geometries that are the focus of this paper.

\subsection{Generalities}

Generalising the idea of \cite{Aganagic:2009zk}, the approach of \cite{Benini:2011cma} was to Kaluza-Klein reduce the M-theory background along a wisely chosen $U(1)_M$ circle action in the $CY_4$, so that the resulting type IIA background is a fibration%
\footnote{More correctly this is a foliation rather than a fibration because, as we will explain, the topology of the ``fibre'' can degenerate and vary as we move along the base.}
of a resolved $CY_3$ $\tilde{Y}$ over a real line $\bR$ parametrised by $r_0$, with RR 2-form fluxes and (anti-)D6-branes \cite{Benini:2011cma}. The $CY_3$ fibre over $r_0$
is the K\"ahler quotient $\tilde{Y}(r_0)\equiv CY_4\kq_{r_0} U(1)_M$.
The K\"ahler volumes $\chi_\alpha (r_0) = \int_{\cC_\alpha} J $ of its 2-cycles $\cC_\alpha$ are piecewise linear functions of $r_0$.
The curvature of the $U(1)_M$ fibration yields RR $F_2$ field strength, whose fluxes $\int_{\cC_\alpha} F_2 = \chi'_\alpha(r_0)$ are piecewise constant in $r_0$. Discontinuities in these fluxes are due to (anti-)D6-branes, which descend from fixed point loci of the circle action (KK monopoles).

One can use toric methods to derive these data, working with the abelian GLSMs whose vacuum moduli spaces are the toric $CY_4$ and $CY_3$ cones. It will be crucial to demand that the GLSM for $CY_4\kq_{r_0} U(1)_M$ is in a geometric phase for any $r_0$. This gives meaning to the geometric description of the previous paragraph and is a necessary condition for the corresponding 2-brane theory to be a toric quiver gauge theory. We expect that it will also be sufficient if the quiver gauge theory is extended to include (anti)fundamental matter coming from massless open string modes stretching between D2-branes and D6-branes along \emph{noncompact} divisors, along the lines of \cite{Benini:2009qs,Jafferis:2009th}.

The previous restriction is most easily stated in terms of toric diagrams.
We start with the toric diagram of the $CY_4$, a 3d convex lattice polytope $\Gamma_3$, and associate by convention the $U(1)_M$ symmetry with the vertical direction in $\bZ^3$.
The 2d toric diagram $\Gamma$ of $Y=CY_4\kq U(1)_M$ is obtained by vertical projection of $\Gamma_3$. Each pair of adjacent vertically aligned points belonging to $\Gamma_3$ leads to a (anti-)D6-brane embedded along the toric divisor of the $CY_3$ associated to the point in $\Gamma$ that the pair of points projects to. Finally, the RR 2-form is determined by the vertical coordinates of the points of the 3d toric diagram, as we will see explicitly in section \ref{subsec:F2_and_D6}.

\begin{figure}[t]
\begin{center}
\subfigure[\small From the bottom.]{\includegraphics[height=5cm]{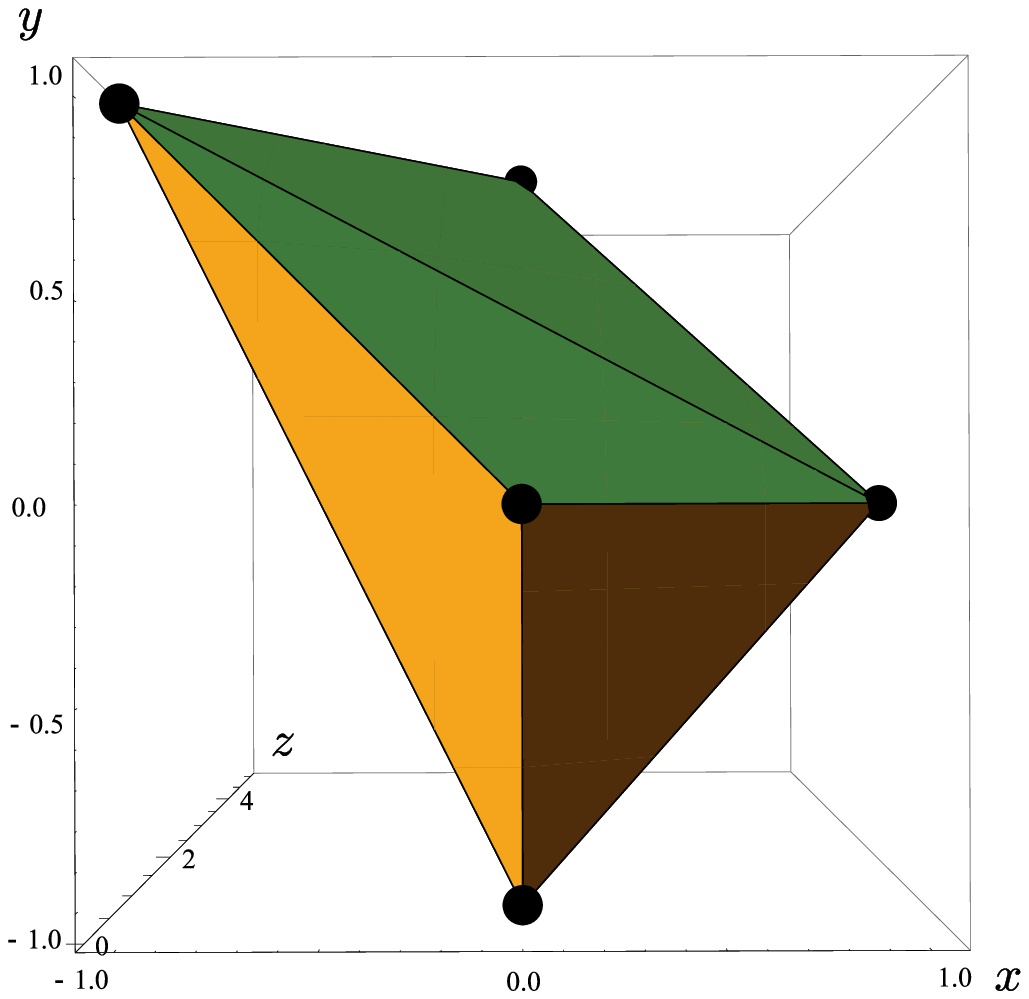}\label{fig: 3d toric bottom}}\qquad
\subfigure[\small From the top.]{\includegraphics[height=5cm]{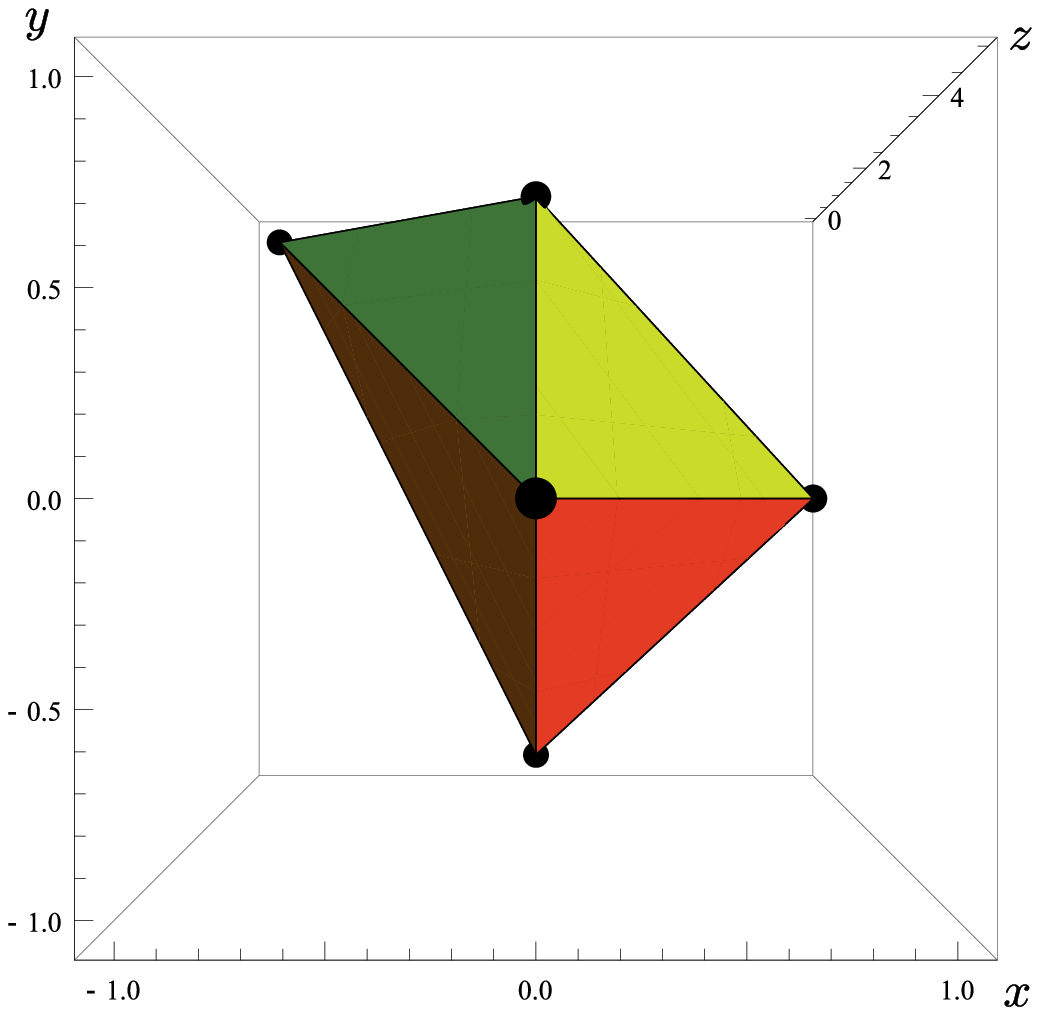}\label{fig: 3d toric top}}\qquad
\caption{\small The 3d toric diagram for the cone over $Y^{p,\,q_1,\,q_2}(dP_1)$, with $(q_1,q_2,p)=(2,1,5)$. Looking at the toric diagram from the top or the bottom, we see the two different triangulations of the 2d toric diagram of $C_\bC(dP_1)$.}\label{Figs: 3d toric diag and triangulations}
\end{center}
\end{figure}

We can initially focus on $CY_4$ metric cones: then the D6-branes wrap toric divisors in the conical $\tilde{Y}(0)=Y$ and the GLSM for $\tilde{Y}(r_0)=CY_4\kq_{r_0}U(1)_M$ is specified by two rays in its FI parameter space, for $r_0<0$ and $r_0>0$ respectively. Up to an  overall dilatation controlled by $|r_0|$, we thus have two toric crepant (partial or complete) resolutions $\tilde{Y}_-$ and $\tilde{Y}_+$  of the singular $CY_3$ $Y=\tilde{Y}(0)$ which lies over $r_0=0$.
It turns out that the triangulated 2d toric diagram $\Gamma_\mp$ for $\tilde{Y}_\mp$ can be found by looking at the 3d toric diagram of the $CY_4$, viewed as a solid lattice polytope, from below ($-$) and above ($+$) respectively --- see Figure \ref{Figs: 3d toric diag and triangulations} for an example.
%%%%%%%%%%%%%%%%%%%%%%%%%%%%%%%%%%%%%%%%%%%%%%
%%%%%%%%%%%%%%%%%%%%%%%%%%%%%%%%%%%%%%%%%%%%%%
Consider $r_0<0$: the $CY_3$ GLSM is in a geometric phase for all $r_0<0$ iff $\Gamma_-$ contains all its lattice points, possibly joined by segments determining a partial or complete (simplicial) triangulation. Similarly for $r_0>0$. Phrased in terms of the original $CY_4$, the necessary and sufficient condition is that the intersection of its toric diagram $\Gamma_3$ with the set of lattice vertical lines $\bZ^2\times \bR$ is a union of vertical segments joining points in $\bZ^3$. Each such vertical segment has then integer length $h\geq 0$ and leads to $h$ D6 on the associated toric divisor in $\tilde{Y}(0)$.
Remark that since $\Gamma_3$ is a convex polytope, if the $CY_3$ $Y$ obtained upon reduction has compact toric divisors then a sufficient number of D6-branes must wrap each of those compact divisors to ensure that the $CY_3$ GLSM is in a geometric phase for any $r_0$.

In the following we will restrict our attention to families of toric $CY_4$ such that the toric $CY_3$ $\tilde{Y}$ obtained upon KK reduction to IIA, in addition to being in a geometric phase for all values of $r_0$, has a single compact toric divisor and no D6-branes along noncompact toric divisors. The former requirement restricts us to $CY_3$ $Y$ which are total spaces of canonical bundles over one of the 16 toric Fano surfaces introduced in section \ref{subsec:_Fano}; the latter guarantees that the resulting 2-brane theory is a quiver gauge theory with only bifundamental matter. The extension of the stringy derivation to M2-brane theories for the entire class of toric $CY_4$ geometries that reduce to geometric $CY_3$ fibrations with both compact and noncompact D6-branes and RR $F_2$ fluxes is an interesting challenge that we leave for future investigation.

\subsection{Cones over toric $\Ypq(B_4)$ Sasaki-Einstein 7-folds}\label{subsec:Ypq(B4)}

Let us then consider a toric $CY_4$ cone whose 3d toric diagram contains a single vertical line of $p\geq 1$ points $s_a=(0,0,a)$, $a=1, \dots, p$, lying above the point $s_0=(0,0,0)$ in the toric diagram. In addition there are $m+2$ points $t_i=(x_i,y_i,z_i)$, $i=1,2,\dots,m+2$, having different horizontal coordinates $(x_i,y_i)$, one for each generator of the toric fan of one of the 16 2-complex-dimensional toric Fano surfaces $B_4$ of Fig. \ref{Figs: the 16 toric diagrams}, with $H_2(B_4,\bZ)=\bZ^m$.
$\{s_0,\dots,s_p,t_1,\dots,t_{m+2}\}$ are all the lattice points belonging to the polytope $\Gamma_3$.
The vertical projection of $\Gamma_3$ is the toric diagram $\Gamma$ of the $CY_3$ $Y$, the total space of $\cO_{B_4}(K)$, which is also the toric fan of $B_4$.

To ensure that the $CY_3$ GLSM is always in a geometric phase and that there are no D6 along noncompact divisors, we need
\begin{inparaenum}[\itshape a\upshape)]
\item $\Gamma_3$ to have no vertical faces nor edges and \label{a}
\item  $s_0$ and $s_p$ to be external points.
\end{inparaenum}
\emph{a}) requires that each external point $v_i=(x_i,y_i)\in \Gamma$ lifts to a single point $t_i=(x_i,y_i,z_i)\in \Gamma_3$. For singular $B_4$ it also requires that the lattice points belonging to an external edge of $\Gamma$ lift to aligned lattice points in $\Gamma_3$, otherwise there would be a vertical face. This imposes some equalities between the heights $z_i$ of lattice points belonging to the same external edge of $\Gamma$. On the other hand \emph{b}) imposes a number of inequalities among linear combinations of the $z_i$ and $0$ or $p$, as we will see explicitly in examples.

Using the subgroup of the $SL(3,\bZ)$ acting on $\Gamma_3$ that leaves $\Gamma$ and $s_0$ invariant, we are free to set the $z_i$ of two lattice points $v_1=(x_1,y_1)$ and $v_2=(x_2,y_2)$ of $\Gamma$ to $0$, as long as $v_1$, $v_2$ and $(0,0)$ form a triangle of minimal area. We then call $z_{i+2}=q_i$ the vertical coordinates of the remaining $m$ external lattice points in $\Gamma$. We made one such choice for each of the 16 toric Fano surfaces: in Fig. \ref{Figs: the 16 toric diagrams} we show next to each point $v_i$ the assignment of $q_i$, $i=1,\dots,m$ which fulfils requirement \emph{a}). We will stick to these conventions in section \ref{sec:_examples}. For each toric Fano $B_4$, we thus have a $(v-1)$-parameter family of toric $CY_4$ cones labelled by $p$ and the set of independent $\bm{q}=(q_i)$, where $v$ is the number of vertices of $\Gamma$.
Generalising the nomenclature of \cite{Gauntlett:2004hh}, we call $Y^{p,\,\bm{q}}(B_4)$ the $SE_7$ base of the $CY_4$ cone.

We still need to impose that $s_0$ and $s_p$ are external in $\Gamma_3$. We will assume that this has been done in the remainder of this section, postponing  to section $\ref{sec:_examples}$ the list of inequalities that this imposes on the geometric parameters $(p,\bm{q})$ for each $B_4$.
\begin{figure}[t]
\begin{center}
\subfigure[\small  $\hat{T}_\Gamma^{1}$.]{\includegraphics[height=3.9cm]{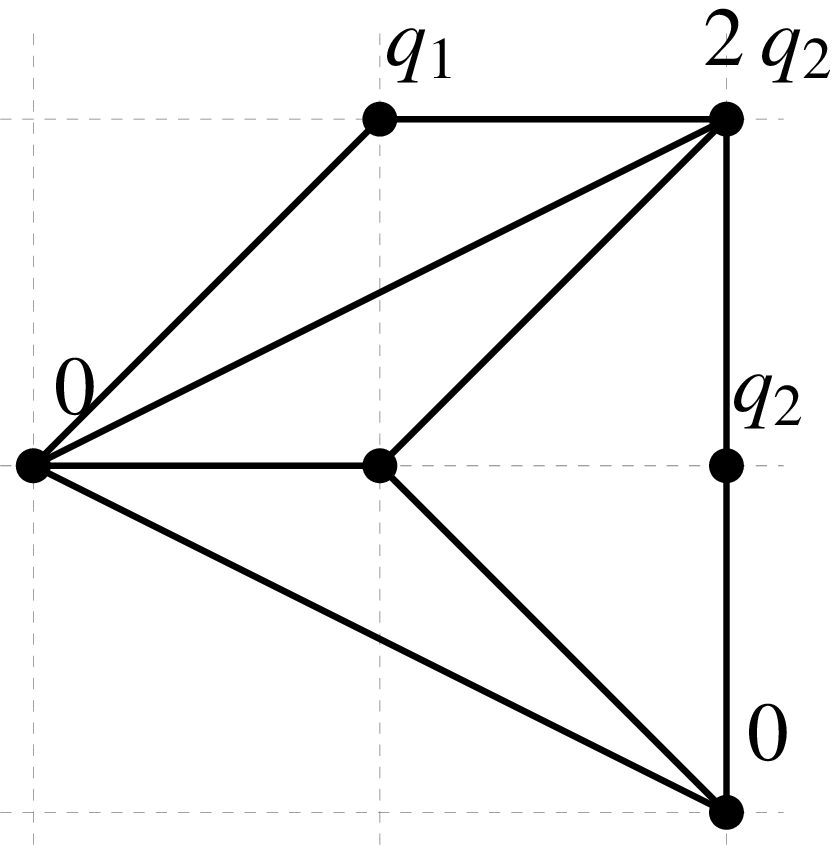}\label{fig:PdP2_part_triang_flopped}}\qquad
\subfigure[\small  $\hat{T}_\Gamma^2$.]{\includegraphics[height=3.9cm]{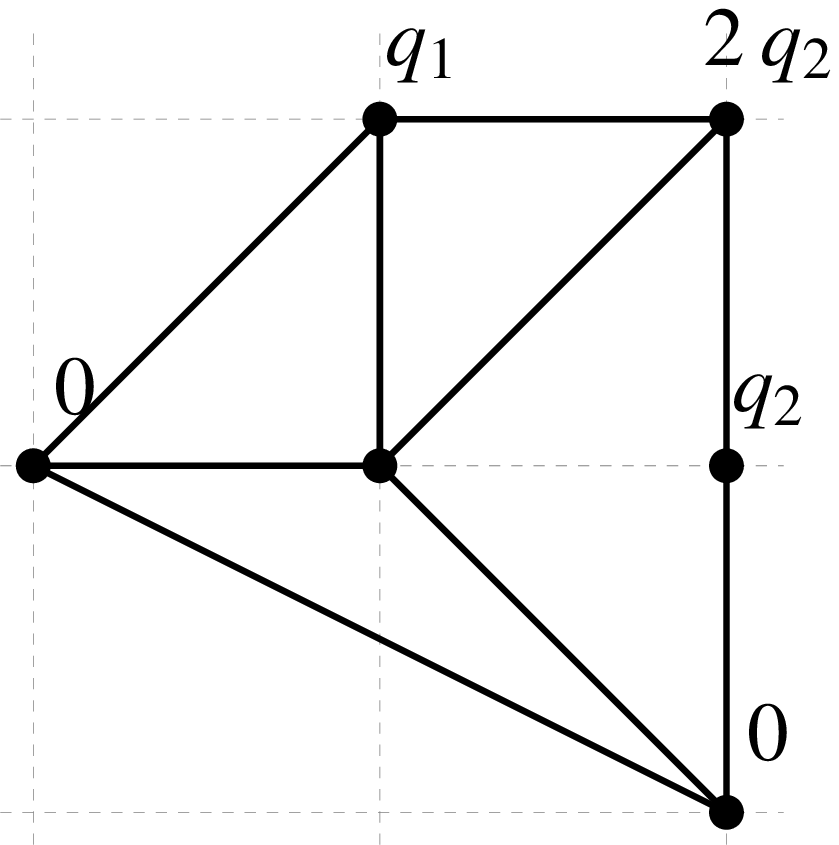}\label{fig:PdP2_part_triang}}\qquad
\subfigure[\small $\hat{T}_\Gamma^{3}$.]{\includegraphics[height=3.9cm]{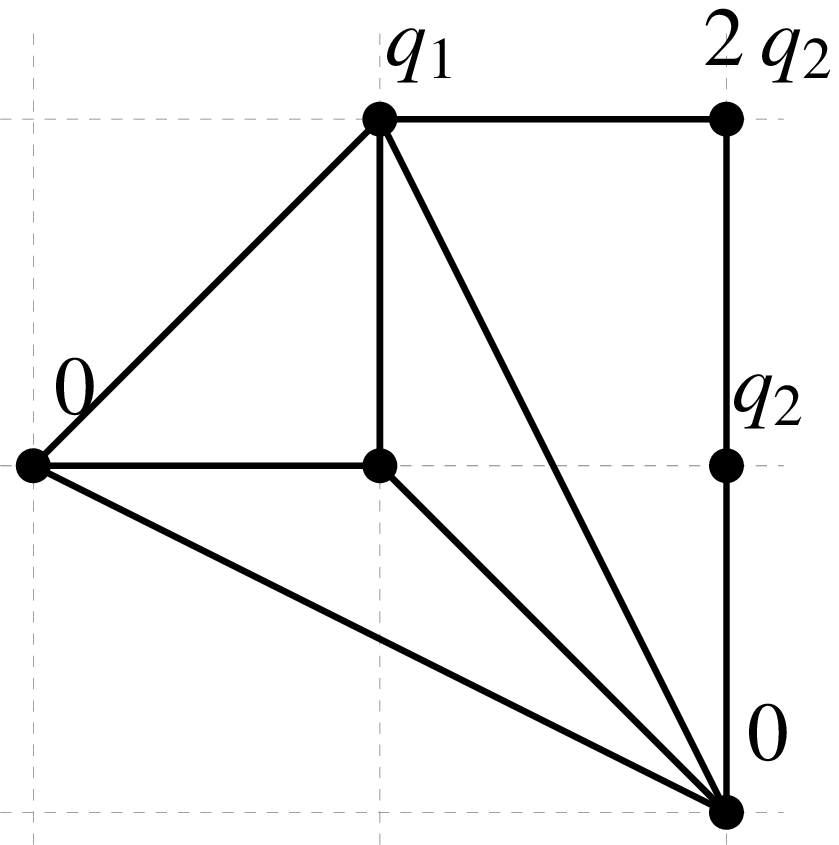}\label{fig:PdP2_part_triang_flopped_like}}\qquad
\caption{\small The three partial triangulations $\hat{T}_\Gamma^{i}$, $i=1,2,3$, of the toric diagram of $PdP_2$ relevant for the reduction of $C(Y^{p,\,q_1,\,q_2}(PdP_2))$.}\label{Figs:PdP2_partial_triangulations}
\end{center}
\end{figure}
Here we consider the illustrative example of $C(Y^{p,\,q_1,\,q_2}(PdP_2))$, which has a non-isolated singularity.  The relevant partial triangulations of the toric diagram are shown in Fig. \ref{Figs:PdP2_partial_triangulations}.
We wrote the $z_i$ coordinates of external points, fulfilling condition \emph{a}) above. We still need to impose that $s_0$ and $s_p$ are external points in $\Gamma_3$: that restricts $q_1$ and $q_2$ by the inequalities
\begin{equation}\label{geom_ineq_PdP2_new_conv}
0 \leq  \frac{q_1}{3}, \frac{q_2}{2}\leq p\;.
\end{equation}
When one of the inequalities is saturated, $s_0$ or $s_p$ lies inside an external face, otherwise they are strictly external. We can further refine the class of geometries into subclasses, depending on which partial triangulation $T_{\Gamma_\pm}$ of $\Gamma_\pm$ results from the reduction:
%\begin{align}
%\tilde{Y}_-:&\;\;T_{\Gamma_-}=\begin{cases}
%\hat{T}_\Gamma^{1} &\mathrm{if}\quad q_1>2q_2\\
%\hat{T}_\Gamma^{2} &\mathrm{if}\quad q_2<q_1<2q_2\\
%\hat{T}_\Gamma^{3} &\mathrm{if}\quad q_1<q_2
%\end{cases}\\
%\tilde{Y}_+:&\;\;T_{\Gamma_+}=\begin{cases}
%\hat{T}_\Gamma^{1} &\mathrm{if}\quad q_1<2q_2-p\\
%\hat{T}_\Gamma^{2} &\mathrm{if}\quad  2q_2-p<q_1<q_2+p \\
%\hat{T}_\Gamma^{3} &\mathrm{if}\quad q_1>q_2+p
%\end{cases}
%\end{align}
\begin{equation}
T_{\Gamma_-}=\begin{cases}
\hat{T}_\Gamma^{1} &\mathrm{if}\quad q_1>2q_2\\
\hat{T}_\Gamma^{2} &\mathrm{if}\quad q_2<q_1<2q_2\\
\hat{T}_\Gamma^{3} &\mathrm{if}\quad q_1<q_2
\end{cases}\;,
\quad
T_{\Gamma_+}=\begin{cases}
\hat{T}_\Gamma^{1} &\mathrm{if}\quad q_1<2q_2-p\\
\hat{T}_\Gamma^{2} &\mathrm{if}\quad  2q_2-p<q_1<q_2+p \\
\hat{T}_\Gamma^{3} &\mathrm{if}\quad q_1>q_2+p
\end{cases}\;.
\end{equation}
When one of the above inequalities becomes an equality the triangulation of $\Gamma$ has fewer edges.
The choices for $T_{\Gamma_-}$ and $T_{\Gamma_+}$ are interchanged by the $\bZ_2$ symmetry $(q_1,q_2)\mapsto (3p-q_1,2p-q_2)$ which identifies $C(Y^{p,\,q_1,\,q_2}(PdP_2))\cong C(Y^{p,\,3p-q_1,\,2p-q_2}(PdP_2))$ and sends $r_0\mapsto -r_0$.
In total there are $7$ possibilities
\begin{equation}
(T_{\Gamma_-},T_{\Gamma_+})=
\begin{cases}
(\hat{T}_\Gamma^{1},\hat{T}_\Gamma^{2}) &\mathrm{if}\quad  2q_2<q_1<q_2+p\\
(\hat{T}_\Gamma^{1},\hat{T}_\Gamma^{3}) &\mathrm{if}\quad  q_1>q_2+p\,,\,2q_2\\
(\hat{T}_\Gamma^{2},\hat{T}_\Gamma^{3}) &\mathrm{if}\quad  q_2+p<q_1<2q_2\\
(\hat{T}_\Gamma^{2},\hat{T}_\Gamma^{2}) &\mathrm{if}\quad  q_2\,,\,2q_2-p<q_1<q_2+p\,,\,2q_2\\
(\hat{T}_\Gamma^{3},\hat{T}_\Gamma^{2}) &\mathrm{if}\quad  2q_2-p<q_1<q_2\\
(\hat{T}_\Gamma^{3},\hat{T}_\Gamma^{1}) &\mathrm{if}\quad  q_1<2q_2-p\,,\,q_2\\
(\hat{T}_\Gamma^{2},\hat{T}_\Gamma^{1}) &\mathrm{if}\quad  q_2<q_1<2q_2-p
\end{cases}
\end{equation}
the last $3$ of which are equivalent to the first $3$ under the $\bZ_2$ identification above.

\subsection{GLSM for the $CY_3$ fibres in type IIA}\label{subsec: CY3 from CY4}
The toric $CY_4$, including all its toric crepant resolutions, can be realised as the moduli space of a supersymmetric abelian gauged linear sigma model for specific choices of the associated FI parameters.
In our examples the GLSM for the $CY_4$ can be written as follows (excluding the last row, which appears for future reference):
\be
\label{general_GLSM_CY4-CY3}
\begin{array}{c|cccccccccc|c}
CY_4 &  t_1 & \cdots & t_{m+2}  & s_0 & s_1 & s_2 & \cdots & s_{p-2} & s_{p-1} & s_p  & \text{FI} \\ \hline
& Q^\alpha_1 &  \cdots& Q^\alpha_{m+2} &  Q^\alpha_{(s_0)} & Q^\alpha_{(s_1)}& 0 & \cdots & 0 & 0 & 0 & \xi^c_\alpha \\
& 0 & \cdots & 0 & 1 & -2 & 1 & \cdots & 0 & 0 & 0 & \xi_2  \\
& 0 & \cdots & 0 & 0 & 1 & -2 & \cdots  & 0 & 0 & 0  & \xi_3  \\
& \vdots & \ddots & \vdots & \vdots & \vdots & \vdots & \ddots  & \vdots & \vdots & \vdots & \vdots\\
& 0 & \cdots& 0 & 0 & 0 & 0 & \cdots & -2 & 1 & 0 & \xi_{p-1}  \\
& 0 & \cdots& 0 & 0 & 0 & 0 & \cdots & 1 & -2 & 1 & \xi_{p}  \\
\hline
U(1)_M & 0 & \cdots& 0 & 1 & -1 & 0 &\cdots & 0  & 0 & 0  & r_0
\end{array}
\ee
The first row denotes the fields of the GLSM, one for each point in the 3d toric diagram. The second row with $\alpha = 1,\dots,m$ denotes their charges under a $U(1)^m$ subgroup. The subset $\{t_1,\dots,t_{m+2}\}$ with $U(1)^m$ charges $Q^\alpha_i$ describes the compact Fano surface $B_4$. The following $p-1$ lines describe the GLSM for a $\bC^2/\bZ_2$ singularity, fibred over $B_4$. The charges $Q^\alpha_{(s_0)}$ and $Q^\alpha_{(s_1)}$ determine how the $\bC^2/\bZ_p$ fibre is twisted over the base $B_4$.
The last column lists FI parameters of the GLSM which control resolutions of the geometry. $\xi_{2,\dots,p}$ have to be non-negative to keep the GLSM of the $CY_4$ in a geometric phase. Similar inequalities involve linear combinations of $\xi^c_\alpha$.

The last row in \eqref{general_GLSM_CY4-CY3} specifies our choice of $U(1)_M$ symmetry acting on the M-theory circle, visualised as the vertical direction in the 3d toric diagram of $CY_4$ \cite{Benini:2009qs}.
Including the last row in \eqref{general_GLSM_CY4-CY3}
yields a GLSM for the K\"ahler quotient $\tilde{Y} (r_0)=CY_4 \kq_{r_0} U(1)_M$. %Its 2d toric diagram is the vertical projection of the 3d toric diagram of the $CY_4$.
The type IIA geometry obtained by KK reduction along the $U(1)_M$ circle  involves a fibration of this $\tilde{Y}(r_0)$ over the real line parametrised by the moment map $r_0$ \cite{Aganagic:2009zk}.
To obtain the precise form of the fibration, we define
\be
\label{def_zeta_in_t_of_xi}
\zeta_0 = -\infty \;,\qquad \zeta_1 = 0 \;,\qquad \zeta_a= \sum_{b=2}^a \xi_b \qquad (a=2,\dots,p) \;, \qquad \zeta_{p+1}= +\infty \;
\ee
and rewrite the GLSM for $CY_4\kq_{r_0} U(1)_M$ in (\ref{general_GLSM_CY4-CY3}), including the last line, as
\be
\label{GLSM_CY3_redundant}
\begin{array}{c|cccccccccc|c}
CY_3 & t_1 & \cdots & t_{m+2} & s_0 & s_1 & s_2 & \cdots & s_{p-2} & s_{p-1} & s_p  & \text{FI} \\
\hline
& Q^\alpha_1 & \cdots & Q^\alpha_{m+2} &  Q^\alpha_{(s_0)} & Q^\alpha_{(s_1)}& 0 & \cdots & 0 &  0 & 0    & \xi^c_\alpha \\
& 0 & \cdots& 0 & 1 & -1  & 0 & \cdots & 0 & 0 & 0 & r_0-\zeta_1 \\
& 0 & \cdots& 0 & 0 & 1 &  -1 & \cdots & 0 & 0 & 0 & r_0-\zeta_2 \\
& 0 & \cdots& 0 & 0 & 0 & 1 & \cdots & 0 & 0 & 0 & r_0-\zeta_3 \\
& \vdots & \ddots & \vdots & \vdots & \vdots & \vdots & \ddots & \vdots & \vdots  & \vdots & \vdots  \\
& 0 & \cdots& 0 & 0 & 0 & 0 & \cdots & 1 & -1 & 0 & r_0-\zeta_{p-1}\\
& 0 & \cdots& 0 & 0 & 0 & 0 & \cdots & 0 & 1 & -1  & r_0-\zeta_{p}\\
\end{array}
\ee
This is a redundant description of $\tilde{Y}(r_0)$:%
\footnote{This redundancy parallels the one encountered in the GLSM for perfect matching variables of section \ref{subsec: toric quiver and resol}. Here it is due to the K\"ahler quotient from the $CY_4$ to the $CY_3$ geometry, rather than from the master space $\cZ$ to the $CY_3$ mesonic moduli space of the abelian toric quiver gauge theory.}
all but one of the $s_{a=0,\dots,p}$ can be eliminated in favour of a remaining unconstrained variable, which depends on the value of $r_0$ as
\be
\label{s_coordinate}
t_0 = s_a  \quad \mathrm{if} \quad \zeta_a \leq r_0 \leq \zeta_{a+1} \;.
\ee
We can rewrite the $\tilde{Y}(r_0)$ GLSM in its minimal form
\be
\label{generic_GLSM_of_CY3}
\begin{array}{c|cccc|c}
CY_3 &  t_1 & \cdots & t_{m+2} & t_0 & \text{FI}\\
\hline
\cC_\alpha & Q^\alpha_1 &  \cdots& Q^\alpha_{m+2} &   Q^\alpha_0 & \chi_\alpha(r_0)
\end{array}
\ee
where
\be\label{def Q0 in term of CY4}
Q^\alpha_0 \equiv Q^\alpha_{(s_0)} + Q^\alpha_{(s_1)}=-\sum_{i=1}^{m+2} Q^\alpha_i = -\int_{\cC_\alpha} c_1(B_4)\;.
\ee
The FI parameters are
\be
\label{resol_param_CY3}
\chi_\alpha(r_0) = \xi^c_\alpha - \Big( \sum_{i=1}^{m+2} z_i Q_i^\alpha \Big) \, (r_0 - \zeta_1) - Q^\alpha_0 \sum_{b=1}^p (r_0 - \zeta_b) \, \Theta(r_0-\zeta_b)\;,
\ee
with $\Theta(x)$ the Heaviside step function. We used the relation
\be\label{relation Qs1 to Qalpha}
-Q^\alpha_{(s_1)} = \sum_{i=1}^{m+2} z_i Q^\alpha_i
\ee
which follows from the toric diagram.
Abusing notation we have identified the $m$ $U(1)$ gauge groups of the GLSM with holomorphic 2-cycles $\cC_\alpha$.

The FI parameters $\chi_\alpha(r_0)$ of the minimal GLSM, which are the K\"ahler parameters in the fibred $\tilde{Y}(r_0)$ if the GLSM is in a geometric phase (as we impose) and the $\cC_\alpha$ are effective curves in the Mori cone, are continuous piecewise linear functions of $r_0$ with first derivatives jumping by $-Q^\alpha_0$ at $r_0 = \zeta_{a=1,\dots,p}$, where the unconstrained coordinate jumps from $s_{a-1}$ to $s_a$. This is due to the presence of an anti-D6-brane wrapping the exceptional toric divisor $D_0$ in $\tilde{Y}(\zeta_a)$.%
\footnote{See \cite{Benini:2011cma} for the explanation of why the object which is mutually BPS with the D2-brane along the quiver locus is a $\overline{\mathrm{D6}}$ rather than a D6.}
If the $CY_4$ is conical, that is $\zeta_a = \xi^c_\alpha = 0$ for all $a$ and $\alpha$, the resolution parameters of $\tilde{Y}(r_0)$ are simply
\be
\label{resol_param_CY3_conical}
\chi_\alpha(r_0) =  \Big[ - \Big( \sum_{i=1}^{m+2} z_i Q^\alpha_i \Big) - p\, Q^\alpha_0 \, \Theta(r_0) \Big] \, r_0  \;.
\ee

\subsection{RR 2-form flux and D6-branes}\label{subsec:F2_and_D6}

Generically the M-theory circle fibration is nontrivial, so its curvature gives a nonvanishing RR 2-form field strength in type IIA. By the same arguments as in \cite{Benini:2011cma}, its cohomology class is
\be
[F_2] = - \sum_{i=1}^{m+2} z_i [D_i] -  \sum_{a=1}^p \, a\,  [D_{(s_a)}] \;,
\ee
where $[D_i]$ is the cohomology class Poincar\'e dual of the toric divisor $D_i = \{t_i = 0\}$.
This expression is still in terms of the redundant GLSM for the $CY_3$. Using the reduced GLSM in (\ref{generic_GLSM_of_CY3}) which minimally describes the resolved $CY_3$ geometry we find
\be
\label{F2 on CY3}
[F_2](r_0) = -\sum_{i=1}^{m+2} z_i [D_i] - [D_0] \sum_{a=1}^p \Theta(r_0 - \zeta_a) \;.
\ee
The flux $[F_2]$ jumps by $-[D_0]$ at $r_0 = \zeta_{a = 1,\dots,p}$: the discontinuity is due to a magnetic source for $F_2$, a $\overline{\mathrm{D6}}$-brane wrapping the toric divisor $D_0$ in $\tilde{Y}(\zeta_a)$. The $\bC^2/\bZ_p$ K\"ahler parameters $\xi_a$ are the separations in the $r_0$ direction between $p$ $\overline{\mathrm{D6}}$-branes wrapping $D_0$. When the $CY_4$ is conical, that is $\xi_\alpha^c=\xi_a=0$, the type IIA background has $p$ coincident $\overline{\mathrm{D6}}$-branes wrapping the collapsed divisor $D_0$ in $\tilde{Y}(0)=Y$.

The fluxes of $F_2$ through the holomorphic 2-cycles $\cC_\alpha$ of $\tilde{Y}(r_0)$ are
\be
\label{relation_F2_flux-Kaehler_par}
\int_{\cC_{\alpha}} F_2 (r_0) = -\sum\nolimits_i z_i Q^\alpha_i - Q^\alpha_0 \sum_{a=1}^p \Theta(r_0 - \zeta_a) = \chi'_\alpha(r_0) \;,
\ee
where recall that $\sum_i z_i Q^\alpha_i = - Q^\alpha_{(s_1)}$.
The equality between 2-form fluxes and derivatives of K\"ahler parameters is a consequence of supersymmetry. In the conical case
\be
\label{F2_fluxes_CY3_conical}
\int_{\cC_{\alpha}} F_2(r_0) = -\sum\nolimits_i z_i Q^\alpha_i - p \, Q^\alpha_0 \, \Theta(r_0) \;.
\ee

\subsection{Torsion $G_4$ and generic type IIA background}
The Sasaki-Einstein seven-manifold $Y^{p, \bm{q}}(B_4)$ has a rather interesting fourth cohomology group which is finite:
\be
H^4(Y^{p, \bm{q}}(B_4), \bZ) = \Gamma\, ,
\ee
with $\Gamma$ given in (\ref{generic sol for Gamma}) in the Appendix. In M-theory, we can turn on discrete torsion $G_4$ flux for any element of $\Gamma$. Equivalently we can wrap an M5-brane on the Poincar\'e dual 3-cycle, known as ``fractional M2-brane'' \cite{Aharony:2008gk}. This gives rise to a large family of $AdS_4\times Y^{p, \bm{q}}(B_4)$ backgrounds which are otherwise undistinguishable. Moving in $\Gamma$ corresponds to changing the ranks and Chern-Simons levels of the dual Chern-Simons quiver theory \cite{Aharony:2008gk, Benini:2011cma}.
In Appendix \ref{sec: Appendix topology} we collect some results on the topology of $Y^{p, \bm{q}}(B_4)$ and of the 6-manifold $M_6$ (an $S^2$ bundle over $B_4$) which appears in the type IIA limit of the $AdS_4$/CFT$_3$ correspondence, in the type IIA background $AdS_4\times_w M_6$.

Torsion $G_4$ flux corresponds to quantised D4-brane Page charges in type IIA, which results in a dynamical quantization of the background
B-field \cite{Aharony:2009fc, Benini:2011cma}. In the conical setup discussed here, we also have explicit D4-branes wrapped on vanishing 2-cycles of $Y$ at $r_0=0$. This type IIA background is characterised by background fluxes measured at $r_0<0$ and $r_0>0$, which we denote
\be\label{def of flux vectors}
\bm{Q}_{\text{flux}, \pm} \,  \equiv\, \left( -Q_{4;\,  \pm} \,|\, Q_{6;\, \alpha \pm} \,|\, 0 \right)\, ,
\ee
and by the explicit D-brane sources: the $p$ $\overline{\mathrm{D6}}$-branes discussed in section \ref{subsec:F2_and_D6}, the D4-branes discussed in the Appendix (section \ref{subsec: flux and Dbranes on CM6}), and of course the $N$ D2-branes corresponding to the $N$ M2-branes we seek to describe. We denote the corresponding brane charges by
\be\label{def of source vector}
\bm{Q}_\text{source} \, \equiv\, \left( Q_{D6} \,|\, (\cI^{-1})^{\alpha\beta} Q_{D4;\, \beta} \,|\, Q_{D2} \right)\, .
\ee
The sources account for the jump of the background fluxes between $r_0<0$ and $r_0>0$, according to
\be\label{rels between sources and fluxes}
 Q_{4;\,  +} - Q_{4;\,  -}  = \cI_{0\alpha} (\cI^{-1})^{\alpha\beta} Q_{D4;\, \beta}\, ,\qquad\qquad
 Q_{6;\, \alpha +}-Q_{6;\, \alpha -}  = \cI_{0\alpha} Q_{D6}    \, ,
\ee
with $\cI_{\alpha\beta}$ defined in (\ref{def 2 cycles and cI}) and $\cI_{0\alpha}=Q_0^{\alpha}$ the intersection number between $B_4$ and $\cC_{\alpha}$ in $Y$.
In Appendix \ref{sec: Appendix topology} we give the explicit form of $\bm{Q}_{\text{flux}, \pm}$, $\bm{Q}_\text{source}$ for the $Y^{p, \bm{q}}(B_4)$ geometry with torsion flux $(n_0, n_{\alpha})\in \Gamma$ --- see equation (\ref{fluxes and sources generic torsion}). In the following we will mainly focus on the torsionless case $(n_0, n_{\alpha})=(0,0)$, in which case (\ref{fluxes and sources generic torsion}) reduces to
\bea\label{fluxes and sources generic torsionless}
 & \bm{Q}_{\text{flux}, -}= (\frac12 s_{\alpha} q_{\beta}  \cI^{\alpha\beta} \, | \, -\cI_{\alpha\beta}  q^{\beta}\, | \, 0  )  \, ,\\
 &\bm{Q}_\text{source} =    ( -p \,|\, -\frac12 s_{\alpha} p    \,|\,  N-\frac18 s_{\alpha}s_{\beta}\cI^{\alpha\beta} p )        \, .
\eea
We will discuss the case of torsion flux in a simple example in section \ref{sec:torsion flux for F0}.

%%%%%%%%%%%%%%%%%%%%%%%%%%%%%%%%%%%%%%%%%%%%%%%%%%%%%%
\section{From type IIA to CS quiver gauge theories and back}\label{section: IIA to quiver}
Once  the type IIA background is understood, the technology of section \ref{section: toric quivers and dictionaries} can be used to derive the low energy worldvolume theory of M2-branes probing the $CY_4$.
The type IIA background obtained from a conical $CY_4$ in M-theory is foliated by $CY_3$ leaves $\tilde{Y}$ along $\bR\cong \{r_0\}$. Algebraically we can characterise it by a choice of \emph{two} partial resolutions of $Y$, $\tilde{Y}_-$ and $\tilde{Y}_+$ at $r_0<0$ and $r_0>0$ respectively. The fluxes on $\tilde{Y}_{\pm}$ are encoded in the flux vectors (\ref{def of flux vectors}), while the D-branes wrapped on vanishing cycles at $r_0=0$ are encoded in the source vector (\ref{def of source vector}).

\subsection{Translating from IIA background to CS quiver}\label{subsec: IIA to CS quiver}
The IIA background is a resolved toric $CY_3$ $\tilde{Y}(r_0)$ fibred along $\bR \cong \{r_0\}$, as described previously. The fibre $\tilde{Y}(r_0)$ can change to a different partial resolution of $Y$ as we cross $r_0=0$, while at $r_0=0$ we have the singular cone $Y$.%
\footnote{If the $CY_4$ has a non-isolated singularity we can have a singular $Y$ on a half-line as well.}
The K\"ahler parameters of $\tilde{Y}(r_0)$ are given by
\be\label{chi of IIA geom}
\bm{\chi}_{\pm} =\, \begin{cases}  -\bm{Q}_{\text{flux}, -}\, (-r_0)\quad &\text{for}\quad r_0<0   \\
                                \bm{Q}_{\text{flux}, +}\, r_0\quad &\text{for}\quad r_0>0
                            \end{cases}
\ee
where $\bm{\chi}$ was defined in (\ref{def big chi}).
This gives us two distinct spaces $\tilde{Y}_+$ and $\tilde{Y}_-$.
To translate this into the quiver language, we need to consider the toric quiver $\cQ$ describing D-branes on the $CY_3$. The background value of the B-field determines in principle which toric phase to use, corresponding to a particular point in the quiver locus. In practice we do not know the exact central charges of all the fractional branes along the quiver locus, and thus we do not know the location of all the Seiberg duality walls. In this paper we will discuss various toric phases for each geometry; it turns out that all the resulting Chern-Simons quivers are 3d Seiberg dual in the sense of \cite{Benini:2011mf, Closset:2012eq}.

D2-branes in the background (\ref{chi of IIA geom}) correspond to $\theta$-stable quiver representations, with $\bm{\theta}$ depending on the sign of $r_0$. According to (\ref{dico xi to chi 0}), we have
\bea\label{definition theta pm}
&\tilde{Y}(r_0<0)\simeq \tilde{Y}_-\, &:& \, \qquad \bm{\theta}_- = - Q^{\vee}_{-} \bm{Q}_{\text{flux}, -}\, ,\\
&\tilde{Y}(r_0>0)\simeq \tilde{Y}_+\, &:& \, \qquad \bm{\theta}_+ =  Q^{\vee}_{+} \bm{Q}_{\text{flux}, +}\, ,
\eea
with $Q^{\vee}_{\pm}$ the relevant dictionaries. We find the correct dictionaries by scanning explicitly over all the K\"ahler chambers (and over large volume monodromies in each chamber), retaining only those dictionaries $Q^{\vee}_{\pm}$ for which $\bm{\theta}_{\pm}$ as defined in (\ref{definition theta pm}) lies in the corresponding open string K\"ahler chambers. We call such dictionaries $Q^{\vee}_{\pm}$ the \emph{consistent dictionaries} for $\tilde{Y}_{\pm}$.

In general there might be several pairs of consistent dictionaries for a given $ \bm{Q}_{\text{flux}, \pm}$, corresponding either to the fact that the type IIA fluxes sets $\tilde{Y_-}$ and/or $\tilde{Y}_+$ on a K\"ahler wall (in which case the different choice of dictionaries lead to the same CS quiver theory), or else to Seiberg-like dualities among different CS gauge theories. For torsionless backgrounds the former situation always occurs, since the $\overline{\mathrm{D6}}$-brane wrapping $B_4$ is mutually BPS with the D2.

Choosing some consistent dictionaries $Q^{\vee}_{\pm}$, the derivation of the field theory is straightforward. Away from the tip $r_0=0$, the mobile D2-brane is a stable bound state of $G$ fractional D2-branes $\mathsf{E}^{\vee}_i$. The $U(1)$ gauge field on $\mathsf{E}^{\vee}_i$ acquires a Chern-Simons interaction from its Wess-Zumino action%
\footnote{We neglect the gravitational coupling in the Wess-Zumino action, because it does not affect our derivation. See \cite{Closset:2012eq} for some comments on that point.}
\be\label{WZ action and CS levels}
%\int_{\bR^{1,2}\times B_4} e^{F_{3d}}\, ch(\mathsf{E}^{\vee}_i)\wedge  C^{(P)} =
 \int_{\bR^{1,2}}A_{3d}\wedge F_{3d} \int_{ B_4}  ch(\mathsf{E}^{\vee}_i)\wedge  F^{(P)}  \, ,
\ee
due to the background fluxes $F^{(P)}$ encoded in $\bm{Q}_{\text{flux}, \pm}$.
Here $F_{3d}=dA_{3d}$ is the worldvolume flux along the $\bR^{1,2}$ directions and $F^{(P)}=e^B F$ the Page current, where $F$ is the improved gauge invariant RR field strength polyform.
From (\ref{WZ action and CS levels}) we read the Chern-Simons levels
\be
\bm{k}_{\pm} = Q^{\vee}_{\pm} \bm{Q}_{\text{flux}, \pm}\, ,
\ee
for a D2-brane at $r_0>0$ or $r_0<0$. Remark that we have $\bm{\theta}_{\pm}= \pm \bm{k}_{\pm}$. At $r_0=0$, the worldvolume gauge theory acquires the CS levels
\be
\bm{k} = \frac12 (\bm{k}_-\, +\, \bm{k}_+)\, .
\ee
The ranks $\bm{N}$ of the CS quiver theory are related to the explicit sources, which we encoded in $\bm{Q}_{\text{source}}$.
In order to use the dictionaries and read the quiver ranks from the branes, we need to split these D-brane sources to the left and right of $r_0=0$:
\be
\bm{Q}_{\text{source}, -}= \bm{\delta Q}_\text{source}\, , \qquad \qquad
\bm{Q}_{\text{source},  +}= \bm{Q}_\text{source} -\bm{\delta Q}_\text{source}\, ,
\ee
in such a way that the bunches $\bm{Q}_{\text{source}, \pm}$ still lie inside the K\"ahler chambers where $Q^{\vee}_{\pm}$ are respectively valid; since these branes affect the background flux, this is a non-trivial constraint. In practice we take an arbitrary splitting $ \bm{\delta Q}_\text{source}$, and compute
\be
\bm{N}_{\text{trial}} = \bm{Q}_{\text{source},-} (Q^{\vee}_-)^{-1} +\bm{Q}_{\text{source},+} (Q^{\vee}_+)^{-1}\, ,
\ee
which depends on some of the unknowns in the arbitrary splitting $ \bm{\delta Q}_\text{source}$. It only depends on the the so-called \emph{anomalous D-branes}, which wrap cycles dual to compact cycles and therefore source the fluxes $\bm{Q}_{\text{flux}, +}$. The anomalous D-branes are the D6-brane wrapped on $\tilde{B}_4$ and the D4-brane on the dual 2-cycle. In term of quiver representations, the distinction between non-anomalous or anomalous D-brane is whether the corresponding dimension vector $\bm{\beta}$ is or not in the kernel of the antisymmetric adjacency matrix $A$: $A\bm{\beta}=0$ for non-anomalous branes.%
\footnote{ Let us stress that there is nothing anomalous about these ``anomalous'' fractional D2-branes: the terminology is inherited from the related setup with fractional D3-branes in type IIB, where a quiver theory with $A\bm{N}\neq 0$ would have a gauge anomaly and the IIB background a RR tadpole.}
The correct $\bm{N}$ is found by requiring that
\be
A\bm{N}_{\text{trial}}= \bm{k}_+\, -  \, \bm{k}_- \, .
\ee
This algorithm gives us a Chern-Simons quiver gauge theory $(\cQ, \bm{N},\bm{k})$ for any choice of consistent dictionaries $Q^{\vee}_-$, $Q^{\vee}_+$. We will show next that the semi-classical moduli space of $(\cQ, \bm{N},\bm{k})$ reproduces by construction the type IIA geometry we started with.

\subsection{Semi-classical moduli space of CS quiver theories and type IIA geometry}\label{subsec:_semiclassical_IIA}
Three dimensional $\cN=2$ supersymmetric quiver gauge theories have complex scalar fields $X_{ij}$ in bifundamental representations and real scalar fields $\sigma_i$ in adjoint representations, leading to a potentially rich semi-classical moduli space. The classical vacuum equations are%
\footnote{To keep formulae simpler, we rescaled $\sigma\to 2\pi \sigma$ with respect to common conventions.}
\bea
\partial_{X} W  \,& = \,  0\, ,\\
\sigma_i X_{ij}- X_{ij} \sigma_j  \,& = \, 0 \, ,\\
\sum_{X_{ij}} X_{ij}^{\dagger}X_{ij} - \sum_{X_{ji}} X_{ji} X_{ji}^{\dagger} \,& = \, \sigma_i k_i\, ,
\eea
whose general solution could be rather intricate. A general analysis of these classical equations was performed in \cite{Jafferis:2008qz, Martelli:2008si, Hanany:2008cd}, whereas the generalization to the one-loop corrected moduli space was given in \cite{Benini:2011cma}. Let us write the ranks as $N_i =  \tilde{N}+N_i$, with $\tilde{N}= \min(N_i)$. We focus on the \emph{geometric branch}, which we define by setting
\be\label{sigma_geom_branch}
\sigma_i = \text{diag}(\sigma_1, \cdots, \sigma_{\tilde{N}}, 0, \cdots, 0)\, ,\qquad  \forall\, i\, .
\ee
In the case $\tilde{N}=1$, the low energy theory at any fixed $\sigma \neq 0$ is Abelian, with vacuum equations
\be\label{CS VMS for Op}
\partial_{X} W  \, = \,  0\, ,\qquad\qquad
\sum_{X_{ij}} |X_{ij}|^2 - \sum_{X_{ji}} |X_{ji}|^2 \, = \, \sigma k_i^{eff}(\sigma)\, ,
\ee
where the effective CS levels $\bm{k}^{eff}$ are given by \cite{Benini:2011cma}
\be\label{kpm defined 00}
\bm{k}^{eff}(\sigma) = \begin{cases}
 \bm{k}_- \, & \quad \text{if}\quad \sigma <0 \\
  \bm{k}_+ \, & \quad \text{if}\quad \sigma >0 \\
\end{cases}\, , \qquad \quad \mathrm{with}\quad \bm{k}_{\pm} = \bm{k} \pm \frac12 A\, \bm{N}\, ,
\ee
due to one-loop corrections upon integrating out massive chiral multiplets.  Remark that $\bm{k}_{\pm}$ are integers.  The equations (\ref{CS VMS for Op}) lead to the  K\"ahler quotient description of a resolved $CY_3$ cone $\tilde{Y}$, with FI parameters $\sigma \bm{k}^{eff}(\sigma)$. The full geometric branch for $\tilde{N}=1$ is a resolved cone $\tilde{Y}$ fibred on a line $\bR\cong \{\sigma \}$ according to (\ref{CS VMS for Op})-(\ref{kpm defined 00}). We have two distinct partial resolutions  $\tilde{Y}_{\pm}$ depending on the sign of $\sigma$.
Equivalently, we can describe the spaces $\tilde{Y}$ by the GIT quotient (\ref{GIT construction 00}),
or in term of semi-stable quiver representations. The $\theta$-stability parameters are given by the effective Chern-Simons levels according to
\be
\bm{\theta}_{\pm}= \pm \bm{k}_{\pm} \, .
\ee
This identity is what makes $\theta$-stability such a natural tool to study Chern-Simons quivers.
For $\tilde{N}>1$, the geometric branch is the $\tilde{N}$-symmetric product of the above result (due to the residual gauge symmetry permuting the non-zero eigenvalues $\sigma_i$).
Therefore we reproduce the type IIA geometry probed by $\tilde{N}$ mobile D2-branes, with the identification $r_0=\sigma$.
 The parameters $\bm{\theta}_{\pm}=\pm \bm{k}_{\pm}$ of the CS quiver determine which open Kahler chambers we sit in at $\sigma$ positive or negative, and which consistent dictionaries we should use.
The K\"ahler parameters $\bm{\chi}$ of $\tilde{Y}_{\pm}$ are found from $\bm{\theta_{\pm}}$ by inverting the relations (\ref{definition theta pm}).

\subsection{Monopole operators and GIT quotient}
The real scalar $\sigma$ is naturally complexified using the dual photon $\varphi$. Good homomorphic coordinates on the Coulomb branch are provided by  the monopole operators $t \sim  \exp(\frac{2\pi}{g^2}\sigma + i \varphi)$. In the conventions of \cite{Benini:2011cma}, we have
\be
t = T \, , \quad \text{for}\quad \sigma <0\, , \qquad \quad t = \tilde{T} \, , \quad \text{for}\quad \sigma >0\, .
\ee
Denoting $t_-= T$ and $t_+=\tilde{T}$, the bare monopole operators have electric charges $\bm{g}(t_{\pm})= -\bm{\theta}_{\pm}= \mp \bm{k}_{\pm}$, respectively, under the torus $\cG= U(1)^G$. On the other hand, in the GIT construction (\ref{GIT construction 00}) we have
the function $t$ on the trivial line bundle which has charges $-\bm{\theta}$ under $\cG$ \cite{Martelli:2008cm}, and the $\cG_{\bC}(\bm{\theta})$-invariant functions are of the form $f_{n \bm{\theta} } t^n$ for $n$ any non-negative integer, with $f_{n \bm{\theta} }$ an homogenous polynomial in the coordinates $z_a$ of degree $n \bm{\theta}$  under (\ref{action of character theta}). Thus we have
\be\label{Ypm in term of Proj}
\tilde{Y}_{\pm} \cong   \cM(\cQ, \bm{\alpha}; \bm{\theta}_{\pm})_{GIT}  \cong \text{Proj}\;  \bigoplus_{n\geq 0}\, \bC[f_{n\bm{\theta}_{\pm}} t^{n}_{\pm}]\, .
\ee
Note that the rings $ \bC[f_{n\bm{\theta}_{\pm}}  t^{n}_{\pm}]$ are in general not freely generated, despite the short-hand notation: there can be \emph{syzygies}, i.e. relations between the generators which follow from their definition in term of the variables $z_a$, $t_{\pm}$; the rings above are thus obtained by further dividing the free ring of gauge invariants by a syzygy ideal which is left implicit.
The invariant functions $f_{n\bm{\theta}_{\pm}}\,  t^{n}_{\pm}$ with $n>0$ are the gauge invariant diagonal monopole operators discussed at length in \cite{Benini:2011cma}, and the ring $\bC[f_{n\bm{\theta}_{\pm}}\,  t^{n}_{\pm}]$ is graded by the magnetic charge $n$. While the singular cone $Y$ corresponds to the spectrum of the $n=0$ subring,
\be
Y = \text{Spec} \; \bC[f_0]\, ,
\ee
the $\text{Proj}$ construction in (\ref{Ypm in term of Proj}) corresponds to a partial resolution $\pi: \tilde{Y}\rightarrow Y$. The local coordinates on the exceptional locus are basically the monopole operators $f_{\bm{\theta}}  t$.

The construction (\ref{Ypm in term of Proj}) ``projectivises'' the affine variety one would obtain from the spectrum of the free ring $\bC[f_{\bm{\theta}}t]$. To obtain the full $CY_4$ geometric branch of the Chern-Simons quiver one would naively replace $\text{Proj}$ with $\text{Spec}$ in (\ref{Ypm in term of Proj}), but the complete story is more subtle. It was shown in various examples \cite{Gaiotto:2009tk, Benini:2009qs, Jafferis:2009th} and conjectured in general in \cite{Benini:2011cma} that the full geometric branch can be obtained as
\be\label{M2 geometric branch algebraically}
\cM_{\text{M2-branes}} = \text{Spec}\; \frac{\bigoplus_{n\geq 0}\, \bC[f_{n\bm{\theta}_{-}}\,  t^{n}_{-},f_{n\bm{\theta}_{+}} t^{n}_{+}]}{\cI_{QR}}  \, \, ,
\ee
and that for $\tilde{N}=1$ this is a conical CY fourfold. The ideal $\cI_{QR}$ corresponds to so-called \emph{quantum relations} involving the monopoles operators. It could not be determined from first principle so far. In the examples which have been worked out, it was enough to conjecture that $\cI_{QR}$ is generated by any binomial%
\footnote{That the relations are of the form ``binomial$=0$'' is necessary for $\cM_{\text{M2-branes}}$ to be a toric space. In non-toric cases such as in \cite{Gaiotto:2009tk} the relations are not binomial.} of the monopoles $f_{n\bm{\theta}_{\pm}}t_{\pm}^n$ homogeneous under all the global symmetries.%
\footnote{It was pointed out to us by Daniel Gulotta that this conjecture fails in some special cases, where one can write down relations amongst monopoles which are allowed by the symmetries but would ruin the identification (\ref{M2 geometric branch algebraically}). In those special cases we should modify the conjecture accordingly.}

\paragraph{Example: The ABJM theory.} As a simple example, consider the ABJM theory, which is the $CY_3$ quiver for D-branes on the conifold. Take the Abelian theory $U(1)_k \times U(1)_{-k}$; there are four bifundamental fields $A_{\alpha}$, $B_{\beta}$ ($\alpha, \beta=1,2$) from node 1 to 2 and from node 2 to 1, respectively. We have
\be\nn
\cZ \cong \bC^4 \cong \{w_1, \cdots, w_4\}\, , \quad \text{with}\quad w_1= A_1B_2, \, w_2= A_2B_1, \, w_3= A_1B_1, \, w_4= A_2B_2\, .
\ee
and (there are no one-loop correction in this non-chiral case)
\be
\bm{\theta}_- = (-k, k)\, , \qquad \bm{\theta}_+ = (k, -k)\,.
\ee
Consider the $k=1$ case for simplicity.
The gauge invariant monopole operators are $B_{\beta}t_-$ and $A_{\alpha}t_+$ at $\sigma<0$ and $\sigma>0$, respectively. Explicitly, the $\text{Proj}$ in (\ref{Ypm in term of Proj}) is obtained by separating the functions $\{w_1, \cdots, w_4\}$ with $n=0$ from the functions $\tilde{w}_l$ with $n>0$, considering $\bC^4\cong \{w\}$ and $\bC^d \cong \{\tilde{w} \}$, taking the zero set of all the relations between the $(w, \tilde{w})$ (syzygies) on $\bC^4\times \bC^d\backslash 0$, and further quotienting by the $\bC^{*}$ action given by the $n$-grading. Let us define
\be
a_{\alpha}= A_{\alpha}t_+\, , \qquad b_{\beta} = B_{\beta}t_{-}\,  .
\ee
We have
\bea
\tilde{Y}_+ & = \text{Proj}\; \bC[w_1, \cdots, w_2]\oplus \bC[a_1, a_2]\oplus \cdots \,\\
& \sim \, \text{Spec}\; \bC[w_1, \cdots, w_2]/(w_1w_2-w_3w_4) \times \frac{\bC[a_1, a_2]}{\bC^{*}}\, ,
\eea
describing the resolved conifold $\cO(-1)\oplus \cO(-1)\rightarrow \CP^1$. The small resolution locus is spanned by the monopoles $[a_1, a_2]$, which are the homogenous coordinates of the $\CP^1$ at the tip. Similarly $\tilde{Y}_-$ is the resolved conifold with the flopped $\CP^1$ described by $[b_1, b_2]$. On the other hand the geometric branch (\ref{M2 geometric branch algebraically}) is given by
\be
\cM_{\text{M2-branes}} =  \text{Spec}\; \frac{\bC[w_{\alpha\beta}, a_{\alpha}, b_{\beta}]}{(a_{\alpha}b_{\beta}-w_{\alpha\beta} )} =  \text{Spec}\; \bC[a_{\alpha}, b_{\beta}] \cong \bC^4\,
\ee
with $w_{\alpha\beta}= A_{\alpha}B_{\beta}$, corresponding to the moduli space of a single M2-brane.

%%%%%%%%%%%%%%%%%%%%%%%%%%%%%%%%%%%%%%%%%%%%%%%%%%%%%%%%%%%%

\section{M2-brane theories for backgrounds without torsion $G_4$ flux}\label{sec:_examples}

In this section we use the type IIA stringy derivation method explained in sections \ref{sec: Mtheory to IIA} and \ref{section: IIA to quiver} to find the low energy worldvolume theory on M2-branes probing toric $CY_4$ cones over the toric $\Ypq(B_4)$ Sasaki-Einstein 7-folds introduced in section \ref{subsec:Ypq(B4)}. We will consider all the 16 2d toric Fano varieties $B_4$ of section \ref{subsec:_Fano}, starting with smooth del Pezzo surfaces and then moving to singular pseudo del Pezzo surfaces including some weighted $\CP^2$'s.

\subsection{$dP_0\equiv\CP^2$}\label{subsec:_dP0}
This example was discussed in great depth in \cite{Benini:2011cma}, to which we refer for more details. We review here some of the results as a warm-up before delving into new examples. We slightly changed conventions with respect to \cite{Benini:2011cma} for later convenience.

\begin{figure}[t]
\begin{center}
\subfigure[\small $dP_0$.]{
\includegraphics[height=3.5cm]{ToricDiag01.eps}
\label{fig:toricDiag01}}
\qquad\qquad
\subfigure[\small Brane tiling for $dP_0$.]{
\includegraphics[height=3.5cm]{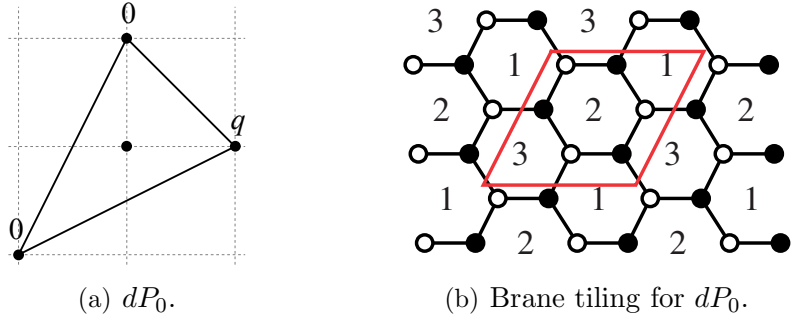}
\label{dP0_tiling_modified}}
\caption{\small Toric diagram and brane tiling for the complex cone over $dP_0$. The toric diagram is a projection on the plane of the 3d toric diagram of $C(Y^{p,q}(\CP^2))$, with the vertical height of the external points indicated on the figure. \label{fig: diagNtiling 01}}
\end{center}
\end{figure}

The toric diagram of the 2-parameter family of toric $CY_4$ cones over $Y^{p,\,q}(\CP^2)$, shown in Fig. \ref{fig:toricDiag01} (or Fig. \ref{fig: torDiag mod 01}), is the convex hull of
\begin{equation}\label{toric_diag_dP0}
(0,0,0), \, (0,0,p),\, (-1,-1,0),\, (0,1,0),\, (1,0,q)\,.
\end{equation}
We are interested in geometric parameters in the range $0\leq q \leq 3p$,
so that all the points \eqref{toric_diag_dP0} are external.
The geometries are identified under the $\bZ_2$ action $q\mapsto 3p-q$.
The metrics for the Sasaki-Einstein bases are explicitly known \cite{Gauntlett:2004hh}.
The minimal GLSM for the $\tilde{Y}(r_0)$ fibre in IIA is
\be
\begin{array}{c|cccc|c}\label{minimal_GLSM_dP0}
CY_3 & p_1 & p_2 & p_3 &  p_0 & \chi(r_0) \\ \hline
\cC & 1 & 1 & 1 & -3 & ( 3p \, \Theta(r_0) - q)\, r_0
\end{array}
\ee
where the K\"ahler volume of the exceptional $\cC=\CP^1\subset \CP^2$ in the $\tilde{Y}(r_0)$ fibre is
\begin{equation}\label{dP0_volume}
\chi(r_0) = \left( 3p \, \Theta(r_0) - q\right)\, r_0\;.
\end{equation}

%\begin{figure}[h]
%\begin{center}
%\includegraphics[width=6cm]{dP0_tiling_modified.eps}
%\caption{\small Brane tiling for $dP_0$.}\label{dP0_tiling_modified}
%\end{center}
%\end{figure}

The toric quiver gauge theory for the complex cone over $dP_0=\CP^2$ is specified by the brane tiling of Fig. \ref{dP0_tiling_modified}, with superpotential
\begin{equation}\label{W_dP0}
W = \eps_{ijh} X_{13}^i X_{32}^j X_{21}^h \;.
\end{equation}
The dimer model has $3$ internal perfect matchings $\{p[4], p[5], p[6]\}$, associated to the open string K\"ahler chambers
\begin{align}
& p[4]:\qquad -\xi_1\geq 0\;,\quad\xi_3\geq 0  \\
& p[5]:\qquad -\xi_3\geq 0\;,\quad \xi_2\geq 0 \\
& p[6]:\qquad -\xi_2\geq 0\;,\quad \xi_1\geq 0
\end{align}
each one with its own dictionary matrix up to large volume monodromies.

A quiver CS theory for the whole class of $C(Y^{p,\,q}(\CP^2))$ geometries can be proposed as follows. Both at $\sigma<0$ and $\sigma>0$ we are on the K\"ahler wall between the maximal dimensional chamber associated to dictionary
\begin{equation}\label{dP0_dict_5}
Q^\vee[dP_0,\,\{p[5]\},\,\{0\}]=
\left(
\begin{array}{ccc}
 -2 & 0 & \frac{3}{4} \\
 1 & \frac{1}{2} & \frac{1}{8} \\
 1 & -\frac{1}{2} & \frac{1}{8}
\end{array}
\right)
\end{equation}
and the one associated to dictionary
\begin{equation}\label{dP0_dict_6}
Q^\vee[dP_0,\,\{p[6]\},\,\{-1\}]=
\left(
\begin{array}{ccc}
 1 & \frac{3}{2} & \frac{9}{8} \\
 1 & \frac{1}{2} & \frac{1}{8} \\
 -2 & -2 & -\frac{1}{4}
\end{array}
\right)
\end{equation}
The wall between these two chambers is given by the cone
\begin{equation}\label{wall_dP0}
\xi_2= 0\;,\;\; \xi_1=-\xi_3\geq 0
\end{equation}
in FI parameter space. Using either one of these dictionaries, both at $\sigma<0$ and $\sigma>0$, we find that the 3d quiver theory has ranks and bare CS levels
\begin{align}
\bm{N} &= (N,\,N-p,\,N)\\
\bm{k} &= (\frac{3}{2}p-q,\,0,\,-\frac{3}{2}p+q)
\end{align}
so that the effective CS levels are
\begin{align} \label{k_-_dP0}
-\bm{k}^- &= (q,\,0,\,-q) \\
+\bm{k}^+ &= (3p-q,\,0,\,-3p+q) \;. \label{k_+_dP0}
\end{align}
The inequality $0\leq q\leq 3$ ensures that the effective FI parameters of this CS toric quiver gauge theory lie precisely on the K\"ahler wall \eqref{wall_dP0} associated to the dictionaries that we used to derive the 3d theory.
Using formula \eqref{dico xi to chi 0}, we find that on this wall the volume of the $\CP^1\subset \CP^2$ in the fibred $CY_3$, computed from the field theory, is
\begin{equation}
\chi(\sigma) = \xi_1^{eff}(\sigma) = (3p\,\Theta(\sigma)-q)\,\sigma
\end{equation}
in agreement with the geometric result \eqref{dP0_volume} of the reduction if $\sigma=r_0$.

\subsection{$\bF_0\equiv \CP^1\times \CP^1$}\label{subsec: F0 geom}

The toric diagram of the 3-parameter family of toric $CY_4$ cones over $Y^{p,\,q_1,\,q_2}(\CP^1\times \CP^1)$, shown in Fig. \ref{fig: torDiag mod 02}, is the convex hull of
\begin{equation}\label{toric_diag_F0}
(0,0,0), \, (0,0,p),\, (0,-1,0),\, (-1,0,0),\, (0,1,q_1),\, (1,0,q_2).
\end{equation}
We are interested in geometric parameters in the range
\begin{equation}\label{geom_ineq_F0}
0 \leq  \frac{q_1}{2}, \frac{q_2}{2} \leq p
\end{equation}
so that all the points \eqref{toric_diag_F0} are external.
The metrics for the Sasaki-Einstein bases are known \cite{Gauntlett:2004hh,Gauntlett:2004hs,Chen:2004nq}.%
\footnote{See also the recent \cite{Tomasiello:2010zz}, which dubbed these Sasaki-Einstein 7-folds $A^{q_1\, q_2\, p}$ and studied Romans mass deformations of the type IIA $AdS_4\times_w M_6$ backgrounds resulting from KK reduction of the $AdS_4\times A^{q_1\, q_2\, p}$ backgrounds of  11d supergravity. We changed notation for the sake of uniformity.}
The geometries are identified under the $\bZ_2\times \bZ_2$ action
\begin{equation}
g: \;\;(q_1,q_2)\mapsto (2p-q_1,2p-q_2)\;, \qquad g': \;(q_1,q_2)\mapsto (q_2,q_1).
\end{equation}
The $CY_4$ singularity is not isolated when at least one of the inequalities \eqref{geom_ineq_F0} is saturated. In that case the $CY_3$ lying over $r_0<0$ or over $r_0>0$ is not completely resolved. Indeed, the minimal GLSM for the $\tilde{Y}(r_0)$ fibre in IIA is
\be
\begin{array}{c|ccccc|c}\label{minimal_GLSM_F0}
CY_3 & p_1 & p_2 & p_3 & p_4 & p_0 & \chi(r_0) \\ \hline
\cC_3 & 0 & 1 & 0 & 1 & -2 & ( 2p \, \Theta(r_0) - q_2)\, r_0 \\
\cC_4 & 1 & 0 & 1 & 0 & -2 & ( 2p \, \Theta(r_0) - q_1) r_0
\end{array}
\ee
with the volumes of the two $\bP^1$'s, $\cC_3$ and $\cC_4$,
\begin{equation}\label{F0_volumes}
\chi_3(r_0) = \left( 2p \, \Theta(r_0) - q_2\right) r_0\;,\qquad\qquad
\chi_4(r_0) = \left( 2p \, \Theta(r_0) - q_1\right) r_0 \;.
\end{equation}

\subsubsection{Phase a of $\bF_0$} \label{subsec:_F0_a}

\begin{figure}[h]
\begin{center}
\includegraphics[width=6cm]{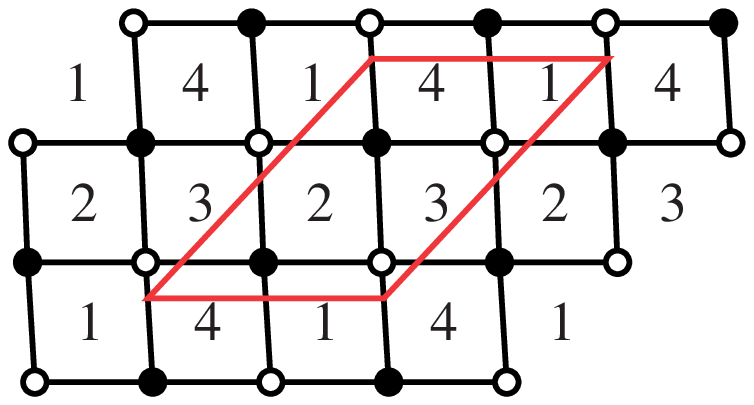}
\caption{\small Brane tiling for toric phase a of $\bF_0$.}\label{F0_a_tiling}
\end{center}
\end{figure}
The toric quiver gauge theory for toric phase a of the complex cone over $F_0$ is specified by the brane tiling of Fig. \ref{F0_a_tiling}, with superpotential
\begin{equation}\label{W_F0_a}
W = \eps_{ij} \eps_{kh} X_{12}^i X_{23}^h X_{34}^j X_{41}^k   \;.
\end{equation}
The dimer model has $4$ internal perfect matchings $\{p[5],\dots, p[8]\}$, associated to the open string K\"ahler chambers
\begin{align}
& p[5]:\qquad -\xi_1\geq 0\;,\quad\xi_2\geq 0\;,\quad \xi_2+\xi_3\geq 0  \label{KC F0a p5} \\
& p[6]:\qquad -\xi_1-\xi_2\geq 0\;,\quad -\xi_2\geq 0\;,\quad\xi_3\geq 0\\
& p[7]:\qquad -\xi_1-\xi_2-\xi_3\geq 0\;,\quad -\xi_2-\xi_3\geq 0\;,\quad -\xi_3\geq 0\\
& p[8]:\qquad \xi_1\geq 0\;,\quad \xi_1+\xi_2\geq 0\;,\quad \xi_1+\xi_2+\xi_3\geq 0 \label{KC F0a p8}
\end{align}
each one with its own dictionary matrix up to large volume monodromies.

A quiver CS theory for the whole class of $C(Y^{p,\,q_1,\,q_2}(F_0))$ geometries based on this toric phase can be proposed as follows. Both at $\sigma<0$ and $\sigma>0$ we are on the K\"ahler wall between the maximal dimensional chamber associated to dictionary
\begin{equation}\label{F0_a_dict_5}
Q^\vee[(\bF_0)_a,\,\{p[5]\},\,\{0, -1\}]=
\left(
\begin{array}{cccc}
 1 & 0 & 0 & 0 \\
 1 & 0 & 1 & 0 \\
 -1 & 1 & -1 & 1 \\
 -1 & -1 & 0 & 0
\end{array}
\right)
\end{equation}
and the one associated to dictionary
\begin{equation}\label{F0_a_dict_8}
Q^\vee[(\bF_0)_a,\,\{p[8]\},\,\{0,0\}]=
\left(
\begin{array}{cccc}
 1 & 0 & 0 & 0 \\
 -1 & 0 & 1 & 0 \\
 -1 & 1 & -1 & 1 \\
 1 & -1 & 0 & 0
\end{array}
\right)
\end{equation}
The wall between these two chambers is given by the cone
\begin{equation}\label{wall_F0_a}
\xi_1= 0\;,\;\; \xi_2\geq 0\;,\;\; \xi_2+\xi_3\geq 0
\end{equation}
in FI parameter space. Using either one of these dictionaries, both at $\sigma<0$ and $\sigma>0$, we find that the 3d quiver theory has ranks and bare CS levels
\begin{align}
\bm{N} &= (N-p,\,N,\,N,\,N)\\
\bm{k} &= (0,\,p-q_1,\,q_1-q_2,\,-p+q_2)
\end{align}
so that the effective CS levels are
\begin{align} \label{k_-_F0_a}
-\bm{k}^- &= (0,\,q_1,\,-q_1+q_2,\,-q_2)\\
+\bm{k}^+ &= (0,\,2p-q_1,\,q_1-q_2 ,\, -2 p+q_2)\;. \label{k_+_F0_a}
\end{align}
It is straightforward to see that the geometric inequalities \eqref{geom_ineq_F0} imply that the effective FI parameters of this  CS toric quiver gauge theory lie precisely on the K\"ahler wall \eqref{wall_F0_a} associated to the dictionaries used to derive the 3d theory.

This guarantees the consistency of the stringy derivation and that the semiclassical computation of the geometric branch of the moduli space reproduces the type IIA geometry, as shown in section \ref{subsec:_semiclassical_IIA}. Let us see it explicitly. In the K\"ahler chamber $p[5]$,
%the solution of D-term equations is in terms of $\sum_i |X_{12}^i|^2$:
%\begin{equation}
%\begin{split}
%\sum_i |X_{23}^i|^2 &= \sum_i |X_{12}^i|^2 + \xi_2 \\
%\sum_i |X_{34}^i|^2 &= \sum_i |X_{12}^i|^2 + \xi_2 + \xi_3 \\
%\sum_i |X_{41}^i|^2 &= \sum_i |X_{12}^i|^2 -\xi_1
%\end{split}
%\end{equation}
% from which we read
using formula \eqref{dico xi to chi 0} with dictionary \eqref{F0_a_dict_5},
the volumes $\chi_{3,\,4}$ of the two $\bP^1$'s are
\begin{equation}
\chi_3 = \xi_2+\xi_3 \;,\qquad \chi_4 = -\xi_1+\xi_2 \;.
\end{equation}
In the K\"ahler chamber $p[8]$,
%the solution of D-term equations is in terms of $\sum_i |X_{41}^i|^2$:
%\begin{equation}
%\begin{split}
%\sum_i |X_{12}^i|^2 &= \sum_i |X_{41}^i|^2 + \xi_1 \\
%\sum_i |X_{23}^i|^2 &= \sum_i |X_{41}^i|^2 + \xi_1 + \xi_2 \\
%\sum_i |X_{34}^i|^2 &= \sum_i |X_{41}^i|^2 + \xi_1 + \xi_2 + \xi_3 \\
%\end{split}
%\end{equation}
using formula \eqref{dico xi to chi 0} with dictionary \eqref{F0_a_dict_8},
the volumes of the two $\bP^1$'s are
\begin{equation}
\chi_3 = 2\xi_1 + \xi_2+\xi_3 \;,\qquad \chi_4 = \xi_1+\xi_2 \;.
\end{equation}
Therefore on the wall $\xi_1=0$ between these two chambers the volumes are
\begin{equation}
\chi_3 = \xi_2+\xi_3 = - \xi_4\;,\qquad \chi_4 = \xi_2\;.
\end{equation}
Plugging in the effective CS levels \eqref{k_-_F0_a}-\eqref{k_+_F0_a}, we find the volumes
\begin{equation}
\chi_3(\sigma) = \left( 2p \, \Theta(\sigma) - q_2\right) \sigma\;,\qquad \qquad
\chi_4(\sigma) = \left( 2p \, \Theta(\sigma) - q_1\right) \sigma \;,
\end{equation}
which reproduce the volumes of the two $\bP^1$'s in the type IIA background \eqref{F0_volumes}, with the identification $\sigma=r_0$.

\subsubsection{Phase b of $\bF_0$} \label{subsec:_F0_b}

\begin{figure}[h]
\begin{center}
\includegraphics[width=4.5cm]{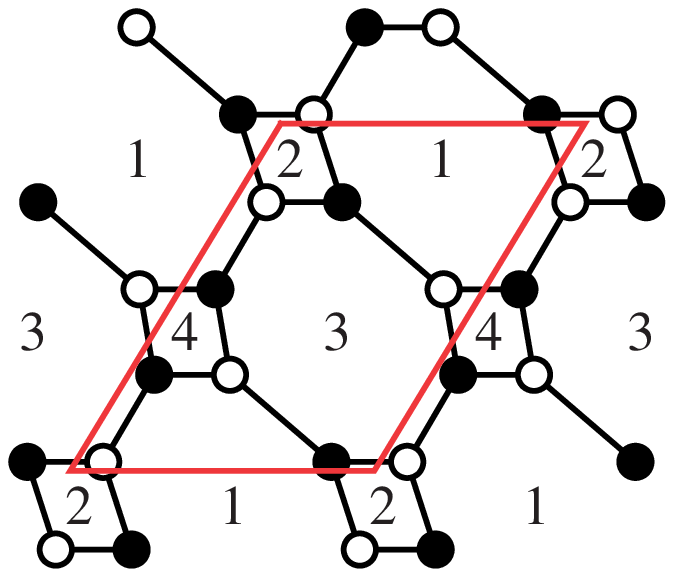}
\caption{\small Brane tiling for toric phase b of $\bF_0$.}\label{F0_b_tiling}
\end{center}
\end{figure}
The toric quiver gauge theory for toric phase b of the complex cone over $F_0$ is specified by the brane tiling of Fig. \ref{F0_b_tiling}, with superpotential
\begin{equation}\label{W_F0_b}
W = \eps_{ij} \eps_{kh} X_{12}^i X_{23}^h X_{31}^{jk} - \eps_{ij} \eps_{kh}  X_{14}^h X_{43}^i X_{31}^{jk}    \;.
\end{equation}
The dimer model has $5$ internal perfect matchings $\{p[5],\dots, p[9]\}$, each one associated to an open string K\"ahler chamber and a dictionary matrix up to large volume monodromies.

A quiver CS theory for the whole class of $C(Y^{p,\,q_1,\,q_2}(F_0))$ geometries based on this toric phase can be obtained by Seiberg duality on gauge group 4 of the theory of phase a. Since the effective FI parameters $\xi_4^\pm\leq 0$, the brane charge dictionaries are obtained by double left mutation $M_{(4;L)}$ of the dictionaries \eqref{F0_a_dict_5} and \eqref{F0_a_dict_8} \cite{Closset:2012eq}, giving
\begin{equation}\label{F0_b_dict_mutation_of_5_in_a}
M_{(4;\,L)}\, Q^\vee[(F_0)_a,\,\{p[5]\},\,\{0,-1\}]=
\left(
\begin{array}{cccc}
 1 & 0 & 0 & 0 \\
 1 & 0 & 1 & 0 \\
-3 & -1 & -1 & 1 \\
 1 & 1 & 0 & 0
\end{array}
\right)
\end{equation}
and
\begin{equation}\label{F0_b_dict_mutation_of_8_in_a}
M_{(4;\,L)}\, Q^\vee[(F_0)_a,\,\{p[8]\},\,\{0,0\}]=
\left(
\begin{array}{cccc}
 1 & 0 & 0 & 0 \\
-1 & 0 & 1 & 0 \\
1 & -1 & -1 & 1 \\
-1 & 1 & 0 & 0
\end{array}
\right)\;.
\end{equation}
Note that these are related to dictionaries
\begin{equation}\label{F0_b_dict_5}
Q^\vee[(\bF_0)_b,\,\{p[5]\},\,\{0, 0\}]=
\left(
\begin{array}{cccc}
 1 & 0 & 0 & 0 \\
 1 & 1 & 0 & 0 \\
 -3 & -1 & -1 & 1 \\
 1 & 0 & 1 & 0
\end{array}
\right)
\end{equation}
and
\begin{equation}\label{F0_b_dict_9}
Q^\vee[(\bF_0)_b,\,\{p[9]\},\,\{0,0\}]=
\left(
\begin{array}{cccc}
 1 & 0 & 0 & 0 \\
 -1 & 1 & 0 & 0 \\
 1 & -1 & -1 & 1 \\
 -1 & 0 & 1 & 0
\end{array}
\right)
\end{equation}
by a quantum $\bZ_2$ monodromy interchanging the role of the two $\bP^1$'s.
The wall between the two chambers is given by the cone
\begin{equation}\label{wall_F0_b}
\xi_1= 0\;,\;\; \xi_2\geq 0\;,\;\; \xi_4\geq 0
\end{equation}
in FI parameter space. On this wall the volumes of the two $\bP^1$'s are
\begin{equation}
\chi_3 = \xi_4 \;,\qquad \chi_4 = \xi_2\;.
\end{equation}
Using either one of the mutated dictionaries \eqref{F0_b_dict_mutation_of_5_in_a} and \eqref{F0_b_dict_mutation_of_8_in_a}, both at $\sigma<0$ and $\sigma>0$, we find the 3d quiver theory with ranks and bare CS levels
\begin{align}
\bm{N} &= (N-p,\,N,\,N,\,N)\\
\bm{k} &= (0,\,p-q_1,\,-2p+q_1+q_2,\,p-q_2)
\end{align}
so that the effective CS levels are
\begin{align} \label{k_-_F0_b}
-\bm{k}^- &= ( 0,\, q_1 ,\,-q_1-q_2 ,\, q_2) \\
+\bm{k}^+ &= ( 0,\, 2 p-q_1 ,\, -4 p+q_1+q_2,\, 2 p-q_2)\;. \label{k_+_F0_b}
\end{align}
The geometric inequalities \eqref{geom_ineq_F0} ensure that the effective FI parameters of this  CS toric quiver gauge theory lie precisely on the K\"ahler wall \eqref{wall_F0_b} associated to the dictionaries used to derive the 3d theory. The volumes of the 2-cycles in the $CY_3$ are again
\begin{equation}
\chi_3(\sigma) = \left( 2p \, \Theta(\sigma) - q_2\right) \sigma\;,\qquad \qquad
\chi_4(\sigma) = \left( 2p \, \Theta(\sigma) - q_1\right) \sigma \;.
\end{equation}

An important remark is in order here: the stringy derivation is subtler if $q_i=0,2p$, which introduces a non-isolated singularity in the $CY_4$ due to the fibration of an isolated singularity of the $CY_3$. Let us consider $q_2=2p$ for simiplicity. There is an extra 1-complex-dimensional Coulomb branch, due to $k^+_4=0$. If this extra branch of the moduli space is parametrised by a monopole operator turning on one unit of flux in gauge group in one of the phases, it is parametrised in the dual phase by an extra singlet coupled in the superpotential to an analogous monopole operator \cite{Aharony:1997gp,Benini:2011mf}. As in simpler brane realizations of 3d Seiberg duality like the type IIB setup of \cite{Cremonesi:2010ae}, it is not known how to account for these extra singlets in terms of branes: the stringy derivation, as developed so far, is not sensitive to these details. It is thus unclear in which of the two toric phases the singlet should be. Similar considerations hold for Seiberg duality on gauge group 2 and $q_1$. In conclusion, the stringy derivation is unambiguous only when the $CY_3$ fibres are completely resolved, so that those extra branches of the moduli space and extra singlets are not there.

\subsection{$dP_1$}\label{subsec:_dP1}

The toric diagram of the 3-parameter family of toric $CY_4$ cones over $Y^{p,\,q_1,\,q_2}(dP_1)$, shown in Fig. \ref{fig: torDiag mod 03}, is the convex hull of
\begin{equation}\label{toric_diag_dP1}
(0,0,0), \, (0,0,p),\, (0,-1,0),\, (-1,1,0),\, (0,1,q_1),\, (1,0,q_2).
\end{equation}
We are interested in geometric parameters in the range
\begin{equation}\label{geom_ineq_dP1}
0 \leq  \frac{q_1}{2}, \frac{q_2}{3} \leq p
\end{equation}
so that all the points \eqref{toric_diag_dP1} are external.
The geometries are identified under the $\bZ_2$ action $(q_1,q_2)\mapsto (2p-q_1,3p-q_2)$.
The minimal GLSM for the $\tilde{Y}(r_0)$ fibre in IIA is
\be\label{minimal_GLSM_dP1}
\begin{array}{c|ccccc|c}
 CY_3 & p_1 & p_2 & p_3 & p_4 & p_0 &  \chi(r_0) \\ \hline
 \cC_3 & 0 & 1 & -1 & 1 & -1 & (p\,\Theta(r_0)+q_1-q_2)\,r_0 \\
 \cC_4 & 1 & 0 & 1 & 0 & -2 & (2 p\,\Theta(r_0)-q_1)\,r_0
\end{array}
\ee
If $0<q_2-q_1<p$, the triangulations $\Gamma_\pm$ are the same: both $\tilde{Y}_\pm$ contain a blown up $dP_1$.
When $q_2-q_1$ crosses $0$ (resp. $p$), the curve $\cC_3\cong\bP^1$ undergoes a flop transition in $\tilde{Y}_-$ (resp. $\tilde{Y}_+$), resulting in a $\bP^1$ intersecting a $\bP^2$.

\begin{figure}[h]
\begin{center}
\includegraphics[width=5.5cm]{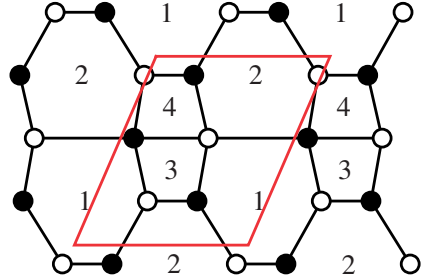}
\caption{\small Brane tiling for $dP_1$.}\label{dP1_tiling}
\end{center}
\end{figure}
The toric quiver gauge theory for the complex cone over $dP_1$ is specified by the brane tiling of Fig. \ref{dP1_tiling}, with superpotential
\begin{equation}\label{W_dP1}
\begin{split}
W &= X_{21}^1 X_{14}^{} X_{42}^1 + X_{21}^2 X_{13}^1 X_{32}^{} + X_{42}^2 X_{21}^3 X_{13}^2 X_{34}^{} +\\
&- X_{13}^2 X_{32}^{} X_{21}^1 - X_{14}^{} X_{42}^2 X_{21}^2 - X_{21}^3 X_{13}^1 X_{34}^{} X_{42}^1 \;.
\end{split}
\end{equation}
The dimer model has $4$ internal perfect matchings $\{p[5],\dots, p[8]\}$ and corresponding open string K\"ahler chambers (before triangulating the toric diagram of the $CY_3$)
\begin{align}
& p[5]:\qquad \xi_1+\xi_2\geq 0\;,\quad\xi_2\geq 0\;,\quad\xi_1+\xi_2+\xi_3\geq 0  \\
& p[6]:\qquad \xi_1\geq 0\;, \quad -\xi_2\geq 0\;,\quad \xi_1+\xi_3\geq 0 \\
& p[7]:\qquad -\xi_1\geq 0\;, \quad -\xi_1-\xi_2\geq 0\;,\quad \xi_3\geq 0 \\
& p[8]:\qquad -\xi_1-\xi_2-\xi_3\geq 0\;,\quad -\xi_1-\xi_3\geq 0\;,\quad -\xi_3\geq 0
\end{align}
each one with its own dictionary matrix up to large volume monodromies.

A quiver CS theory for the whole class of $C(Y^{p,\,q_1,\,q_2}(dP_1))$ geometries (\emph{i.e.} also for any triangulations of $\tilde{Y}_\pm)$ can be proposed as follows. Both at $\sigma<0$ and $\sigma>0$ we are on the K\"ahler wall between the maximal dimensional chamber associated to dictionary
\begin{equation}\label{dP1_dict_5}
Q^\vee[dP_1,\,\{p[5]\},\,\{0, 0\}]=
\left(
\begin{array}{cccc}
 -2 & 1 & 0 & 1 \\
 1 & 0 & \frac{1}{2} & 0 \\
 0 & -1 & 0 & 0 \\
 1 & 0 & -\frac{1}{2} & 0
\end{array}
\right)
\end{equation}
and the one associated to dictionary
\begin{equation}\label{dP1_dict_6}
Q^\vee[dP_1,\,\{p[6]\},\,\{-1, -1\}]=
\left(
\begin{array}{cccc}
 1 & 1 & \frac{3}{2} & 1 \\
 1 & 0 & \frac{1}{2} & 0 \\
 -1 & -1 & -\frac{1}{2} & 0 \\
 -1 & 0 & -\frac{3}{2} & 0
\end{array}
\right)
\end{equation}
The wall between these two chambers is given by the cone
\begin{equation}\label{wall_dP1}
\xi_1\geq 0\;,\;\; \xi_2= 0\;,\;\; \xi_1+\xi_3\geq 0
\end{equation}
in FI parameter space.%
\footnote{This cone can be further refined into two cones with $\xi_3\leq 0$ and $\xi_3\geq 0$ respectively, which the dictionaries translate to $\chi_3\geq 0$ and $\chi_3\leq 0$. This subdivision is sensitive to the triangulation of the toric diagram: $\tilde{Y}$ undergoes a flop transition at the common boundary of the two subcones.}
Using either one of these dictionaries, both at $\sigma<0$ and $\sigma>0$, we find that the 3d quiver theory has ranks and bare CS levels
\begin{align}
\bm{N} &= (N,\,N-p,\,N,\,N)\\
\bm{k} &= \left(\frac{3p}{2}-q_2,\,0,\,-\frac{p}{2}-q_1+q_2,\,-p+q_1\right)
\end{align}
so that the effective CS levels are
\begin{align}
-\bm{k}^- &= (q_2,\,0,\,q_1-q_2,\,-q_1)\\
+\bm{k}^+ &= (3 p-q_2,\,0,\,-p-q_1+q_2 ,\, -2 p+q_1)\;.
\end{align}
The geometric inequalities \eqref{geom_ineq_dP1} imply that the effective FI parameters of this  CS toric quiver gauge theory lie precisely on the K\"ahler wall \eqref{wall_dP1} associated to the dictionaries used to derive the 3d theory. This guarantees the consistency of the derivation and that the semiclassical computation of the geometric branch of the moduli space reproduces the type IIA geometry: plugging the effective FI parameters and any of the dictionaries \eqref{dP1_dict_5}-\eqref{dP1_dict_6} into formula \eqref{dico xi to chi 0}, the volumes of 2-cycles of $\tilde{Y}(\sigma)$ computed in field theory match the IIA data \eqref{minimal_GLSM_dP1} with $r_0=\sigma$.

\subsection{$dP_2$}

The toric diagram of the 4-parameter family of toric $CY_4$ cones over $Y^{p,\,q_1,\,q_2,\,q_3}(dP_2)$, shown in Fig. \ref{fig: torDiag mod 05}, is the convex hull of
\begin{equation}\label{toric_diag_dP2}
(0,0,0), \, (0,0,p),\, (1,-1,0) ,\,  (-1,0,0),\, (-1,1,q_1),\, (0,1,q_2),\, (1,0,q_3).
\end{equation}
We are interested in geometric parameters in the range
\begin{equation}\label{geom_ineq_dP2}
0 \leq  \frac{q_1}{2} , \frac{q_2}{3}, \frac{q_3}{2} \leq p
\end{equation}
so that all the points \eqref{toric_diag_dP2} are external.
The geometries are identified under the $\bZ_2$ action $(q_1,q_2,q_3)\mapsto (2p-q_1,3p-q_2,2p-q_3)$.
The minimal GLSM for the $\tilde{Y}(r_0)$ fibre in IIA is
\be\label{minimal_GLSM_dP2}
\begin{array}{c|cccccc|c}
 CY_3 & p_1 & p_2 & p_3 & p_4 & p_5 & p_0 & \chi(r_0)  \\
\hline
 \cC_3 & 0 & 1 & -1 & 1 & 0 & -1 & (p\,\Theta(r_0)+q_1-q_2)\,r_0 \\
 \cC_4 & 0 & 0 & 1 & -1 & 1 & -1 & (p\,\Theta(r_0)-q_1+q_2-q_3)\,r_0 \\
 \cC_5 & 1 & 0 & 0 & 1 & -1 & -1 & (p\,\Theta(r_0)-q_2+q_3)\,r_0
\end{array}
\ee

\subsubsection{Phase a of $dP_2$} \label{subsec:_dP2_phase_a}

\begin{figure}[h]
\begin{center}
\includegraphics[width=5.5cm]{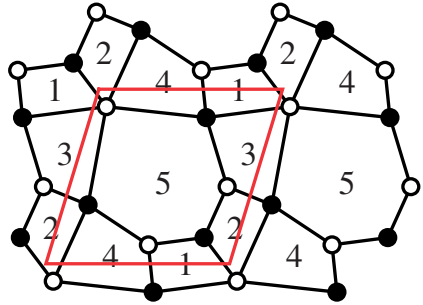}
\caption{\small Brane tiling for toric phase a of $dP_2$.}\label{dP2_a_tiling_modified}
\end{center}
\end{figure}
The brane tiling for toric phase a of $dP_2$ is in Fig. \ref{dP2_a_tiling_modified}. The superpotential is
\begin{equation}
\begin{split}
W &= X_{25}^{} X_{53}^2 X_{32}^{} + X_{51}^{} X_{14}^{} X_{45}^2
+ X_{31}^{} X_{12}^{} X_{24}^{} X_{45}^1 X_{53}^1 + \\
& - X_{25}^{} X_{51}^{} X_{12}^{} - X_{53}^1 X_{32}^{} X_{24}^{} X_{45}^2 - X_{53}^2 X_{31}^{} X_{14}^{} X_{45}^1 \;.
\end{split}
\end{equation}
The dimer model has $5$ internal perfect matchings $\{p[6],\dots, p[10]\}$.

The quiver CS theory for M2-branes at $C(Y^{p,\,q_1,\,q_2,\,q_3}(dP_2))$ in the absence of torsion $G_4$ flux is on the wall between the chamber of dictionary
\begin{equation}\label{dP2_a_dict_7}
Q^\vee[(dP_2)_a,\,\{p[7]\},\,\{0,0,1\}]=
\left(
\begin{array}{ccccc}
 -1 & 0 & -\frac{1}{2} & 1 & \frac{9}{8} \\
 0 & -1 & 0 & 0 & 0 \\
 -1 & 1 & \frac{1}{2} & 0 & \frac{1}{8} \\
 1 & 0 & -\frac{1}{2} & -1 & -\frac{1}{8} \\
 1 & 0 & \frac{1}{2} & 0 & -\frac{1}{8}
\end{array}
\right)
\end{equation}
and the one of dictionary
\begin{equation}\label{dP2_a_dict_8}
Q^\vee[(dP_2)_a,\,\{p[8]\},\,\{-1,-1,0\}]=
\left(
\begin{array}{ccccc}
 0 & 0 & 0 & 1 & 1 \\
 -1 & -1 & -\frac{1}{2} & 0 & \frac{1}{8} \\
 1 & 1 & \frac{3}{2} & 0 & -\frac{1}{8} \\
 -1 & 0 & -\frac{3}{2} & -1 & \frac{1}{8} \\
 1 & 0 & \frac{1}{2} & 0 & -\frac{1}{8}
\end{array}
\right)
\;.
\end{equation}
In FI parameter space the wall is the cone
\begin{equation}\label{wall_dP2_a_from_dP3_c}
\xi_1+\xi_3\geq 0\;,\;\; \xi_3\geq 0\;,\;\;
-\xi_4 \geq 0\;,\;\; \xi_5=0\;.
\end{equation}
The gauge ranks and bare CS levels of the M2-brane theory are
\begin{align}
\bm{N} &= (N,\,N,\,N,\,N,\,N-p)\\
\bm{k} &= (\frac{1}{2}p-q_2+q_3,\,-\frac{1}{2}p-q_1+q_2,\,p-q_3,\,-p+q_1,\,0)
\end{align}
The effective CS levels
\begin{align}
-\bm{k}^- &= (q_2-q_3,\, q_1-q_2,\,q_3,\,-q_1,\,0)\\
+\bm{k}^+ &= (p-q_2+q_3,\,-p-q_1+q_2,\,2p-q_3,\,-2p+q_1,\,0)
\end{align}
are such that the effective FI parameters lie in the cone \eqref{wall_dP2_a_from_dP3_c} for geometric parameters in the window \eqref{geom_ineq_dP2}. Then the dictionary matrices translate the effective FI parameters $\bm{\xi}(\sigma)$ of the gauge theory into the GLSM FI parameters $\bm{\chi}(r_0)$ of $\tilde{Y}(r_0)$, with $r_0=\sigma$.

\subsubsection{Phase b of $dP_2$} \label{subsec:_dP2_phase_b}

\begin{figure}[h]
\begin{center}
\includegraphics[width=5.5cm]{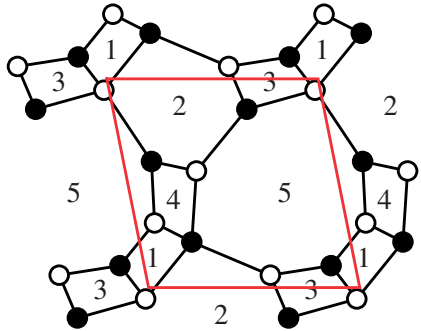}
\caption{\small Brane tiling for toric phase b of $dP_2$.}\label{dP2_b_tiling_modified}
\end{center}
\end{figure}
The brane tiling for toric phase b of $dP_2$ is in Fig. \ref{dP2_b_tiling_modified}. The superpotential is
\begin{equation}
\begin{split}
W &= X_{25}^2 X_{53}^2 X_{32}^{} + X_{15}^{} X_{54}^2 X_{41}^{} + X_{25}^3 X_{54}^1 X_{42}^{} + X_{12}^{} X_{25}^1 X_{53}^1 X_{31}^{} +\\
& - X_{25}^1 X_{54}^2 X_{42}^{} - X_{15}^{} X_{53}^2 X_{31}^{} - X_{25}^3 X_{53}^1 X_{32}^{} - X_{12}^{} X_{25}^2 X_{54}^1 X_{41}^{} \;.
\end{split}
\end{equation}
The dimer model has $6$ internal perfect matchings $\{p[6],\dots, p[11]\}$.

We can propose a quiver CS theory for M2-branes probing  $C(Y^{p,\,q_1,\,q_2,\,q_3}(dP_2))$, for the entire class of geometries specified by $q_1\in[0,2p]$, $q_2\in[0,3p]$ and $q_3\in[0,2p]$: it is on the wall between the chamber associated to dictionary
\begin{equation}\label{dP2_b_dict_9}
Q^\vee[(dP_2)_b,\,\{p[9]\},\,\{0,0,1\}]=\left(
\begin{array}{ccccc}
 0 & 0 & -1 & 0 & 1 \\
 1 & -1 & -\frac{1}{2} & -1 & -\frac{1}{8} \\
 -1 & 1 & \frac{1}{2} & 0 & \frac{1}{8} \\
 -1 & 0 & \frac{1}{2} & 1 & \frac{1}{8} \\
 1 & 0 & \frac{1}{2} & 0 & -\frac{1}{8}
\end{array}
\right)
\end{equation}
and the one associated to
\begin{equation}\label{dP2_b_dict_10}
Q^\vee[(dP_2)_b,\,\{p[10]\},\,\{0,-1,0\}]=
\left(
\begin{array}{ccccc}
 -1 & 0 & -\frac{3}{2} & 0 & \frac{9}{8} \\
 -2 & -1 & -2 & -1 & \frac{1}{4} \\
 1 & 1 & \frac{3}{2} & 0 & -\frac{1}{8} \\
 1 & 0 & \frac{3}{2} & 1 & -\frac{1}{8} \\
 1 & 0 & \frac{1}{2} & 0 & -\frac{1}{8}
\end{array}
\right)\;,
\end{equation}
which in FI parameter space is given by the cone
\begin{equation}\label{wall_dP2_b}
-\xi_2\geq 0\;,\;\; \xi_3\geq 0\;,\;\;
\xi_4 \geq 0\;,\;\; \xi_5=0\;.
\end{equation}
The gauge ranks and bare CS levels are
\begin{align}
\bm{N} &= (N,\,N,\,N,\,N,\,N-p)\\
\bm{k} &= (-\frac{1}{2}p+q_1-q_2+q_3,\,-\frac{3}{2}p+q_2,\,p-q_3,\,p-q_1,\,0)\;,
\end{align}
and the effective CS levels are
\begin{align}
-\bm{k}^- &= (-q_1+q_2-q_3,\, -q_2,\,q_3,\,q_1,\,0)\\
+\bm{k}^+ &= (-p+q_1-q_2+q_3,\,-3p+q_2,\,2p-q_3,\,2p-q_1,\,0)\;,
\end{align}
so that the effective FI parameters belong to the cone \eqref{wall_dP2_b} thanks to the geometric inequalities \eqref{geom_ineq_dP2}. This quiver CS theory is nothing but the dual of the theory in phase a of section \ref{subsec:_dP2_phase_a} under a maximally chiral Seiberg duality  of gauge group 4 \cite{Benini:2011mf}. The dictionaries \eqref{dP2_b_dict_9} and \eqref{dP2_b_dict_10} are obtained by left mutation $M_{(4;L)}$ of the dictionaries \eqref{dP2_a_dict_7} and \eqref{dP2_a_dict_8} of phase a respectively, with no need of quantum monodromies.

\subsection{$dP_3$}

The toric diagram of the 5-parameter family of toric $CY_4$ cones over $Y^{p,\,q_1,\,q_2,\,q_3,\,q_4}(dP_3)$, shown in Fig. \ref{fig: torDiag mod 07}, is the convex hull of
\begin{equation}\label{toric_diag_dP3}
(0,0,0), \, (0,0,p),\, (1,-1,0) ,\,  (0,-1,0),\, (-1,0,q_1),\, (-1,1,q_2),\, (0,1,q_3),\, (1,0,q_4).
\end{equation}
We require that the points \eqref{toric_diag_dP3} are all external, which means
\begin{equation}\label{geom_ineq_dP3}
0 \leq  \frac{q_1+q_4}{2} , \frac{q_2}{2}, \frac{q_3}{2}, \frac{q_1+q_3}{3}, \frac{q_2+q_4}{3} \leq p\;.
\end{equation}
The geometries are identified under the $\bZ_2$ action $(q_1,q_2,q_3,q_4)\mapsto (p-q_1,2p-q_2,2p-q_3,p-q_4)$.
The minimal GLSM for the $\tilde{Y}(r_0)$ fibre in IIA is
\be\label{minimal_GLSM_dP3}
\begin{array}{c|ccccccc|cc}
CY_3 & p_1 & p_2 & p_3 & p_4 & p_5 & p_6 & p_0 & \chi(r_0)  \\ \hline
 \cC_3 & 0 & 1 & -1 & 1 & 0 & 0 & -1 & (p\,\Theta(r_0)+q_1-q_2)\,r_0 \\
 \cC_4 & 0 & 0 & 1 & -1 & 1 & 0 & -1 & (p\,\Theta(r_0)-q_1+q_2-q_3)\,r_0 \\
 \cC_5 & 0 & 0 & 0 & 1 & -1 & 1 & -1 & (p\,\Theta(r_0)-q_2+q_3-q_4)\,r_0 \\
 \cC_6 & 1 & 0 & 0 & 0 & 1 & -1 & -1 & (p\,\Theta(r_0)-q_3+q_4)\,r_0
\end{array}
\ee

\subsubsection{Phase d of $dP_3$} \label{subsec:_dP3_phase_d}

\begin{figure}[h]
\begin{center}
\includegraphics[width=6cm]{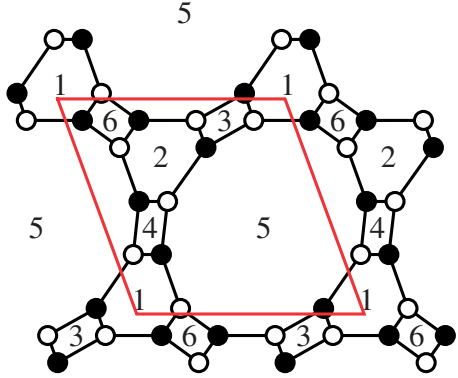}
\caption{\small Brane tiling for toric phase d of $dP_3$.}\label{dP3_d_tiling}
\end{center}
\end{figure}
Toric phase d of $dP_3$ is specified by the brane tiling of Fig. \ref{dP3_d_tiling}. The superpotential is
\begin{equation}\label{W_dP3d}
\begin{split}
W &= X_{32}^{} X_{25}^2 X_{53}^2 + X_{31}^{} X_{15}^1 X_{53}^1 + X_{41}^{} X_{15}^3 X_{54}^2 + X_{42}^{} X_{25}^3 X_{54}^1 +\\
&+ X_{56}^1 X_{62}^{} X_{25}^1 + X_{56}^2 X_{61}^{} X_{15}^2 - X_{31}^{} X_{15}^3 X_{53}^2 - X_{32}^{} X_{25}^3 X_{53}^1 +\\
&- X_{42}^{} X_{25}^1 X_{54}^2 - X_{41}^{} X_{15}^2 X_{54}^1 - X_{56}^1 X_{61}^{} X_{15}^1 - X_{56}^2 X_{62}^{} X_{25}^2 \;.
\end{split}
\end{equation}
The dimer model has $11$ internal perfect matchings $\{p[7],\dots, p[17]\}$ and corresponding open string K\"ahler chambers (before triangulating the toric diagram of the $CY_3$), each one with its own dictionary matrix up to large volume monodromies.

A quiver CS theory for the whole class of $C(Y^{p,\,q_1,\,q_2,\,q_3,\,q_4}(dP_3))$ geometries can be proposed as follows. Both at $\sigma<0$ and $\sigma>0$ we are on the K\"ahler wall between the maximal dimensional chamber associated to dictionary
\begin{equation}\label{dP3_d_dict_13}
Q^\vee[(dP_3)_d,\,\{p[13]\},\,\{1, 0, 0, 0\}]=
\left(
\begin{array}{cccccc}
 1 & \frac{1}{2} & -1 & -1 & -\frac{1}{2} & -\frac{1}{4} \\
 1 & -\frac{1}{2} & -1 & -1 & \frac{1}{2} & -\frac{1}{4} \\
 -1 & \frac{1}{2} & 1 & 0 & -\frac{1}{2} & \frac{1}{4} \\
 -1 & -\frac{1}{2} & 1 & 1 & -\frac{1}{2} & \frac{5}{4} \\
 1 & \frac{1}{2} & 0 & 0 & \frac{1}{2} & -\frac{1}{4} \\
 -1 & -\frac{1}{2} & 0 & 1 & \frac{1}{2} & \frac{1}{4}
\end{array}
\right)
\end{equation}
and the one associated to dictionary
\begin{equation}\label{dP3_d_dict_14}
Q^\vee[(dP_3)_d,\,\{p[14]\},\,\{1, 0, 0, -1\}]=
\left(
\begin{array}{cccccc}
 -2 & -1 & -1 & -1 & -2 & \frac{1}{2} \\
 -2 & -2 & -1 & -1 & -1 & \frac{1}{2} \\
 1 & \frac{3}{2} & 1 & 0 & \frac{1}{2} & -\frac{1}{4} \\
 1 & \frac{1}{2} & 1 & 1 & \frac{1}{2} & \frac{3}{4} \\
 1 & \frac{1}{2} & 0 & 0 & \frac{1}{2} & -\frac{1}{4} \\
 1 & \frac{1}{2} & 0 & 1 & \frac{3}{2} & -\frac{1}{4}
\end{array}
\right)\;.
\end{equation}
The wall between these two K\"ahler chambers is given by the cone
\begin{equation}\label{wall_dP3_d}
-\xi_1\geq 0\;,\;\; -\xi_2\geq 0\;,\;\; \xi_3\geq 0\;,\;\;
\xi_4 \geq 0\;,\;\; \xi_5=0\;,\;\;\xi_6\geq 0
\end{equation}
in FI parameter space. Using either one of these dictionaries, both at $\sigma<0$ and $\sigma>0$, we find that the 3d quiver theory has ranks and bare CS levels
\begin{align}
\bm{N} &= (N,\,N,\,N,\,N,\,N-p,\,N)\\
\bm{k} &= (-\frac{3}{2}p+q_1+q_3,\,-\frac{3}{2}p+q_2+q_4,\,p-q_3,\,p-q_1-q_4,\,0,\,p-q_2)
\end{align}
so that the effective CS levels are
\begin{align}
-\bm{k}^- &= (-q_1-q_3,\,-q_2-q_4,\,q_3,\,q_1+q_4,\,0,\,q_2)\\
+\bm{k}^+ &= (-3p+q_1+q_3,\,-3p+q_2+q_4,\,2p-q_3,\,2p-q_1-q_4,\,0,\,2p-q_2)\;.
\end{align}
It is straightforward to see that the geometric inequalities \eqref{geom_ineq_dP3} ensure that the effective FI parameters of this  CS toric quiver gauge theory lie on the K\"ahler wall \eqref{wall_dP3_d} associated to the dictionaries used to derive the 3d theory. This guarantees the consistency of the stringy derivation and that the semiclassical computation of the geometric branch of the moduli space reproduces the type IIA geometry \eqref{minimal_GLSM_dP3}.

\subsubsection{Phase c of $dP_3$}

We next move to phase c of the $dP_3$ quiver, which is obtained upon a ``maximally chiral'' Seiberg duality on gauge group $4$ \cite{Benini:2011mf}.  In D-brane terms \cite{Closset:2012eq} it is a double right mutation $M_{(4;\,R)}$ on the dictionary matrices \eqref{dP3_d_dict_13}-\eqref{dP3_d_dict_14}, giving the dictionaries
\begin{equation}\label{dP3_c_dict_mutation_of_13_in_d}
M_{(4;\,R)}\, Q^\vee[(dP_3)_d,\,\{p[13]\},\,\{1, 0, 0, 0\}]=\left(
\begin{array}{cccccc}
 0 & 0 & 0 & 0 & -1 & 1 \\
 0 & -1 & 0 & 0 & 0 & 1 \\
 -1 & \frac{1}{2} & 1 & 0 & -\frac{1}{2} & \frac{1}{4} \\
 1 & \frac{1}{2} & -1 & -1 & \frac{1}{2} & -\frac{5}{4} \\
 1 & \frac{1}{2} & 0 & 0 & \frac{1}{2} & -\frac{1}{4} \\
 -1 & -\frac{1}{2} & 0 & 1 & \frac{1}{2} & \frac{1}{4}
\end{array}
\right)
\end{equation}
and
\begin{equation}\label{dP3_c_dict_mutation_of_14_in_d}
M_{(4;\,R)}\, Q^\vee[(dP_3)_d,\,\{p[14]\},\,\{1, 0, 0, -1\}]=
\left(
\begin{array}{cccccc}
 -1 & -\frac{1}{2} & 0 & 0 & -\frac{3}{2} & \frac{5}{4} \\
 -1 & -\frac{3}{2} & 0 & 0 & -\frac{1}{2} & \frac{5}{4} \\
 1 & \frac{3}{2} & 1 & 0 & \frac{1}{2} & -\frac{1}{4} \\
 -1 & -\frac{1}{2} & -1 & -1 & -\frac{1}{2} & -\frac{3}{4} \\
 1 & \frac{1}{2} & 0 & 0 & \frac{1}{2} & -\frac{1}{4} \\
 1 & \frac{1}{2} & 0 & 1 & \frac{3}{2} & -\frac{1}{4}
\end{array}
\right)\;.
\end{equation}

\begin{figure}[h]
\begin{center}
\includegraphics[width=6cm]{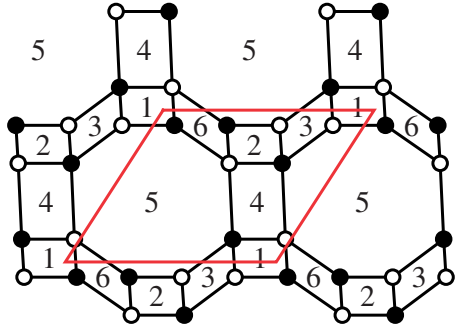}
\caption{\small Brane tiling for toric phase c of $dP_3$.}\label{dP3_c_tiling}
\end{center}
\end{figure}
The brane tiling for toric phase c is in Fig. \ref{dP3_c_tiling}, with superpotential
\begin{equation}\label{W_dP3c}
\begin{split}
W &= X_{25}^{} X_{53}^2 X_{32}^{} +X_{31}^{} X_{15}^{} X_{53}^1 +X_{56}^1 X_{62}^{} X_{24}^{} X_{45}^1+X_{56}^2 X_{61}^{} X_{14}^{} X_{45}^2 +\\
& -X_{25}^{} X_{56}^2 X_{62}^{}-X_{61}^{} X_{15}^{} X_{56}^1 -X_{53}^1 X_{32}^{} X_{24}^{} X_{45}^2-X_{53}^2 X_{31}^{} X_{14}^{} X_{45}^1  \;.
\end{split}
\end{equation}
The dimer model has $8$ internal perfect matchings $\{p[7],\dots, p[14]\}$ and corresponding open string K\"ahler chambers. From the D6-charges we can infer that  \eqref{dP3_c_dict_mutation_of_13_in_d} is valid in the K\"ahler chamber of perfect matching $p[9]$ and that \eqref{dP3_c_dict_mutation_of_14_in_d} is valid in the K\"ahler chamber of perfect matching $p[10]$.
%
%\footnote{   \textbf{... but we still need to reproduce the dictionary with a quantum monodromy on the dictionaries of HHV.}}
The wall in FI parameter space between the corresponding maximal dimensional chambers is
\begin{equation}\label{wall_dP3_c}
-\xi_1-\xi_4\geq 0\;, \;\; -\xi_2-\xi_4\geq 0\;, \;\; \xi_3\geq 0\;,\;\;
-\xi_4\geq 0\;, \;\; \xi_5= 0\;, \;\; \xi_6\geq 0\;.
\end{equation}

Using either one of these mutated dictionaries, both at $\sigma<0$ and $\sigma>0$, or applying the Seiberg duality rules of \cite{Benini:2011mf} to the theory in phase d, we find a 3d quiver theory in phase c with gauge ranks and bare CS levels
\begin{align}
\bm{N} &= (N,\,N,\,N,\,N,\,N-p,\,N)\\
\bm{k} &= (-\frac{1}{2}p+q_3-q_4,\,-\frac{1}{2}p+q_2-q_1,\,p-q_3,\,-p+q_1+q_4,\,0,\,p-q_2)
\end{align}
so that the effective CS levels are
\begin{align} \label{k_eff_dP3_c}
-\bm{k}^- &= (-q_3+q_4,\,-q_2+q_1,\,q_3,\,-q_1-q_4,\,0,\,q_2)\\
+\bm{k}^+ &= (-p+q_3-q_4,\,-p+q_2-q_1,\,2p-q_3,\,-2p+q_1+q_4,\,0,\,2p-q_2)\;.
\end{align}
%A \textbf{subtlety} appears when $k_4^-$ or $k_4^+$ vanish so that there is an extra 1-complex dimensional Coulomb branch: the field theory duality of \cite{Benini:2011mf} implies that there is a singlet in one of the two dual phases considered so far, but it is unclear which. More generally we expect this subtlety to appear whenever the $CY_3^-$ or the $CY_3^+$ is not completely resolved. \textbf{[Can we make this precise?]}
Once again \eqref{geom_ineq_dP3} ensures that the  effective FI parameters of the gauge theory are in the cone \eqref{wall_dP3_c} and therefore that the type IIA geometry \eqref{minimal_GLSM_dP3} is reproduced.

\subsubsection{Phase b of $dP_3$}

\begin{figure}[h]
\begin{center}
\includegraphics[width=5.5cm]{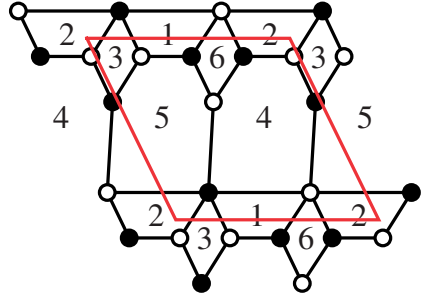}
\caption{\small Brane tiling for toric phase b of $dP_3$.}\label{dP3_b_tiling}
\end{center}
\end{figure}

We next move to phase b of the $dP_3$ quiver by means of a Seiberg duality on gauge group $2$ of the quiver of phase c. We see from the effective CS levels \eqref{k_eff_dP3_c} that the Seiberg duality in question is maximally chiral if $0\leq q_2-q_1\leq p$ (a double $M_{(2;\,L)}$ left mutation), whereas it is minimally chiral if $q_2-q_1<0$ ($M_{(2;\,R)}$/$M_{(2;\,L)}$ mutation at $\sigma<0/\sigma<0$ resp.) or $q_2-q_1>p$ ($M_{(2;\,L)}$/$M_{(2;\,R)}$ mutation).
As a consequence, three windows will be needed to cover the full class of geometries using toric phase b.

The brane tiling of toric phase b is in figure \ref{dP3_b_tiling}, with superpotential
\begin{equation}\label{W_dP3b}
\begin{split}
W &= X_{31}^{} X_{15}^{} X_{53}^{} +X_{42}^{} X_{23}^{} X_{34}^{} +X_{56}^{} X_{64}^{} X_{45}^1 +X_{52}^{} X_{26}^{} X_{61}^{} X_{14}^{} X_{45}^2+\\
& -X_{42}^{} X_{26}^{} X_{64}^{} -X_{53}^{} X_{34}^{} X_{45}^2 -X_{56}^{} X_{61}^{} X_{15}^{} -X_{14}^{} X_{45}^1 X_{52}^{} X_{23}^{} X_{31}^{}
  \;.
\end{split}
\end{equation}
The dimer model has $7$ internal perfect matchings $\{p[7],\dots, p[13]\}$ and corresponding open string K\"ahler chambers.

In the window $0\leq q_2-q_1\leq p$, we are both at $\sigma<0$ and $\sigma>0$ on the K\"ahler wall between dictionaries $\{\{p[7]\},\,\{-1, 0, 1, 0\}\}$ and $\{\{p[13]\},\,\{-1, 0, 1, 1\}\}$, which gives the cone
\begin{equation}\label{wall_dP3_b_1}
\xi_2\geq 0\;, \;\; \xi_2+\xi_3\geq 0\;, \;\; \xi_2+\xi_6\geq 0\;,\;\;
\xi_1+\xi_4\leq 0\;,\;\; \xi_4\leq 0\;, \;\; \xi_5= 0
\end{equation}
in FI parameter space. The ranks and bare levels of the CS theory are
\begin{align}
\bm{N} &= (N,\,N,\,N,\,N,\,N-p,\,N)\\
\bm{k} &= (-\frac{p}{2}+q_3-q_4,\,\frac{p}{2}+q_1-q_2,\,\frac{p}{2}-q_1+q_2-q_3,\, -p+q_1+q_4,\,0,\,\frac{p}{2}-q_1)
\end{align}
so that the effective CS levels are
\begin{align} \label{k_eff_dP3_b_1}
-\bm{k}^- &= (-q_3+q_4,\,-q_1+q_2,\,q_1-q_2+q_3,\,-q_1-q_4,\,0,\,q_1)\\
+\bm{k}^+ &= (-p+q_3-q_4,\,p+q_1-q_2,\,p-q_1+q_2-q_3,\,-2p+q_1+q_4,\,0,\,p-q_1)\;.
\end{align}
In order for the dictionaries we used in the derivation to be consistent, the effective FI parameters must lie in the cone \eqref{wall_dP3_b_1}: this indeed requires that $0\leq q_2-q_1\leq p$.

We can leave the K\"ahler cone \eqref{wall_dP3_b_1} either by going to $q_2-q_1\leq 0$ or to $q_2-q_1\geq p$, changing sign to the effective $\xi_2$ at $\sigma<0$ or $\sigma>0$: we end up on the wall between
dictionaries $\{\{p[8]\},\,\{0, 0, 0, -1\}\}$ and $\{\{p[13]\},\,\{0, 0, 1, 1\}\}$, which gives the cone
\begin{equation}\label{wall_dP3_b_2}
\xi_2\leq 0\;, \;\; \xi_3\geq 0\;, \;\; \xi_6\geq 0\;,\;\;
\xi_1+\xi_4\leq 0\;,\;\; \xi_4\leq 0\;, \;\; \xi_2+\xi_5= 0
\end{equation}
in FI parameter space.

In the window $ q_2-q_1\leq 0$, the field theory is on the wall \eqref{wall_dP3_b_2} at $\sigma<0$ and and on the wall \eqref{wall_dP3_b_1} at $\sigma>0$.
The ranks and bare levels of the CS theory are
\begin{align}
\bm{N} &= (N,\,N+q_1-q_2,\,N,\,N,\,N-p,\,N)\\
\bm{k} &= (-\frac{p}{2}+q_3-q_4,\,\frac{p}{2}+q_1-q_2,\, \frac{p}{2}-\frac{q_1}{2}+\frac{q_2}{2}-q_3,\,\\ & \quad\;\; -p+\frac{q_1}{2}+\frac{q_2}{2}+q_4,\,-\frac{q_1}{2}+\frac{q_2}{2},\, \frac{p}{2}-\frac{q_1}{2}-\frac{q_2}{2})
\end{align}
so that the effective CS levels are
\begin{align} \label{k_eff_dP3_b_2}
-\bm{k}^- &= (-q_3+q_4,\,-q_1+q_2,\,q_3,\,-q_2-q_4,\,q_1-q_2,\,q_2)\\
+\bm{k}^+ &= (-p+q_3-q_4,\,p+q_1-q_2,\,p-q_1+q_2-q_3,\,-2p+q_1+q_4,\,0,\,p-q_1)\;.
\end{align}
The dictionaries we used are consistent when $ q_2-q_1\leq 0$ in addition to \eqref{geom_ineq_dP3}.

Conversely, in the window $ q_2-q_1\geq p$, the field theory is on the wall \eqref{wall_dP3_b_1} at $\sigma<0$ and and on the wall \eqref{wall_dP3_b_2} at $\sigma>0$.
The ranks and levels of the CS theory are
\begin{align}
\bm{N} &= (N,\,N-p-q_1+q_2,\,N,\,N,\,N-p,\,N)\\
\bm{k} &= (-\frac{p}{2}+q_3-q_4,\,\frac{p}{2}+q_1-q_2,\, p-\frac{q_1}{2}+\frac{q_2}{2}-q_3,\,\\ & \quad\;\; -\frac{3}{2}p+\frac{q_1}{2}+\frac{q_2}{2}+q_4,\,-\frac{p}{2}-\frac{q_1}{2}+\frac{q_2}{2},\, p-\frac{q_1}{2}-\frac{q_2}{2})
\end{align}
so that the effective CS levels are
\begin{align} \label{k_eff_dP3_b_3}
-\bm{k}^- &= (-q_3+q_4,\,-q_1+q_2,\,q_1-q_2+q_3,\,-q_1-q_4,\,0,\,q_2)\\
+\bm{k}^+ &= (-p+q_3-q_4,\,p+q_1-q_2,\,2p-q_3,\,-3p+q_2+q_4,\,-p-q_1+q_2,\,2p-q_2)\;.
\end{align}
The dictionaries we used are consistent when $ q_2-q_1\geq p$ in addition to \eqref{geom_ineq_dP3}.

Joining the three windows that we described in this subsection, we have provided M2-brane theories for the full class of $\Ypq(dP_3)$ geometries with \eqref{geom_ineq_dP3}.

One can similarly Seiberg dualise to toric phase a, where again several windows associated to pairs of consistent dictionaries are needed. We leave that as an exercise to the readers and move instead to singular toric Fano surfaces.

\subsection{$W\bC\bP^2_{[1,1,2]}$} \label{subsec_WP2_112}

The toric diagram of the 2-parameter family of toric $CY_4$ cones over $Y^{p,\,q}(W\bC\bP^2_{[1,1,2]})$, shown in Fig. \ref{fig: torDiag mod 04}, is the convex hull of
\begin{equation}\label{toric_diag_P2_112}
(0,0,0), \, (0,0,p),\, (0,1,2q),\, (1,0,q),\,(2,-1,0).
\end{equation}
We are interested in geometric parameters in the range
\begin{equation}\label{geom_ineq_WP2_112}
0 \leq \frac{q}{2}\leq p
\end{equation}
so that the points \eqref{toric_diag_P2_112} are all external. The geometries are identified under the $\bZ_2$ action $q\mapsto 2p-q$.
We further restricted the $CY_4$ geometries to avoid the presence in the IIA reduction of D6-branes along the exceptional divisor $\bP^1\times \bC$ corresponding to the toric point $(1,0)$ That is achieved by picking its vertical coordinate to be $q$ and the ones of $(1,-1)$ and $(1,1)$ to be $0$ and $2q$ respectively. The $CY_3$ fibre is only partially resolved because that exceptional $\bP^1$ vanishes.

\begin{figure}[h]
\begin{center}
\includegraphics[width=5.5cm]{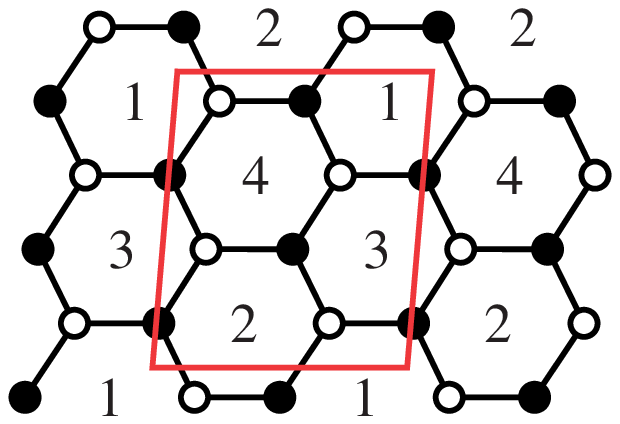}
\caption{\small Brane tiling for $W\bC\bP^2_{[1,1,2]}$.}\label{WP2_112_tiling}
\end{center}
\end{figure}
The brane tiling for D-branes at the complex cone over $W\bC\bP^2_{[1,1,2]}$ is in Fig. \ref{WP2_112_tiling} and has superpotential
\begin{equation}\label{W_WP2_112}
\begin{split}
W &= X_{12}^1 X_{24}^{} X_{41}^1 +X_{31}^{} X_{12}^2 X_{23}^1  +X_{34}^2 X_{42}^{} X_{23}^2 +X_{41}^2 X_{13}^{} X_{34}^1+   \\
& -X_{12}^1 X_{23}^2 X_{31}^{} -X_{13}^{} X_{34}^2 X_{41}^1 -X_{34}^1 X_{42}^{} X_{23}^1-X_{41}^2 X_{12}^2 X_{24}^{}   \;.
\end{split}
\end{equation}
The dimer model has 3 strictly external perfect matchings, $2$ internal-external perfect matchings $\{p[4],p[5]\}$ and $4$ strictly internal perfect matchings $\{p[6],\dots, p[9]\}$. There are 8 open string K\"ahler chambers, but we will not need to choose between $p[4]$ and $p[5]$ since the associated exceptional $\bP^1$ will remain blown down in the $CY_3$ fibre.

A quiver CS theory for the whole class of $C(Y^{p,\,q}(W\bC\bP^2_{[1,1,2]}))$ geometries can be proposed as follows. Both at $\sigma<0$ and $\sigma>0$ we are on the K\"ahler wall between the maximal dimensional chambers associated to perfect matchings $\{p[4],p[6]\}$, $\{p[4],p[7]\}$, $\{p[5],p[6]\}$ and $\{p[5],p[7]\}$, with suitable large volume monodromies.
The intersection of these 4 K\"ahler chambers is the cone
\begin{equation}\label{wall_WP2_112}
\xi_2=\xi_4 = 0\;,\;\; \xi_1=-\xi_3\geq 0
\end{equation}
in FI parameter space. Using any of these dictionaries, both at $\sigma<0$ and $\sigma>0$, we find that the 3d quiver theory has ranks and bare CS levels
\begin{align}
\bm{N} &= (N,\,N,\,N,\,N-p)\\
\bm{k} &= (p-q,\,0,-p+q,\,0 )
\end{align}
so that the effective CS levels are
\begin{align}
-\bm{k}^- &= (q,\,0,\, -q,\, 0)\\
+\bm{k}^+ &= (2p-q,\,0,\, -2p+q,\, 0)\;.
\end{align}
The geometric inequalities \eqref{geom_ineq_WP2_112} imply that the effective FI parameters belong to the cone \eqref{wall_WP2_112} where all the four dictionaries that can be used to derive the 3d theory are valid. This guarantees that the semiclassical computation of the geometric branch of the moduli space reproduces the type IIA geometry.

\subsection{$PdP_2$}\label{subsec:_PdP2}

The toric diagram of the 3-parameter family of toric $CY_4$ cones over $Y^{p,\,q_1,\,q_2}(PdP_2)$, shown in Fig. \ref{fig: torDiag mod 06}, is the convex hull of
\begin{equation}\label{toric_diag_PdP2}
(0,0,0), \, (0,0,p),\, (-1,0,0),\,  (0,1,q_1),\, (1,1,2q_2),\, (1,0,q_2),\,(1,-1,0).
\end{equation}
We are interested in geometric parameters in the range
\begin{equation}\label{geom_ineq_PdP2}
0 \leq  \frac{q_1}{3}, \frac{q_2}{2}\leq p
\end{equation}
so that the points \eqref{toric_diag_PdP2} are all external. The geometries are identified under $(q_1,q_2)\mapsto (3p-q_1,2p-q_2)$.
We restricted the $CY_4$ geometries to avoid D6-branes along noncompact toric divisors in the IIA background. The $CY_3$ fibre is only partially resolved.

The quiver diagram and brane tiling for D-branes at the complex cone over $PdP_2$ were shown in figures \ref{fig: PdP2 quiver0}-\ref{fig: PdP2 tiling}. The superpotential is \eqref{W_PdP2}. As reviewed in section \ref{subsec: toric quiver and resol}, the dimer model has 4 strictly external perfect matchings, $2$ internal-external perfect matchings $\{p[5],p[6]\}$ and $5$ strictly internal perfect matchings $\{p[7],\dots, p[11]\}$. There are 10 open string K\"ahler chambers, but we will not need to choose between $p[5]$ and $p[6]$ since the associated exceptional $\bP^1$ will remain blown down.

A quiver CS theory for the whole class of $C(Y^{p,\,q_1,\,q_2}(PdP_2))$ geometries can be proposed as follows. Both at $\sigma<0$ and $\sigma>0$ we are on the K\"ahler wall between the maximal dimensional chambers associated to perfect matchings $\{p[5],p[10]\}$, $\{p[6],p[10]\}$, $\{p[5],p[11]\}$ and $\{p[6],p[11]\}$ with suitable LVM's.
%&=
%\left(
%\begin{array}{ccccc}
% -1 & 1 & \frac{1}{2} & 0 & \frac{1}{8} \\
% 1 & -1 & -\frac{1}{2} & -1 & -\frac{1}{8} \\
% -1 & 1 & -\frac{1}{2} & 1 & \frac{9}{8} \\
% 0 & -1 & 0 & 0 & 0 \\
% 1 & 0 & \frac{1}{2} & 0 & -\frac{1}{8}
%\end{array}
%\right)
%\\
%\label{PdP2_dict_6-10}
%Q^\vee[PdP_2,\,\{p[6],p[10]\},\,\{0,0,0\}]&=
%\left(
%\begin{array}{ccccc}
% -1 & 1 & \frac{1}{2} & 1 & \frac{1}{8} \\
% 1 & -1 & -\frac{1}{2} & 0 & -\frac{1}{8} \\
% -1 & 1 & -\frac{1}{2} & 0 & \frac{9}{8} \\
% 0 & -1 & 0 & -1 & 0 \\
% 1 & 0 & \frac{1}{2} & 0 & -\frac{1}{8}
%\end{array}
%\right)
%\end{align}
%and
%\begin{align}
%\label{PdP2_dict_5-11}
%Q^\vee[PdP_2,\,\{p[5],p[11]\},\,\{-1,-1,0\}] &=
%\left(
%\begin{array}{ccccc}
% 1 & 1 & \frac{3}{2} & 0 & -\frac{1}{8} \\
% -1 & -1 & -\frac{3}{2} & -1 & \frac{1}{8} \\
% 0 & 1 & 0 & 1 & 1 \\
% -1 & -1 & -\frac{1}{2} & 0 & \frac{1}{8} \\
% 1 & 0 & \frac{1}{2} & 0 & -\frac{1}{8}
%\end{array}
%\right)
%\\
%\label{PdP2_dict_6-11}
%Q^\vee[PdP_2,\,\{p[6],p[11]\},\,\{-1,-1,-1\}]&=
%\left(
%\begin{array}{ccccc}
% 1 & 1 & \frac{3}{2} & 1 & -\frac{1}{8} \\
% -1 & -1 & -\frac{3}{2} & 0 & \frac{1}{8} \\
% 0 & 1 & 0 & 0 & 1 \\
% -1 & -1 & -\frac{1}{2} & -1 & \frac{1}{8} \\
% 1 & 0 & \frac{1}{2} & 0 & -\frac{1}{8}
%\end{array}
%\right)
%\;.
%\end{align}
The intersection of these 4 chambers is the cone
\begin{equation}\label{wall_PdP2}
\xi_1=-\xi_2\geq 0\;,\;\; \xi_1+\xi_3\geq 0\;,\;\; \xi_5= 0
\end{equation}
in FI parameter space. Using any of these dictionaries, both at $\sigma<0$ and $\sigma>0$, we find that the 3d quiver theory has ranks and bare CS levels
\begin{align}
\bm{N} &= (N,\,N,\,N,\,N,\,N-p)\\
\bm{k} &= (p-q_2,\,-p+q_2,\,\frac{1}{2}p-q_1+q_2,\,-\frac{1}{2}p+q_1-q_2,\,0   )
\end{align}
so that the effective CS levels are
\begin{align}
-\bm{k}^- &= (q_2,\,-q_2,\, q_1-q_2,\, -q_1+q_2,\, 0)\\
+\bm{k}^+ &= (2p-q_2,\,-2p+q_2,\, p-q_1+q_2,\, -p+q_1-q_2,\, 0)\;.
\end{align}
The geometric inequalities \eqref{geom_ineq_PdP2} imply that the effective FI parameters belong to the cone \eqref{wall_PdP2} where all the four dictionaries that can be used to derive the 3d theory are valid. This guarantees that the semiclassical computation of the geometric branch of the moduli space reproduces the type IIA geometry.

\subsection{$PdP_{3b}$}

The toric diagram of the 4-parameter family of toric $CY_4$ cones over $Y^{p,\,q_1,\,q_2,\,q_3}(PdP_{3b})$, shown in Fig. \ref{fig: torDiag mod 08}, is the convex hull of
\begin{equation}\label{toric_diag_PdP3b}
(0,0,0), \, (0,0,p),\, (2,-1,0), \, (1,-1,0) ,\, (-1,0,q_1),\, (-1,1,q_2),\, (0,1,2q_3),\, (1,0,q_3).
\end{equation}
We are interested in geometric parameters in the range
\begin{equation}\label{geom_ineq_PdP3b}
0 \leq  \frac{q_1+q_3}{2} , \frac{q_2}{2}, \frac{q_1+q_2}{3}, \frac{q_1+2q_3}{3} \leq p
\end{equation}
so that the points \eqref{toric_diag_PdP3b} are all external. We avoided D6-branes along noncompact toric divisors in type IIA. The $CY_3$ fibre is only partially resolved.

\subsubsection{Phase c of $PdP_{3b}$} \label{subsec:_PdP3b_phase_c}

\begin{figure}[h]
\begin{center}
\includegraphics[width=5.5cm]{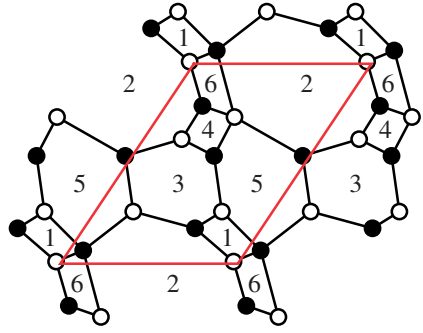}
\caption{\small Brane tiling for toric phase c of $PdP_{3b}$.}\label{PdP3b_c_tiling}
\end{center}
\end{figure}
Toric phase c of $PdP_{3b}$ is specified by the brane tiling of Fig. \ref{PdP3b_c_tiling}, with superpotential
\begin{equation}\label{W_PdP3b_c}
\begin{split}
W &= X_{21}^{} X_{16}^{} X_{62}^1 + X_{24}^{} X_{43}^{} X_{32}^1 + X_{25}^1 X_{53}^{} X_{32}^2 + X_{51}^{} X_{13}^{} X_{35}^{} +X_{54}^{} X_{46}^{} X_{62}^2 X_{25}^2   \\
& - X_{21}^{} X_{13}^{} X_{32}^2 - X_{24}^{} X_{46}^{} X_{62}^1 - X_{25}^2 X_{53}^{} X_{32}^1 -X_{54}^{} X_{43}^{} X_{35}^{} -  X_{51}^{} X_{16}^{} X_{62}^2 X_{25}^1
 \;.
\end{split}
\end{equation}
The dimer model has 5 strictly external perfect matchings, $2$ internal-external perfect matchings $\{p[6],p[7]\}$ and $8$ strictly internal perfect matchings $\{p[8],\dots, p[15]\}$. Correspondingly there are 32 open string K\"ahler chambers, but we will not need to choose between $p[6]$ and $p[7]$ since the associated exceptional $\bP^1$ will remain blown down.

A quiver CS theory for the whole class of $C(Y^{p,\,q_1,\,q_2,\,q_3}(PdP_{3b}))$ geometries can be proposed as follows. Both at $\sigma<0$ and $\sigma>0$ we are on the K\"ahler wall between the maximal dimensional chambers associated to internal perfect matchings $\{p[6],p[14]\}$, $\{p[7],p[14]\}$, $\{p[6],p[15]\}$ and $\{p[7],p[15]\}$, with suitable large volume monodromies in the dictionaries.
The intersection of these four K\"ahler chambers is the cone
\begin{equation}\label{wall_PdP3b_c}
\xi_2=\xi_3+\xi_5=0\;,\;\; \xi_5\geq 0\;,\;\;\xi_1+\xi_5\geq 0\;,\;\;\xi_4+\xi_5\geq 0\;,\;\;\xi_1+\xi_4\geq 0
\end{equation}
in FI parameter space. Using any of these dictionaries, both at $\sigma<0$ and $\sigma>0$, we find that the 3d quiver theory has ranks and bare CS levels
\begin{align}
\bm{N} &= (N,\,N-p,\,N,\,N,\,N,\,N)\\
\bm{k} &= (\frac{1}{2}p-q_3,\,0,\,-p+q_1+q_3,\,\frac{1}{2}p-q_2+q_3,\,p-q_1-q_3,\,-p+q_2)
\end{align}
so that the effective CS levels are
\begin{align}
-\bm{k}^- &= (q_3,\,0,\, -q_1-q_3,\, q_2-q_3,\, q_1+q_3, \,-q_2)\\
+\bm{k}^+ &= (p-q_3,\,0,\, -2p+q_1+q_3,\, p-q_2+q_3,\, 2p-q_1-q_3, \,-2p+q_2)\;.
\end{align}
The geometric inequalities \eqref{geom_ineq_PdP3b} imply that the effective FI parameters belong to the cone \eqref{wall_PdP3b_c} where all the four dictionaries that can be used to derive the 3d theory are valid. This guarantees that the semiclassical computation of the geometric branch of the moduli space reproduces the type IIA geometry.

A maximally chiral Seiberg duality on node 6 (a double $M_{(6;L)}$ mutation on the dictionaries) leads to a quiver CS theory which is just the CP-conjugate of the original one, up to a relabelling $(4,3)\leftrightarrow (1,5)$: without relabelling, ranks and levels are
\begin{align}
\bm{N} &= (N,\,N-p,\,N,\,N,\,N,\,N)\\
\bm{k} &= (-\frac{1}{2}p+q_2-q_3,\,0,\,-p+q_1+q_3,\,-\frac{1}{2}p+q_3,\,p-q_1-q_3,\,p-q_2)
\end{align}
and the superpotential is
\begin{equation}\label{W_PdP3b_c_dual}
\begin{split}
W &= x_{26}^{2} x_{64}^{} X_{43} X_{32}^1 + X_{25}^1 X_{53} X_{32}^2 + X_{51} X_{13} X_{35} + X_{54} M_{42}^2 X_{25}^2 +x_{61}^{} M_{12}^2 x_{26}^1 +   \\
& - x_{26}^2 x_{61} X_{13}^{} X_{32}^2 - X_{25}^2 X_{53}^{} X_{32}^1 -X_{54}^{} X_{43}^{} X_{35}^{} -  X_{51}^{} M_{12}^2 X_{25}^1 - x_{64} M_{42}^2 x_{26}^1
 \;.
\end{split}
\end{equation}

We can also reach toric phase b by a Seiberg duality on node 1 or 4. Let us consider duality on node 1 for definiteness. The duality is maximally chiral if $0\leq q_3\leq p$ (double right mutation $M_{(1;R)}$), whereas it is minimally chiral if $p < q_3\leq \frac{3}{2}p$ (mutations $M_{(1;R)}$ at $r_0<0$ and $M_{(1;L)}$ at $r_0>0$), so the resulting theory in phase b needs more than one window to cover the whole class of geometries \eqref{geom_ineq_PdP3b}. We leave this and further duality to phase a as an exercise to the interested reader.

\subsection{$PdP_{3c}$}\label{subsec:_PdP3c}

The toric diagram of the 3-parameter family of toric $CY_4$ cones over $Y^{p,\,q_1,\,q_2}(PdP_{3c})$, shown in Fig. \ref{fig: torDiag mod 09}, is the convex hull of
\begin{equation}\label{toric_diag_PdP3c}
\begin{split}
& (0,0,0), \, (0,0,p),\, (1,1,0),\, (1,0,0),\, (1,-1,0),\, \\
& (-1,0,q_1),\, (-1,1,2q_2),\, (0,1,q_2).
\end{split}
\end{equation}
We are interested in geometric parameters in the range
\begin{equation}\label{geom_ineq_PdP3c}
0 \leq  \frac{q_1}{2} ,  q_2  \leq p
\end{equation}
so that the points \eqref{toric_diag_PdP3c} are all external. We avoided D6-branes along noncompact toric divisors in the type IIA background.

\subsubsection{Phase b of $PdP_{3c}$} \label{subsec:_PdP3c_phase_b}
\begin{figure}[h]
\begin{center}
\includegraphics[width=5cm]{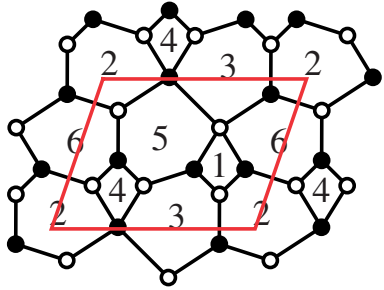}
\caption{\small Brane tiling for toric phase b of $PdP_{3c}$.}\label{PdP3c_b_tiling}
\end{center}
\end{figure}
Toric phase b of $PdP_{3c}$ is specified by the brane tiling of Fig. \ref{PdP3c_b_tiling}, with superpotential
\begin{equation}\label{W_PdP3c_b}
\begin{split}
W &= X_{31}^{} X_{12}^{} X_{23}^{} +X_{34}^{} X_{45}^{} X_{53}^2 +X_{56}^{} X_{62}^{} X_{25}^{} +X_{64}^{} X_{42}^{} X_{26}^{}  +X_{61}^{} X_{15}^{} X_{53}^1 X_{36}^{} + \\
& -X_{31}^{} X_{15}^{} X_{53}^2 -X_{36}^{} X_{62}^{} X_{23}^{} -X_{56}^{} X_{64}^{} X_{45}^{} -X_{61}^{} X_{12}^{} X_{26}^{}  -X_{25}^{} X_{53}^1 X_{34}^{} X_{42}^{}
 \;.
\end{split}
\end{equation}
The dimer model has 4 strictly external perfect matchings, $2$ pairs of internal-external perfect matchings $\{p[5],p[6]\}$ and $\{p[7],p[8]\}$, and $7$ strictly internal perfect matchings $\{p[9],\dots, p[15]\}$. Correspondingly there are 28 open string K\"ahler chambers, but we will not need to choose between internal-external perfect matching variables since the associated exceptional $\bP^1$'s will remain blown down.

A quiver CS theory for the whole class of $C(Y^{p,\,q_1,\,q_2}(PdP_{3c}))$ geometries can be proposed as follows.
Both at $\sigma<0$ and $\sigma>0$ we are on the K\"ahler wall between the maximal dimensional chambers associated to strictly internal perfect matchings
$p[11]$ and $p[15]$, as well as on the wall between $p[5]$ and $p[6]$ and between $p[7]$ and $p[8]$, with suitable large volume monodromies in the dictionaries.
The intersection of these eight K\"ahler chambers is the cone
\begin{equation}\label{wall_PdP3c_b}
\begin{split}
&\xi_3=\xi_1+\xi_4+\xi_5=\xi_2+\xi_4=0\;,\\
&\xi_1+\xi_5\leq 0\;,\;\;\xi_5\leq 0 \;,\;\;\xi_1+\xi_2+\xi_5\leq 0\;,\;\;\xi_2+\xi_5\leq 0\;.
\end{split}
\end{equation}
in FI parameter space. The 3d quiver theory has ranks and bare CS levels
\begin{align}
\bm{N} &= (N,\,N,\,N-p,\,N,\,N,\,N)\\
\bm{k} &= (\frac{p}{2}-q_1+q_2,\,-\frac{p}{2}+q_2,\,0,\,\frac{p}{2}-q_2,\,-p+q_1,\,\frac{p}{2} - q_2)
\end{align}
so that the effective CS levels are
\begin{align}
-\bm{k}^- &= (q_1-q_2,\,-q_2,\, 0,\, q_2,\, -q_1, \,q_2)\\
+\bm{k}^+ &= (p-q_1+q_2,\,-p+q_2,\, 0,\, p-q_2,\, -2p+q_1, \,p-q_2)\;.
\end{align}
The geometric inequalities \eqref{geom_ineq_PdP3c} ensure that the effective FI parameters belong to the cone \eqref{wall_PdP3c_b} where all the eight dictionaries that can be used to derive the 3d theory are valid. Consequently the semiclassical computation of the geometric branch of the moduli space reproduces the type IIA geometry.

A single proposal for all the geometries can be made in toric phase a too: it is obtained by a maximally chiral Seiberg duality on node 4.

\subsection{$W\bC\bP^2_{[1,2,3]}$} \label{subsec:_WP2_123}

The toric diagram of the 2-parameter family of toric $CY_4$ cones over $Y^{p,\,q}(W\bC\bP^2_{[1,2,3]})$, shown in Fig. \ref{fig: torDiag mod 10}, is the convex hull of
\begin{equation}\label{toric_diag_P2_123}
(0,0,0), \, (0,0,p),\, (-1,0,0),\, (-1,1,0),\,(-1,2,0),\,
(0,1,q),\, (1,0,2q),\,(2,-1,3q).
\end{equation}
We impose that all the points \eqref{toric_diag_P2_123} are external:
\begin{equation}\label{geom_ineq_WP2_123}
0 \leq q \leq p\;.
\end{equation}
The geometries are identified under the $\bZ_2$ action $q\mapsto p-q$.
We restricted the $CY_4$ geometries to avoid the presence in the reduction to type IIA of D6-branes along noncompact toric divisors.

\begin{figure}[h]
\begin{center}
\includegraphics[width=6cm]{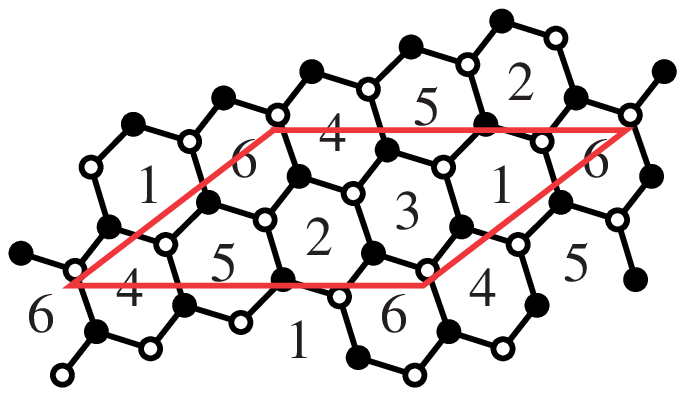}
\caption{\small Brane tiling for $W\bC\bP^2_{[1,2,3]}$.}\label{WP2_123_tiling}
\end{center}
\end{figure}
The brane tiling for D-branes at the complex cone over $W\bC\bP^2_{[1,2,3]}$ is in Fig. \ref{WP2_123_tiling} and has superpotential
\begin{equation}\label{W_WP2_123}
\begin{split}
W &= X_{12}^{} X_{26}^{} X_{61}^{} +X_{24}^{} X_{43}^{} X_{32}^{} +X_{35}^{} X_{51}^{} X_{13}^{} + X_{41}^{} X_{15}^{} X_{54}^{} + X_{56}^{} X_{62}^{} X_{25}^{} +\\
&+ X_{63}^{} X_{34}^{} X_{46}^{} -X_{12}^{} X_{25}^{} X_{51}^{} -X_{24}^{} X_{46}^{} X_{62}^{} -X_{35}^{} X_{54}^{} X_{43}^{} -X_{41}^{} X_{13}^{} X_{34}^{} +\\
&-X_{56}^{} X_{61}^{} X_{15}^{} -X_{63}^{} X_{32}^{} X_{26}^{}     \;.
\end{split}
\end{equation}
The dimer model has 3 strictly external perfect matchings, internal-external perfect matchings $\{p[4],p[5]\}$, $\{p[6],p[7],p[8]\}$ and $\{p[9],p[10],p[11]\}$, and $6$ strictly internal perfect matchings $\{p[12],\dots, p[17]\}$. There are 108 open string K\"ahler chambers, but we will not need to choose between internal-external perfect matching variables associated to the same lattice point, since the associated exceptional $\bP^1$ will remain blown down in the $CY_3$ fibre.

A quiver CS theory for the whole class of $C(Y^{p,\,q}(W\bC\bP^2_{[1,2,3]}))$ geometries has ranks and bare CS levels
\begin{align}
\bm{N} &= (N,\,N,\,N,\,N,\,N,\,N-p)\\
\bm{k} &= (\frac{p}{2}-q,\,0,\,\frac{p}{2}-q,\,-\frac{p}{2}+q,\,-\frac{p}{2}+q,\,0)
\end{align}
so that the effective CS levels are
\begin{align}
-\bm{k}^- &= (q,\,0,\,q,\, -q,\, -q, \, 0) \\
+\bm{k}^+ &= (p-q,\,0,\,p-q,\, -p+q,\, -p+q, \, 0)   \;.
\end{align}
The geometric inequalities \eqref{geom_ineq_WP2_123} guarantee that the semiclassical computation of the geometric branch of the moduli space reproduces the type IIA geometry.

\subsection{$PdP_4$}\label{subsec:_PdP4}

The toric diagram of the 4-parameter family of toric $CY_4$ cones over $Y^{p,\,q_1,\,q_2,\,q_3}(PdP_4)$, shown in Fig. \ref{fig: torDiag mod 11}, is the convex hull of
\begin{equation}\label{toric_diag_PdP4}
\begin{split}
& (0,0,0), \, (0,0,p),\, (1,1,0),\, (1,0,0),\, (1,-1,0),\, \\
& (0,-1,q_1),\, (-1,0,q_2),\, (-1,1,2q_3),\, (0,1,q_3).
\end{split}
\end{equation}
We are interested in geometric parameters in the range
\begin{equation}\label{geom_ineq_PdP4}
0 \leq  \frac{q_1+q_3}{2} , \frac{q_2}{2}, q_3 , \frac{q_1+q_2}{3}  \leq p
\end{equation}
so that the points \eqref{toric_diag_PdP4} are all external. We avoided D6-branes along noncompact toric divisors in the type IIA background.

\subsubsection{Phase c of $PdP_4$} \label{subsec:_PdP4_phase_c}

\begin{figure}[h]
\begin{center}
\includegraphics[width=4.6cm]{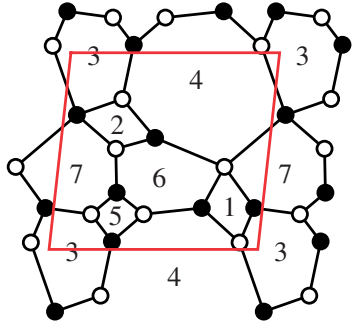}
\caption{\small Brane tiling for toric phase c of $PdP_4$.}\label{PdP4_c_tiling}
\end{center}
\end{figure}
Toric phase c of $PdP_4$ is specified by the brane tiling of Fig. \ref{PdP4_c_tiling}, with superpotential
\begin{equation}\label{W_PdP4}
\begin{split}
W &= X_{41}^{} X_{13}^{} X_{34}^1 + X_{42}^{} X_{23}^{} X_{34}^2  +X_{45}^{} X_{56}^{} X_{64}^1 +X_{67}^{} X_{72}^{} X_{26}^{} +X_{75}^{} X_{53}^{} X_{37}^{} +  \\
& -X_{41}^{} X_{16}^{} X_{64}^1 -X_{42}^{} X_{26}^{} X_{64}^2 -X_{45}^{} X_{53}^{} X_{34}^2 -X_{67}^{} X_{75}^{} X_{56}^{} -X_{71}^{} X_{13}^{} X_{37}^{} +\\
& + X_{47}^{} X_{71}^{} X_{16}^{} X_{64}^2 -X_{47}^{} X_{72}^{} X_{23}^{}X_{34}^1
 \;.
\end{split}
\end{equation}

The dimer model has 5 strictly external perfect matchings, $2$ pairs of internal-external perfect matchings $\{p[6],p[7]\}$ and $\{p[8],p[9]\}$, and $12$ strictly internal perfect matchings $\{p[10],\dots, p[21]\}$. Correspondingly there are 48 open string K\"ahler chambers, but we will not need to choose between internal-external perfect matching variables since the associated exceptional $\bP^1$'s will remain blown down.

A quiver CS theory for the whole class of $C(Y^{p,\,q_1,\,q_2,\,q_3}(PdP_4))$ geometries can be proposed as follows.
Both at $\sigma<0$ and $\sigma>0$ we are on the K\"ahler wall between the maximal dimensional chambers associated to strictly internal perfect matchings $p[10]$ and $p[21]$, as well as on the wall between $p[6]$ and $p[7]$ and between $p[8]$ and $p[9]$, with suitable large volume monodromies in the dictionaries.
The intersection of these eight K\"ahler chambers is the cone
\begin{equation}\label{wall_PdP4_c}
\begin{split}
&\xi_4=\xi_1+\xi_5+\xi_6=\xi_2+\xi_3+\xi_5=0\;,\qquad\xi_3\leq 0\;,\;\;\xi_6\leq 0 \;,\\
&\xi_1+\xi_2+\xi_3+\xi_6\leq 0\;,\;\;\xi_1+\xi_3+\xi_6\leq 0\;,\;\;\xi_3+\xi_5+\xi_6\leq 0\;,\;\;\xi_2+\xi_3\leq 0   \;.
\end{split}
\end{equation}
in FI parameter space. The 3d quiver theory has ranks and bare CS levels
\begin{align}
\bm{N} &= (N,\,N,\,N,\,N-p,\,N,\,N,\,N)\\
\bm{k} &= (\frac{1}{2}p-q_2+q_3,\, \frac{1}{2}p-q_1 ,\,-p+q_1+q_3,\,0,\,\frac{1}{2}p-q_3,\,-p+q_2,\,\frac{1}{2}p-q_3)
\end{align}
so that the effective CS levels are
\begin{align}
-\bm{k}^- &= (q_2-q_3,\,q_1,\, -q_1-q_3,\,0,\, q_3,\, -q_2, \,q_3)\\
+\bm{k}^+ &= (p-q_2+q_3,\,p-q_1,\, -2p+q_1+q_3,\,0,\, p-q_3,\, -2p+q_2, \,p-q_3)\;.
\end{align}
The geometric inequalities \eqref{geom_ineq_PdP4} ensure that the effective FI parameters belong to the cone \eqref{wall_PdP4_c} where all the eight dictionaries that can be used to derive the 3d theory are valid. Consequently the semiclassical computation of the geometric branch of the moduli space reproduces the type IIA geometry.

A single proposal for all the geometries can be made in toric phase a too: it is obtained by a maximally chiral Seiberg duality on node 5. More windows are needed in phase b.

\subsection{$PdP_{4b}$} \label{subsec:_model_V}

The toric diagram of the 3-parameter family of toric $CY_4$ cones over $Y^{p,\,q_1,\,q_2}(PdP_{4b})$, shown in Fig. \ref{fig: torDiag mod 12}, is the convex hull of
\begin{equation}\label{toric_diag_V}
\begin{split}
&(0,0,0), \, (0,0,p),\, (-1,2,0),\,(0,1,0),\,(1,0,0),\,(2,-1,0),\,\\
& (-1,1,q_1),\,(-1,0,2q_2),\,(-1,1,q_2)\,.
\end{split}
\end{equation}
We impose that all the points \eqref{toric_diag_V} are external:
\begin{equation}\label{geom_ineq_V}
0 \leq \frac{q_1+q_2}{2}, q_2 \leq p\;.
\end{equation}
The geometries are identified under the $\bZ_2$ action $(q_1,q_2)\mapsto (p-q_1,p-q_2)$.
We avoided D6-branes along noncompact toric divisors in IIA.

\begin{figure}[h]
\begin{center}
\includegraphics[width=4.5cm]{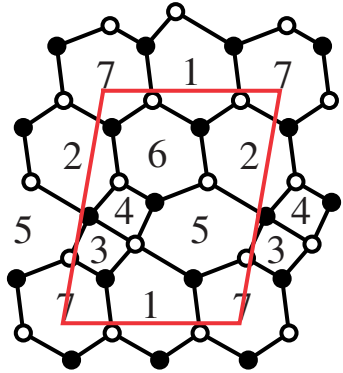}
\caption{\small Brane tiling for $PdP_{4b}$.}\label{V_tiling}
\end{center}
\end{figure}
The brane tiling for D-branes at the complex cone over $PdP_{4b}$ is in Fig. \ref{V_tiling} and has superpotential
\begin{equation}\label{W_V}
\begin{split}
W &= X_{21}^{} X_{17}^{} X_{72}^{} +X_{42}^{} X_{26}^{} X_{64}^{} +X_{56}^{} X_{62}^{} X_{25}^{}  +X_{67}^{} X_{71}^{} X_{16}^{} +  \\
&  +X_{75}^{} X_{53}^{} X_{37}^{} +X_{13}^{} X_{34}^{} X_{45}^{} X_{51}^{}
-X_{13}^{} X_{37}^{} X_{71}^{} -X_{16}^{} X_{62}^{} X_{21}^{} +\\
& -X_{56}^{} X_{64}^{} X_{45}^{} -X_{67}^{} X_{72}^{} X_{26}^{} -X_{75}^{} X_{51}^{} X_{17}^{}-X_{25}^{} X_{53}^{} X_{34}^{} X_{42}^{}
  \;.
\end{split}
\end{equation}
The dimer model has $4$ strictly external perfect matchings, $2+3+3$ internal-external perfect matchings and $9$ strictly internal perfect matchings.

The quiver CS theory for the whole class of $C(Y^{p,\,q_1,\,q_2}(PdP_{4b}))$ geometries  has ranks and bare CS levels
\begin{align}
\bm{N} &= (N,\,N,\,N,\,N,\,N-p,\,N,\,N)\\
\bm{k} &= (\frac{p}{2} - q_2 ,\,-\frac{p}{2}+q_2,\,\frac{p}{2} - q_1,\,-\frac{p}{2}+q_1,\,0,\,\frac{p}{2} - q_2,\,-\frac{p}{2}+q_2)
\end{align}
so that the effective CS levels are
\begin{align}
-\bm{k}^- &= (q_2,\,-q_2,\,q_1,\, -q_1,\,0,\, q_2, \, -q_2) \\
+\bm{k}^+ &= (p-q_2,\,-p+q_2,\,p-q_1,\, -p+q_1,\,0,\, p-q_2, \, -p+q_2)  \;.
\end{align}
The geometric inequalities \eqref{geom_ineq_V} guarantee that the semiclassical computation of the geometric branch of the moduli space reproduces the type IIA geometry.

\subsection{$PdP_5$} \label{subsec:_PdP5}

The toric diagram of the 3-parameter family of toric $CY_4$ cones over $Y^{p,\,q_1,\,q_2}(PdP_5)$, shown in Fig. \ref{fig: torDiag mod 13}, is the convex hull of
\begin{equation}\label{toric_diag_PdP5}
\begin{split}
&(0,0,0), \, (0,0,p),\, (1,1,0),\,(1,0,0),\,(1,-1,0),\,(0,-1,q_1),\,\\
& (-1,-1,2q_1),\,(-1,0,q_1+q_2),\,(-1,1,2q_2),\,(0,1,q_1).
\end{split}
\end{equation}
We impose that all the points \eqref{toric_diag_PdP5} are external:
\begin{equation}\label{geom_ineq_PdP5}
0 \leq q_1, q_2 \leq p\;.
\end{equation}
The $CY_4$ geometries are identified under the $\bZ_2$ action $(q_1,q_2)\mapsto (p-q_1,p-q_2)$.
We restricted them to avoid D6-branes along noncompact toric divisors in IIA.

There are $4$ toric phases for the D-brane quiver gauge theory. Using each of them it is possible to find a single proposal for a CS quiver gauge theory that is valid for the whole class of $CY_4$ geometries. In the following we will present only toric phase a.

\begin{figure}[h]
\begin{center}
\includegraphics[width=4.5cm]{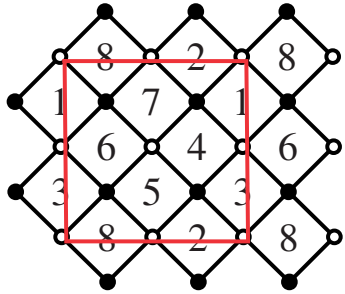}
\caption{\small Brane tiling for toric phase a of $PdP_5$.}\label{PdP5_a_tiling}
\end{center}
\end{figure}
The brane tiling in phase a for D-branes at the complex cone over $PdP_5$, which is a $\bZ_2\times\bZ_2$ orbifold of the conifold, is in Fig. \ref{PdP5_a_tiling} and has superpotential
\begin{equation}\label{W_PdP5_a}
\begin{split}
W &= X_{23}^{} X_{38}^{} X_{81}^{} X_{12}^{} +X_{41}^{} X_{16}^{} X_{63}^{} X_{34}^{} +X_{67}^{} X_{74}^{} X_{45}^{} X_{56}^{} +X_{85}^{} X_{52}^{} X_{27}^{} X_{78}^{} +\\
& -X_{27}^{}X_{74}^{}X_{41}^{}X_{12}^{}
-X_{45}^{}X_{52}^{}X_{23}^{}X_{34}^{}
-X_{63}^{}X_{38}^{}X_{85}^{}X_{56}^{}
-X_{81}^{}X_{16}^{}X_{67}^{}X_{78}^{}
 \;.
\end{split}
\end{equation}
The dimer model has $4$ strictly external perfect matchings, $2+2+2+2$ internal-external perfect matchings and $12$ strictly internal perfect matchings.
The quiver CS theory for the whole class of $C(Y^{p,\,q_1,\,q_2}(PdP_5))$ geometries  has ranks and bare CS levels
\begin{align}
\bm{N} &= (N,\,N,\,N,\,N,\,N,\,N,\,N,\,N-p)\\
\bm{k} &= (\frac{p}{2}- q_1 ,\,q_1-q_2,\,-\frac{p}{2}+q_2,\,0,\,\frac{p}{2}-q_1,\,q_1-q_2,\,-\frac{p}{2}+q_2, \,0)
\end{align}
so that the effective CS levels are
\begin{align}
-\bm{k}^- &= (q_1,\,-q_1+q_2,\,-q_2,\,0,\,q_1,\,-q_1+q_2,\,-q_2,\,0 ) \\
+\bm{k}^+ &= (p-q_1,\,q_1-q_2,\,-p+q_2,\,0,\,p-q_1,\,q_1-q_2,\,-p+q_2,\,0)  \;.
\end{align}
The geometric inequalities \eqref{geom_ineq_PdP5} guarantee that the semiclassical computation of the geometric branch of the moduli space reproduces the type IIA geometry.

\subsection{$PdP_{5b}$} \label{subsec:_PdP5b}

The toric diagram of the 3-parameter family of toric $CY_4$ cones over $Y^{p,\,q_1,\,q_2}(PdP_{5b})$, shown in Fig. \ref{fig: torDiag mod 14}, is the convex hull of
\begin{equation}\label{toric_diag_PdP5b}
\begin{split}
&(0,0,0), \, (0,0,p),\, (-1,1,0),\,(0,1,0),\,(1,1,0),\,(2,1,0),\,\\
& (1,0,q_1),\,(0,-1,2q_1),\,(-1,-1,2q_2),\,(-1,0,q_2).
\end{split}
\end{equation}
We impose that all the points \eqref{toric_diag_PdP5b} are external:
\begin{equation}\label{geom_ineq_PdP5b}
0 \leq q_1, q_2 \leq p\;.
\end{equation}
The $CY_4$ geometries are identified under the $\bZ_2$ action $(q_1,q_2)\mapsto (p-q_1,p-q_2)$.
We restricted them to avoid D6-branes along noncompact toric divisors in IIA.

The D-brane quiver gauge theory has 2 toric phases. Here we only introduce phase b, which allows us to propose a theory that is valid for the whole class of geometries.

\begin{figure}[h]
\begin{center}
\includegraphics[width=4.5cm]{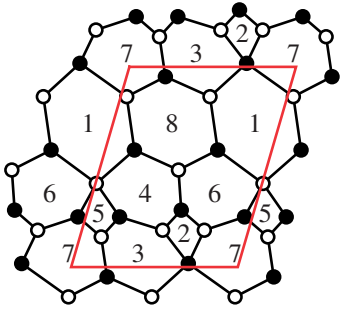}
\caption{\small Brane tiling for toric phase b of $PdP_{5b}$.}\label{PdP5b_b_tiling}
\end{center}
\end{figure}
The brane tiling in phase a for D-branes at the complex cone over $PdP_{5b}$, which is a $\bZ_2$ orbifold of the real cone over the $SE_5$ $L^{1,\,3,\,1}$, is in Fig. \ref{PdP5b_b_tiling} and has superpotential
\begin{equation}\label{W_PdP5b_b}
\begin{split}
W &= X_{31}^{}X_{18}^{}X_{83}^{} +X_{42}^{}X_{23}^{}X_{34}^{} +X_{53}^{}X_{37}^{}X_{75}^{} +X_{67}^{}X_{72}^{}X_{26}^{} +X_{78}^{}X_{81}^{}X_{17}^{} +\\
& +X_{86}^{}X_{64}^{}X_{48}^{} +X_{14}^{}X_{45}^{}X_{56}^{}X_{61}^{} -X_{14}^{}X_{48}^{}X_{81}^{} -X_{42}^{}X_{26}^{}X_{64}^{} -X_{53}^{} X_{34}^{} X_{45}^{} \\
&  -X_{67}^{}X_{75}^{}X_{56}^{} -X_{78}^{}X_{83}^{}X_{37}^{} -X_{86}^{}X_{61}^{}X_{18}^{} -X_{17}^{}X_{72}^{}X_{23}^{}X_{31}^{}
 \;.
\end{split}
\end{equation}
The dimer model has $4$ strictly external perfect matchings, $3+3+2+2$ internal-external perfect matchings and $14$ strictly internal perfect matchings.
The quiver CS theory for the whole class of $C(Y^{p,\,q_1,\,q_2}(PdP_{5b}))$ geometries  has ranks and bare CS levels
\begin{align}
\bm{N} &= (N-p,\,N,\,N,\,N,\,N,\,N,\,N,\,N)\\
\bm{k} &= (0,\,-q_1+q_2,\,-\frac{p}{2}+q_1,\,\frac{p}{2}- q_1, \,q_1-q_2,\,-\frac{p}{2}+q_2, \,\frac{p}{2}-q_2,\,0)
\end{align}
so that the effective CS levels are
\begin{align}
-\bm{k}^- &= (0,\, q_1-q_2,\, -q_1,\, q_1,\, -q_1+q_2,\, -q_2,\, q_2,\, 0) \\
+\bm{k}^+ &= (0,\,-q_1=q_2,\,-p+q_1,\,p-q_1,\,q_1-q_2,\,-p+q_2,\,p-q_2,\,0)  \;.
\end{align}
The geometric inequalities \eqref{geom_ineq_PdP5b} guarantee that the semiclassical computation of the geometric branch of the moduli space reproduces the type IIA geometry.

\subsection{$W\bC\bP^2_{[2,2,4]}$} \label{subsec:_WP2_224}

The toric diagram of the 2-parameter family of toric $CY_4$ cones over $Y^{p,\,q}(W\bC\bP^2_{[2,2,4]})$, shown in Fig. \ref{fig: torDiag mod 15}, is the convex hull of
\begin{equation}\label{toric_diag_P2_224}
\begin{split}
&(0,0,0), \, (0,0,p),\, (-1,1,0),\,(0,1,0),\, (1,1,0),\,(2,1,0),\,(3,1,0),\\
& (1,0,q),\, (-1,-1,2q),\, (-1,0,q)\,.
\end{split}
\end{equation}
We impose that all the points \eqref{toric_diag_P2_224} are external:
\begin{equation}\label{geom_ineq_WP2_224}
0 \leq q \leq p\;.
\end{equation}
The geometries are identified under the $\bZ_2$ action $q\mapsto p-q$.
We restricted the $CY_4$ geometries to avoid D6-branes along noncompact toric divisors in type IIA.

\begin{figure}[h]
\begin{center}
\includegraphics[width=4.5cm]{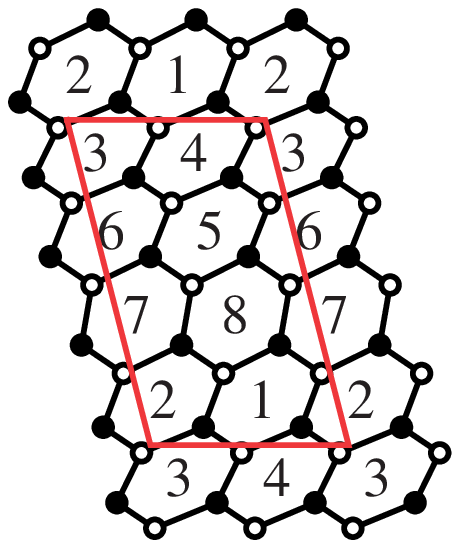}
\caption{\small Brane tiling for $W\bC\bP^2_{[2,2,4]}$.}\label{WP2_224_tiling}
\end{center}
\end{figure}
The brane tiling for D-branes at the complex cone over $W\bC\bP^2_{[2,2,4]}$ is in Fig. \ref{WP2_224_tiling} and has superpotential
\begin{equation}\label{W_WP2_224}
\begin{split}
W &= X_{17}^{}X_{72}^{}X_{21}^{} +X_{28}^{}X_{81}^{}X_{12}^{} +X_{31}^{}X_{14}^{}X_{43}^{} +X_{42}^{}X_{23}^{}X_{34}^{} +\\ &+X_{53}^{}X_{36}^{}X_{65}^{} +X_{64}^{}X_{45}^{}X_{56}^{} +X_{75}^{}X_{58}^{}X_{87}^{} +X_{86}^{} X_{67}^{} X_{78}^{}+\\
&-X_{17}^{}X_{78}^{}X_{81}^{} -X_{28}^{}X_{87}^{}X_{72}^{} -X_{31}^{}X_{12}^{}X_{23}^{} -X_{42}^{}X_{21}^{}X_{14}^{} +\\
&-X_{53}^{}X_{34}^{}X_{45}^{} -X_{64}^{}X_{43}^{}X_{36}^{} -X_{75}^{}X_{56}^{}X_{67}^{} -X_{86}^{}X_{65}^{}X_{58}^{}
    \;.
\end{split}
\end{equation}
A quiver CS theory for the $C(Y^{p,\,q}(W\bC\bP^2_{[2,2,4]}))$ geometries has ranks and bare CS levels
\begin{align}
\bm{N} &= (N,\,N,\,N-p,\,N,\,N,\,N,\,N,\,N)\\
\bm{k} &= (\frac{p}{2}-q,\,-\frac{p}{2}+q,\,0,\,0,\,-\frac{p}{2}+q, \,\frac{p}{2}-q,\,0,\,0)
\end{align}
so that the effective CS levels are
\begin{align}
-\bm{k}^- &= ( q,\, -q,\, 0 , \,0,\, - q ,\, q ,\, 0 ,\,0) \\
+\bm{k}^+ &= ( p-q,\, -p+q,\, 0 , \,0,\, -p+q ,\, p-q ,\, 0 ,\,0)   \;.
\end{align}
The geometric inequalities \eqref{geom_ineq_WP2_224} guarantee that the semiclassical computation of the geometric branch of the moduli space reproduces the type IIA geometry.

\subsection{$W\bC\bP^2_{[3,3,3]}$} \label{subsec:_WP2_333}

The toric diagram of the 2-parameter family of toric $CY_4$ cones over $Y^{p,\,q}(W\bC\bP^2_{[3,3,3]})$, shown in Fig. \ref{fig: torDiag mod 16}, is the convex hull of
\begin{equation}\label{toric_diag_P2_333}
\begin{split}
&(0,0,0), \, (0,0,p),\, (-1,-1,0),\,(-1,0,0),\, (-1,1,0),\,(-1,2,0),\,\\
& (0,1,q),\, (1,0,2q),\,(2,-1,3q),\,(1,-1,2q),\,(0,-1,q)\,.
\end{split}
\end{equation}
We impose that all the points \eqref{toric_diag_P2_333} are external:
\begin{equation}\label{geom_ineq_WP2_333}
0 \leq q \leq p\;.
\end{equation}
The geometries are identified under the $\bZ_2$ action $q\mapsto p-q$.
We restricted the $CY_4$ geometries to avoid D6-branes along noncompact toric divisors in type IIA.

\begin{figure}[h]
\begin{center}
\includegraphics[width=5.5cm]{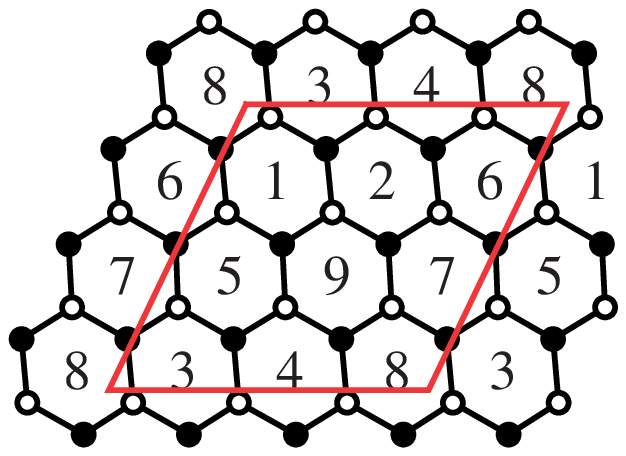}
\caption{\small Brane tiling for $W\bC\bP^2_{[3,3,3]}$.}\label{WP2_333_tiling}
\end{center}
\end{figure}
The brane tiling for D-branes at the complex cone over $W\bC\bP^2_{[3,3,3]}$ is in Fig. \ref{WP2_333_tiling} and has superpotential
\begin{equation}\label{W_WP2_333}
\begin{split}
W &= X_{15}^{} X_{56}^{} X_{61}^{}  +X_{29}^{} X_{91}^{} X_{12}^{}  +X_{31}^{} X_{18}^{} X_{83}^{} +X_{42}^{} X_{23}^{} X_{34}^{} + X_{53}^{} X_{37}^{} X_{75}^{}+  \\
& +X_{67}^{} X_{72}^{} X_{26}^{}  +X_{78}^{} X_{89}^{} X_{97}^{}+X_{86}^{}X_{64}^{}X_{48}^{} +X_{94}^{}X_{45}^{}X_{59}^{} -X_{15}^{}X_{59}^{}X_{91}^{}+      \\
&  -X_{29}^{}X_{97}^{}X_{72}^{} -X_{31}^{}X_{12}^{}X_{23}^{} -X_{42}^{}X_{26}^{} X_{64}^{} -X_{53}^{}X_{34}^{}X_{45}^{}-X_{67}^{}X_{75}^{}X_{56}^{} +\\
& -X_{78}^{} X_{83}^{} X_{37}^{}-X_{86}^{} X_{61}^{} X_{18}^{} -X_{94}^{} X_{48}^{} X_{89}^{}  \;.
\end{split}
\end{equation}
A quiver CS theory for the whole class of $C(Y^{p,\,q}(W\bC\bP^2_{[3,3,3]}))$ geometries has ranks and bare CS levels
\begin{align}
\bm{N} &= (N,\,N,\,N,\,N,\,N,\,N,\,N,\,N-p,\,N)\\
\bm{k} &= (-\frac{p}{2}+q,\,0,\,\frac{p}{2}-q,\,-\frac{p}{2}+q,\,0,\,\frac{p}{2} -q,\,-\frac{p}{2}+q,\,0,\,\frac{p}{2}-q)
\end{align}
so that the effective CS levels are
\begin{align}
-\bm{k}^- &= ( -q,\, 0 ,\, q ,\, -q ,\, 0 ,\, q ,\,-q ,\,0 ,\, q) \\
+\bm{k}^+ &= ( -p+q,\, 0 ,\,p-q ,\,-p+q ,\, 0 ,\, p-q ,\,-p+q ,\,0 ,\,p-q)   \;.
\end{align}
The geometric inequalities \eqref{geom_ineq_WP2_333} guarantee that the semiclassical computation of the geometric branch of the moduli space reproduces the type IIA geometry.

\section{Partial resolutions and Higgsing}\label{sec: par resol and Higgsing}

In this section we study partial or complete resolutions of the $CY_4$ cones over $\Ypq(B_4)$ from the perspective of the moduli space of the M2-brane field theories that we proposed. Since for any toric Fano $B_4$ we found at least a toric phase in which the conformal field theory could be derived using the same dictionaries at $r_0$ negative and positive, we restrict our attention to such models. Then the ranks of the gauge groups all equal $N$, except for the group associated to the D6 wrapped on $B_4$, whose rank is $N-p$, and which we relabel to be node 1 in the quiver. Following \cite{Benini:2011cma}, we can generalise the analysis of the geometric branch of the moduli space in section  \ref{section: IIA to quiver}, allowing bare FI parameters $\xi_i$ and common extra $p$ eigenvalues for the $G-1$ $U(N)$ gauge groups:
\be
\begin{split}
\sigma_1 &= \text{diag}(\sigma_1, \cdots, \sigma_{N-p})\\
\sigma_i &= \text{diag}(\sigma_1, \cdots, \sigma_{N-p}, \tilde{\sigma}_1, \cdots, \tilde{\sigma}_p)\, ,\qquad  \forall\, i=2,\dots,p\, .
\end{split}
\ee
The extra eigenvalues $\tilde{\sigma}_{\tilde{a}}$ of $\sigma_i$ parametrise the location of the $p$ wrapped $\overline{\mathrm{D6}}$-branes along the $\sigma=r_0$ real line and  modify the real masses of bifundamental matter fields, low energy modes of open strings stretched between D-branes separated in the $\bR$ direction.
Integrating out these massive chiral multiplets shifts the CS levels at one loop and affects the effective FI parameters. The 1-loop effective FI parameters for the Cartan $U(1)$'s are
\be
\begin{split}
\xi^{eff}_1(\sigma_a) &= \xi_1 \\
\xi^{eff}_i(\sigma_a) &=
\xi_i + k_i \sigma_a + \frac{1}{2} A_{i1} \sum_{\tilde{b}=1}^p
|\sigma_a-\tilde{\sigma}_{\tilde{b}}|
\qquad  (i=2,\dots,p) \\
\xi^{eff}_i(\tilde{\sigma}_{\tilde{a}}) &= \xi^{eff}_i(\sigma_a)|_{\;\sigma_a\to \tilde{\sigma}_{\tilde{a}}} \qquad\qquad  (i=2,\dots,p)\;.
\end{split}
\ee
These effective FI parameters appear in D-term equations involving chiral superfields with vanishing real mass. The D-term equations split into two sets: those for the first $N-p$ eigenvalues, each one giving a $U(1)^G$ quiver of massless fields which describes a regular D2-brane probing the type IIA geometry; those for the extra $p$ eigenvalues, each one giving a $U(1)^{G-1}$ quiver of massless fields, lacking node 1 and the fields charged under it, which describes a D2-$\overline{\mathrm{D6}}$ bound state with a mobile D2 on the $\overline{\mathrm{D6}}$ \cite{Benini:2011cma}.

Let us consider the $U(1)^{G-1}$ quiver first. Since no matter is charged under its diagonal $U(1)$, a necessary conditions for having D-flat solutions is
\be
0 = \sum_{i=2}^G \xi_i^{eff}(\tilde{\sigma}_{\tilde{a}}) = \sum_{i=2}^G \xi_i= \sum_{i=2}^G \xi_i^{eff}(\sigma_a) \quad \Longrightarrow \quad \xi_1=0\;.
\ee
Demanding that the D-terms of the $U(1)^{G-1}$ subquiver can be solved restricts the effective FI parameters $\xi^{eff}_i(\tilde{\sigma}_{\tilde{a}})$, $i=2,\dots,G$, to belong to its K\"ahler cone, a subcone of the K\"ahler cone of the full $U(1)^G$ quiver theory.

We leave a general analysis of the K\"ahler cone of the $U(1)^{G-1}$ subquivers and of the full nonabelian theory to future work, and we content ourselves here with a simple example of such an analysis.
Let us then consider the field theory for M2-branes probing $C(Y^{p,\,q_1,\,q_2}(\bF_0))$ of section \ref{subsec:_F0_a} as a concrete example. The D-term equations for the $U(1)^3$ subquiver for a D2-$\overline{\mathrm{D6}}$ bound state read
\begin{align}
\xi_2^{eff}(\tilde{\sigma}_{\tilde{a}}) &= \sum_{i=1}^2 |\tilde{x}_{23}^j|^2 \\
\xi_3^{eff}(\tilde{\sigma}_{\tilde{a}}) &= -\sum_{i=1}^2 |\tilde{x}_{23}^j|^2 + \sum_{i=1}^2 |\tilde{x}_{34}^j|^2 \\
\xi_4^{eff}(\tilde{\sigma}_{\tilde{a}}) &= -\sum_{i=1}^2 |\tilde{x}_{34}^j|^2
\end{align}
therefore we need the effective CS levels $\xi_i^{eff}(\tilde{\sigma}_{\tilde{a}})$ to belong to the cone
\be\label{cone_F0}
\xi_2^{eff}(\tilde{\sigma}_{\tilde{a}})\geq 0\;,\qquad \quad \xi_4^{eff}(\tilde{\sigma}_{\tilde{a}})\leq 0 \qquad \qquad \forall\, \tilde{\sigma}_{\tilde{a}}, \quad \tilde{a}=1,\dots,p\;.
\ee
The effective FI parameters $\xi^{eff}_i(\sigma)$ of the $U(1)^4$ D2-brane quiver are
\begin{align}
\xi_1^{eff}(\sigma) &= \xi_1 = 0\\
\xi_2^{eff}(\sigma) &= \xi_2 + (p-q_1) \sigma + \sum_{\tilde{a}=1}^p
|\sigma-\tilde{\sigma}_{\tilde{a}}| \\
\xi_3^{eff}(\sigma) &= \xi_3 + (q_1-q_2) \sigma \\
\xi_4^{eff}(\sigma) &= \xi_4 + (q_2-p) \sigma - \sum_{\tilde{a}=1}^p
|\sigma-\tilde{\sigma}_{\tilde{a}}| \;.
\end{align}
$\xi^{eff}_2(\sigma)$ and $\xi^{eff}_4(\sigma)$ take their minimum and maximum respectively at $\sigma=\tilde{\sigma}_{\tilde{a}}$ for some special $\tilde{a}$, therefore \eqref{cone_F0} implies the same inequalities for $\xi^{eff}_i(\sigma)$: we have learned that the effective FI parameters are forced to remain on the K\"ahler wall \eqref{wall_F0_a} associated to the dictionaries used to derive the superconformal theory.
Consequently, the semiclassical moduli space of the gauge theory matches the type IIA geometry, as long as the GLSM of the fibred $CY_3$ $\tilde{Y}(r_0)$ is in a geometric phase.

This last comment deserves a more detailed explanation. Recall that in the derivation of the field theory we had to require the type IIA background to be in a geometric phase of the GLSM. That constrained the conical $CY_4$ geometry in M-theory to satisfy inequalities like \eqref{geom_ineq_F0}, and also its partial resolutions. For instance, consider sending to infinity the volume of an exceptional $\bP^1$'s in the $\bC^2/\bZ_p$ fibre to reduce $p\to p-1$ and remove the point $s_p$ from the toric diagram. Iterating this process, at some point  the inequality \eqref{geom_ineq_F0} will be violated. In type IIA, after sending too many $\overline{\mathrm{D6}}$-branes to infinity, the GLSM of the fibred $CY_3$ $\tilde{Y}(r_0)$ is no longer in a geometric phase for all $r_0$.
This has a neat counterpart in field theory: if we try to increase the $\tilde{\sigma}$ eigenvalues accordingly, the effective FI parameters $\xi^{eff}_i(\tilde{\sigma}_{\tilde{a}})$ at some point leave the K\"ahler cone of the $U(1)^{G-1}=U(1)^3$ subquiver \eqref{cone_F0} and the gauge theory is no longer in a supersymmetric vacuum.
Hence the quiver gauge theory only probes partial resolutions of the $CY_4$ that keep the $\tilde{Y}(r_0)$ GLSM in a gometric phase, due to the extra $p$ components of the D-term equations for the $U(1)^{G-1}$ subquiver.%
\footnote{This corrects the claim made in \cite{Benini:2011cma} that the quiver gauge theory can describe in its supersymmetric moduli space all the toric crepant resolutions of the $CY_4$ cone. We thank Francesco Benini for discussions on this point.}

\subsection{Higgsings}

In the remainder of this section we will keep the $p$ $\overline{\mathrm{D6}}$-branes on top of each other, $\tilde{\sigma}_{\tilde{a}}=0$, and consider instead partial resolutions of the exceptional $B_4$ surface. In field theory this is achieved by turning on a VEV for a bifundamental which is uncharged under the $\overline{\mathrm{D6}}$ gauge group, together with bare FI parameters so that the D-term equations are solved. We will only consider some representative Higgsings among models studied in section \ref{sec:_examples}. More general Higgsings leading to noncompact D6-branes in type IIA will be analysed elsewhere \cite{in:progress}.

\subsubsection*{Higgsing phase d of $dP_3$ to phase b of $dP_2$}

Consider the theory for phase d of $dP_3$ with bare FI parameters $\xi_6=-\xi_1\equiv \xi >0$, and the VEV $\langle X_{61}\rangle = \xi I_{N\times N}$ so that D-term equations are solved. Integrating out massive matter leads to a low energy CS quiver theory for phase b of $dP_2$, with the brane tiling in figure \ref{dP2_b_tiling_modified}, gauge ranks and bare CS levels
\begin{align}
\bm{N} &= (N,\,N,\,N,\,N,\,N-p)\\
\bm{k} &= (-\frac{1}{2}p+q_1-q_2+q_3,\,-\frac{3}{2}p+q_2+q_4,\,p-q_3,\,p-q_1-q_4,\,0)\;.
\end{align}
We can get rid of one of the $q$ parameters by the redefinition
\begin{equation}
q_1+q_4=q'_1\in[0,2p]\;,\;\; q_2+q_4=q_2'\in[0,3p]\;,
\end{equation}
so that the bare CS levels become
\begin{equation}
\bm{k} = (-\frac{1}{2}p+q'_1-q'_2+q_3,\,-\frac{3}{2}p+q'_2,\,p-q_3,\,p-q'_1,\,0)\;.
\end{equation}
This is nothing but the quiver CS theory for M2-branes at $C(Y^{p,\,q'_1,\,q'_2,\,q_3}(dP_2))$  presented in section \ref{subsec:_dP2_phase_b}, that uses toric phase b of $dP_2$.

\subsubsection*{Higgsing phase c of $dP_3$ to phase a of $dP_2$}

Consider the theory for phase c of $dP_3$ with bare FI parameters $\xi_6=-\xi_1\equiv \xi >0$, and the VEV $\langle X_{61}\rangle = \xi I_{N\times N}$ so that D-term equations are solved. Integrating out massive matter leads to a low energy CS quiver theory for phase a of $dP_2$, with the brane tiling in figure \ref{dP2_a_tiling_modified}, gauge ranks and bare CS levels
\begin{align}
\bm{N} &= (N,\,N,\,N,\,N,\,N-p)\\
\bm{k} &= (\frac{1}{2}p-q_2+q_3-q_4,\,-\frac{1}{2}p-q_1+q_2,\,p-q_3,\,-p+q_1+q_4,\,0)\;.
\end{align}
As in the (Seiberg dual) previous section, we redefine
\begin{equation}
q_1+q_4= q'_1\in[0,2p]\;,\;\; q_2+q_4= q'_2\in[0,3p]\;,
\end{equation}
so that the bare CS levels become
\begin{equation}
\bm{k} = (\frac{1}{2}p-q'_2+q_3,\,-\frac{1}{2}p-q'_1+q'_2,\,p-q_3,\,-p+q'_1,\,0)\;.
\end{equation}
This is nothing but the quiver CS theory of section \ref{subsec:_dP2_phase_a}
for M2-branes probing the real cone over $Y^{p,\,q'_1,\,q'_2,\,q_3}(dP_2)$, that uses toric phase a of $dP_2$.

\subsubsection*{Higgsing phases a and b of $dP_2$ to $dP_1$}

Consider the theory for phase a of $dP_2$ of section \ref{subsec:_dP2_phase_a}, with bare FI parameters $\xi_3=-\xi_1\equiv \xi >0$, and the VEV $\langle X_{31}\rangle = \xi I_{N\times N}$ so that D-term equations are solved. Integrating out massive matter and relabelling gauge groups $(1/3,2,4,5)\to(1,3,4,2)$ leads to the CS quiver theory for $C(Y^{p,\,q_1,\,q_2}(dP_1))$ of section \ref{subsec:_dP1}, with ranks and levels
\begin{align}
\bm{N} &= (N,\,N-p,\,N,\,N)\\
\bm{k} &= \left(\frac{3p}{2}-q_2,\,0,\,-\frac{p}{2}-q_1+q_2,\,-p+q_1\right)\;.
\end{align}

Similarly, one can consider as a starting point the (Seiberg dual) theory of section \ref{subsec:_dP2_phase_b} in phase b of $dP_2$. Following the same Higgsing one gets to a quiver for $dP_1$, the Seiberg dual (for node 4) of the previous one. That is the same quiver theory up to relabelling and a CP transformation that reverses arrows and changes sign to CS levels.

\subsubsection*{Higgsing phase a of $dP_2$ to phase a of $\bF_0$}

Consider again the theory for phase a of $dP_2$, now with bare FI parameters $\xi_1=-\xi_2\equiv \xi >0$, and the VEV $\langle X_{12}\rangle = \xi I_{N\times N}$ so that D-term equations are solved. Integrating out massive matter and relabelling gauge groups $(1/2,3,4,5)\to(3,2,4,1)$ leads to the CS quiver theory for $C(Y^{p,\,q_3,\,q_1}(\bF_0))$ in phase a of section \ref{subsec:_F0_a},
with ranks and levels
\begin{align}
\bm{N} &= (N-p,\,N,\,N,\,N)\\
\bm{k} &= \left(0,\,p-q_3,\,q_3-q_1,\,-p+q_1\right)\;.
\end{align}

\subsubsection*{Higgsing phase b of $dP_2$ to phase b of $\bF_0$}

We can follow the same Higgsing in the Seiberg dual quiver, in phase b of $dP_2$, with bare FI parameters $\xi_1=-\xi_2\equiv \xi >0$ and the VEV $\langle X_{12}\rangle = \xi I_{N\times N}$ solving D-term equations. Relabelling gauge groups $(1/2,3,4,5)\to(3,2,4,1)$ leads to the CS quiver theory for $C(Y^{p,\,q_3,\,q_1}(\bF_0))$ in phase b of section \ref{subsec:_F0_b},
with ranks and levels
\begin{align}
\bm{N} &= (N-p,\,N,\,N,\,N)\\
\bm{k} &= \left(0,\,p-q_3,\,-2p+q_1+q_3,\,p-q_1\right)\;.
\end{align}

\subsubsection*{Higgsing the $dP_1$ quiver to the $dP_0$ quiver}

Finally consider the theory for M2-branes at $C(Y^{p,\,q_1,\,q_2}(dP_1))$ of section \ref{subsec:_dP1}, with bare FI parameters $\xi_3=-\xi_4\equiv \xi >0$, and the D-flat VEV $\langle X_{34}\rangle = \xi I_{N\times N}$. The resulting low energy theory is the CS quiver theory for M2-branes at $C(Y^{p,\,q_2}(dP_0))$ of section \ref{subsec:_dP0}, with the brane tiling of figure \ref{dP0_tiling_modified} and ranks and bare CS levels
\begin{align}
\bm{N} &= (N,\,N-p,\,N)\\
\bm{k} &= (\frac{3}{2}p-q_2,\,0,-\frac{3}{2}p+q_2)\;.
\end{align}

\subsubsection*{Higgsing phase c of $PdP_4$ to phase c of $PdP_{3b}$}

Consider the theory for phase c of $PdP_4$ of section \ref{subsec:_PdP4}, with bare FI parameters $\xi_7=-\xi_1\equiv \xi >0$ and the VEV $\langle X_{71}\rangle = \xi I_{N\times N}$ that solves D-term equations. After integrating out massive matter,  relabelling gauge groups $(1/7,2,3,4,5,6)\to(5,4,6,2,1,3)$ and redefining
\begin{equation}
q_2=q'_1+q'_3\in[0,2p]\;,\;\; q_1+q_3=q'_2\in[0,2p]\;,\;\; q_1=q'_2-q'_3\in[0,2p]
\end{equation}
we get to the CS quiver theory for phase c of $PdP_{3b}$ of section \ref{subsec:_PdP3b_phase_c}, the worldvolume theory on M2-branes probing $C(Y^{p,\,q'_1,\,q'_2,\,q'_3}(PdP_{3b}))$.

\subsubsection*{Higgsing phase c of $PdP_4$ to phase b of $PdP_{3c}$}

Consider again the theory for phase c of $PdP_4$ of section \ref{subsec:_PdP4}, now with bare FI parameters $\xi_2=-\xi_3\equiv \xi >0$ and the VEV $\langle X_{23}\rangle = \xi I_{N\times N}$ solving D-term equations. After integrating out massive matter,  relabelling gauge groups $(1,2/3,4,5,6,7)\to(1,2,3,4,5,6)$ and redefining
\begin{equation}
q_2=q'_1\in[0,2p]\;,\;\; q_3=q'_2\in[0,p]\;,
\end{equation}
we get to the CS quiver theory for phase b of $PdP_{3c}$ of section \ref{subsec:_PdP3c_phase_b}, the worldvolume theory on M2-branes probing $C(Y^{p,\,q'_1,\,q'_2}(PdP_{3c}))$.

\subsubsection*{Higgsing phase c of $PdP_{3b}$ to phases a and b of $dP_2$}

Consider the theory for phase c of $PdP_{3b}$ of section \ref{subsec:_PdP3b_phase_c}, with bare FI parameters $\xi_1=-\xi_3\equiv \xi >0$ and the VEV $\langle X_{13}\rangle = \xi I_{N\times N}$ that solves D-term equations. After integrating out massive matter,  relabelling gauge groups $(1/3,2,4,5,6)\to(1,5,2,4,3)$ and redefining
\begin{equation}
q_1+q_3=q'_1\in[0,2p]\;,\;\; q_1+q_2=q'_2\in[0,3p]\;,\;\; q_2=q'_3\in[0,3p]
\end{equation}
we get to the CP-conjugate of the CS quiver theory for phase a of $dP_2$ of section \ref{subsec:_dP2_phase_a}, which is the worldvolume theory on M2-branes probing $C(Y^{p,\,q'_1,\,q'_2,\,q'_3}(dP_2))$. This is nothing but the Seiberg dual of the CS theory for phase b of $dP_2$ of section \ref{subsec:_dP2_phase_b}, as can be seen by dualising on the new node 3 and relabelling some nodes.

Similarly we can turn on bare FI parameters $\xi_5=-\xi_1\equiv \xi >0$ in the theory for phase c of $PdP_{3b}$: the D-term equations are solved if we turn on the VEV $\langle X_{51}\rangle = \xi I_{N\times N}$. After integrating out massive matter,  relabelling gauge groups $(1/5,2,3,4,6)\to(2,5,4,1,3)$ and redefining
\begin{equation}
q_1+q_3=q'_1\in[0,2p]\;,\;\; q_1+2q_3=q'_2\in[0,3p]\;,\;\; q_2=q'_3\in[0,3p]
\end{equation}
we get to the CP-transform of the CS quiver theory for phase b of $dP_2$ of section \ref{subsec:_dP2_phase_a}, the worldvolume theory of M2-branes probing $C(Y^{p,\,q'_1,\,q'_2,\,q'_3}(dP_2))$. This is again dual to the CS theory for phase a of $dP_2$ of section \ref{subsec:_dP2_phase_a}, as can be seen by dualising on the new node 3 and relabelling some nodes.

\subsubsection*{Higgsing phase c of $PdP_{3b}$ to $PdP_2$}

Consider again the theory for phase c of $PdP_{3b}$ of section \ref{subsec:_PdP3b_phase_c}, with bare FI parameters $\xi_4=-\xi_6\equiv \xi >0$ and the VEV $\langle X_{46}\rangle = \xi I_{N\times N}$ that solves D-term equations. After integrating out massive matter, relabelling gauge groups $(1,2,3,4/6,5)\to(3,5,2,4,1)$ and redefining
\begin{equation}
q_1+2q_3=q'_1\in[0,3p]\;,\;\; q_1+q_3=q'_2\in[0,2p]
\end{equation}
we get the CS quiver theory for $PdP_2$ of section \ref{subsec:_PdP2}, the worldvolume theory on M2-branes probing $C(Y^{p,\,q'_1,\,q'_2}(PdP_2))$.

\subsubsection*{Higgsing $PdP_2$ to $dP_1$}

Consider the theory for $PdP_2$ of section \ref{subsec:_PdP2}, with bare FI parameters $\xi_1=-\xi_3\equiv \xi >0$ and VEV $\langle X_{13}\rangle = \xi I_{N\times N}$. Integrating out massive matter and relabelling gauge groups $(1/3,2,4,5)\to(1,4,3,2)$ leads to the low energy CS quiver theory for $N$ M2-branes probing $Y^{p,\,q_2,\,q_1}(dP_1)$, see section \ref{subsec:_dP1}.

\subsubsection*{Higgsing $PdP_2$ to $W\bC\bP^2_{[1,1,2]}$}

Consider again the $PdP_2$ theory, now with bare FI parameters $\xi_3=-\xi_4\equiv \xi >0$ and VEV $\langle X_{34}\rangle = \xi I_{N\times N}$. Relabelling gauge groups $(1,2,3/4,5)\to(1,3,2,4)$, the low energy theory is the CS quiver theory for $N$ M2-branes probing $Y^{p,\,q_2}(W\bC\bP^2_{[1,1,2]})$ of section \ref{subsec_WP2_112}.

%%%%%%%%%%%%%%%%%%%%%%%%%%%%%%%%%%%%%%%%%%%%%%%%%%%%%%%%%%%%%%%%%%%%

\section{Adding torsion $G_4$ flux: The $Y^{p,\,q}(\bF_0)$ case}\label{sec:torsion flux for F0}

All of our $\Ypq(B_4)$ geometries are associated to rather large families of M-theory background, corresponding to turning on $G_4$ torsion flux in $H^4(\Ypq, \bZ)$.
We leave a completely general analysis of the field theories dual to the $\Ypq(B_4)$ background with any value of the torsion for future work. In the following we work out in some detail the next simplest example after the $dP_0= \CP^2$ case worked out in detail in \cite{Benini:2011cma}, which is $B_4= \bF_0=\CP^1\times \CP^1$.  The $AdS/$CFT correspondence for the  $Y^{p,\,q_1,\, q_2}(\CP^1\times \CP^1)$ geometries has been studied in \cite{Martelli:2008rt, Tomasiello:2010zz}.
For simplicity we will also set $q_1=q_2$ in most of the following.

In \cite{Tomasiello:2010zz} a Chern-Simons quiver with ranks $\bm{N}=(N,N,N,N)$ based on the (phase $a$) $\bF_0$ quiver was proposed which describes half of the $Y^{p,\,q_1,\, q_2}(\bF_0)$ geometries. In section \ref{sec: explaining TZ theories} below we will derive this theory from our formalism, showing that it corresponds to non-zero $G_4$ torsion and explaining from $\theta$-stability why it cannot cover the whole $Y^{p,\,q_1,\, q_2}(\bF_0)$ family.

Our toric conventions for the $Y^{p,\,q_1,\, q_2}(\bF_0)$ geometry were introduced in section \ref{subsec: F0 geom}.
In the basis $\{\cC_1, \cC_2\}=\{D_3, D_4 \}$  of toric divisors of $\bF_0$, the $H^4$ torsion group (\ref{generic sol for Gamma}) is
\be
\bZ^3/\langle v_0, v_1, v_2 \rangle \, , \quad v_0=(2q_1+2q_2, q_2,q_1), \, v_1= (q_1,p,0),\, v_2=(q_2,0,p) ,
\ee
while the flux and source vectors (\ref{fluxes and sources generic torsion}) are given by
\bea\label{fluxes and sources generic torsion for F0}
 & \bm{Q}_{\text{flux}, -}= (- n_0   \, | \, -q_2 \, , \,  -q_1   \, | \, 0  )  \, ,\\
  & \bm{Q}_{\text{flux}, +}= (- n_0 + 2n_1 +2n_2   \, | \, -q_2+2p \, , \,  -q_1+2p   \, | \, 0  )  \, ,\\
 &\bm{Q}_\text{source} =    ( -p \,|\,n_2   \,,\, n_1   \,|\,  N )        \, .
\eea
Remark that there is no Freed-Witten anomaly.
It will be convenient to consider a different (non-toric) basis for the 2-cycles of $\bF_0$:
\be\label{new basis H2 F0}
\cC_1' = \cC_1+\cC_2\, ,\qquad \cC_2'= \cC_2\,.
\ee
In this basis, we have $v_0'= (2q_1+2q_2, q_1+q_2,q_1)$, $ v_1'= (q_1, p, 0)$,$v_2'= (-q_1+q_2, 0, p)$.
In following we set $q_1=q_2\equiv q$. In that case, the torsion group is determined by the periodicity vectors
\be \label{peridocity vector q1 eq q2 eq q for FO}
v_0'= (4q, 2q, q)\, ,\; v_1'= (q, p, 0)\, ,\; v_2'= (0, 0, p)\, .
\ee
These periodicities are realised by large gauge transformations of the B-field in type IIA. The B-fields periods are
\be
b_0= \frac{p n_0-q n}{2q(2p-q)} \, ,\quad\qquad  b_1^+{}' = \frac{n}{p}  -\frac{2q}{p}b_0\, ,\qquad\quad b_2^+{}'= b_2^+= \frac{n_2}{p}-\frac{q}{p}b_0 \, ,
\ee
where we defined $n\equiv n_1+n_2$. Another important period is
\be\label{def btilde for F0}
\tilde{b} \equiv b_2^+ - b_1^+ = b_2^- - b_1^- = \frac{2n_2-n}{p}\, .
\ee
The periodicities $v_0'$, $v_{\alpha}'$ are related to the shift of the B-periods as
\bea\label{shift of v and B forF0}
& \delta(n_0, n ,n_2) = v_0'\, &:&\quad \delta (b_0,\,b_1^+{}',\, b_2^+{}',\, \tilde{b}) = (1, 0,0,0)\, , \\
& \delta(n_0, n ,n_2) =v_1'\, &:&\quad \delta (b_0,\, b_1^+{}',\, b_2^+{}',\, \tilde{b}) = (0, 1,0,-1)\, , \\
& \delta(n_0, n ,n_2) =v_2'\, &:&\quad \delta (b_0,\, b_1^+{}',\, b_2^+{}',\, \tilde{b}) = (0, 0,1,2)\, .
\eea
The central charge of a D4-brane on any 2-cycle $\cC$ is given exactly by
\be
Z(D4_{\cC}) = t_{\cC} = b_{\cC} +  i\chi_{\cC}\, ,
\ee
and in particular its real part is the corresponding B-field period. Of particular interest are the ``non-anomalous'' D4-branes wrapped on $\tilde{\cC}\equiv \cC_2-\cC_1$. We denote by $D4^{NA}_l$ the brane wrapped on $\tilde{\cC}$ with $l$ units of worldvolume flux, and by $\overline{D4}^{NA}_{-l+1}$ the brane wrapped with opposite orientation with $-l+1$ units of worldvolume flux. In term of Chern characters on $\bF_0$,
\be\label{ch of D4NA states}
ch(D4^{NA}_l)=(0,-1,1, l)\, , \qquad ch(\overline{D4}^{NA}_{-l+1})=(0,1,-1,-l+1)\, .
\ee
From the central charge, we find the inverse gauge couplings
\be\label{inverse gauge coupling}
g^{-2}(D4^{NA}_l)= \tilde{b}+l \, ,\qquad g^{-2}(\overline{D4}^{NA}_{-l+1})= -\tilde{b}-l+1 \, .
\ee
We will see shortly that the states (\ref{ch of D4NA states}) occur in the quiver spectrum: they correspond to $\cQ$ representations of positive dimension. When $\tilde{b}\in \bZ$, one of the gauge couplings (\ref{inverse gauge coupling}) diverges, and we should change the basis of fractional branes accordingly, as we will explain momentarily. The locus $g^{-2}=0$ defines a wall of the second kind, or Seiberg duality wall \cite{2008arXiv0811.2435K, Aganagic:2010qr}.
In term of $(n_0, n, n_2)$, this occurs at
\be\label{F0, location of SD walls}
n = 2 n_2 + l p \, , \qquad \forall l \in \bZ\, .
\ee
We call the locus (\ref{F0, location of SD walls}) for a given $l$ the ``Seiberg duality wall of level $l$''. Crossing the wall corresponds to doing \emph{two} simultaneous Seiberg dualities on two different quiver nodes. The relevant Seiberg dualities for Chern-Simons quivers have been elucidated in \cite{Benini:2011mf, Closset:2012eq}.

\begin{figure}[t]
\begin{center}
\includegraphics[width=14cm]{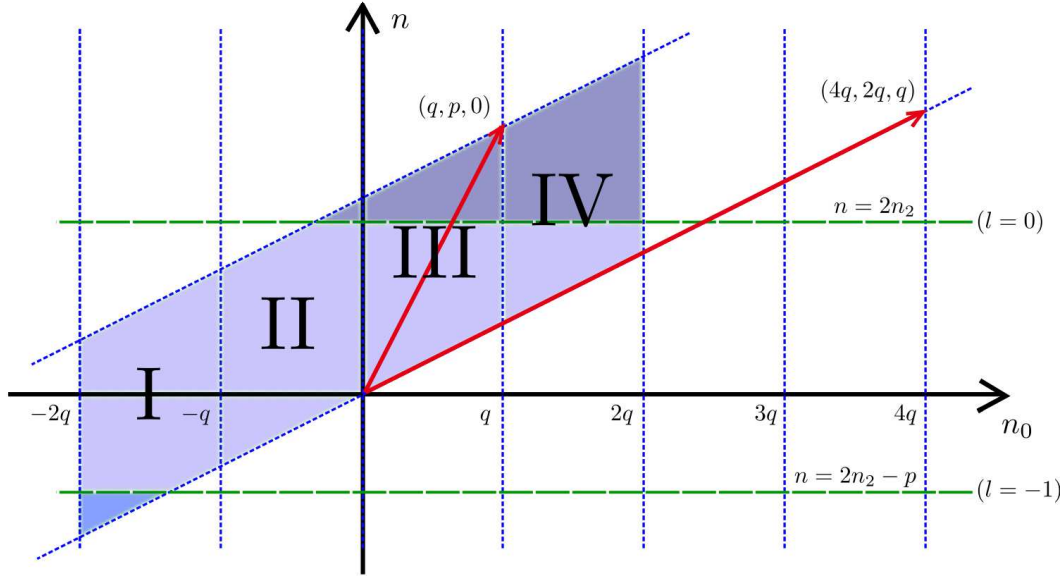}
\caption{\small The $(n_0, n)$ plane for $Y^{p,\,q}(\bF_0)$, at some arbitrary value for $n_2$.  The red vectors are the periodicities $v_0'=(4q,2q,q)$ and $v_1'=(q,p,0)$ projected on the plane. The last periodicity $v_2'=(0,0,p)$ goes perpendicular to that plane. The walls of the first kind are shown in blue and the walls of the second kind (Seiberg duality walls) in green. The shaded area covers the fundamental domain once, and is subdivided into four windows $I-IV$ delimited by walls of the first kind. \label{fig: torsion domain YpqF0}}
\end{center}
\end{figure}

In addition, we also have the marginal stability walls --- or walls of the first kind --- for the fractional branes, which have been thoroughly discussed in this paper. Changing the value of the torsion flux $(n_0, n, n_2)$, we can cross the two kinds of walls; this is shown for instance in Figure \ref{fig: torsion domain YpqF0}.

%%%%%%%%%%%%%%%%%%%%%%%%%5
%%%%%%%%%%%%%%%%%%%%
\subsection{Theories covering the full torsion group}
Let us consider the toric quiver for $\cO_{\bF_0}(K)$ in phase a, and let us start with type IIA background fluxes near the torsionless point $(n_0, n, n_2)=(0,0,0)$. We saw in section \ref{subsec:_F0_a}  that we could use either dictionary $Q^{\vee}[\{p_5\}, \{0,-1\}]$ or $Q^{\vee}[\{p_8\}, \{0,0\}]$ at the torsionless point. Using first the dictionaries $Q^{\vee}_-=Q^{\vee}_+=Q^{\vee}[\{p_8\}, \{0,0\}]$ with generic torsion, we find that
\bea
&\bm{\theta_-}\,=\, (n_0, q-n_0, -n_0, -q+n_0)\, , \\
&\bm{\theta_+}\,=\, (2n-n_0, 2p-q-2n+n_0, -2n +n_0, -2p+q+2n-n_0)\, .
\eea
Looking at the inequalities (\ref{KC F0a p8}) defining the open string Kahler chamber $\{p_8\}$, one finds that this choice of dictionaries $Q^{\vee}_{\pm}$ is consistent for
\be\label{thetas F0 p8p8}
0 \leq n_0 \leq q\, , \quad 0 \leq 2n-n_0 \leq 2p- q\, .
\ee
This corresponds to the window III shown in Figure \ref{fig: torsion domain YpqF0}.  Leaving window III towards window II, we see that $\bm{\theta_{+}}$ in (\ref{thetas F0 p8p8}) stays in the K\"ahler chamber  $\{p_8\}$, while $\bm{\theta_{-}}$ crosses the wall to the K\"ahler chamber $\{p_5\}$. The correct LVM is obtained by checking consistency of the new pair of dictionaries, or equivalently by requiring that the theories glue along the wall $n_0=0$ between windows III and II. Crossing the walls to the successive windows, we find that we need to use the dictionaries:
\bea \label{four dictionaries torsion group F0 level-1}
\text{window} \, I &:&\quad & Q^{\vee}_- =Q^{\vee}[\{p_6\},\{-1,-1\}]\, , &\quad & Q^{\vee}_+ =Q^{\vee}[\{p_8\},\{0,0\}]\, , \\
\text{window} \, II &:&\quad & Q^{\vee}_- =Q^{\vee}[\{p_5\},\{0,-1\}]\, , &\quad & Q^{\vee}_+ =Q^{\vee}[\{p_8\},\{0,0\}]\, , \\
\text{window} \, III &:&\quad & Q^{\vee}_- =Q^{\vee}[\{p_8\},\{0,0\}]\, , &\quad  & Q^{\vee}_+ =Q^{\vee}[\{p_8\},\{0,0\}]\, , \\
\text{window} \, IV &:&\quad & Q^{\vee}_- =Q^{\vee}[\{p_7\},\{1,0\}]\, , &\quad  & Q^{\vee}_+ =Q^{\vee}[\{p_8\},\{0,0\}]\, ,
\eea
The field theories for these four windows are given in Table \ref{tab: F0 theories at level -1}. The four dictionaries appearing in (\ref{four dictionaries torsion group F0 level-1}) are given explicitly by
\bea\label{show 4 dictionaires level 0 explicitly}
&
 Q^{\vee}_{(0)}[\{p_6\},\{-1,-1\}] &=
\tiny{\mat{ -1 & 0 & -2 & 0 \\
 1 & 0 & 1 & 0 \\
 1 & 1 & 1 & 1 \\
 -1 & -1 & 0 & 0}}\, , \quad\quad
&Q^{\vee}_{(0)}[\{p_5\},\{0,-1\}] &=
\tiny{\mat{ 1 & 0 & 0 & 0 \\
 1 & 0 & 1 & 0 \\
 -1 & 1 & -1 & 1 \\
 -1 & -1 & 0 & 0}}\, , \\
&  Q^{\vee}_{(0)}[\{p_8\},\{0,0\}] &=
\tiny{\mat{ 1 & 0 & 0 & 0 \\
 -1 & 0 & 1 & 0 \\
 -1 & 1 & -1 & 1 \\
 1 & -1 & 0 & 0}}\, , \quad \quad
& Q^{\vee}_{(0)}[\{p_7\},\{1,0\}]&=
\tiny{\mat{-1 & 2 & 0 & 0 \\
 -1 & 0 & 1 & 0 \\
 1 & -1 & -1 & 1 \\
 1 & -1 & 0 & 0}}\, .
\eea

So far we did not take into account the presence of the walls of the second kind (Seiberg duality walls) at $n=2n_2 + lp$, $l \in \bZ$. The four dictionaries (\ref{show 4 dictionaires level 0 explicitly}) have in common that
\be
ch(\mathsf{E}_1^{\vee}+\mathsf{E}_3^{\vee})=(0,1,-1,1)\, , \qquad ch(\mathsf{E}_2^{\vee}+\mathsf{E}_4^{\vee})=(0,-1,1,0)\, .
\ee
Therefore $\mathsf{E}_1^{\vee}+\mathsf{E}_3^{\vee}$ is the D4-brane $\overline{D4}^{NA}_1$, and $\mathsf{E}_2^{\vee}+\mathsf{E}_4^{\vee}$ is the D4-brane $D4^{NA}_0$, in the notation of (\ref{ch of D4NA states}). Requiring that these two states have real gauge coupling, we find that
\be\label{range of n-2n2 at level 0}
-p \leq n-2n_2 \leq 0 \, ,
\ee
corresponding to being between the Seiberg dualities walls (\ref{F0, location of SD walls}) of levels $l=-1$ and $l=0$. The subscript $(0)$ in (\ref{show 4 dictionaires level 0 explicitly}) is to remind us that these dictionaries are valid only in the range (\ref{range of n-2n2 at level 0}).
If the real part of $Z(\mathsf{E}_1^{\vee}+\mathsf{E}_3^{\vee})$ goes negative, we should perform a double Seiberg duality on nodes 1 and 3, corresponding to crossing a wall (\ref{F0, location of SD walls}) at level $l=-1$. Similarly, if $\text{Re}\,Z(\mathsf{E}_2^{\vee}+\mathsf{E}_4^{\vee})=0$ we are crossing the Seiberg duality wall of level $l=0$, and we perform a double duality on nodes 2 and 4.
%%%%%%%%%%%%%%%%%%%%%%%%%%%%%%%%%
\begin{table}[h]
\bea\nn
&\qquad  \text{Conditions}: \qquad -2q \leq n_0 \leq -q\,\; ,\; \qquad\qquad  0 \leq 2n - n_0 \leq 2p - q \\
&\boxed{ I \,\; : \,\; \begin{cases}
&\bm{N} \,\;=\,\; (N+n-n_0-p,\;N+n-n_0-n_2-q,\;N,\;N-n_2 ) \\
&\bm{k} \,\;=\,\; (n+q,\;-n+p-q,\;-n-q,\;n-p+q )
\end{cases}}\\
\\
&\qquad  \text{Conditions}: \qquad -q \leq n_0 \leq  0\,\; ,\; \qquad\qquad 0 \leq 2n - n_0 \leq 2p - q \\
&\boxed{ II \,\; : \,\; \begin{cases}
&\bm{N} \,\;=\,\; (N+n-n_0-p,\;N+n-n_2,\;N,\;N-n_2 ) \\
&\bm{k} \,\;=\,\; ( n-n_0,\;-n+p-q,\;n_0-n,\;n-p+q)
\end{cases}}\\
\\
&\qquad  \text{Conditions}: \qquad 0 \leq n_0 \leq  q\,\; ,\; \qquad\qquad 0 \leq 2n - n_0 \leq 2p - q\\
&\boxed{ III \,\; : \,\; \begin{cases}
&\bm{N} \,\;=\,\; (N+n-p,\;N+n-n_2,\;N,\;N-n_2) \\
&\bm{k} \,\;=\,\; (n-n_0,\;-n+n_0+p-q,\;n_0-n,\;n-n_0-p+q )
\end{cases}}\\
\\
&\qquad  \text{Conditions}: \qquad  q \leq n_0 \leq  2q\,\; ,\; \qquad\qquad 0 \leq 2n - n_0 \leq 2p - q \\
&\boxed{ IV \,\; : \,\; \begin{cases}
&\bm{N} \,\;=\,\; (N+n-p,\;N+n-n_2,\;N,\;N+n_0-n_2-q) \\
&\bm{k} \,\;=\,\; (n-q,\;-n+n_0+p-q,\;q-n,\;n-n_0-p+q )
\end{cases}}\\
\eea
\caption{\small Theories for $F_0$ covering the torsion domain. As explained in the text, these theories are valid only between the Seiberg duality walls of level $l=-1$ and $l=0$, which means for $2n_2-p \leq n \leq 2n_2$. }\label{tab: F0 theories at level -1}
\end{table}
%%%%%%%%%%%%%%%%%%%%%%%%%%%%%%%%%%%%%%%%%%%%%
\subsubsection{Crossing the Seiberg duality walls: mutated dictionaries}
Consider for instance crossing the wall $l=0$. We should perform a 3d Seiberg duality on node 2 and 4. The new dictionaries are determined by mutations as explained in \cite{Closset:2012eq}. For instance, in window $I$ --- first line of (\ref{four dictionaries torsion group F0 level-1}) ---, we have
\bea\label{example mutation F0: theta params}
& \theta_{2-}= n_0+q \leq 0 \, , &\qquad\quad  & \theta_{4-}= -n_0-q \geq 0\, ,   \\
 & \theta_{2+}= -2n+n_0+2p-q \geq 0\, , &\qquad\quad & \theta_{4+}= 2n-n_0-2p+q \leq 0\, .
\eea
Starting with node 2, we should perform a left mutation on $Q^{\vee}_-$ and a right mutation on $Q^{\vee}_+$. This gives us a Chern-Simons theory based on the quiver of phase b%
\footnote{With different conventions for labelling the notes with respect to section \ref{subsec:_F0_b}.},
 whose $\theta_{4\pm}$ parameters are still the same as in (\ref{example mutation F0: theta params}). Therefore we proceed with a duality on node 4, corresponding to a right mutation at $r_0<0$ and a left mutation at $r_0 >0$. The resulting quiver is in phase $a$ again, but with the arrows reversed. We reverse to our original conventions by relabelling the nodes $(1,2,3,4)\rightarrow (1,4,3,2)$. With these mutations and relabelling, the new dictionaries we should use beyond the wall $l=0$ (and up to the wall $l=1$) are
\bea\label{show 4 dictionaires level 1 explicitly}
&
 Q^{\vee}_{(1)}[\{p_6\},\{-1,-1\}] &=
\tiny{\mat{ -1 & -2 & 0 & 0 \\
 1 & 1 & 0 & 0 \\
 1 & 1 & 1 & 1 \\
 -1 & 0 & -1 & 0}}\, , \quad\quad
&Q^{\vee}_{(1)}[\{p_5\},\{0,-1\}] &=
\tiny{\mat{ 1 & 0 & 0 & 0 \\
 1 & 1 & 0 & 0 \\
 -1 & -1 & 1 & 1 \\
 -1 & 0 & -1 & 0}}\, , \\
&  Q^{\vee}_{(1)}[\{p_8\},\{0,0\}] &=
\tiny{\mat{ 1 & 0 & 0 & 0 \\
 -1 & 1 & 0 & 0 \\
 -1 & -1 & 1 & 1 \\
 1 & 0 & -1 & 0}}\, , \quad \quad
& Q^{\vee}_{(1)}[\{p_7\},\{1,0\}]&=
\tiny{\mat{-1 & 0 & 2 & 0 \\
 -1 &1 & 0 & 0 \\
 1 & -1 & -1 & 1 \\
 1 & 0 & -1 & 0}}\, .
\eea
The walls of the first kinds remain unchanged in this new basis. The dictionaries (\ref{show 4 dictionaires level 1 explicitly}) are valid for $2n_2 \leq n \leq 2n_2 +p$.

A similar analysis can be performed at the Seiberg duality wall of level $l=-1$. The resulting dictionaries after mutation on nodes 1 and 3 (and relabelling of the nodes $(1,2,3,4)\rightarrow (3,2,1,4)$) are
\bea\label{show 4 dictionaires level -1 explicitly}
&
 Q^{\vee}_{(-1)}[\{p_6\},\{-1,-1\}] &=
\tiny{\mat{-1 & -1 & -1 & -1 \\
 1 & 0 & 1 & 0 \\
 1 & 0 & 2 & 0 \\
 -1 & 1 & -2 & 2}}\, , \quad\quad
&Q^{\vee}_{(-1)}[\{p_5\},\{0,-1\}] &=
\tiny{\mat{ 1 & -1 & 1 & -1 \\
 1 & 0 & 1 & 0 \\
 -1 & 0 & 0 & 0 \\
 -1 & 1 & -2 & 2}}\, , \\
&  Q^{\vee}_{(-1)}[\{p_8\},\{0,0\}] &=
\tiny{\mat{1 & -1 & 1 & -1 \\
 -1 & 2 & -1 & 2 \\
 -1 & 0 & 0 & 0 \\
 1 & -1 & 0 & 0}}\, , \quad \quad
& Q^{\vee}_{(-1)}[\{p_7\},\{1,0\}]&=
\tiny{\mat{-1 & 1 & 1 & -1 \\
 -1 & 2 & -1 & 2 \\
 1 & -2 & 0 & 0 \\
 1 & -1 & 0 & 0}}\, .
\eea
They are valid for $2n_2-2p \leq n \leq 2n_2 -p$.

We can easily generalise these results to any wall (\ref{F0, location of SD walls}).
At the Seiberg duality wall of level $l$, the corresponding mutation acts on dictionaries according to
\be\label{mutated dictionary at wall l}
 Q^{\vee}_{l+ 1}\ = \, Q^ {\vee}_l M_l\, , \qquad \quad \text{with}
\quad M_l = \mat{1 & l &-l &-l^ 2 \\0 &0 &1 &l \\0 &1 &0 &-l \\0 & 0& 0&1 }\, ,
\ee
while the walls of the first kind remain unchanged.  Remark that $M_l^{-1}= M_l$ and therefore we can cross the wall in the opposite direction with $Q^{\vee}_{l}\ = \, Q^ {\vee}_{l+1} M_l$. The dictionaries $Q^{\vee}_l$ are valid for
\be\label{validity of Qvl}
(l-1)p \leq n-2n_2 \leq l p\, .
\ee
 Since the ranks and Chern-Simons levels are given by $\bm{N}= Q^{\vee\, T -1}\bm{Q}_{\text{source}}$ and $\bm{k}=Q^{\vee} \bm{Q}_{\text{flux}}$%(up to the fact that we use two dictionaries, as explained in section \ref{subsec: IIA to CS quiver})
, we can account for the mutation (\ref{mutated dictionary at wall l}) by a ficticious change of the type IIA parameters $\bm{Q}_{\text{source}}\rightarrow M^{T}_l\bm{Q}_{\text{source}}$ and $\bm{Q}_{\text{flux}}\rightarrow M_l\bm{Q}_{\text{flux}}$. We thus find that crossing a wall of level $l$ amounts to changing the IIA parameters according to%
\footnote{The simplicity of these rules comes from the fact that we chose $q_1=q_2=q$, which implies that only $\bm{N}$ changes as we cross the wall, while $\bm{k}$ is invariant.}
\be\label{rules of replacement for  Seiberg duality walls}
%n_0\rightarrow n_0\, , \quad n\rightarrow n\, ,\quad
 n_2\rightarrow n-n_2 - l p\, , \qquad\quad N\rightarrow N + l(2n_2-n+l p)\, ,
\ee
keeping $q$, $p$, $n_0$ and $n$ fixed. These rules make it obvious that the field theories change continously as we cross the wall at $n=2n_2+lp$. They are  reminisent of the rules for non-chiral Seiberg dualities obtained from Hanany-Witten setups in \cite{Aharony:2009fc}. We should insist, however, that this change of the IIA parameters is ficticious: The parameters of the background geometry stay what they are, while what changes are the dictionaries $Q^{\vee}$, and therefore the field theory parameters $(\bm{N}, \bm{k})$. The rules (\ref{rules of replacement for  Seiberg duality walls}) are just a convenient summary of the effect of these Seiberg dualities on the Chern-Simons quivers. Hence the field theories for $0 \leq n-2n_2 \leq p$ are found from the field theories of Table \ref{tab: F0 theories at level -1} by replacing $n_2\rightarrow n-n_2$, $N\rightarrow N$, and so on and so forth to attain any of the regions (\ref{validity of Qvl}), as long as we are still inside the walls of the first kind delimiting the regions $I-IV$ in Figure \ref{fig: torsion domain YpqF0}.

\subsection{Periodicities and Seiberg duality}
As we perform a shift of $(n_0,n, n_2)$ by any of the periodicity vectors (\ref{peridocity vector q1 eq q2 eq q for FO}), the field theory should stay invariant, up to a shift of $N$ expected from the corresponding shift (\ref{general formula for shift of Q2}) of the D2-brane Page charge.
For the case at hand, (\ref{general formula for shift of Q2}) gives
\bea\label{expected shift of N for F0}
& \delta(n_0, n ,n_2) = v_0'\, &:&\quad \delta N= -n_0-2q \, , \\
& \delta(n_0, n ,n_2) =v_1'\, &:&\quad \delta N= - n_2 \, ,\\
& \delta(n_0, n ,n_2) =v_2'\, &:&\quad \delta N=  2n_2 - n + p \,.
\eea
We should check these type IIA expectations against the field theories we derived above.

First of all, we can check that the $v_0'$ periodicity is realised explictly in Table \ref{tab: F0 theories at level -1}: the theories on the left boundary of window $I$ are the same as the theories on the right boundary of window $IV$, without any shift of $N$ (this agrees with the first line of (\ref{expected shift of N for F0}), since we start at $n_0=-2q$).

From (\ref{shift of v and B forF0}), it is apparent that a shift by $v_1'$ necessitates the crossing of one Seiberg duality wall in the negative direction ($\delta \tilde{b}=-1$) while a shift by $v_2'$ implies a crossing  of two such walls in the positive direction ($\delta \tilde{b}=2$).%
\footnote{Remark that $v_0'$ does \emph{not} cross any Seiberg duality wall, unlike what is suggested by a naive reading of Fig.\ref{fig: torsion domain YpqF0}, because $v_0'=(4q,2q,q)$ also goes in the $n_2$ direction perpendicular to the $(n_0,n)$ plane shown in that figure.}
Consider first the $v_1'$ periodicity. Starting with any theory of Table \ref{tab: F0 theories at level -1} on the bottom of the fundamental domain in Fig.\ref{fig: torsion domain YpqF0} (hence such that we are between the Seiberg duality walls $l=-1$ and $l=0$), we can go to the upper boundary of the fundamental domain with a $v_1'=(q,p,0)$ shift, crossing the wall $l=0$ in the process. Using the rules (\ref{rules of replacement for  Seiberg duality walls}) for Seiberg duality, we can check that these two CS quivers are the same, up to a shift $\delta N = -n_2$ which agrees with the expectation (\ref{expected shift of N for F0}).

Finally, let us consider the periodicity $v_2'= (0,0,p)$, which shifts $n_2$ by $p$. According to the rules (\ref{rules of replacement for  Seiberg duality walls}) for Seiberg duality, crossing two succesive Seiberg duality walls shifts $n_2$ according to
\be
n_2\qquad \rightarrow\qquad n-n_2- lp \qquad\rightarrow\qquad n-(n-n_2-(l-1)p) -lp = n_2 -p\, .
\ee
This effective shift of $n_2$ from Seiberg dualities cancels the shifts from $v_2'$. The shift $\delta N= 2n_2-n +p$ expected from (\ref{expected shift of N for F0}) is also recovered from (\ref{rules of replacement for  Seiberg duality walls}).

We thus found that the periodicities of the torsion group $H_4(Y^{p,q}(\bF_0))$ are reproduced by the field theories we derived, providing a non-trivial consistency check on their correctness.

\subsection{Remark on $Y^{p, \,q_1, \,q_2}(\bF_0)$ and quiver with equal ranks}\label{sec: explaining TZ theories}
For general $q_1$, $q_2$, the structure of walls of the first and second kind is more complicated than for the $q_1=q_2$ case of Figure \ref{fig: torsion domain YpqF0}, and we will not present a full analysis here. Nevertheless we can make some preliminary remarks allowing us to check the conjecture of \cite{Tomasiello:2010zz} concerning some Chern-Simons quiver describing those geometries. The theory proposed in \cite{Tomasiello:2010zz} is a Chern-Simons quiver with equal ranks, like in most of the heuristic proposals in the literature. We stress that from the string theory point of view there is nothing special about having equal ranks: it just corresponds to some particular value of the torsion flux.

Consider first the $q_1=q_2$ theories of Table \ref{tab: F0 theories at level -1}. We can find a theory with ranks $\bm{N}=(N,N,N,N)$ in window $IV$ if we choose $n=n_2=p$ and $n_0=p+q$. However, requiring that this theory actually sits in window $IV$ we must also set $q=p$. This results in a theory with
\be\label{Aganagic theory for Q111/Zp}
\bm{N}=(N,N,N,N)\, , \qquad \quad \bm{k}= (0, p, 0, -p)\, ,
\ee
which moreover sits on the Seiberg duality wall $l=-1$. The corresponding geometry is $Y^{p,p,p}(\bF_0) = Q^{1,1,1}/\bZ_p$, and the theory (\ref{Aganagic theory for Q111/Zp}) was first proposed in \cite{Aganagic:2009zk} for that geometry.%
\footnote{We see that, ironically, the type IIA derivation first proposed in that same paper \cite{Aganagic:2009zk} was imprecisely applied to this $Q^{1,\,1,\,1}$ case, which needed the formalism of the present paper to be fully understood.}
Similarly, there is a theory at $n=n_2=0$ and $n_0=-p$, with $q=p$, sitting in window $II$, corresponding to $\bm{k}=(-p, 0, p, 0)$.

A third possibility from Table \ref{tab: F0 theories at level -1} is the equal rank theory in Window $I$, which again occurs only if $q=p$, at torsion flux  $n=n_0+p$ and $n_2=0$. The field theory is
\be\label{Q222 theories etc}
\bm{N}=(N,N,N,N)\, , \qquad \quad \bm{k}= (2p +n_0, -p-n_0, -2p-n_0, p+n_0)\, ,
\ee
for $-2 p\leq n_0\leq -p$. This theory again describes $Q^{1,1,1}/\bZ_p$. For $p=2$, $n_0=-3$, such that $\bm{k}=(1,1,-1,-1)$, this theory was proposed in \cite{Hanany:2008cd, Davey:2009sr} to describe $Q^{2,2,2}\cong Q^{1,1,1}/\bZ_2$, and we have thus derived this proposal and shown how it fits in a larger family of CS quiver theories. 

Consider next the $Y^{p, \,q_1, \,q_2}(\bF_0)$ geometry. The various chambers are modified with respect to Fig.\ref{fig: torsion domain YpqF0} but there is still a generalization of window $IV$ where the consistent dictionaries are $Q^{\vee}_- =Q^{\vee}[\{p_7\},\{1,0\}]$ and $ Q^{\vee}_+ =Q^{\vee}[\{p_8\},\{0,0\}]$. Using these dictionaries with the charges (\ref{fluxes and sources generic torsion for F0}), we find the Chern-Simons quiver
\bea
&\bm{N}= (N+ n_1+n_2-p, \, N+ n_1,\, N,\, N+ n_0-n_2-q_2 )   \, , \\
& \bm{k}= (n-q_2,\, -n+n_0+p-q_1,\, q_1-n,\,  n-n_0-p+q_2)  \, .
\eea
This includes a theory with equal ranks, for the choice of fluxes $(n_0, n_1, n_2)=(p+q_2, 0, p)$. With this choice, we have the theory
\be\label{TZ theory derived}
\bm{N}=(N,N,N,N)\, , \qquad \quad \bm{k}= (p-q_2,\, p-q_1+q_2,\, q_1-p,\, -p)\, ,
\ee
which is the theory conjectured in \cite{Tomasiello:2010zz}. Moreover, $\theta$-stability for this theory constraints the values of the Chern-Simons levels, and we find that the string theory derivation is consistent only if
\be
p \leq q_1 \leq 2p\, , \qquad 0\leq q_2 \leq p\, .
\ee
This explains why only in those ranges does the quiver theory (\ref{TZ theory derived}) reproduce the cone $C(Y^{p, \,q_1, \,q_2}(\bF_0))$ as its Abelian moduli space \cite{Tomasiello:2010zz}.
Remark that the theory (\ref{TZ theory derived}) is not CP invariant, which is explained by the fact that it is  dual to an M-theory background with non-zero torsion flux (M-theory parity acts as $(n_0, n_1, n_2)\rightarrow -(n_0, n_1, n_2)$ on our D4 Page charges); the special case (\ref{Aganagic theory for Q111/Zp}) is CP invariant and the M-theory parity action $(n_0, n_1, n_2)=(2p,0, p)\rightarrow (-2p, 0, -p)$ corresponds to a periodicity of $\Gamma$.

%%%%%%%%%%%%%%%%%%%%%%%%%%%%%%%%%%%%%%%%%%%%%
\section{Conclusion and outlook}
We provided a type IIA string theory derivation of Chern-Simons quiver gauge theories describing the low energy dynamics of M2-branes on the $CY_4$ cone over the seven-manifold $\Ypq(B_4)$, for any of the 16 toric Fano varieties $B_4$.

While we focused on the type IIA description, either in term of branes or in term of the CS quiver, we only briefly commented on how the CS gauge theory probes the full CY$_4$ geometry in M-theory. We recalled the conjecture that the quantum chiral ring of the CS theory encodes the coordinate ring of the CY$_4$, but we still lack a first principle approach to deal with the chiral ring of the IR SCFT. We believe that the conjecture (\ref{M2 geometric branch algebraically}) deserves further study, and we hope that the link we stressed to the GIT desciption of quiver moduli spaces can be fruitful in this respect.

There are a number of further directions for research one could follow. First of all, we can allow for D6-branes along noncompact toric divisors in some $\cO_{B_4}(K)$ fibres in type IIA, still keeping the GLSM of $\tilde{Y}$ geometric over the whole $\bR$ base so that the reduction to type IIA is understood.
It would be interesting to pursue this generalisation and combine the results of this paper with the flavouring procedure of \cite{Benini:2009qs}, allowing for even more general toric $CY_4$ geometries in term of CS quivers with fundamental matter.

Secondly, we mostly focused on the case of zero $G_4$ torsion flux in $\Ypq$. The case of general torsion has been solved for a special case in section \ref{sec:torsion flux for F0} (and before that in \cite{Benini:2011cma} for $B_4= \CP^2$), but remains challenging in general. To find the CS quiver theories dual to  $\Ypq$ with any value of the torsion flux,  one has to understand in general the interplay between the two kinds of walls in K\"ahler moduli space, and in its discretised version spanned by the type IIA quantised fluxes. The second kind of wall is a Seiberg duality wall, and the relevant 3d Seiberg dualities have been understood in some detail only recently  \cite{Benini:2011mf, Closset:2012eq}.

More generally, it would be desirable to carry on the type IIA derivation to much more general cases, corresponding to generic toric $CY_3$ in type IIA. One of the main challenges to do so is also of general interest for D-brane physics: we would need a generalization of the dictionaries $Q^{\vee}$ to any $CY_3$. While the result of \cite{2009arXiv0909.2013B} reviewed in section \ref{subsec: tilting collection of line bdl} provides a tilting collection of line bundles $\cE$ for any crepant partial resolution of a toric $CY_3$ cone, it is not known in general how to define a ``dual'' collection $\cE^{\vee}$ corresponding to the fractional D-branes.

The final step to cover all toric $CY_4$ geometries would then involve generalising to cases where the $U(1)_M$ circle action degenerates to $U(1)_M/Z_h$ over certain loci and the $CY_3$ GLSM in type IIA is not in a geometric phase over the whole $\bR$ base. It was suggested that in such cases the M2-brane quiver contains non-Lagrangian sectors \cite{Jafferis:2009th}. It would be interesting to pursue this proposal further.

Last but not least, the quiver-based approach we followed to study the ``open string K\"ahler chambers'' (delimited by marginal stability walls for fractional branes) in the K\"ahler moduli space of any toric $CY_3$ is of more general interest. For instance these methods might be applied to baryon counting problems in D-brane theories \cite{Butti:2006au}.

\section*{Acknowledgments}
We thank Noppadol Mekareeya for collaboration at the beginning of this project, Francesco Benini for collaboration on earlier related work, and Jurgis Pasukonis for allowing us to use and develop his Mathematica package.
CC is grateful to Ofer Aharony for interesting discussions.
SC thanks Amihay Hanany and Rak-Kyeong Seong for sharing their database and results prior to publication, and Paola Cognigni for help with figures.
CC is a Feinberg Postdoctoral Fellow at the Weizmann Institute of Science.
The work of SC is supported by the STFC Consolidated Grant ST/J000353/1.

%%%%%%%%%%%%%%%%%%%%%%%%%%%%%%%%%%%%%%%%%%%%%%%%%%%%%%%%%%%%%%%%%%%%%%%%%%%%%%%%%%%%%%%%%%%%%%%%%%%%
%%%%%%%%%%%%%%%%%%%%%%%%%%%%%%%%%%%%%%%%%%%%%%%%%%%%%%%%%%%%%%%%%%%%%%%%%%%%%%%%%%%%%%%%%%%%%%%%%%%%
%%%%%%%%%%%%%%%%%%%%%%%%%%%%%%%%%%%%%%%%%%%%%%%%%%%%%%%%%%%%%%%%%%%%%%%%%%%%%%%%%%%%%%%%%%%%%%%%%%%%
%%%%%%%%%%%%%%%%%%%%%%%%%%%%%%%%%%%%%%%%%%%%%%%%%%%%%%%%%%%%%%%%%%%%%%%%%%%%%%%%%%%%%%%%%%%%%%%%%%%%
%%%%%%%%%%%%%%%%%%%%%%%%%%%%%%%%%%%%%%%%%%%%%%%%%%%%%%%%%%%%%%%%%%%%%%%%%%%%%%%%%%%%%%%%%%%%%%%%%%%%
\appendix

\section{$\Ypq(B_4)$ geometry and type IIA background}\label{sec: Appendix topology}

For any smooth toric $CY_4$ cone, there exist a Sasaki-Einstein metric on the seven dimensional base $Y_7$ of the cone \cite{Futaki:2006cc} and a corresponding $AdS_4\times Y_7$ background of M-theory. In the simplest cases when $B_4$ is $\CP^2$ or $\bF_0$, the $\Ypq(B_4)$ metrics have been explicitly constructed in \cite{Gauntlett:2004hh, Chen:2004nq, Gauntlett:2004hs}. In this paper we are not explicitly interested in the metrics nor in the supergravity limit, but we do need to understand the topology of $\Ypq$.

Following \cite{Gauntlett:2004hh, Tomasiello:2010zz}, it is useful to realise the $SE_7$ geometry $\Ypq(B_4)$ as a circle bundle over a six-manifold $M_6$, which in turn is an $S^2$ bundle over the Fano variety $B_4$.
Such representation is physically sensible, because the circle fiber is the very M-theory circle we choose in section \ref{sec: Mtheory to IIA} to reduce to type IIA.
Therefore $M_6$ is the manifold that appears transverse to $AdS_4$ in the IIA limit of the $AdS_4$/CFT$_3$ duality.

In the following subsection we discuss the topology of $M_6$, which is intimately related to the $CY_3$ $\tilde{Y}= \cO_{B_4}(K)$ we discussed in detail in section \ref{section: toric quivers and dictionaries}.
In subsection \ref{subsec:_SE7_and_third_homology} we discuss the topology of $\Ypq$.
In the remaining subsections we give more details on the IIA fluxes obtained from the reduction of M-theory on  $AdS_4\times \Ypq$ with non-zero torsion $G_4$ flux.

%%%%%%%%%%%%%%%%%%%%%%%%%%%%%%%%%%%%%%%%%%%%%%%%%%%%%%%%%%%%%%%%%%%%%%%%%%%%%%%%%%%%%%%%%%%%%%%%%%%%
\subsection{Topology of $M_6$}\label{subsec:_M6}

The manifold $M_6$ is defined as $\Ypq/U(1)_M$, where $\Ypq$ is the seven manifold at the base of the $CY_4$ cone, and the $U(1)_M$ action is the one discussed in section \ref{subsec: CY3 from CY4}.
In the following we infer the topology of $M_6$ from the GLSM of the CY fourfold, following closely section 5.3 of \cite{Martelli:2004wu}.  To discuss the conical $CY_4$, it is convenient to rewrite (\ref{general_GLSM_CY4-CY3}) as a minimal GLSM%
\footnote{For simplicity we consider the case where $p$ and $q_\alpha$ are coprime.} (using that $Q^{\alpha}_{(s_0)}= Q_0^{\alpha}+\cI^{\alpha\beta}q_{\beta}$ and $Q^{\alpha}_{(s_1)}= -\cI^{\alpha\beta}q_{\beta}$, as follows from (\ref{def Q0 in term of CY4}) and  (\ref{relation Qs1 to Qalpha})):
\be\label{CY4 GLSM reduced}
\begin{array}{c|ccccc}
\text{CY}_4 & t_1 & \cdots & t_{m+2}  & s_0 &  s_p  \\
\hline
U(1)_{\alpha}& p\, Q^\alpha_1 &  \cdots& p\, Q^\alpha_{m+2} &  p\, Q_0^\alpha + \tilde{q}^{\alpha} & \,- \tilde{q}^{\alpha} \\
\hline
U(1)_M& 0 & \cdots& 0 & 1 & -1
\end{array}
\ee
which only takes into account the external points of the toric diagram. The last line denotes the charges under the $U(1)_M$ of the M-theory circle, and we have defined $\tilde{q}^{\alpha}\equiv \cI^{\alpha\beta} q_\beta$ for ease of notation.
The coordinates $(s_0, s_p)$ span the covering space $\bC^2$ of the $\bC^2/\bZ_p$ fiber over $B_4$. To see this, notice that the locus $s_0=s_p=0$ is an orbifold locus, left invariant by the subgroup $\bZ_p \subset U(1)_{\alpha}$, $\forall\;  U(1)_\alpha$. The action of the $\bZ_p$ on the fiber is
\be
\bZ_p \; : \; (s_0, s_p) \quad \mapsto\quad   (\omega_p^{\tilde q^{\alpha}} s_0, \omega_p^{-\tilde q^{\alpha}} s_p)\, .
\ee
Hence, when all $q^{\alpha}$ are coprime with $p$ we truly have a $\bZ_p$ action embedded in $U(1)_M$. This $\bZ_p$ acting on the fiber $(s_0, s_p)$ is a residual gauge symmetry once we have gauge fixed the $U(1)_{\alpha}$ acting on $\{t_i\}$.

Topologically, the compact $SE_7$ $\Ypq$ is described by the equations
\bea\label{Y7 from GLSM}
\sum_i Q_i^{\alpha} \, |t_i|^2 + (Q_0^{\alpha} p +\tilde{q}^{\alpha})|s_0|^2 -\tilde{q}^{\alpha} |s_p|^2 &= 0\, , \qquad \alpha=1, \cdots m\, ,\\
\sum_i |t_i|^2 + |s_0|^2 + |s_p|^2 &= L\, .
\eea
modulo the $U(1)^m$ gauge equivalence of the GLSM (\ref{CY4 GLSM reduced}) -- without the last line; $L$ is real and positive, and otherwise arbitrary. The locus $s_0= s_p=0$ does not intersect the manifold $Y^{p,\bm{q}}$ cut out by (\ref{Y7 from GLSM}); in fact the fixed point $s_0=s_p=0$ is the $B_4$ of vanishing size which lives at the tip of the $CY_3$ (and at $r_0=0$) in the $CY_3 \times\bR$ description of section \ref{sec: Mtheory to IIA}.  Therefore $\bZ_p$ acts freely on $\Ypq$. We have a $S^3/\bZ_p$ bundle over $B_4$, with $S^3/\bZ_p$ a Lens space. We want to quotient this Lens space by $U(1)_M$, to obtain a $S^2$ bundle over $B_4$, called $M_6$. To do that at the level of the GLSM description, we would have to construct new coordinates  invariant under $U(1)_M$. The only holomorphic choice, $s_0 s_p$, will not do, because such a coordinate can reach the origin $s_0=s_p=0$. Instead, since $s_p$ and $s_0$ can never vanish together, we consider local patches on the $S^3/\bZ_p$ fiber, with either $s_0$ or $s_p$ different from zero. Consider first the patch for which $s_0 \neq 0$. To go to the $S^2$ bundle, we introduce a coordinate
\be\label{t+ from CY4}
t^+ = \frac{s_p^*}{s_0}\, ,
\ee
which is invariant under $U(1)_M$. To summarise the remaining $U(1)_{\alpha}$ charges of the fields $t_i$ and $t^+$, we can write an auxiliary GLSM:
\be
\begin{array}{c|ccccc}
 & t_1 &          \dots    & t_{m+2}  &      t_0^+ \\
\hline
U(1)_{\alpha} & Q^\alpha_1 &  \dots& Q^\alpha_{m+2} &    -Q_0^\alpha  \\
\end{array}
\ee
This GLSM describes the anti-canonical bundle over $B_4$, $\cO_{B_4}(-K)$. In other words, the local patch $s_0\neq 0$ on the fiber $S^3/U(1)_M \cong S^2$ is fibered over $B_4$ exactly like the line bundle $\cO_{B_4}(-K)$.
Similarly, on the patch $s_p \neq 0$ we define the coordinate
\be\label{t- from CY4}
t^- = \frac{s_0}{s_p^*}
\ee
In this case we have the canonical bundle $\cO_{B_4}(K)$, with GLSM
\be
\begin{array}{c|ccccc}
 & t_1 &          \dots    & t_{m+2}  &      t_0^- \\
\hline
U(1)_{\alpha} & Q^\alpha_1 &  \dots& Q^\alpha_{m+2} &    Q_0^\alpha  \\
\end{array}
\ee
The two line bundles are patched together into a Riemann sphere, with $t^+= 1/t^-$ on the overlap, giving us the sought-after $M_6$ manifold.
The full $\CP^1$ bundle is just the anti-canonical line bundle with the point at infinity on the $\bC$ fiber added. This can be written
\be\label{global def of M6}
M_6 \cong \mathbb{P} ( \cO_{B_4}(-K) \oplus \cO_{B_4} )\, .
\ee
Equivalently, we could also write $M_6$ in term of the canonical bundle,  $M_6 \cong \mathbb{P} ( \cO_{B_4}(K) \oplus \cO_{B_4} )$.
We can describe both cases by the following GLSM's :
\be
\label{M6 toric: description as GLSM}
\begin{array}{c|ccccc}
M_6 & t_1 &          \dots    & t_{m+2}  &       t_0^- & t_0^+ \\
\hline
\cC^\alpha_+ & Q^\alpha_1 &  \dots& Q^\alpha_{m+2} &   0 & -Q_0^\alpha \\
\cC_0 &  0 &  \dots& 0 &   1 & 1
\end{array}
\qquad\qquad \quad\quad
\begin{array}{c|ccccc}
M_6 & t_1 &          \dots    & t_{m+2}  &       t_0^- & t_0^+\\ \hline
 \cC_\alpha^- & Q^\alpha_1 &  \dots& Q^\alpha_{m+2} &   Q_0^\alpha & 0 \\
  \cC_0 &  0 &  \dots& 0 &   1 & 1
\end{array}
\ee
We should remark that the complex structure apparent in this toric description of $M_6$ is \emph{not} inherited from the complex structure of the CY fourfold, due to the non-holomorphic relations (\ref{t+ from CY4}) and (\ref{t- from CY4}). In fact, the physical metric on $M_6$ which preserves $\cN=2$ supersymmetry might not even be K\"ahler, in general. The toric description with its unphysical complex structure is however quite useful as a tool to describe the topology of $M_6$.

Since $M_6$ is a sphere bundle over $B_4$, $S^2 \rightarrow M_6 \rightarrow B_4$, one can compute its cohomology in term of the cohomology of $B_4$ through the Gysin sequence, and its homology by Poincar\'e duality. We find
\be\label{homology of M6}
\begin{tabular}{c|ccccccc}
 $M_6$ & $H_0$ &$H_1$ & $H_2$ & $H_3$ & $H_4$ & $H_5$ & $H_6$    \\
 \hline
  & $\bZ$ & $0$ & $\bZ^{m+1}$ & $0$ & $\bZ^{m+1}$ & $0$ &  $\bZ $
\end{tabular}
\ee
In particular, we have
\be\label{homology M6}
H_2(M_6,\bZ) \cong H_2(B_4,\bZ) \oplus H_0(B_4,\bZ) \,,\quad
H_4(M_6,\bZ) \cong H_2(B_4,\bZ) \oplus H_4(B_4,\bZ)   \,.
\ee
To be more concrete in the construction of $H_2(M_6)$ and $H_4(M_6)$ we need to specify how to embed cycles of $B_4$ in $M_6$, that is we need to specify global sections of the $S^2$ bundle. The  $S^2$ fiber is twisted along $B_4$ by the $U(1)$ which rotates its azimuthal angle. There are two fixed points of this $U(1)$ action, the north and the south pole of $S^2$ ($t^+=0$ and $t^-=0$, respectively), which can be used to embed submanifolds of $B_4$ (and $B_4$ itself) in $M_6$ as global sections \cite{Martelli:2008rt}.
Let $\sigma_N$ and $\sigma_S$ denote the push-forward maps $B_4 \xrightarrow{\pi_*} M_6$ which fix the north pole or the south pole of the  $S^2$ fiber, respectively. One can define nice representatives of the 2-cycles (\ref{homology M6}) by
\be
\label{Set of 2 cycles}
-\cC_0 = S^2 \;,\qquad\qquad \cC_\alpha^+ = \sigma_N \cC_\alpha \;,\qquad\qquad \cC_\alpha^- = \sigma_S \cC_\alpha \;,
\ee
where $-\cC_0$ is the $S^2$ fiber over an arbitrary point on the base.%
\footnote{It will be convenient to take a minus sign in the definition $S^2= -\cC_0$, because the $\cC_0$ as defined by (\ref{M6 toric: description as GLSM}) has an orientation opposite to the ``natural'' one, due to the complex conjugation in (\ref{t+ from CY4}).}
We anticipated these definitions in (\ref{M6 toric: description as GLSM}). All such 2-cycles are not independent: as a basis for $H_2$ we could pick $\{ -\cC_0, \cC_\alpha^+ \}$, but it will be convenient to work with the redundant set (\ref{Set of 2 cycles}). Similarly we define the 4-cycles $D_\alpha$ and $D^\pm$:
\be\label{Set of 4 cycles}
S^2 \hookrightarrow D_\alpha \to \cC_\alpha \;,\qquad\qquad D^+ = \sigma_N B_4 \;,\qquad\qquad D^- = \sigma_S B_4 \;,
\ee
where $D_\alpha$ are restrictions of the $S^2$-bundle to $\cC_\alpha$. The 4-cycles $D^+$ and $D^{-}$ corresponds to the toric divisors $\{t^+=0\}$ and $\{t^-=0\}$, while the 4-cycles $D_{\alpha}$ and the toric divisors $D_i = \{t_i=0\}$ are related by
\be
D_i = Q_i^{\alpha} (\cI^{-1})_{\alpha\beta}\,  D^{\beta}
\ee
We can also invert this relation, meaning that we can take $D_{\alpha}= X_{\alpha}^i D_i$ for any $X^i_{\alpha}$ such that $X_{\alpha}^i Q_{i}^{\beta} = \cI^{\beta}_{\alpha}$. The homology relations among the representatives (\ref{Set of 2 cycles})-(\ref{Set of 4 cycles}) are
\bea
\label{relations in homology}
\cC_\alpha^+ &= \cC_\alpha^- - Q_0^\alpha \, \cC_0\; , \\
D^+ &= D^- - Q^\alpha_0 (\cI^{-1})_{\alpha\beta} \, D^\beta\; ,
\eea
as follows from (\ref{M6 toric: description as GLSM}).
Moreover, using the toric description it is straightforward to compute the intersections among 2-cycles and 4-cycles:
\be\label{intersections 2-4cycles}
\begin{array}{c|ccc|ccc}
& \cC_0 & \cC_\alpha^+ & \cC_\alpha^- & D_\alpha & D^+ & D^- \\
\hline
D_\beta & 0 & \cI_{\alpha\beta} & \cI_{\alpha\beta} & \cI_{\alpha\beta} \, \cC_0 & \cC_\beta^+ & \cC_\beta^- \\
D^+ & 1 & -Q_0^\alpha & 0 & \cC_\alpha^+ & - Q_0^\rho (\cI^{-1})_\rho^\sigma \, \cC^+_\sigma & 0 \\
D^- & 1 & 0 & Q_0^\alpha & \cC_\alpha^- & 0 & Q_0^\rho (\cI^{-1})_\rho^\sigma \, \cC_\sigma^-
\end{array}
\ee
Equipped with the understanding of $M_6$, we turn back to the Sasaki-Einstein seven-manifold which appears in M-theory.

\subsection{Topology of $\Ypq(B_4)$}\label{subsec:_SE7_and_third_homology}
Consider the seven-manifold $\Ypq$ given by a circle fibration over $M_6$:
\be
S^1 \,\rightarrow\, \Ypq \,\rightarrow\, M_6 \;.
\ee
The circle bundle is fully characterised by its first Chern class $c_1 \in H^2(M_6, \bZ)$, which equals the type IIA RR 2-form flux in $M_6$, or equivalently by the $m+1$ Chern numbers $(p, q_\alpha)$. We can compute the cohomology of $\Ypq$ from $H_*(M_6,\bZ)$ (\ref{homology of M6}) and the Gysin sequence.
One can show that
\be
\pi_1(Y^{p,\bm{q}}) = \bZ_{\mathrm{gcd}(p, q_1, \cdots, q_m)} \;,
\ee
so that $\Ypq$ is simply connected  if and only if the Chern numbers are co-prime.
The homology $H_*(\Ypq,\bZ)$ is
\be
\begin{tabular}{c|cccccccc}
$\Ypq$ & $H_0$ &$H_1$ & $H_2$ & $H_3$ & $H_4$ & $H_5$ & $H_6$ & $H_7$   \\
\hline
& $\bZ$ & $\bZ_{\mathrm{gcd}(p, q_{\alpha})}$ & $\bZ^m$ & $\Gamma$ & $0$ & $\bZ^m \oplus \bZ_{\mathrm{gcd}(p, q_{\alpha})}$ & $0$ & $\bZ$
\end{tabular} \;,
\ee
and similarly for cohomology by Poincar\'e duality.
The most interesting group is
\be
H^4(\Ypq, \bZ) \cong H_3(Y^{p,\bm{q}}, \bZ) \cong \Gamma \;,
\ee
which we now explain. From the Gysin sequence one finds that
\be
\label{identity for H4 of X7}
H^4 (\Ypq) \cong  H^4(M_6) / \mathrm{Im}(c_1) \;,\qquad \text{with} \;\quad c_1: H^2(M_6) \stackrel{\wedge c_1}{\longrightarrow} H^4(M_6) \;.
\ee
Let us consider $(-D^-, D_{\alpha} )$ as a basis of $H_4(M_6)$; in that case, the dual basis of $H^4(M_6)$ is given by $(-\cC_0,(\cI^{-1})^{\alpha\beta}  \cC^+_\beta)$.
The image of the map $c_1$ in (\ref{identity for H4 of X7}) is computed from (\ref{intersections 2-4cycles}):
\bea
\label{c1 act 001}
& c_1(-D^-) &= &\;\; q_{\alpha} Q_0^{\alpha}\, \cC_0  + q^{\alpha} \, \cC_{\alpha}^+ \\
& c_1(D_{\alpha}) &= &\;\; - \cI_{{\alpha\beta}}q^{\beta} \, \cC_0 \, + \, p \, \cC_{\alpha}^+ \;.
\eea
 We find that $\Gamma$ is the finite Abelian group
\be
\label{generic sol for Gamma}
\Gamma = \bZ^{m+1} / <v_0, v_1, \cdots, v_{m}>
\ee
where the vectors $v= (v_0, v_\alpha)$ are read from (\ref{c1 act 001}):
\be
\label{generic v}
v_0 = (-q_{\gamma} Q_0^{\gamma} \,,\,  \cI^{\beta\gamma} q_{\gamma}) \;,\qquad\qquad
v_{\alpha} = (q_{\alpha} \,,\, \delta_\alpha^\beta \, p ) \;,
\ee
and the index $\beta$ parameterises the coordinates (but the first one) of the vectors.
As a simple example, let us consider  $Y^{p,q}(\CP^2)$. We have $m=1$, $\cI=1$, $Q_0=-3$ and $F_2 = p\, D^+ - q\, D$, therefore
\be
\Gamma = \bZ^2/ < (3q, q) \,,\,  (q, p)> \;,
\ee
which was computed in \cite{Martelli:2008rt} and explained from the field theory point of view in \cite{Benini:2011cma}.

%%%%%%%%%%%%%%%%%%%%%%%%%
\subsection{Type IIA background and the IIA dual of torsion flux}
\label{subsec:_IIA_dual_of_torsion}

Given a Sasaki-Einstein metric on $\Ypq(B_4)$, we have $\cN=2$ supersymmetric solution of 11d supergravity, given by
\be\label{Mth metric and flux}
\begin{split}
ds^2 \,&=  \,  R^2\left( \frac{1}{4}\,ds^2(AdS_4)   \, +\, ds^2(Y^{p, q_{\alpha}}(B_4)) \right)\, , \\
G_4 \, &= \, \frac{3}{8}\, R^3 \, d\vol{(AdS_4)} \, .
\end{split}
\ee
We have $N$ units of M2-brane charge on $\Ypq$, where $N$ is related to the radius $R$ according to:
\be
\frac{1}{(2\pi l_p)^6} \int_{Y^{p,\bm{q}}} \ast G_4 = N\, = \,   \, \frac{6 R^6}{(2 \pi l_p)^6} \mathrm{Vol}(Y^{p,\bm{q}})   \,  .
\ee
The manifold $Y^{p, \bm{q}}$ has a fourth cohomology $H_3(Y^{p,\bm{q}},\bZ) \cong \Gamma$ (\ref{generic sol for Gamma}) which is purely torsion. A torsion $G_4$ flux does not affect the supergravity equations of motion, and we therefore have a distinct M-theory background for each element $[G_4] \in \Gamma$.

We are interested in the corresponding type IIA solutions $AdS_4 \times M_6$. The $M_6$ metric is of the type introduced in \cite{Gauntlett:2004hh}. A nice account of this class of IIA reduction can be found in \cite{Martelli:2009ga}. As we already stated, we are not interested in the metric nor in the supegravity limit in particular. For this reason we will only discuss the RR fluxes that are present on $M_6$, as we go from the M-theory background (\ref{Mth metric and flux}) with generic $[G_4] \in \Gamma$ to the type IIA dual. The charges so-defined are conserved quantities of great use to derive the dual CS quiver gauge theory.

One can straightforwardly generalise the analysis of \cite{Benini:2011cma} (section 3.5) of the D-brane charges to any $Y^{p, \bm{q}}(B_4)$.
The D6-brane Page charges are directly related to the Chern numbers, according to
\bea\label{D6 Page charge}
&Q_{6;\, 0}\equiv \int_{-\cC_0} F_2 = -p  \,,\quad\\
&Q_{6;\, \alpha-}\equiv \int_{\cC_\alpha^-} F_2 =   -\cI^{\alpha\beta}q_{\beta}  \,,\quad\qquad
Q_{6;\, \alpha+} \equiv \int_{\cC_\alpha^+} F_2=  -\cI^{\alpha\beta} q_{\beta} -  Q^\alpha_0\, p  \,.
\eea
The D4-brane Page charges are the integral of $B_2\wedge F_2$ over the 4-cycles (\ref{Set of 4 cycles}). They are given by
\be\label{def Q4 Page charges for M6}
\begin{array}{llll}
Q_{4;\, -} & = {\displaystyle -\int_{D^- \rule[-.5em]{0pt}{1em}} F_2 \wedge B_2} &=    \;
 n_0 - \frac12 q^\alpha\,  \cI_{\alpha\beta}  s^\beta\, ,  \\
Q_{4;\, \alpha} &= {\displaystyle \int_{D_\alpha \rule[-.5em]{0pt}{1em}} F_2 \wedge B_2} &= \; n_\alpha - \frac12 p \,\cI_{\alpha\beta} s^\beta \, , \\
Q_{4;\, +} & = {\displaystyle -\int_{D^+} F_2 \wedge B_2} &= Q_{4;\, -} + Q^\alpha_0 (\cI^{-1})_{\alpha\beta} Q_{4;\, \beta} \;.
\end{array}
\ee
with
\be\label{integer values torsion group}
(n_0 \,,\, n_\alpha) \in \bZ^{m+1}
\ee
the integers parameterizing the torsion group $\Gamma$ (\ref{generic sol for Gamma}). The parameters $s_{\alpha}$ are the Freed-Witten anomaly parameters, which are $0$ or $1$ depending on whether the 4-cycle $D_{\alpha}$ in $M_6$ is spin or only spin$^c$ --- see equation (\ref{wv flux for FW and def of s}) below.
Finally, the D2-brane Page charge is
\be\label{D2 Page charge in our background}
Q_2 = N - \frac{p}{8} s_\alpha s_\beta \cI^{\alpha\beta} \;.
\ee

The quantization of the D4 Page charge results in a quantised flat background B-field. Defining the periods
\be
b_0 \equiv \int_{-\cC_0} B_2 \;,\qquad\qquad b_\alpha^- \equiv \int_{\cC_\alpha^-} B_2 \;,
\qquad\qquad b_\alpha^+ \equiv \int_{\cC_\alpha^+} B_2  = b_\alpha^- + Q^0_\alpha b_0 \; ,
\ee
we find
\be\label{b fields with FW anomaly}
b_0 = \frac{-p\, n_0 + q^\alpha n_\alpha}{q_\alpha \cI^{\alpha\beta} q_\beta + p Q_0^\alpha q_\alpha}\, , \qquad \qquad
b_\alpha^+ = - \cI_{\alpha\beta} \frac{s^\beta}2 + \frac{n_\alpha}p - \frac{\cI^{\alpha\beta}q_\beta }p \, b_0 \;.
\ee
The periodicities (\ref{generic v}) defining $\Gamma$ are realised as the following large gauge transformations of the B-field:
\bea
\label{large gauge transfo and n shift}
\delta (n_0,\, n_\beta) &= v_0 \qquad &&\longleftrightarrow\qquad & \delta(b_0,\, b_\beta^+) &= (1,0) \\
\delta (n_0,\, n_\beta) &= v_\alpha \qquad &&\longleftrightarrow\qquad & \delta(b_0,\, b_\beta^+) &= (0, \delta_{\beta\alpha}) \;.
\eea

\subsection{D6-branes and Freed-Witten anomaly}\label{subsec: D6s and FW}
\label{subsec:FW_and_D4_Page_charge}
To obtain the $F_2$ fluxes and the corresponding Page D6-charges (\ref{D6 Page charge}), we can wrap D6-branes on various 4-cycles of $M_6$ at some radial position in $AdS_4$%
\footnote{Or at some radial position in the cone $C(M_6)$, if we do not wish to consider the decoupling limit nor the supergravity limit, like in most of this paper.}
where they are not stable, and let them fall inside. At a fixed radial position such a wrapped D6-brane acts as a domain wall between two regions with different RR flux.

When a 4-cycle is not spin but only spin$^c$, the wrapped D6-brane must carry some half-integer worldvolume flux to cancel the Freed-Witten anomaly \cite{Freed:1999vc}. This induces some extra background D4 Page charge, giving rise to the half-integer shifts in (\ref{def Q4 Page charges for M6}). It also induces the shift of the D2 Page charge in (\ref{D2 Page charge in our background}).

Let us consider the various 4-cycles (\ref{Set of 4 cycles}) in $M_6$. The topology of $D^\pm$ is $B_4$. We have
\be\label{B4 and FW 001}
\int_{\cC_{\alpha}} c_1(B_4) \, = \, -Q_0^{\alpha}\, ,
\ee
so that $B_4$ fails to be spin whenever some of the $Q_0^{\alpha}$ is odd.  The other 4-cycles are $D_{\alpha}$. These are $S^2$ bundles over $\cC_{\alpha}$%
\footnote{In $B_4$ (or rather in $\tilde{B}_4$) all the $\cC_{\alpha}$ are topologically 2-spheres as well. $S^2$ bundle over $S^2$ are classified by $\pi_1 (SO(3))= \bZ_2$, so there are only two distinct topologies, trivial or not. In term of the description (\ref{global def of M6}) we have $D_{\alpha} \cong  \bP(\cO(Q_0^{\alpha})\oplus \cO)$. When $Q^{\alpha}_0$ is odd the corresponding $D_{\alpha}$ is not spin (and it corresponds to the non-trivial topology).},
which are spin or spin$^c$ depending on whether $Q_0^{\alpha}$ is odd or even.
One can show that all these Freed-Witten anomalies can be cancelled by turning on the pull-back of a common bulk 2-form
\be\label{wv flux for FW and def of s}
F =  \frac12 \sum_\alpha s^\alpha D_\alpha \, , \qquad \quad \text{with}\qquad s_{\alpha}=(\cI^ {-1})_{\alpha\beta}Q_0^{\beta} \quad \text{mod}\; 2\, .
\ee
on every D6-brane.
Let us also remark  that at the torsionless point $(n_0, n_{\alpha})=(0,0)$, we have a non-zero flat $B$-field
\be
B_2 = -\frac12 \sum\nolimits_{\alpha} s_{\alpha} \, D^{\alpha}\, ,
\ee
turned on whenever there is a Freed-Witten anomaly. This B-field induces the non-zero half-integer periods  $b^+_{\alpha}$ in (\ref{b fields with FW anomaly}), and it is such that the gauge invariant worldvolume flux $\cF=  B_2 + F $ on any probe D6-brane vanishes.
That the B-field is non-zero even in the torsionless case was first argued for the ABJM theory in \cite{Aharony:2009fc};  the present results generalise that understanding to our family of geometries. How this subtelty in charge quantization translates in the M-theory language is not well understood, to the best of our present knowledge.

\subsection{Large gauge transformations and shift of the D2 Page charge}
Under a large gauge transformation of the B-field, Page charges shift by integers \cite{Marolf:2000cb, Benini:2007gx}. We have already seen this in (\ref{large gauge transfo and n shift}) for the D4-brane Page charges: Large gauge transformations of $B$ are in one to one correspondence with periodicities of the torsion group $\Gamma$ in M-theory.

It is important to note that the D2-brane Page charges shifts as well along these periodicities. The D2 Page charge computed on $M_6$ is
\be
\label{D2 brane charge}
Q_2 = \int_{M_6}\left( \ast F_4 - \frac{1}{2} B_2\wedge B_2 \wedge F_2   \right) \;.
\ee
Remark that $Q_2 = Q^{\mathrm{Maxwell}}_2 + \tilde{Q}_2$, where we defined
\be
\tilde{Q}_2 \equiv -\frac{1}{2 }\int_{M_6} B_2\wedge B_2\wedge F_2 \;.
\ee
Computing $\tilde{Q}_2$ explcitly in term of $(n_0, n_\alpha)$, we find
\be
\tilde Q_2(n_0, n_{\alpha}) = - \frac p8 s_\alpha \cI^{\alpha\beta} s_\beta + \frac{n_\alpha s^\alpha}2 - \frac{n_\alpha (\cI^{-1})^{\alpha\beta} n_\beta}{2p} + \frac{(p n_0 - q^\alpha n_\alpha)^2}{2p(q_\alpha I^{\alpha\beta} q_\beta + p Q_0^\alpha q_\alpha)} \;.
\ee
At the torsionless point we have%
\footnote{There are further corrections from gravitational effects similarly to \cite{Bergman:2009zh}, but  such contributions will not be studied in this work.}
 $Q^{\mathrm{Maxwell}}_2 =N$, while the Page charge is given by (\ref{D2 Page charge in our background}). As we move in the torsion group,  the Page charge $Q_2$ is invariant, while the Maxwell charge varies accordingly. On the other hand, under a large gauge transformation (\ref{large gauge transfo and n shift}) the Page charge $Q_2$ shifts according to
\be\label{general formula for shift of Q2}
\delta_v Q_2 = \tilde{Q}_2(n_0+\delta n_0, n_{\alpha}+\delta n_{\alpha}) -\tilde{Q}_2(n_0, n_{\alpha})\, ,
\ee
where $v=(\delta n_0, \delta n_{\alpha})$ is any periodicity vector of $\Gamma$. One can show that the shift $\delta_v Q_2$ is always an integer.

\subsection{Fluxes and D-branes in $C(M_6)$}\label{subsec: flux and Dbranes on CM6}
In the bulk of this paper we are not interested in $M_6$ \emph{per se}, but rather in the cone $C(M_6)= CY_4/U(1)_M$ seen as a folliation of $\tilde{Y}_{\pm}\cong \cO_{B_4}(K)$ along a line $\bR\cong \{r_0\}$. We showed in section \ref{subsec:_M6} that $M_6$ and $\tilde{Y}$ are closely related. The north pole (resp. south pole) of the $S^2$ fiber of $M_6$ is the exceptional locus $B_4$ of $\tilde{Y}_+$ at $r_0 > 0$ (resp. $\tilde{Y}_-$ at $r_0 < 0$), and cycles living there can be compared. We have:
\bea\label{rel btw cycles M6 CY3}
\cC_\alpha^+ &= \cC_\alpha \text{ at } r_0 > 0 \qquad\qquad &
D^+ &=  D_0 \text{ at } r_0 > 0 \\
\cC_\alpha^- &= \cC_\alpha \text{ at } r_0 < 0 \qquad\qquad &
D^- &= D_0 \text{ at } r_0 < 0 \;.
\eea
These are the only cycles which are common to the $CY_3$ and $M_6$; for instance the divisors $D_\alpha$, are different in the $CY_3$ and in $M_6$ despite having the same name: they are non-compact in the $CY_3$ and compact in $M_6$.

The fluxes through $\cC_{\alpha}^{\pm}$ and $B_4$ in $\tilde{Y}_{\pm}$ are the same as measured on the corresponding cycles in $M_6$. We collect them into two $(m+2)$-covectors of charges
\be
\bm{Q}_{\text{flux}, \pm} \,  \equiv\, \left( -Q_{4;\,  \pm} \,|\, Q_{6;\, \alpha \pm} \,|\, 0 \right)\, .
\ee
The last entry is zero because we do not allow for D8-brane charge $F_0$, thus allowing a M-theory uplift \cite{Aharony:2010af}; see \cite{Closset:2012eq} for a sketch of how to generalise the present formalism to the case of $F_0\neq 0$.

On the other hand, the fluxes through the remaining 2- and 4- cycles $\cC_0$ and $D_{\alpha}$ of $M_6$ correspond to explicit D-brane sources wrapped on the dual 4-cycles $B_4$ and $(\cI^{-1})^{\alpha\beta}\cC_{\beta} \subset B_4$ at $r_0=0$. For the (anti)D6-brane wrapped on $B_4$ this was shown in section \ref{subsec:F2_and_D6}. Similarly, the jump of the D4 Page charge across $r_0=0$ denotes explicit sources through the 2-cycles:
\be\label{jump D4 flux}
 Q_{4;\,  +} - Q_{4;\,  -}  = Q_0^{\alpha} (\cI^{-1})^{\alpha}_{\beta} Q_{D4;\, \beta}\, .
\ee
In the presence of torsion flux $(n_0, n_{\alpha})\in \Gamma$, we have $(\cI)^{\beta\alpha} n_{\alpha}$ D4-branes wrapped on $\cC_{\beta}$ at $r_0=0$. Remark that only the combination $Q_0^{\alpha} (\cI^{-1})_{\alpha\beta} n^{\beta}$ appears in (\ref{jump D4 flux}); the remaining $m-1$ choices $n_{\alpha}$ orthogonal to $(\cI^{-1})_{\alpha\beta} Q_0^{\beta}$ are ``non-anomalous'' D4-branes. We collect the information about the D-brane sources in a covector of Page charges
\be
\bm{Q}_\text{source} \, \equiv\, \, \left( Q_{6;0} \,|\, (\cI^{-1})^{\alpha\beta} Q_{4; \beta} \,|\, Q_2 \right)\, .
\ee
In this notation, the results (\ref{D6 Page charge}), (\ref{def Q4 Page charges for M6}) and (\ref{D2 Page charge in our background}) are summarised by
\bea\label{fluxes and sources generic torsion}
 & \bm{Q}_{\text{flux}, -}= (- n_0 +\frac12 s_{\alpha} q_{\beta}  \cI^{\alpha\beta} \, | \, -\cI_{\alpha\beta} q^{\beta}\, | \, 0  )  \, ,\\
 &\bm{Q}_\text{source} =    ( -p \,|\, (\cI^{-1})^{\alpha\beta} n_{\beta} -\frac12 s_{\alpha} p    \,|\,  N-\frac18 s_{\alpha}s_{\beta}\cI^{\alpha\beta} p )        \, .
\eea
%%%%%%%%%%%%%%%%%%%%%%%%%%%%%%%%%%%%%%%%%%%%%%%%%%%%%%%%%%%%%%%%%%%%%%%%%%%%%%%%%%%
%%%%%%%%%%%%%%%%%%%%%%%%%%%%%%%%%%%%%%%%%%%%%%%%%%%%%%%%%%%%%%%%%%%%%%%%%%%%%%%%%%%
\bibliographystyle{utphys}
\bibliography{bibM2Fano}{}

\providecommand{\href}[2]{#2}\begingroup\raggedright\begin{thebibliography}{10}

\bibitem{Aharony:2008ug}
O.~Aharony, O.~Bergman, D.~L. Jafferis, and J.~Maldacena, ``{N=6 superconformal
  Chern-Simons-matter theories, M2-branes and their gravity duals},''
  \href{http://dx.doi.org/10.1088/1126-6708/2008/10/091}{{\em JHEP} {\bfseries
  10} (2008) 091},
\href{http://arxiv.org/abs/0806.1218}{{\ttfamily arXiv:0806.1218 [hep-th]}}.
%%CITATION = 0806.1218;%%.

\bibitem{Maldacena:1997re}
J.~M. Maldacena, ``{The Large N limit of superconformal field theories and
  supergravity},'' \href{http://dx.doi.org/10.1023/A:1026654312961}{{\em
  Adv.Theor.Math.Phys.} {\bfseries 2} (1998) 231--252},
\href{http://arxiv.org/abs/hep-th/9711200}{{\ttfamily arXiv:hep-th/9711200
  [hep-th]}}.
%%CITATION = HEP-TH/9711200;%%.

\bibitem{Acharya:1998db}
B.~S. Acharya, J.~Figueroa-O'Farrill, C.~Hull, and B.~J. Spence, ``{Branes at
  conical singularities and holography},'' {\em Adv.Theor.Math.Phys.}
  {\bfseries 2} (1999) 1249--1286,
\href{http://arxiv.org/abs/hep-th/9808014}{{\ttfamily arXiv:hep-th/9808014
  [hep-th]}}.
%%CITATION = HEP-TH/9808014;%%.

\bibitem{Jafferis:2008qz}
D.~L. Jafferis and A.~Tomasiello, ``{A Simple class of N=3 gauge/gravity
  duals},'' \href{http://dx.doi.org/10.1088/1126-6708/2008/10/101}{{\em JHEP}
  {\bfseries 0810} (2008) 101},
\href{http://arxiv.org/abs/0808.0864}{{\ttfamily arXiv:0808.0864 [hep-th]}}.
%%CITATION = ARXIV:0808.0864;%%.

\bibitem{Martelli:2008si}
D.~Martelli and J.~Sparks, ``{Moduli spaces of Chern-Simons quiver gauge
  theories and AdS(4)/CFT(3)},''
  \href{http://dx.doi.org/10.1103/PhysRevD.78.126005}{{\em Phys. Rev.}
  {\bfseries D78} (2008) 126005},
\href{http://arxiv.org/abs/0808.0912}{{\ttfamily arXiv:0808.0912 [hep-th]}}.
%%CITATION = 0808.0912;%%.

\bibitem{Hanany:2008cd}
A.~Hanany and A.~Zaffaroni, ``{Tilings, Chern-Simons Theories and M2 Branes},''
  \href{http://dx.doi.org/10.1088/1126-6708/2008/10/111}{{\em JHEP} {\bfseries
  0810} (2008) 111},
\href{http://arxiv.org/abs/0808.1244}{{\ttfamily arXiv:0808.1244 [hep-th]}}.
%%CITATION = ARXIV:0808.1244;%%.

\bibitem{Franco:2008um}
S.~Franco, A.~Hanany, J.~Park, and D.~Rodriguez-Gomez, ``{Towards M2-brane
  Theories for Generic Toric Singularities},''
  \href{http://dx.doi.org/10.1088/1126-6708/2008/12/110}{{\em JHEP} {\bfseries
  0812} (2008) 110},
\href{http://arxiv.org/abs/0809.3237}{{\ttfamily arXiv:0809.3237 [hep-th]}}.
%%CITATION = ARXIV:0809.3237;%%.

\bibitem{Ueda:2008hx}
K.~Ueda and M.~Yamazaki, ``{Toric Calabi-Yau four-folds dual to
  Chern-Simons-matter theories},''
  \href{http://dx.doi.org/10.1088/1126-6708/2008/12/045}{{\em JHEP} {\bfseries
  0812} (2008) 045},
\href{http://arxiv.org/abs/0808.3768}{{\ttfamily arXiv:0808.3768 [hep-th]}}.
%%CITATION = ARXIV:0808.3768;%%.

\bibitem{Imamura:2008qs}
Y.~Imamura and K.~Kimura, ``{Quiver Chern-Simons theories and crystals},''
  \href{http://dx.doi.org/10.1088/1126-6708/2008/10/114}{{\em JHEP} {\bfseries
  0810} (2008) 114},
\href{http://arxiv.org/abs/0808.4155}{{\ttfamily arXiv:0808.4155 [hep-th]}}.
%%CITATION = ARXIV:0808.4155;%%.

\bibitem{Hanany:2008fj}
A.~Hanany, D.~Vegh, and A.~Zaffaroni, ``{Brane Tilings and M2 Branes},''
  \href{http://dx.doi.org/10.1088/1126-6708/2009/03/012}{{\em JHEP} {\bfseries
  0903} (2009) 012},
\href{http://arxiv.org/abs/0809.1440}{{\ttfamily arXiv:0809.1440 [hep-th]}}.
%%CITATION = ARXIV:0809.1440;%%.

\bibitem{Aganagic:2009zk}
M.~Aganagic, ``{A Stringy Origin of M2 Brane Chern-Simons Theories},''
\href{http://arxiv.org/abs/0905.3415}{{\ttfamily arXiv:0905.3415 [hep-th]}}.
%%CITATION = 0905.3415;%%.

\bibitem{Jafferis:2009th}
D.~L. Jafferis, ``{Quantum corrections to N=2 Chern-Simons theories with flavor
  and their AdS(4) duals},'' \href{http://arxiv.org/abs/0911.4324}{{\ttfamily
  arXiv:0911.4324 [hep-th]}}.

\bibitem{Benini:2009qs}
F.~Benini, C.~Closset, and S.~Cremonesi, ``{Chiral flavors and M2-branes at
  toric CY4 singularities},''
  \href{http://dx.doi.org/10.1007/JHEP02(2010)036}{{\em JHEP} {\bfseries 1002}
  (2010) 036}, \href{http://arxiv.org/abs/0911.4127}{{\ttfamily arXiv:0911.4127
  [hep-th]}}.

\bibitem{Benini:2011cma}
F.~Benini, C.~Closset, and S.~Cremonesi, ``{Quantum moduli space of
  Chern-Simons quivers, wrapped D6-branes and AdS4/CFT3},''
  \href{http://dx.doi.org/10.1007/JHEP09(2011)005}{{\em JHEP} {\bfseries 1109}
  (2011) 005}, \href{http://arxiv.org/abs/1105.2299}{{\ttfamily arXiv:1105.2299
  [hep-th]}}.

\bibitem{Gang:2011jj}
D.~Gang, C.~Hwang, S.~Kim, and J.~Park, ``{Tests of AdS$_4$/CFT$_3$
  correspondence for $\mathcal{N}=2$ chiral-like theory},''
\href{http://arxiv.org/abs/1111.4529}{{\ttfamily arXiv:1111.4529 [hep-th]}}.
%%CITATION = ARXIV:1111.4529;%%.

\bibitem{Jafferis:2010un}
D.~L. Jafferis, ``{The Exact Superconformal R-Symmetry Extremizes Z},''
\href{http://arxiv.org/abs/1012.3210}{{\ttfamily arXiv:1012.3210 [hep-th]}}.
%%CITATION = 1012.3210;%%.

\bibitem{Hama:2010av}
N.~Hama, K.~Hosomichi, and S.~Lee, ``{Notes on SUSY Gauge Theories on
  Three-Sphere},'' \href{http://dx.doi.org/10.1007/JHEP03(2011)127}{{\em JHEP}
  {\bfseries 03} (2011) 127},
\href{http://arxiv.org/abs/1012.3512}{{\ttfamily arXiv:1012.3512 [hep-th]}}.
%%CITATION = 1012.3512;%%.

\bibitem{Imamura:2011su}
Y.~Imamura and S.~Yokoyama, ``{Index for three dimensional superconformal field
  theories with general R-charge assignments},''
  \href{http://dx.doi.org/10.1007/JHEP04(2011)007}{{\em JHEP} {\bfseries 04}
  (2011) 007},
\href{http://arxiv.org/abs/1101.0557}{{\ttfamily arXiv:1101.0557 [hep-th]}}.
%%CITATION = 1101.0557;%%.

\bibitem{Cheon:2011vi}
S.~Cheon, H.~Kim, and N.~Kim, ``{Calculating the partition function of N=2
  Gauge theories on $S^3$ and AdS/CFT correspondence},''
  \href{http://dx.doi.org/10.1007/JHEP05(2011)134}{{\em JHEP} {\bfseries 1105}
  (2011) 134},
\href{http://arxiv.org/abs/1102.5565}{{\ttfamily arXiv:1102.5565 [hep-th]}}.
%%CITATION = ARXIV:1102.5565;%%.

\bibitem{Martelli:2011qj}
D.~Martelli and J.~Sparks, ``{The large N limit of quiver matrix models and
  Sasaki-Einstein manifolds},''
  \href{http://dx.doi.org/10.1103/PhysRevD.84.046008}{{\em Phys.Rev.}
  {\bfseries D84} (2011) 046008},
\href{http://arxiv.org/abs/1102.5289}{{\ttfamily arXiv:1102.5289 [hep-th]}}.
%%CITATION = ARXIV:1102.5289;%%.

\bibitem{Cheon:2011th}
S.~Cheon, D.~Gang, S.~Kim, and J.~Park, ``{Refined test of AdS4/CFT3
  correspondence for N=2,3 theories},''
  \href{http://dx.doi.org/10.1007/JHEP05(2011)027}{{\em JHEP} {\bfseries 1105}
  (2011) 027},
\href{http://arxiv.org/abs/1102.4273}{{\ttfamily arXiv:1102.4273 [hep-th]}}.
%%CITATION = ARXIV:1102.4273;%%.

\bibitem{Jafferis:2011zi}
D.~L. Jafferis, I.~R. Klebanov, S.~S. Pufu, and B.~R. Safdi, ``{Towards the
  F-Theorem: N=2 Field Theories on the Three-Sphere},''
  \href{http://dx.doi.org/10.1007/JHEP06(2011)102}{{\em JHEP} {\bfseries 1106}
  (2011) 102},
\href{http://arxiv.org/abs/1103.1181}{{\ttfamily arXiv:1103.1181 [hep-th]}}.
%%CITATION = ARXIV:1103.1181;%%.

\bibitem{Amariti:2011uw}
A.~Amariti, C.~Klare, and M.~Siani, ``{The Large N Limit of Toric Chern-Simons
  Matter Theories and Their Duals},''
\href{http://arxiv.org/abs/1111.1723}{{\ttfamily arXiv:1111.1723 [hep-th]}}.
%%CITATION = ARXIV:1111.1723;%%.

\bibitem{Hanany:2005ve}
A.~Hanany and K.~D. Kennaway, ``{Dimer models and toric diagrams},''
\href{http://arxiv.org/abs/hep-th/0503149}{{\ttfamily arXiv:hep-th/0503149
  [hep-th]}}.
%%CITATION = HEP-TH/0503149;%%.

\bibitem{Franco:2005rj}
S.~Franco, A.~Hanany, K.~D. Kennaway, D.~Vegh, and B.~Wecht, ``{Brane dimers
  and quiver gauge theories},''
  \href{http://dx.doi.org/10.1088/1126-6708/2006/01/096}{{\em JHEP} {\bfseries
  0601} (2006) 096},
\href{http://arxiv.org/abs/hep-th/0504110}{{\ttfamily arXiv:hep-th/0504110
  [hep-th]}}.
%%CITATION = HEP-TH/0504110;%%.

\bibitem{Gauntlett:2004hh}
J.~P. Gauntlett, D.~Martelli, J.~F. Sparks, and D.~Waldram, ``{A New infinite
  class of Sasaki-Einstein manifolds},'' {\em Adv.Theor.Math.Phys.} {\bfseries
  8} (2006) 987--1000,
\href{http://arxiv.org/abs/hep-th/0403038}{{\ttfamily arXiv:hep-th/0403038
  [hep-th]}}.
%%CITATION = HEP-TH/0403038;%%.

\bibitem{Martelli:2008rt}
D.~Martelli and J.~Sparks, ``{Notes on toric Sasaki-Einstein seven-manifolds
  and AdS(4) / CFT(3)},''
  \href{http://dx.doi.org/10.1088/1126-6708/2008/11/016}{{\em JHEP} {\bfseries
  0811} (2008) 016}, \href{http://arxiv.org/abs/0808.0904}{{\ttfamily
  arXiv:0808.0904 [hep-th]}}.

\bibitem{Futaki:2006cc}
A.~Futaki, H.~Ono, and G.~Wang, ``{Transverse Kahler geometry of Sasaki
  manifolds and toric Sasaki-Einstein manifolds},''
  \href{http://arxiv.org/abs/math/0607586}{{\ttfamily arXiv:math/0607586
  [math-dg]}}.

\bibitem{Hanany:2012hi}
A.~Hanany and R.-K. Seong, ``{Brane Tilings and Reflexive Polygons},''
\href{http://arxiv.org/abs/1201.2614}{{\ttfamily arXiv:1201.2614 [hep-th]}}.
%%CITATION = ARXIV:1201.2614;%%.

\bibitem{King1994}
A.~D. King, ``{Moduli of representations of finite-dimensional algebras},''
  {\em Quart. J. Math. Oxford Ser. (2), Vol. 45, No. 180.} (1994) .

\bibitem{Tomasiello:2010zz}
A.~Tomasiello and A.~Zaffaroni, ``{Parameter spaces of massive IIA
  solutions},'' \href{http://dx.doi.org/10.1007/JHEP04(2011)067}{{\em JHEP}
  {\bfseries 1104} (2011) 067},
\href{http://arxiv.org/abs/1010.4648}{{\ttfamily arXiv:1010.4648 [hep-th]}}.
%%CITATION = ARXIV:1010.4648;%%.

\bibitem{Benini:2011mf}
F.~Benini, C.~Closset, and S.~Cremonesi, ``{Comments on 3d Seiberg-like
  dualities},'' \href{http://dx.doi.org/10.1007/JHEP10(2011)075}{{\em JHEP}
  {\bfseries 1110} (2011) 075},
\href{http://arxiv.org/abs/1108.5373}{{\ttfamily arXiv:1108.5373 [hep-th]}}.
%%CITATION = ARXIV:1108.5373;%%.

\bibitem{Closset:2012eq}
C.~Closset, ``{Seiberg duality for Chern-Simons quivers and D-brane
  mutations},''
\href{http://arxiv.org/abs/1201.2432}{{\ttfamily arXiv:1201.2432 [hep-th]}}.
%%CITATION = ARXIV:1201.2432;%%.

\bibitem{Mathematica7}
I.~Wolfram~Research, {\em {Mathematica Edition: Version 7.0}}.
\newblock Wolfram Research, Inc., 2008.

\bibitem{Davey:2009bp}
J.~Davey, A.~Hanany, and J.~Pasukonis, ``{On the Classification of Brane
  Tilings},'' \href{http://dx.doi.org/10.1007/JHEP01(2010)078}{{\em JHEP}
  {\bfseries 1001} (2010) 078},
\href{http://arxiv.org/abs/0909.2868}{{\ttfamily arXiv:0909.2868 [hep-th]}}.
%%CITATION = ARXIV:0909.2868;%%.

\bibitem{Douglas:1996sw}
M.~R. Douglas and G.~W. Moore, ``{D-branes, quivers, and ALE instantons},''
\href{http://arxiv.org/abs/hep-th/9603167}{{\ttfamily arXiv:hep-th/9603167
  [hep-th]}}.
%%CITATION = HEP-TH/9603167;%%.

\bibitem{Diaconescu:1997br}
D.-E. Diaconescu, M.~R. Douglas, and J.~Gomis, ``{Fractional branes and wrapped
  branes},'' {\em JHEP} {\bfseries 9802} (1998) 013,
  \href{http://arxiv.org/abs/hep-th/9712230}{{\ttfamily arXiv:hep-th/9712230
  [hep-th]}}.

\bibitem{Lawrence:1998ja}
A.~E. Lawrence, N.~Nekrasov, and C.~Vafa, ``{On conformal field theories in
  four-dimensions},''
  \href{http://dx.doi.org/10.1016/S0550-3213(98)00495-7}{{\em Nucl.Phys.}
  {\bfseries B533} (1998) 199--209},
\href{http://arxiv.org/abs/hep-th/9803015}{{\ttfamily arXiv:hep-th/9803015
  [hep-th]}}.
%%CITATION = HEP-TH/9803015;%%.

\bibitem{Klebanov:1998hh}
I.~R. Klebanov and E.~Witten, ``{Superconformal field theory on three-branes at
  a Calabi-Yau singularity},''
  \href{http://dx.doi.org/10.1016/S0550-3213(98)00654-3}{{\em Nucl.Phys.}
  {\bfseries B536} (1998) 199--218},
\href{http://arxiv.org/abs/hep-th/9807080}{{\ttfamily arXiv:hep-th/9807080
  [hep-th]}}.
%%CITATION = HEP-TH/9807080;%%.

\bibitem{Morrison:1998cs}
D.~R. Morrison and M.~Plesser, ``{Nonspherical horizons. 1.},'' {\em
  Adv.Theor.Math.Phys.} {\bfseries 3} (1999) 1--81,
  \href{http://arxiv.org/abs/hep-th/9810201}{{\ttfamily arXiv:hep-th/9810201
  [hep-th]}}.
Revised.
%%CITATION = HEP-TH/9810201;%%.

\bibitem{Beasley:1999uz}
C.~Beasley, B.~R. Greene, C.~Lazaroiu, and M.~Plesser, ``{D3-branes on partial
  resolutions of Abelian quotient singularities of Calabi-Yau threefolds},''
  \href{http://dx.doi.org/10.1016/S0550-3213(99)00646-X}{{\em Nucl.Phys.}
  {\bfseries B566} (2000) 599--640},
\href{http://arxiv.org/abs/hep-th/9907186}{{\ttfamily arXiv:hep-th/9907186
  [hep-th]}}.
%%CITATION = HEP-TH/9907186;%%.

\bibitem{Feng:2000mi}
B.~Feng, A.~Hanany, and Y.-H. He, ``{D-brane gauge theories from toric
  singularities and toric duality},''
  \href{http://dx.doi.org/10.1016/S0550-3213(00)00699-4}{{\em Nucl.Phys.}
  {\bfseries B595} (2001) 165--200},
\href{http://arxiv.org/abs/hep-th/0003085}{{\ttfamily arXiv:hep-th/0003085
  [hep-th]}}.
%%CITATION = HEP-TH/0003085;%%.

\bibitem{Feng:2002fv}
B.~Feng, S.~Franco, A.~Hanany, and Y.-H. He, ``{UnHiggsing the del Pezzo},''
  {\em JHEP} {\bfseries 0308} (2003) 058,
\href{http://arxiv.org/abs/hep-th/0209228}{{\ttfamily arXiv:hep-th/0209228
  [hep-th]}}.
%%CITATION = HEP-TH/0209228;%%.

\bibitem{Kennaway:2007tq}
K.~D. Kennaway, ``{Brane Tilings},''
  \href{http://dx.doi.org/10.1142/S0217751X07036877}{{\em Int. J. Mod. Phys.}
  {\bfseries A22} (2007) 2977--3038},
\href{http://arxiv.org/abs/0706.1660}{{\ttfamily arXiv:0706.1660 [hep-th]}}.
%%CITATION = 0706.1660;%%.

\bibitem{Marolf:2000cb}
D.~Marolf, ``{Chern-Simons terms and the three notions of charge},''
\href{http://arxiv.org/abs/hep-th/0006117}{{\ttfamily arXiv:hep-th/0006117}}.
%%CITATION = HEP-TH/0006117;%%.

\bibitem{Hanany:2005ss}
A.~Hanany and D.~Vegh, ``{Quivers, tilings, branes and rhombi},''
  \href{http://dx.doi.org/10.1088/1126-6708/2007/10/029}{{\em JHEP} {\bfseries
  10} (2007) 029},
\href{http://arxiv.org/abs/hep-th/0511063}{{\ttfamily arXiv:hep-th/0511063}}.
%%CITATION = HEP-TH/0511063;%%.

\bibitem{Gulotta:2008ef}
D.~R. Gulotta, ``{Properly ordered dimers, R-charges, and an efficient inverse
  algorithm},'' \href{http://dx.doi.org/10.1088/1126-6708/2008/10/014}{{\em
  JHEP} {\bfseries 0810} (2008) 014},
  \href{http://arxiv.org/abs/0807.3012}{{\ttfamily arXiv:0807.3012 [hep-th]}}.

\bibitem{Franco:2006gc}
S.~Franco and D.~Vegh, ``{Moduli spaces of gauge theories from dimer models:
  Proof of the correspondence},''
  \href{http://dx.doi.org/10.1088/1126-6708/2006/11/054}{{\em JHEP} {\bfseries
  0611} (2006) 054},
\href{http://arxiv.org/abs/hep-th/0601063}{{\ttfamily arXiv:hep-th/0601063
  [hep-th]}}.
%%CITATION = HEP-TH/0601063;%%.

\bibitem{Luty:1995sd}
M.~A. Luty and W.~Taylor, ``{Varieties of vacua in classical supersymmetric
  gauge theories},'' \href{http://dx.doi.org/10.1103/PhysRevD.53.3399}{{\em
  Phys. Rev.} {\bfseries D53} (1996) 3399--3405},
\href{http://arxiv.org/abs/hep-th/9506098}{{\ttfamily arXiv:hep-th/9506098}}.
%%CITATION = HEP-TH/9506098;%%.

\bibitem{Martelli:2008cm}
D.~Martelli and J.~Sparks, ``{Symmetry-breaking vacua and baryon condensates in
  AdS/CFT},'' \href{http://dx.doi.org/10.1103/PhysRevD.79.065009}{{\em
  Phys.Rev.} {\bfseries D79} (2009) 065009},
  \href{http://arxiv.org/abs/0804.3999}{{\ttfamily arXiv:0804.3999 [hep-th]}}.

\bibitem{2009arXiv0908.3475M}
S.~{Mozgovoy}, ``{Crepant resolutions and brane tilings I: Toric
  realization},'' \href{http://arxiv.org/abs/0908.3475}{{\ttfamily
  arXiv:0908.3475 [math.AG]}}.

\bibitem{2009arXiv0909.2013B}
M.~{Bender} and S.~{Mozgovoy}, ``{Crepant resolutions and brane tilings II:
  Tilting bundles},'' \href{http://arxiv.org/abs/0909.2013}{{\ttfamily
  arXiv:0909.2013 [math.AG]}}.

\bibitem{Hanany:2006nm}
A.~Hanany, C.~P. Herzog, and D.~Vegh, ``{Brane tilings and exceptional
  collections},'' \href{http://dx.doi.org/10.1088/1126-6708/2006/07/001}{{\em
  JHEP} {\bfseries 0607} (2006) 001},
  \href{http://arxiv.org/abs/hep-th/0602041}{{\ttfamily arXiv:hep-th/0602041
  [hep-th]}}.

\bibitem{Aspinwall:2008jk}
P.~S. Aspinwall, ``{D-Branes on Toric Calabi-Yau Varieties},''
\href{http://arxiv.org/abs/0806.2612}{{\ttfamily arXiv:0806.2612 [hep-th]}}.
%%CITATION = 0806.2612;%%.

\bibitem{Cox:book}
D.~Cox, J.~Little, and H.~Schenck, {\em {Toric Varieties}}.
\newblock American Mathematical Society, 2011.

\bibitem{Carqueville:2009xu}
N.~Carqueville and A.~Quintero~Velez, ``{Remarks on quiver gauge theories from
  open topological string theory},''
  \href{http://dx.doi.org/10.1007/JHEP03(2010)129}{{\em JHEP} {\bfseries 1003}
  (2010) 129},
\href{http://arxiv.org/abs/0912.4699}{{\ttfamily arXiv:0912.4699 [hep-th]}}.
%%CITATION = ARXIV:0912.4699;%%.

\bibitem{sage}
W.~Stein {\em et al.}, {\em {S}age {M}athematics {S}oftware ({V}ersion 4.7.2)}.
\newblock The Sage Development Team, 2011.
\newblock {\tt http://www.sagemath.org}.

\bibitem{Minasian:1997mm}
R.~Minasian and G.~W. Moore, ``{K theory and Ramond-Ramond charge},'' {\em
  JHEP} {\bfseries 9711} (1997) 002,
  \href{http://arxiv.org/abs/hep-th/9710230}{{\ttfamily arXiv:hep-th/9710230
  [hep-th]}}.

\bibitem{Aspinwall:2004jr}
P.~S. Aspinwall, ``{D-branes on Calabi-Yau manifolds},''
  \href{http://arxiv.org/abs/hep-th/0403166}{{\ttfamily arXiv:hep-th/0403166
  [hep-th]}}.

\bibitem{Herzog:2004qw}
C.~P. Herzog, ``{Seiberg duality is an exceptional mutation},''
  \href{http://dx.doi.org/10.1088/1126-6708/2004/08/064}{{\em JHEP} {\bfseries
  0408} (2004) 064}, \href{http://arxiv.org/abs/hep-th/0405118}{{\ttfamily
  arXiv:hep-th/0405118 [hep-th]}}.

\bibitem{Freed:1999vc}
D.~S. Freed and E.~Witten, ``{Anomalies in string theory with D-branes},''
  \href{http://arxiv.org/abs/hep-th/9907189}{{\ttfamily arXiv:hep-th/9907189
  [hep-th]}}.

\bibitem{Douglas:2000ah}
M.~R. Douglas, B.~Fiol, and C.~Romelsberger, ``{Stability and BPS branes},''
  \href{http://dx.doi.org/10.1088/1126-6708/2005/09/006}{{\em JHEP} {\bfseries
  0509} (2005) 006}, \href{http://arxiv.org/abs/hep-th/0002037}{{\ttfamily
  arXiv:hep-th/0002037 [hep-th]}}.

\bibitem{Aspinwall:2004mb}
P.~S. Aspinwall, ``{D-branes, Pi-stability and theta-stability},''
  \href{http://arxiv.org/abs/hep-th/0407123}{{\ttfamily arXiv:hep-th/0407123
  [hep-th]}}.

\bibitem{Aspinwall:2004vm}
P.~S. Aspinwall and I.~V. Melnikov, ``{D-branes on vanishing del Pezzo
  surfaces},'' \href{http://dx.doi.org/10.1088/1126-6708/2004/12/042}{{\em
  JHEP} {\bfseries 0412} (2004) 042},
  \href{http://arxiv.org/abs/hep-th/0405134}{{\ttfamily arXiv:hep-th/0405134
  [hep-th]}}.

\bibitem{Aharony:2008gk}
O.~Aharony, O.~Bergman, and D.~L. Jafferis, ``{Fractional M2-branes},''
  \href{http://dx.doi.org/10.1088/1126-6708/2008/11/043}{{\em JHEP} {\bfseries
  11} (2008) 043},
\href{http://arxiv.org/abs/0807.4924}{{\ttfamily arXiv:0807.4924 [hep-th]}}.
%%CITATION = 0807.4924;%%.

\bibitem{Aharony:2009fc}
O.~Aharony, A.~Hashimoto, S.~Hirano, and P.~Ouyang, ``{D-brane Charges in
  Gravitational Duals of 2+1 Dimensional Gauge Theories and Duality
  Cascades},'' \href{http://dx.doi.org/10.1007/JHEP01(2010)072}{{\em JHEP}
  {\bfseries 01} (2010) 072},
\href{http://arxiv.org/abs/0906.2390}{{\ttfamily arXiv:0906.2390 [hep-th]}}.
%%CITATION = 0906.2390;%%.

\bibitem{Gaiotto:2009tk}
D.~Gaiotto and D.~L. Jafferis, ``{Notes on adding D6 branes wrapping RP$^3$ in
  AdS(4) $\times$ CP$^3$},''
\href{http://arxiv.org/abs/0903.2175}{{\ttfamily arXiv:0903.2175 [hep-th]}}.
%%CITATION = 0903.2175;%%.

\bibitem{Gauntlett:2004hs}
J.~P. Gauntlett, D.~Martelli, J.~Sparks, and D.~Waldram, ``{Supersymmetric AdS
  backgrounds in string and M-theory},''
\href{http://arxiv.org/abs/hep-th/0411194}{{\ttfamily arXiv:hep-th/0411194
  [hep-th]}}.
%%CITATION = HEP-TH/0411194;%%.

\bibitem{Chen:2004nq}
W.~Chen, H.~Lu, C.~Pope, and J.~F. Vazquez-Poritz, ``{A Note on Einstein Sasaki
  metrics in D $\geq$ 7},''
  \href{http://dx.doi.org/10.1088/0264-9381/22/17/004}{{\em Class.Quant.Grav.}
  {\bfseries 22} (2005) 3421--3430},
\href{http://arxiv.org/abs/hep-th/0411218}{{\ttfamily arXiv:hep-th/0411218
  [hep-th]}}.
%%CITATION = HEP-TH/0411218;%%.

\bibitem{Aharony:1997gp}
O.~Aharony, ``{IR duality in d = 3 N = 2 supersymmetric USp(2N(c)) and U(N(c))
  gauge theories},''
  \href{http://dx.doi.org/10.1016/S0370-2693(97)00530-3}{{\em Phys. Lett.}
  {\bfseries B404} (1997) 71--76},
\href{http://arxiv.org/abs/hep-th/9703215}{{\ttfamily arXiv:hep-th/9703215}}.
%%CITATION = HEP-TH/9703215;%%.

\bibitem{Cremonesi:2010ae}
S.~Cremonesi, ``{Type IIB construction of flavoured ABJ(M) and fractional M2
  branes},'' \href{http://dx.doi.org/10.1007/JHEP01(2011)076}{{\em JHEP}
  {\bfseries 1101} (2011) 076},
\href{http://arxiv.org/abs/1007.4562}{{\ttfamily arXiv:1007.4562 [hep-th]}}.
%%CITATION = ARXIV:1007.4562;%%.

\bibitem{in:progress}
S.~Cremonesi, {\em {work in progress}}.

\bibitem{2008arXiv0811.2435K}
M.~{Kontsevich} and Y.~{Soibelman}, ``{Stability structures, motivic
  Donaldson-Thomas invariants and cluster transformations},'' {\em ArXiv
  e-prints} (Nov., 2008) , \href{http://arxiv.org/abs/0811.2435}{{\ttfamily
  arXiv:0811.2435 [math.AG]}}.

\bibitem{Aganagic:2010qr}
M.~Aganagic and K.~Schaeffer, ``{Wall Crossing, Quivers and Crystals},''
  \href{http://arxiv.org/abs/1006.2113}{{\ttfamily arXiv:1006.2113 [hep-th]}}.

\bibitem{Davey:2009sr}
J.~Davey, A.~Hanany, N.~Mekareeya, and G.~Torri, ``{Phases of M2-brane
  Theories},'' \href{http://dx.doi.org/10.1088/1126-6708/2009/06/025}{{\em
  JHEP} {\bfseries 0906} (2009) 025},
\href{http://arxiv.org/abs/0903.3234}{{\ttfamily arXiv:0903.3234 [hep-th]}}.
%%CITATION = ARXIV:0903.3234;%%.

\bibitem{Butti:2006au}
A.~Butti, D.~Forcella, and A.~Zaffaroni, ``{Counting BPS baryonic operators in
  CFTs with Sasaki-Einstein duals},''
  \href{http://dx.doi.org/10.1088/1126-6708/2007/06/069}{{\em JHEP} {\bfseries
  0706} (2007) 069},
\href{http://arxiv.org/abs/hep-th/0611229}{{\ttfamily arXiv:hep-th/0611229
  [hep-th]}}.
%%CITATION = HEP-TH/0611229;%%.

\bibitem{Martelli:2004wu}
D.~Martelli and J.~Sparks, ``{Toric geometry, Sasaki-Einstein manifolds and a
  new infinite class of AdS/CFT duals},''
  \href{http://dx.doi.org/10.1007/s00220-005-1425-3}{{\em Commun.Math.Phys.}
  {\bfseries 262} (2006) 51--89},
  \href{http://arxiv.org/abs/hep-th/0411238}{{\ttfamily arXiv:hep-th/0411238
  [hep-th]}}.

\bibitem{Martelli:2009ga}
D.~Martelli and J.~Sparks, ``{AdS$_4$/CFT$_3$ duals from M2-branes at
  hypersurface singularities and their deformations},''
\href{http://arxiv.org/abs/0909.2036}{{\ttfamily arXiv:0909.2036 [hep-th]}}.
%%CITATION = 0909.2036;%%.

\bibitem{Benini:2007gx}
F.~Benini, F.~Canoura, S.~Cremonesi, C.~Nunez, and A.~V. Ramallo,
  ``{Backreacting Flavors in the Klebanov-Strassler Background},''
  \href{http://dx.doi.org/10.1088/1126-6708/2007/09/109}{{\em JHEP} {\bfseries
  09} (2007) 109},
\href{http://arxiv.org/abs/0706.1238}{{\ttfamily arXiv:0706.1238 [hep-th]}}.
%%CITATION = 0706.1238;%%.

\bibitem{Bergman:2009zh}
O.~Bergman and S.~Hirano, ``{Anomalous radius shift in AdS(4)/CFT(3)},''
  \href{http://dx.doi.org/10.1088/1126-6708/2009/07/016}{{\em JHEP} {\bfseries
  0907} (2009) 016}, \href{http://arxiv.org/abs/0902.1743}{{\ttfamily
  arXiv:0902.1743 [hep-th]}}.

\bibitem{Aharony:2010af}
O.~Aharony, D.~Jafferis, A.~Tomasiello, and A.~Zaffaroni, ``{Massive type IIA
  string theory cannot be strongly coupled},''
  \href{http://dx.doi.org/10.1007/JHEP11(2010)047}{{\em JHEP} {\bfseries 1011}
  (2010) 047},
\href{http://arxiv.org/abs/1007.2451}{{\ttfamily arXiv:1007.2451 [hep-th]}}.
%%CITATION = ARXIV:1007.2451;%%.

\end{thebibliography}\endgroup

\end{document}